\documentclass[11pt,letterpaper,oneside,openany]{book}

\title{Bayesian parameter estimation for relativistic heavy-ion collisions}
\author{Jonah E.\ Bernhard}

\usepackage[utf8]{inputenc}
\usepackage[T1]{fontenc}
\usepackage{lmodern}
\usepackage[light]{raleway}
\usepackage{lato}

\usepackage{mathtools}
\usepackage{amssymb}

\usepackage{graphicx}
\graphicspath{{figures/}}

\usepackage{wrapfig}
\usepackage{tikz}
\usetikzlibrary{shapes,calc,matrix}
\usepackage{xcolor}

\definecolor{theblue}{RGB}{22,79,149}
\usepackage[pdfusetitle,colorlinks=true,allcolors=theblue,bookmarksdepth=2,linktoc=section,bookmarksopen=false]{hyperref}

\usepackage{titlesec}
\titleformat{\chapter}[display]{\Huge\sffamily\color[RGB]{83,94,113}\filcenter}{\thechapter}{1ex}{}
\titleformat*{\section}{\Large\bfseries\color{theblue}}
\titleformat*{\subsection}{\large\bfseries\color{theblue}}
\titleformat*{\subsubsection}{\bfseries\color{theblue}}
\titleformat*{\paragraph}{\bfseries\color[gray]{.1}}
\titlespacing{\paragraph}{0pt}{*1.4}{*2.5}

\usepackage[font=small,labelfont={bf,color=theblue},labelsep=quad]{caption}
\newcommand{\graphicsandcaption}[3]{
  \centering
  \includegraphics[width=#1\textwidth]{#2}
  \captionsetup{width=#1\textwidth}
  \caption{#3}
}

\usepackage{lettrine}

\usepackage{enumitem}
\newcounter{enumisave}
\setlist[itemize,1]{label=\raisebox{.25ex}{\color{theblue}\tiny\textbullet}}

\usepackage{booktabs}
\usepackage{multirow}

\usepackage[style=phys,biblabel=brackets,eprint=true,url=true,maxcitenames=3]{biblatex}
\newcommand{\parbreakboldstart}[1]{\medskip\noindent\textbf{#1}}
\newcommand{\fmc}{\ensuremath{\text{fm}/c}}
\newcommand{\Nch}{N_\text{ch}}
\newcommand{\Np}{N_\text{part}}
\newcommand{\avg}[1]{\langle #1 \rangle}
\newcommand{\llangle}{{\langle\!\langle}}
\newcommand{\rrangle}{{\rangle\!\rangle}}
\newcommand{\davg}[1]{\llangle #1 \rrangle}
\newcommand{\vnk}[2]{v_{#1}\{#2\}}
\newcommand{\order}[1]{\ensuremath{\mathcal{O}(#1)}}
\newcommand{\del}{\nabla}
\newcommand{\mn}{^{\mu\nu}}
\newcommand{\trento}{T\raisebox{-0.5ex}{R}ENTo}
\newcommand{\T}{\tilde{T}}
\newcommand{\dmin}{d_\text{min}}
\newcommand{\tfs}{\tau_\text{fs}}
\newcommand{\Tsw}{T_\text{switch}}
\newcommand{\sumsp}{\sum_\text{sp}}
\newcommand{\shear}{_\text{shear}}
\newcommand{\bulk}{_\text{bulk}}
\newcommand{\dv}{\mathbf d}
\newcommand{\kv}{\mathbf k}
\newcommand{\vv}{\mathbf v}
\newcommand{\xv}{\mathbf x}
\newcommand{\yv}{\mathbf y}
\newcommand{\zv}{\mathbf z}
\newcommand{\zerov}{\mathbf 0}
\newcommand{\alphav}{\boldsymbol\alpha}
\newcommand{\epsv}{\boldsymbol\epsilon}
\newcommand{\muv}{\boldsymbol\mu}
\newcommand{\thetav}{\boldsymbol\theta}
\newcommand{\tran}{^\mathrm{T}}
\newcommand{\N}{\mathcal{N}}
\newcommand{\D}{\mathcal{D}}
\newcommand{\stat}{^\text{stat}}
\newcommand{\sys}{^\text{sys}}

\DeclareMathOperator{\diag}{diag}
\DeclareMathOperator{\cov}{cov}
\DeclareMathOperator{\SC}{SC}

\begin{document}

\frontmatter

\makeatletter
\begin{titlepage}
  \centering
  \sffamily
  \vspace*{.25\textheight}
  \huge\@title \\
  \flafamily
  \vspace{.05\textheight}
  \Large\@author \\
  \vspace{.05\textheight}
  \large Ph.D.\ dissertation \\[.25ex]
  Advisor: Steffen A.\ Bass \\[.25ex]
  Department of Physics, Duke University \\
  \vspace{.05\textheight}
  \today
\end{titlepage}
\makeatother

\chapter{Abstract}
\label{ch:abstract}

I develop and apply a Bayesian method for quantitatively estimating properties of the quark-gluon plasma (QGP), an extremely hot and dense state of fluid-like matter created in relativistic heavy-ion collisions.

The QGP cannot be directly observed---it is extraordinarily tiny and ephemeral, about $10^{-14}$ meters in size and living $10^{-23}$ seconds before freezing into discrete particles---but it can be indirectly characterized by matching the output of a computational collision model to experimental observations.
The model, which takes the QGP properties of interest as input parameters, is calibrated to fit the experimental data, thereby extracting a posterior probability distribution for the parameters.

In this dissertation, I construct a specific computational model of heavy-ion collisions and formulate the Bayesian parameter estimation method, which is based on general statistical techniques.
I then apply these tools to estimate fundamental QGP properties, including its key transport coefficients and characteristics of the initial state of heavy-ion collisions.

Perhaps most notably, I report the most precise estimate to date of the temperature-dependent specific shear viscosity $\eta/s$, the measurement of which is a primary goal of heavy-ion physics.
The estimated minimum value is $\eta/s = 0.085_{-0.025}^{+0.026}$ (posterior median and 90\% uncertainty), remarkably close to the conjectured lower bound of $1/4\pi \simeq 0.08$.
The analysis also shows that $\eta/s$ likely increases slowly as a function of temperature.

Other estimated quantities include the temperature-dependent bulk viscosity $\zeta/s$, the scaling of initial state entropy deposition, and the duration of the pre-equilibrium stage that precedes QGP formation.

\titlespacing{\chapter}{0pt}{-15pt}{30pt}
\setcounter{tocdepth}{1}
\tableofcontents
\titlespacing{\chapter}{0pt}{50pt}{40pt}

\mainmatter

\chapter{Introduction}
\label{ch:intro}

\lettrine{I}{magine:} One day, strolling through the countryside, you happen upon the ruins of an old, burned-down building.
Little remains of the original structure; the charred rubble is haphazardly piled, the wood battered by decades of wind and rain.
Yet, the foundation is mostly intact.
A fraction of a wall still stands.
You can't help but wonder: What did the building look like, what was its purpose?
And how---and why---did the fire start?

This is the allegory I often use when attempting to convey the difficulty of a much different problem:
How can we measure the properties of the quark-gluon plasma, a highly excited and transient state of matter that cannot be observed directly?
Quark-gluon plasma (QGP) can only be created on Earth in ultra-relativistic collisions of heavy nuclei, and even then only in microscopic droplets that almost instantly disintegrate into a shower of particles, which are detected eons (relatively speaking) after the original collision.
From those particles---the ``ashes''---we wish to infer not only that a QGP was, in fact, the source, but also its precise properties.

What connects these mysteries is that we can only observe the final state of the system, not its original state nor its transformation.
How can we turn back the clock?
In the case of the burned-down building, perhaps we could ignite some other structures, observe how they burn, and compare the results to the discovered ruins.
That could get expensive, but the basic strategy is sound:
We need a way to replay history, then we can match the outcomes to our limited observations.

Enter the modern computational model.
The generic setup is as follows:
We have a physical system with an observed final state (experimental data), a set of undetermined parameters which characterize the system, and a computer model of said system.
The model takes the parameters as input, simulates the full time evolution of the system, and produces output analogous to the observations.
We tune the parameters and run the model until its output optimally matches the experimental data, thereby determining the true values of the parameters.

In this way, a realistic computational model can be an invaluable tool for inferring the properties of physical systems.
We could likely solve our first puzzle via simulations of burning buildings under different conditions, and extract even more information while we're at it, such as how the fire spread through the building, and how hot it burned.

Computational models of relativistic heavy-ion collisions simulate the entire time evolution of the collision, from the moment of impact, to the formation of the quark-gluon plasma, through its disintegration back into particles, and the final interactions of those particles before they are detected.
By varying the parameters of the model and matching the simulation output to experimental observations, we can characterize the initial state of the collision and the properties of the QGP, such as its transport coefficients.

One of the most sought after QGP coefficients is the specific shear viscosity $\eta/s$---the dimensionless ratio of the shear viscosity to entropy density.
Early observations pointed towards a small $\eta/s$, meaning that the QGP is nearly a ``perfect'' fluid (which would have zero viscosity).
Meanwhile, a purely theoretical calculation posited that the minimum specific shear viscosity is $1/4\pi$, or approximately 0.08.
Subsequent studies, comparing viscous relativistic hydrodynamics models (which take $\eta/s$ as an input parameter) to experimental measurements of collective behavior, showed that the specific shear viscosity is likely nonzero, and within roughly a factor of three of the conjectured $1/4\pi$ limit.

Model-to-data comparison is broadly applicable in a variety of scientific disciplines.
We can acquire stunning images of distant galaxies, but given the timescale of most galactic processes, these images are effectively frozen snapshots.
Galactic evolution models help unravel the life cycle of galaxies, all the way back to their formation.

Recently, the LIGO Scientific Collaboration used observations of gravitational waves combined with numerical simulations of general relativity to quantify properties of binary black hole mergers, such as the masses of the original black holes, the amount of energy radiated, and the distance to the event.
(More recently, a binary neutron star merger was also detected.)
These quantitative conclusions are highly nontrivial and, in my opinion, underappreciated.
After all, the observed gravitational waveforms---while astonishingly impressive in their own right---do not directly indicate that the source was a black hole merger, much less provide its precise properties.
Only by comparing the signal to numerical relativity simulations can that information be decoded.

Of course, the ``generic setup'' described above does not proceed so simply.
Experimental measurements always contain some noise, precluding exact measurement of the model input parameters.
Multiple parameters often have complex interrelationships, so they cannot be tuned independently, and their correlations may contribute additional uncertainty.
And computational models are seldom perfect representations of reality.
Due to these factors (and others), parameters determined from model-to-data comparison are inevitably uncertain and should be viewed as \emph{estimates} of the true values.
This is the terminology I'll most often use for the remainder of this work (``parameter estimation'' is even in the title!).

Given this inherent inexactness, it is crucial to strive for not only the ``best-fit'' parameters but also faithful assessments of their associated uncertainties.

All of this calls for a rigorous, systematic approach.
To this end, we frame the problem in terms of Bayes' theorem.

In the Bayesian interpretation, our complete knowledge of the parameters is contained in the \emph{posterior} probability distribution on the model parameters.
To obtain the posterior distribution, we first encode our initial knowledge as the \emph{prior} distribution, for example we probably know a reasonable range for each parameter.
Then, for any given parameter values, we compute the \emph{likelihood} by evaluating the model and calculating the fit to data, folding in any sources of uncertainty from the model calculation and experimental measurements.
Finally, we invoke Bayes' theorem, which states that the posterior is proportional to the product of the likelihood and the prior:
\begin{equation*}
  \text{posterior} \propto \text{likelihood} \times \text{prior}.
\end{equation*}
From the posterior, we can extract quantitative estimates of each parameter, their uncertainties, and any other statistical metrics;
visualizations can reveal detailed structures and relationships among parameters.

Let us step back for a moment and consider what happened here.
The likelihood and posterior have quite different meanings, even when they are mathematically equivalent (for example if the prior is constant).
The likelihood is the probability of observing the evidence given a proposed set of parameters; in other words, if we assume certain values of the parameters are true, then what is the probability of a universe where the resulting model calculations and experimental data exist together?
The posterior incorporates our prior knowledge and reverses the conditionality:
it is the probability of the parameters, given the observed universe and the prior.
The former is what we can compute directly, the latter is the quantity we're after.

Armed with a dataset, a computational model, and Bayes' theorem, we can compute the posterior probability at any point in parameter space.
To extract estimates of each parameter, we must now construct their \emph{marginal} distributions,
obtained for any given parameter by marginalizing over (integrating out) all the rest.
Importantly, this folds in the remaining uncertainty of the marginalized parameters, for instance if the estimates of several parameters are correlated, the uncertainty in each parameter contributes to the uncertainty of the others.

Marginalization necessitates calculating multidimensional numerical integrals, for which Monte Carlo techniques usually perform best.
Markov chain Monte Carlo methods are the canonical choice for sampling posterior distributions;
this entails roughly a million evaluations of the posterior, plus or minus a few orders of magnitude, depending on the problem at hand and which quantities are desired.
So unless the model runs rather quickly, the required computation time is prohibitive.
A model that runs in a tenth of a second would take a little over a day for a million evaluations;
heavy-ion collision models need at least a few thousand hours (consisting of tens of thousands of individual events, each of which runs in a few minutes on average), which translates to a total time of over a hundred thousand years!
(In practice, some degree of parallelization would reduce this, but not enough.)

One strategy to dramatically reduce the required computation time is to use a surrogate model, or emulator, that predicts the output of the true model in much less time than a full calculation.
The surrogate is trained on the input-output behavior of the true model, then used as a stand-in during Monte Carlo sampling.
Gaussian process emulators are a common choice, since they perform well in high dimensions, do not require any assumptions about the parametric form of the model, and naturally provide the uncertainty of their predictions.

\medskip

\lettrine{I}{n} this dissertation, I develop a complete framework for applying Bayesian parameter estimation methods to quantitatively estimate the properties of the quark-gluon plasma created in ultra-relativistic heavy-ion collisions.

I begin by laying the groundwork in chapter \ref{ch:overview}, reviewing the history of heavy-ion physics and surveying its current status, focusing on the aspects most relevant to this work.
In chapter \ref{ch:models}, I go in depth on the computational modeling of heavy-ion collisions.
I present several original contributions to the modeling landscape, and assemble a specific set of models to be used later for parameter estimation.
Then, in chapter \ref{ch:param-est}, I describe the Bayesian parameter estimation method, building on existing techniques and tailoring them to heavy-ion physics.

Finally, in chapter \ref{ch:quant-qcd-props}, I present a sequence of case studies that I have performed over the past several years.
The first study is a proof of concept; a demonstration that parameter estimation can succeed in heavy-ion physics.
The subsequent studies progress from there, improving various features of both the computational model and the statistical analysis.
It all culminates in a state-of-the-art analysis with the most precise estimates of QGP properties to date, including the temperature dependence of the highly sought-after specific shear viscosity.

\chapter{A pragmatic overview of heavy-ion collisions}
\chaptermark{A pragmatic overview}
\label{ch:overview}

\lettrine{A}{} pair of lead nuclei hurtle towards each other, circling in opposite directions around the 27 kilometer ring of the Large Hadron Collider (LHC) near Geneva, Switzerland.
Stripped of their electrons, the positively-charged ions have been accelerated to virtually the speed of light by the collider's powerful electromagnetic field.
They are nearly-flat discs due to relativistic length contraction.

The discs collide, depositing their kinetic energy in a nucleus-sized area and creating temperatures $T \sim 300$ MeV, or about $3 \times 10^{12}$ K, over 100,000 times hotter than the core of the Sun ($1.6 \times 10^7$ K \cite{Williams:2016sun}).
By around 1 \fmc\ after the collision,%
\footnote{
  A \fmc\ is the time it takes light to travel a femtometer ($10^{-15}$ meter, usually abbreviated fermi or fm), which works out to approximately $3 \times 10^{-24}$ seconds.
  This is a convenient unit of time in the context of heavy-ion collisions, since most particles move at a significant fraction of the speed of light and the typical length scale is several fermi.
}
the quarks and gluons that made up the original protons and neutrons in the lead nuclei have escaped and formed an extremely hot and dense state of fluid-like matter known as quark-gluon plasma (QGP) \cite{Collins:1974ky,Shuryak:1980tp}.

Quarks and gluons are elementary particles, the constituents of composite particles called hadrons, of which protons and neutrons are examples.
In normal matter, quarks and gluons exist \emph{only} in hadrons, confined by the strong nuclear force, which also binds protons and neutrons into nuclei.
The theory of the strong force, quantum chromodynamics (QCD), stipulates that---under normal conditions---particles must have neutral ``color'' charge, the QCD analog of electric charge (the prefix \emph{chromo} means color).
There are two main ways to form a color-neutral hadron:
a quark and antiquark of the same color charge, called a meson, and three (anti)quarks of different color charges, known as a (anti)baryon.

QCD further predicts that the constituents become deconfined at sufficiently high temperature and density.
Such conditions materialized in the early universe, microseconds after the Big Bang, suggesting that the entire universe was once a large progenitor QGP;
in the present day, superdense celestial objects such as neutron stars may contain a QGP-like phase, although their distance from Earth makes them difficult to characterize.
High-energy nuclear collisions are the only way to create similarly extreme conditions in the laboratory.
The notion of performing these collisions in search of a hot and dense phase of free quarks and gluons dates back to the 1970s \cite{Chapline:1974zf}.

\begin{figure}[h]
  \makebox[\textwidth]{
    \includegraphics[height=.2\paperwidth]{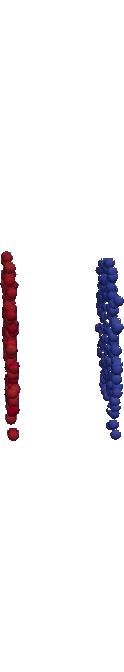} \hspace{.04\paperwidth}
    \includegraphics[height=.2\paperwidth]{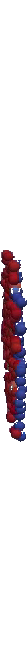} \hspace{.04\paperwidth}
    \includegraphics[height=.2\paperwidth]{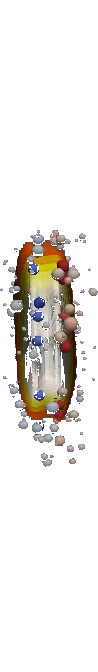} \hspace{.04\paperwidth}
    \includegraphics[height=.2\paperwidth]{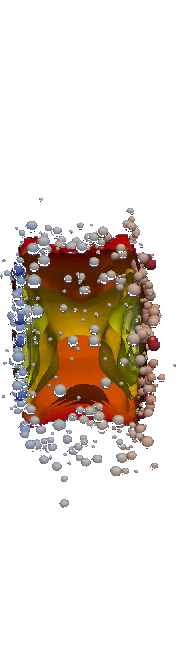} \hspace{.04\paperwidth}
    \includegraphics[height=.2\paperwidth]{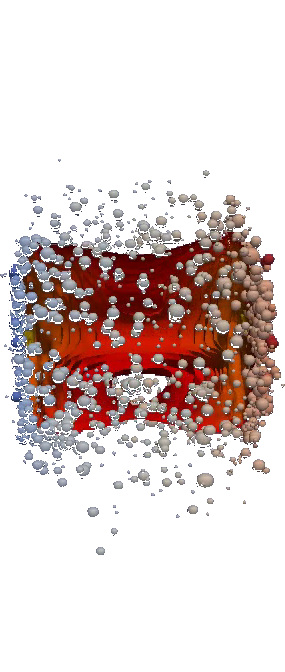} \hspace{.04\paperwidth}
    \includegraphics[height=.2\paperwidth]{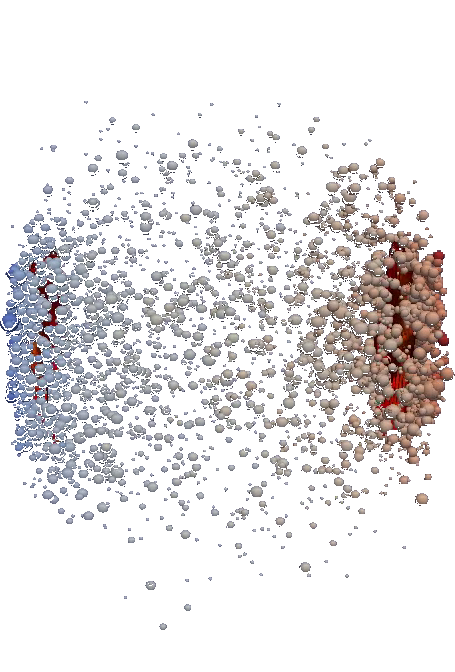} \hspace{.04\paperwidth}
    \includegraphics[height=.2\paperwidth]{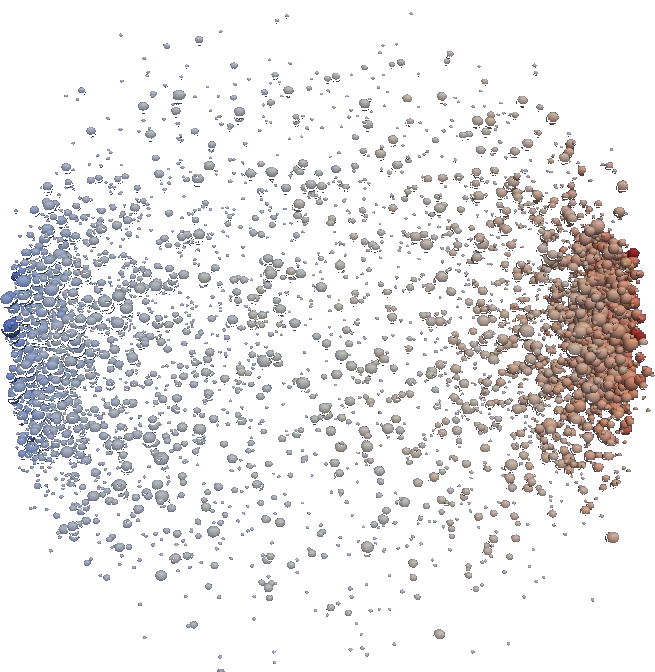}
  }
  \centering
  $\xrightarrow{\hspace{4em}\text{Time}\hspace{4em}}$
  \caption{
    A rendering of the stages of a heavy-ion collision.
    From left to right, nuclei approach each other and collide, the QGP medium forms and expands while particles are emitted, and the QGP dissipates as the hadron gas expands.
    Visualization originally created by Hannah Petersen and modified by the author for this work.
  }
  \label{fig:event-frames}
\end{figure}

\noindent\textbf{Back to the lead-lead collision}, where the Lorentz-contracted nuclei are receding along the $z$-axis with the created droplet of QGP between them.
Bjorken outlined the basic collision spacetime evolution in 1982 \cite{Bjorken:1982qr}.
The QGP is located near the origin, expanding hydrodynamically in both the transverse ($x$-$y$) plane and the longitudinal ($z$) direction;
at any given $z$ position, the fluid has approximate longitudinal velocity $z/t$.
As the nuclei continue to recede, the fluid forms at later times further from $z = 0$, roughly on a spacetime hyperbola defined by a constant ``proper time''
\begin{equation}
  \tau \equiv \sqrt{t^2 - z^2} \sim 1\ \fmc.
\end{equation}
The partner variable to proper time is the spacetime rapidity
\begin{equation}
  \eta_s \equiv \frac{1}{2} \log \frac{t + z}{t - z},
\end{equation}
which specifies the position along proper-time hyperbolas.
This is a convenient kinematic variable since Lorentz boosts simply add as a function of rapidity, i.e.\ a boost of $\eta_s^A$ followed by a second of $\eta_s^B$ is equivalent to a single boost of $\eta_s^A + \eta_s^B$.

\begin{figure}[h]
  \centering
  \includegraphics{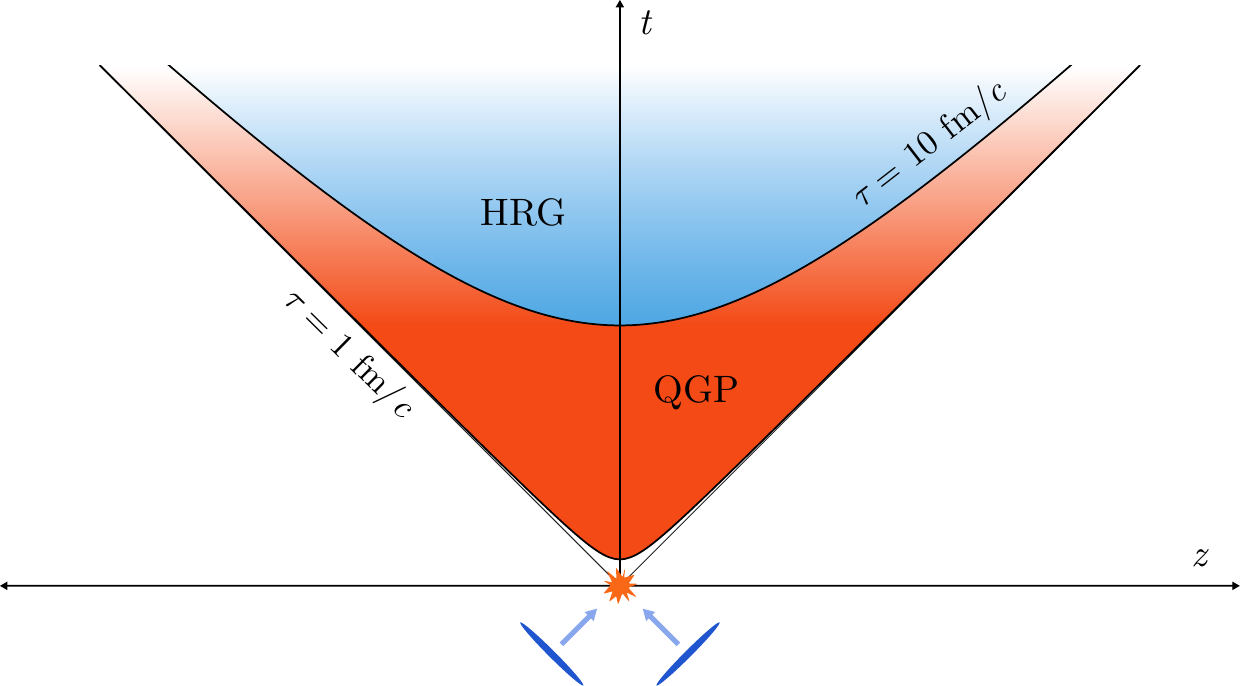}
  \caption{
    Spacetime diagram of a heavy-ion collision.
    The nuclei (blue discs) propagate along the $z$-axis at the speed of light and collide at $z = t = 0$.
    The quark-gluon plasma (QGP, orange region) medium forms at proper time $\tau \sim 1\ \fmc$ and converts to a hadron resonance gas (HRG, blue region) around $\tau \sim 10\ \fmc$.
  }
  \label{fig:spacetime-diagram}
\end{figure}

The system is approximately invariant under Lorentz boosts near central rapidity ($\eta_s \sim 0$), because, as Bjorken argues, the nuclei are so extremely boosted (Lorentz factor $\gamma > 1000$ at the LHC) that the collision dynamics appear similar in all near-center-of-mass frames.
This implies a central plateau structure in the density distribution as a function of rapidity, and that particle production is constant per unit rapidity within the plateau region.
This approximation, ``boost-invariance'', is substantiated by experimental data, as shown in the next section, and is an important simplification for hydrodynamic models.

As the QGP expands and cools, the strong force quickly reasserts itself and the quarks and gluons recombine into hadrons.
The latest calculations show that this conversion occurs as a crossover phase transition around $T \sim 145$--165 MeV at zero net baryon density \cite{Bazavov:2014pvz,Borsanyi:2013bia}, i.e.\ equal parts matter and antimatter.
Meanwhile, at zero temperature and large baryon density, there is a first-order phase transition from normal nuclear matter to a color superconductor \cite{Rapp:1997zu,Alford:1997zt}.

Based on these insights, we can draw a schematic phase diagram of QCD matter:

\begin{figure}[h]
  \graphicsandcaption{.75}{etc/phase_diagram}{Schematic of the QCD phase diagram \cite{NSAC:2008lrp}.}
  \label{fig:phase-diagram}
\end{figure}

\noindent Following convention, the diagram is a function of temperature $T$ and baryon chemical potential $\mu_B$, which quantifies the net baryon density, where positive $\mu_B$ means more baryons than antibaryons.
Given the crossover at zero $\mu_B$ and the first-order transition at zero temperature, it is logical to draw a first-order phase boundary terminating in a critical point at some $(T, \mu_B)$ combination \cite{Stephanov:1998dy}, however, current experimental evidence for the existence of a QCD critical point is inconclusive.

Our quantitative knowledge of the crossover phase transition at zero $\mu_B$ derives from lattice QCD calculations of the equation of state, which connects the system's various thermodynamic quantities: temperature, pressure, energy density, etc.
At sufficiently high collision energy, $\mu_B$ is small enough that it may be approximated as zero---this is the case at the LHC, for example.
See subsection \ref{subsec:eos} for more on the equation of state.

Heavy-ion collisions trace various trajectories through the phase diagram, beginning as a QGP at high temperature and eventually cooling into a hadron gas, undergoing either a crossover or first-order phase transition depending on the value of $\mu_B$.
Higher energy collisions have a larger initial temperature and smaller baryon chemical potential, thus, different energy collisions probe different regions of the phase diagram.

The conversion back to particles (hadronization) completes by proper time $\tau \sim 10\ \fmc$.
The system is now a hadron resonance gas (HRG), consisting of mostly pions---the lightest hadron---but also protons, neutrons, and a slew of other species, including many unstable resonances.
The gas continues to expand and cool as particles scatter off each other and resonances decay into stable species.
Soon after hadronization, the decays and other chemical interactions complete, freezing the composition of the system (``chemical freeze-out'').
Around temperature $T \sim 120$ MeV, the system is dilute enough that scatterings cease, freezing all particle momenta (``kinetic freeze-out'').
A few nanoseconds later, the particles stream into the experimental detector, where they are recorded as tracks to be processed into observable quantities.

\parbreakboldstart{This is the broad picture} of ultra-relativistic heavy-ion collisions.
Of course, none of it can be observed directly---the system is far too miniscule and ephemeral, and free quarks and gluons cannot be detected directly due to QCD color confinement.
Much of what we know is inferred by matching computational collision models to experimental observations.
The primary goal of the present work is to perform this model-to-data comparison in systematic fashion, and make quantitative statements on the physical properties of the QGP and precisely what transpires in heavy-ion collisions.
For the remainder of this chapter, I introduce the experimental observations key to this comparison and describe the properties we wish to measure.

\section{Experimental observations}
\label{sec:expt-obs}

In this section, I review the current heavy-ion collision experiments and the primary experimental signatures of the strongly-interacting quark-gluon plasma.

\subsection{Ongoing experiments}

There are two particle accelerators with ongoing heavy-ion programs:
the Relativistic Heavy-ion Collider (RHIC) at Brookhaven National Lab in Upton, NY and the aforementioned Large Hadron Collider (LHC), operated by the European Organization for Nuclear Research (CERN) near Geneva, Switzerland (the accelerator ring intersects the French-Swiss border).

RHIC\footnote{
The acronym RHIC is colloquially pronounced like the name ``Rick'', and as a result, is used in speech like a name, e.g.\ people say ``at RHIC'' instead of ``at \emph{the} RHIC'', even though the latter formally makes more sense.
Meanwhile, the acronym LHC is pronounced simply as its letters spelled out, and so people usually say ``at the L-H-C''.
I will use the acronyms here as they are colloquially spoken.
}
has been operational since 2000, colliding assorted combinations of nuclear species including gold, uranium, copper, aluminum, protons, deuterons, and helium-3 at center-of-mass energies ranging from $\sqrt s = 7.7$ to 200 GeV per nucleon-nucleon pair.
The LHC, which turned on in 2009, runs proton-proton, proton-lead, and lead-lead collisions.
Although the LHC focuses on proton-proton, it is the lead-lead collisions, at energies of $\sqrt s = 2.76$ and 5.02 TeV, that are the most relevant to this work.
The two facilities are complementary:
While the LHC achieves higher energy, RHIC can run more collisions systems over a wide energy range---crucial for exploring the QCD phase diagram.

For the purposes of heavy-ion collisions, size is the principal difference among the various projectile species; the larger the nucleus, the larger the produced QGP.
Most nuclei used in collisions are approximately spherical, the notable exception being uranium, whose deformed spheroidal shape has some interesting consequences for the collision dynamics (see related discussion starting on page \pageref{loc:trento-uranium}).

\begin{figure}[t]
  \makebox[\textwidth]{
    \includegraphics[width=1.2\textwidth]{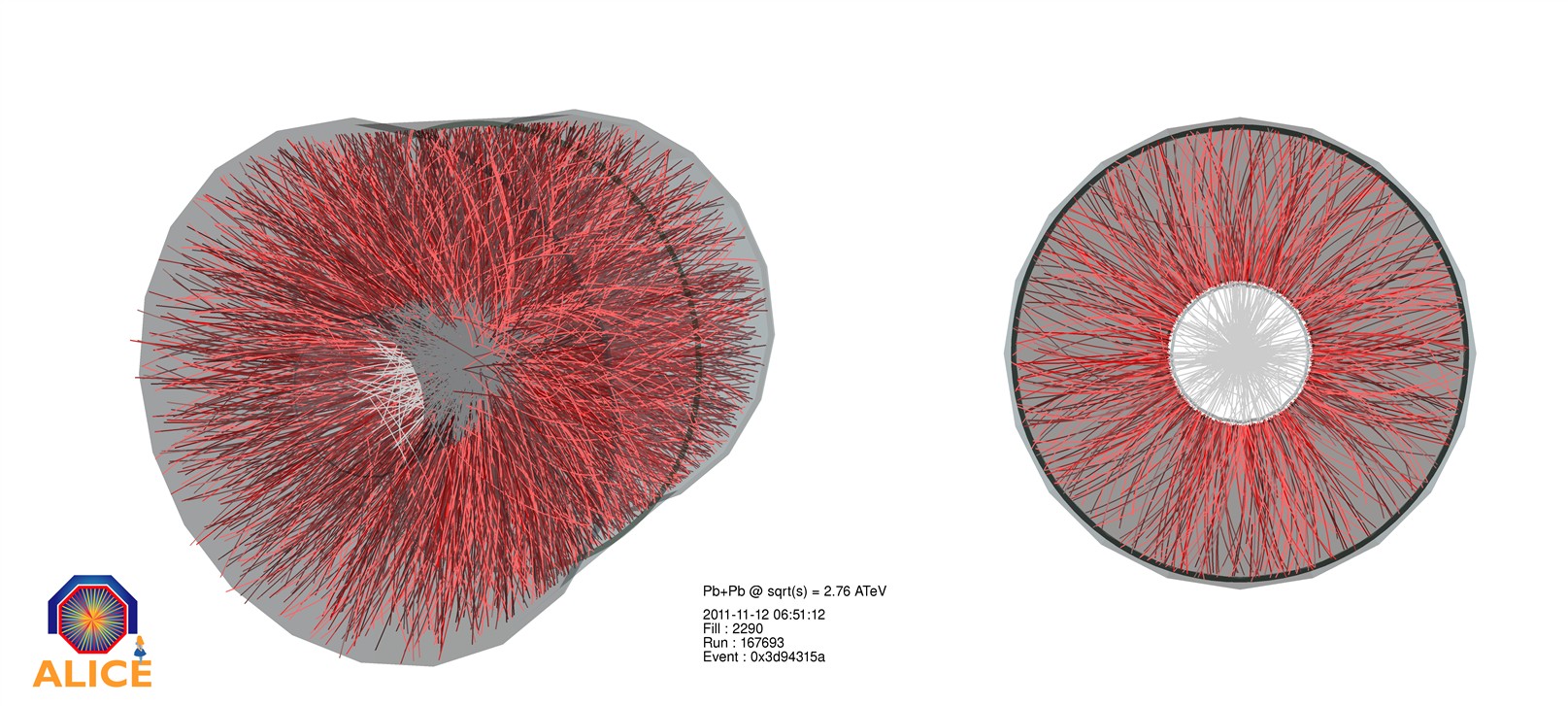}
  }
  \caption{
    Event display of a Pb-Pb collision in the ALICE detector \cite{ALICE:2011hic}, which has a toroidal shape measuring $16 \times 16 \times 26\ \text{m}^3$ \cite{Aamodt:2008zz}.
    The beam coincides with the central axis of the toroid and collisions occur in the center.
    As particles are emitted, they propagate through the various layers of the experimental apparatus and are recorded as tracks, represented here as lines.
    Left: perspective view, right: beam-axis view.
  }
\end{figure}

Both colliders have several experimental detectors distributed around their accelerator rings, each optimized for studying certain phenomena of high-energy collisions.
In this work, I use data from ALICE (A Large Ion Collider Experiment) at the LHC \cite{Aamodt:2008zz}, which specializes in heavy-ion collisions.
The ALICE Collaboration has published consistent data at both $\sqrt s = 2.76$ and 5.02 TeV \cite{Aamodt:2010cz,ALICE:2011ab,Abelev:2013vea,Abelev:2013xaa,Abelev:2013zaa,Abelev:2014ckr,Adam:2015ptt,Adam:2016thv,Adam:2016izf,ALICE:2016kpq} suitable for direct comparison with computational models.
The other heavy-ion experiments at the LHC are ATLAS and CMS;
at RHIC there is STAR, PHENIX, PHOBOS, and BRAHMS (although these all stand for something, most are fairly contrived and the acronyms are used almost exclusively).

\parbreakboldstart{Nearly two decades} into the RHIC era, there is unequivocal evidence that a strongly-interacting phase of QCD matter is created in heavy-ion collisions \cite{Adcox:2004mh,Adams:2005dq,Back:2004je,Arsene:2004fa,Muller:2012zq}.
In the following subsections, I review the experimental signatures of the QGP, emphasizing the observables that I will later use to estimate QGP properties.

\subsection{Particle and energy production}
\label{subsec:particle-energy-production}

Among the most straightforward observable quantities from high-energy collisions are the number of produced particles (multiplicity) and the amount of produced energy.
But they should not be overlooked, for despite (and perhaps because of) their simplicity, these observables connect to the basic thermal properties of the QGP, and serve as important constraints for computational models.

Particle and energy yields are typically reported per unit rapidity $y$ or pseudorapidity $\eta$.
Not to be confused with the spacetime rapidity $\eta_s$, these quantities have similar form but operate on the energy-momentum vector rather than the spacetime position.
The rapidity is defined as \cite{Patrignani:2016xqp}
\begin{equation}
  y \equiv \frac{1}{2} \log{\frac{E + p_z}{E - p_z}}.
\end{equation}
However, since this expression contains the energy, it requires direct measurement of the particle's total energy or its mass, which is not always experimentally available.
The pseudorapidity
\begin{equation}
  \eta \equiv -\log\bigl[ \tan(\theta/2) \bigr] = \frac{1}{2} \log{\frac{|\mathbf p| + p_z}{|\mathbf p| - p_z}}
\end{equation}
is sometimes more accessible, since it only depends on the polar angle of the momentum vector relative to the beam axis ($\cos\theta = p_z/|\mathbf p|$).
The rapidity and pseudorapidity are equal in the ultra-relativistic limit, $p \gg m$.

\subsubsection{Midrapidity yields}

\begin{figure}[b!]
  \centering
  \begin{tikzpicture}
    \node (img) {\includegraphics[width=.7\textwidth]{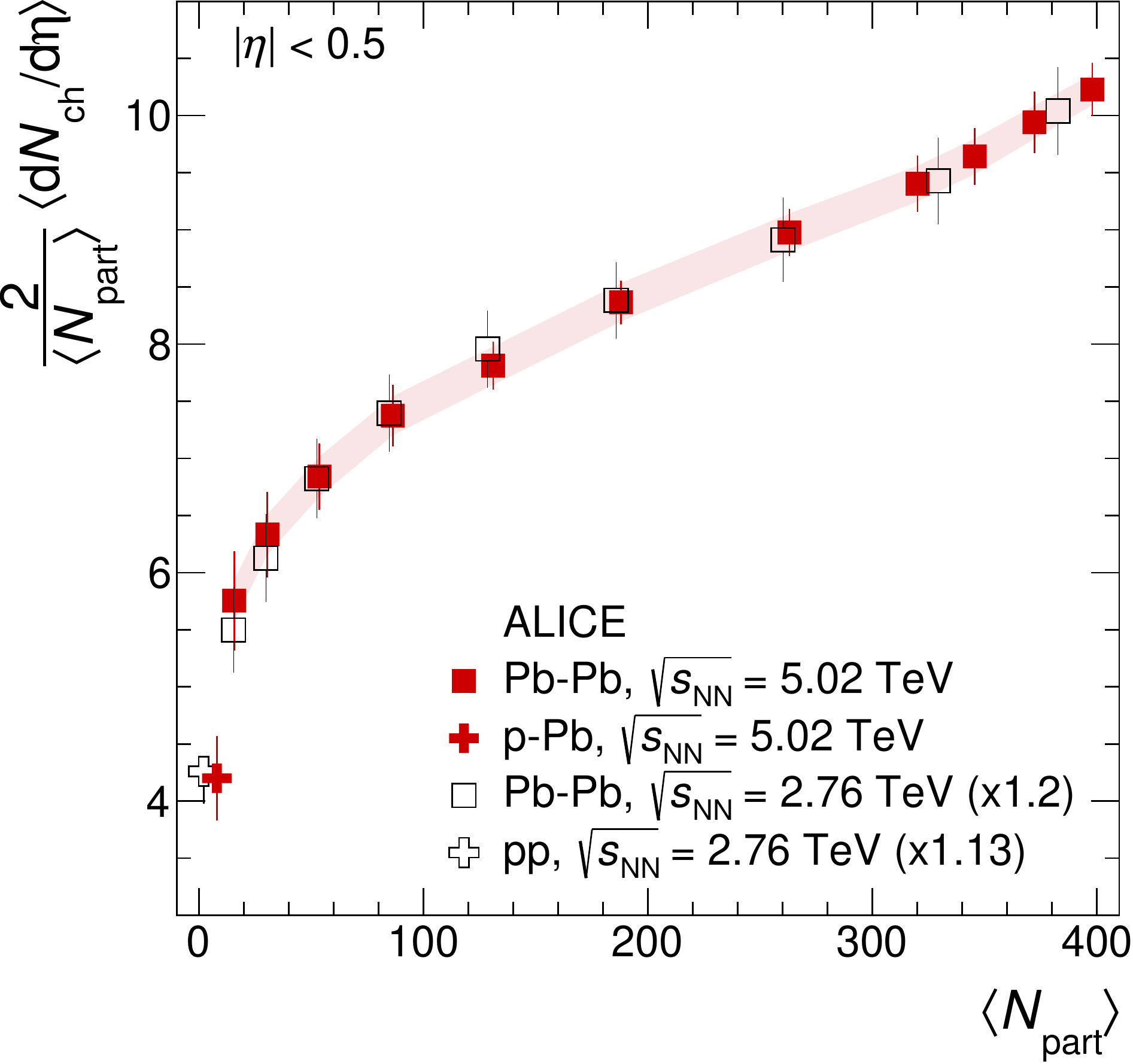}};
    \foreach \anchor/\xshift in {
      south west/40, south west/53,
      south/15, south/22,
      south east/1, south east/4
    } \fill[opacity=.3] (img.\anchor) circle [radius=.3, xshift=\xshift];
  \end{tikzpicture}
  \captionsetup{width=.75\textwidth}
  \caption{
    Charged-particle multiplicity at midrapidity per participant pair as a function of the number of participants \cite{Adam:2015ptt,Aamodt:2010cz,Adam:2015gka,ALICE:2012xs}.
    The circle diagrams show the approximate nuclear overlap of the collision depending on $\Np$.
  }
  \label{fig:dNdeta-midrap}
\end{figure}

Particles emitted transverse to the beam, i.e.\ at midrapidity (near $\eta = 0$), are the purest sample of matter produced in the collision.

Figure \ref{fig:dNdeta-midrap} shows the charged particle multiplicity per unit pseudorapidity, $d\Nch/d\eta$, in the central rapidity unit, $|\eta| < 0.5$, from ALICE measurements of lead-lead collisions at $\sqrt s = 2.76$ and 5.02 TeV and proton-proton and proton-lead collisions for comparison \cite{Adam:2015ptt,Aamodt:2010cz,Adam:2015gka,ALICE:2012xs}.
The multiplicities are shown as a function of the number of participating nucleons, $\Np$, and scaled by participant pair, $\Np/2$.
A ``participant'' is a nucleon that engages in inelastic collision processes, as opposed to a spectator, which continues down the beam pipe unaffected.
More ``central'' collisions, i.e.\ those with small impact parameter and more complete nuclear overlap, have more participants;
``peripheral'' collisions with large impact parameter have fewer participants.
The maximum number of participants for a collision of $^{208}$Pb nuclei is 416.

Due to the high energy of the collision, many more particles are produced than the original number of nucleons.
In the most central collisions with the most participants, $\Np \sim 400$, about 10 charged particles are produced per participant pair---or about $d\Nch/d\eta \sim 2000$ total particles---in the central rapidity unit alone at 5.02 TeV.
Yet, particle production is not simply proportional to the number of participants:
Central collisions create particles more efficiently per participant than peripheral collisions.
The shape of this trend is almost identical at both beam energies, with 5.02 TeV collisions uniformly producing about 20\% more particles than 2.76 TeV.

\begin{figure}[b!]
  \graphicsandcaption{.83}{expt-data/dETdeta_dNdeta}{
    Average transverse energy per charged particle at midrapidity as a function of the number of participants
    \cite{Adam:2016thv,Adams:2004cb,Adcox:2001ry,Adler:2004zn}.
  }
  \label{fig:dETdeta-dNdeta}
\end{figure}

Another common measurement of the amount of produced matter is the transverse energy
\begin{equation}
  E_T = \sum_i E_i \, \sin\theta_i,
\end{equation}
where $E_i$ and $\theta_i$ are the total energy and angle with respect to the beam, respectively, of particle $i$.
Transverse energy is closely related to charged-particle production and has a similar trend as a function of $\Np$.
Figure \ref{fig:dETdeta-dNdeta} shows the average transverse energy per charged particle at midrapidity for lead-lead collisions at 2.76 TeV and RHIC gold-gold collisions at 200 GeV \cite{Adam:2016thv,Adams:2004cb,Adcox:2001ry,Adler:2004zn};
the ratio is constant within uncertainty as a function of $\Np$, but clearly higher-energy collisions produce more transverse energy per particle.

\subsubsection{Centrality}

\begin{figure}[b!]
  \graphicsandcaption{.75}{expt-data/centrality_glauber}{
    Centrality determination of Pb-Pb collisions at 2.76 TeV by ALICE \cite{Aamodt:2010cz,Abelev:2013qoq}.
    The histogram is the distribution of the VZERO amplitude, apportioned into centrality percentile bins, and the red line is the Glauber model fit.
  }
  \label{fig:centrality-glauber}
\end{figure}

I have already mentioned the concept of centrality and its relation to $\Np$, but, as the primary classifier of heavy-ion collision events, it warrants a dedicated discussion.
Centrality categorizes events based on a final-state observable that quantifies the amount of matter produced in the collision, such as $\Nch$ or $E_T$.
Geometric properties of the initial state, such as $\Np$ and the impact parameter $b$---which are not directly measurable---can then be connected to centrality and estimated using a geometric initial condition model.

\newcommand{\Nchthr}{\Nch^\text{THR}}

Formally, centrality is the fraction of the nuclear interaction cross section $\sigma$ above some threshold of particle or energy production, for example
\begin{equation}
  c(\Nchthr) \approx \frac{1}{\sigma} \int_{\Nchthr}^\infty \frac{d\sigma}{d\Nch'} d\Nch',
\end{equation}
where $\Nchthr$ is a threshold number of charged particles.
Thus, if $\Nchthr$ is close to the maximum number of particles that an event can produce, then only a small fraction of the differential cross section $d\sigma/d\Nch$ will be above the threshold, so the centrality fraction will be small.
Inversely, if the threshold is low, then most of the cross section will be above it, so the centrality fraction will be large.

To construct centrality bins, experiments run a large number of events, sort them by the chosen observable, and then apportion the events into percentile bins.
Figure \ref{fig:centrality-glauber} shows the centrality bins determined by ALICE, which defines centrality by the VZERO amplitude \cite{Abelev:2013qoq}.
The ALICE VZERO detector covers the forward and backward pseudorapidity ranges $2.8 < \eta < 5.1$ and $-3.7 < \eta < -1.7$ \cite{Aamodt:2008zz};
the ``amplitude'' is proportional to the number of detected particles, i.e.\ multiplicity.

The multiplicity distribution is fit to a Monte Carlo Glauber model (a geometric initial condition model---see subsection \ref{subsec:properties-initial-state}) convolved with multiplicity fluctuations from a negative binomial distribution, then the average number of participants is determined from the fit for each centrality bin, as shown in table \ref{tab:centrality} \cite{Aamodt:2010cz}.
Thus, the data points shown as functions of $\Np$ in figures \ref{fig:dNdeta-midrap} and \ref{fig:dETdeta-dNdeta} are in fact centrality bins in disguise.

\begin{table}[t]
  \centering
  \captionsetup{width=.62\textwidth}
  \caption{Average charged-particle multiplicity at midrapidity and estimated average number of participants for the centrality bins in figure \ref{fig:centrality-glauber} \cite{Aamodt:2010cz}.}
  \label{tab:centrality}
  \small
  \begin{tabular}{lcc}
    \toprule
    Centrality \% & $\avg{d\Nch/d\eta}$ & $\avg{\Np}$ \\
    \midrule
    0--5   & $1601 \pm 60$ & $382.8 \pm 3.1$ \\
    5--10  & $1294 \pm 49$ & $329.7 \pm 4.6$ \\
    10--20 & $966  \pm 37$ & $260.5 \pm 4.4$ \\
    20--30 & $649  \pm 23$ & $186.4 \pm 3.9$ \\
    30--40 & $426  \pm 15$ & $128.9 \pm 3.3$ \\
    40--50 & $261  \pm 9 $ & $85.0  \pm 2.6$ \\
    50--60 & $149  \pm 6 $ & $52.8  \pm 2.0$ \\
    60--70 & $76   \pm 4 $ & $30.0  \pm 1.3$ \\
    70--80 & $35   \pm 2 $ & $15.8  \pm 0.6$ \\
    \bottomrule
  \end{tabular}
\end{table}

The following graphic summarizes the relationship between centrality and collision geometry ($\Np$ and impact parameter $b$), where $R$ is the nuclear radius and $A$ the mass number (number of nucleons):
\begin{center}
  \begin{tikzpicture}
    \draw[semithick,<->]
      (0, 0)
        node (L) {}
        node[above right] {100\% centrality}
        node[below right, align=left] {$\Np \sim 2$ \\ $b \sim 2R$} --
      (.5\textwidth, 0)
        node(R) {}
        node[above left] {0\% centrality}
        node[below left, align=right] {$\Np \sim 2A$ \\ $b \sim 0$};
    \foreach \pos/\xshift in {
      L/-25, L/-55,
      R/25, R/27
    } \fill[opacity=.3] (\pos) circle [radius=.6, xshift=\xshift];
  \end{tikzpicture}
\end{center}

\subsubsection{Pseudorapidity distributions}

In addition to midrapidity yields, experiments have measured particle production as a function of pseudorapidity, for example:
\begin{figure}[h]
  \includegraphics[width=\textwidth]{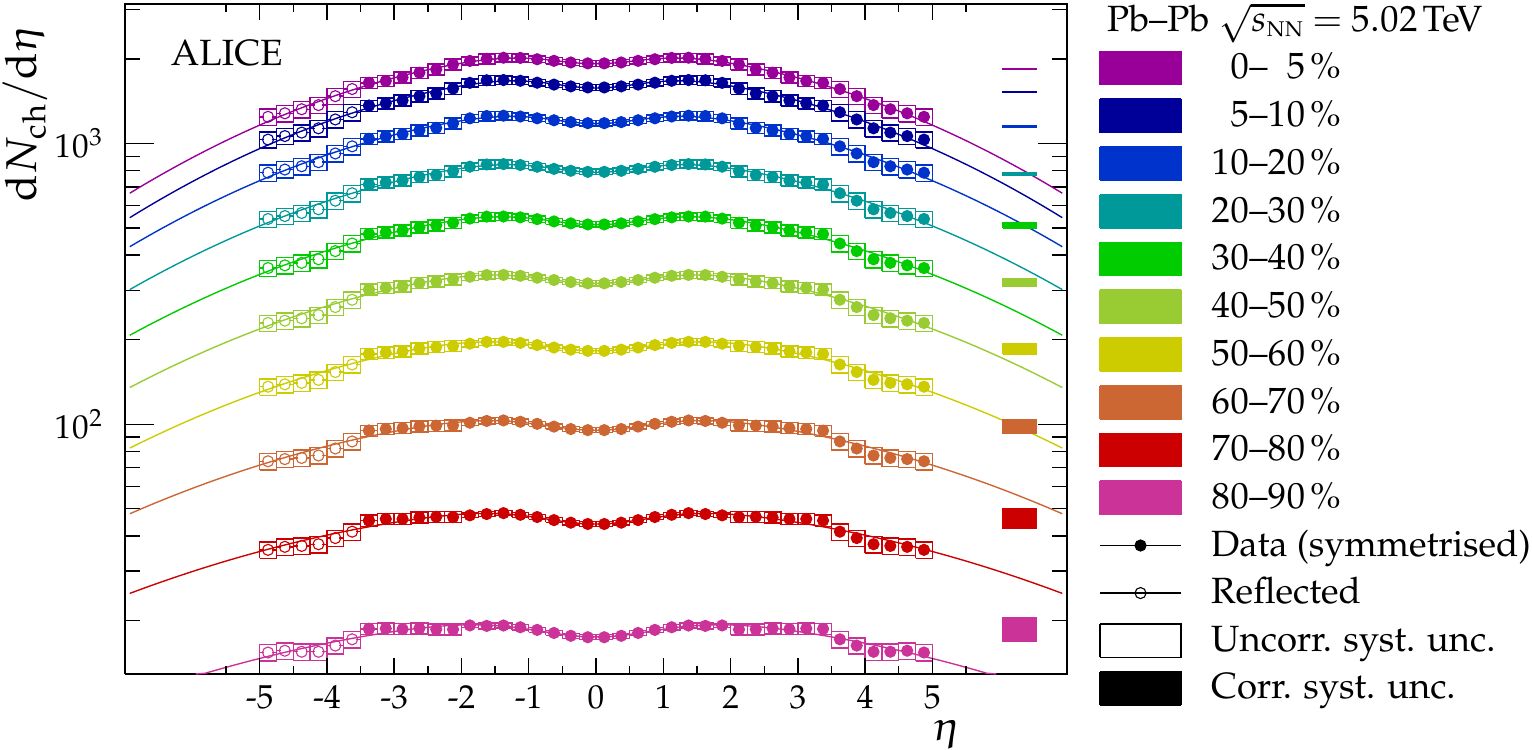}
  \caption{Charged-particle pseudorapidity density in several centrality bins measured by ALICE for Pb-Pb collisions at 5.02 TeV \cite{Adam:2016ddh}.}
  \label{fig:dNdeta-dists}
\end{figure} \\
Boost invariance asserts that these distributions should be flat near midrapidity, which is the case for the central rapidity unit $|\eta| < 0.5$ out to intermediate centrality.
Note that pseudorapidity is a highly nonlinear function of the physical angle; each successive rapidity unit away from $\eta = 0$ has less angular coverage.
For example the central unit $|\eta| < 0.5$ covers the central $55^\circ$, or about 30\% of the total angular space, while $2 < |\eta| < 2.5$ covers only about $12^\circ$.

\subsubsection{Identified particle yields}

The observables discussed to this point do not differentiate among the various hadronic species created in heavy-ion collisions, such as pions, kaons, and protons.
Yields of specific identified particles are conventionally reported per unit rapidity (not pseudorapidity), $dN/dy$.

The ratios of various identified particle yields provide insights on chemical freeze-out, expected to occur shortly after the QGP medium hadronizes.
A simple description of particle production is the statistical hadronization model \cite{BraunMunzinger:2003zd,Becattini:2009sc}, which assumes that particles are thermally produced in the grand canonical ensemble, so each species's yield is controlled by its Boltzmann factor $e^{-m/T}$ and spin degeneracy.

\begin{figure}[h]
  \graphicsandcaption{.8}{expt-data/dNdy_shm}{
    Statistical hadronization model fit to identified particle yields in central Pb-Pb collisions at 2.76 TeV \cite{Stachel:2013zma}.
    Data from ALICE \cite{Abelev:2013vea,Abelev:2013xaa,Abelev:2013zaa,Abelev:2014uua,Adam:2015vda}.
  }
  \label{fig:dNdy-shm}
\end{figure}

Figure \ref{fig:dNdy-shm} shows statistical model fits to hadron yields in central lead-lead collisions at 2.76 TeV \cite{Stachel:2013zma}.
The primary fit has two free parameters: the temperature and effective system volume (to normalize overall particle production), with the baryon chemical potential fixed to zero.
In this model, the various yields are well-described---with the possible exception of protons, which are somewhat overpredicted---and the best-fit temperature $T = 156$ MeV is within the QCD transition region, consistent with a prompt chemical freeze-out.
The alternate fit has $\mu_B = 1$ MeV and a higher temperature, but the description of the data is inferior, suggesting near matter-antimatter symmetry.

It is quite remarkable that such a simple model, with a single meaningful free parameter (perhaps two, if the chemical potential counts), is able to quantitatively describe such a wide variety of hadron yields.
This is in contrast to high-energy collisions of smaller projectiles, such as proton-proton and electron-positron, where hadrons containing strange quarks are underproduced relative to their thermal ratios.
Fits to proton-proton and electron-positron data require an artificial strangeness suppression fugacity parameter $\gamma_s \sim 0.6$, while for nucleus-nucleus collisions $\gamma_s \sim 1$, implying full chemical equilibrium \cite{Becattini:2010sk,Blume:2011sb}.
Interestingly, the best-fit temperature is consistently 155--170 MeV for all collision systems.

The total chemical equilibrium in heavy-ion collisions---as opposed to strangeness suppression in other systems---is an important signal of QGP formation, since, in the plasma, strange-antistrange pairs can be produced directly from pairs of free quarks and gluons, and these processes equilibrate within the timescale of heavy-ion collisions \cite{Rafelski:1982pu,Koch:1986ud}.
These avenues are not available in hadronic systems, so small collision systems (that don't create QGP) cannot produce as much strangeness.

\subsubsection{Transverse momentum distributions}

A standard measurement in high-energy collisions is the distribution of particle production as a function of the transverse momentum, $p_T = \sqrt{p_x^2 + p_y^2}$.
These distributions, often called $p_T$ spectra, are usually reported as something like $d^2N/(N_\text{ev} \, 2\pi p_T \, dp_T \, dy)$, meaning a histogram of particle counts binned by $p_T$, per unit rapidity, averaged over the events in a centrality bin.
The factor $1/2\pi p_T$ corrects for the phase space density $d^2p_T = 2\pi p_T \, dp_T \, d\phi$, since $p_T$ is effectively a polar or cylindrical radius.

\begin{figure}[t]
  \centering
  \includegraphics{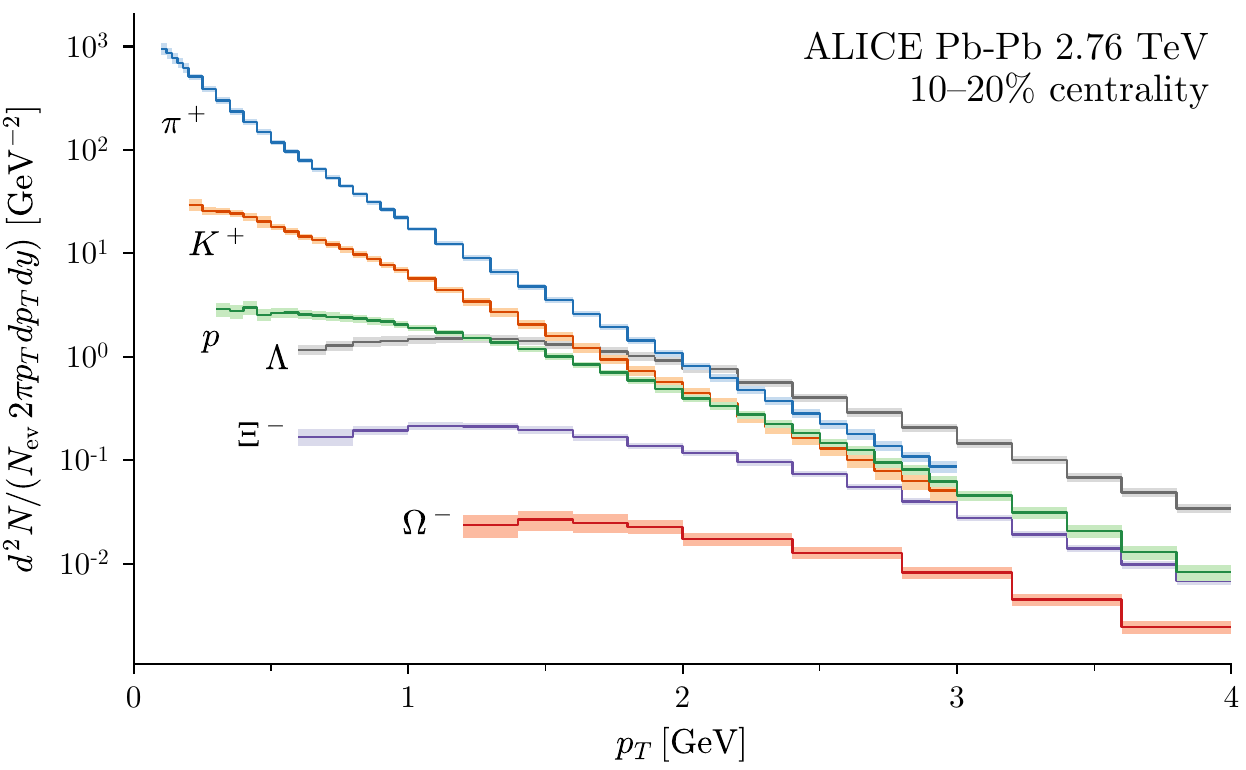}
  \caption{Transverse momentum distributions (histograms) for the labeled identified particles at midrapidity measured by ALICE \cite{Abelev:2013vea,Abelev:2013xaa,Abelev:2013zaa,Abelev:2014uua}.}
  \label{fig:pT-spectra}
\end{figure}

\begin{figure}[b!]
  \makebox[\textwidth]{
    \includegraphics[width=.53\textwidth]{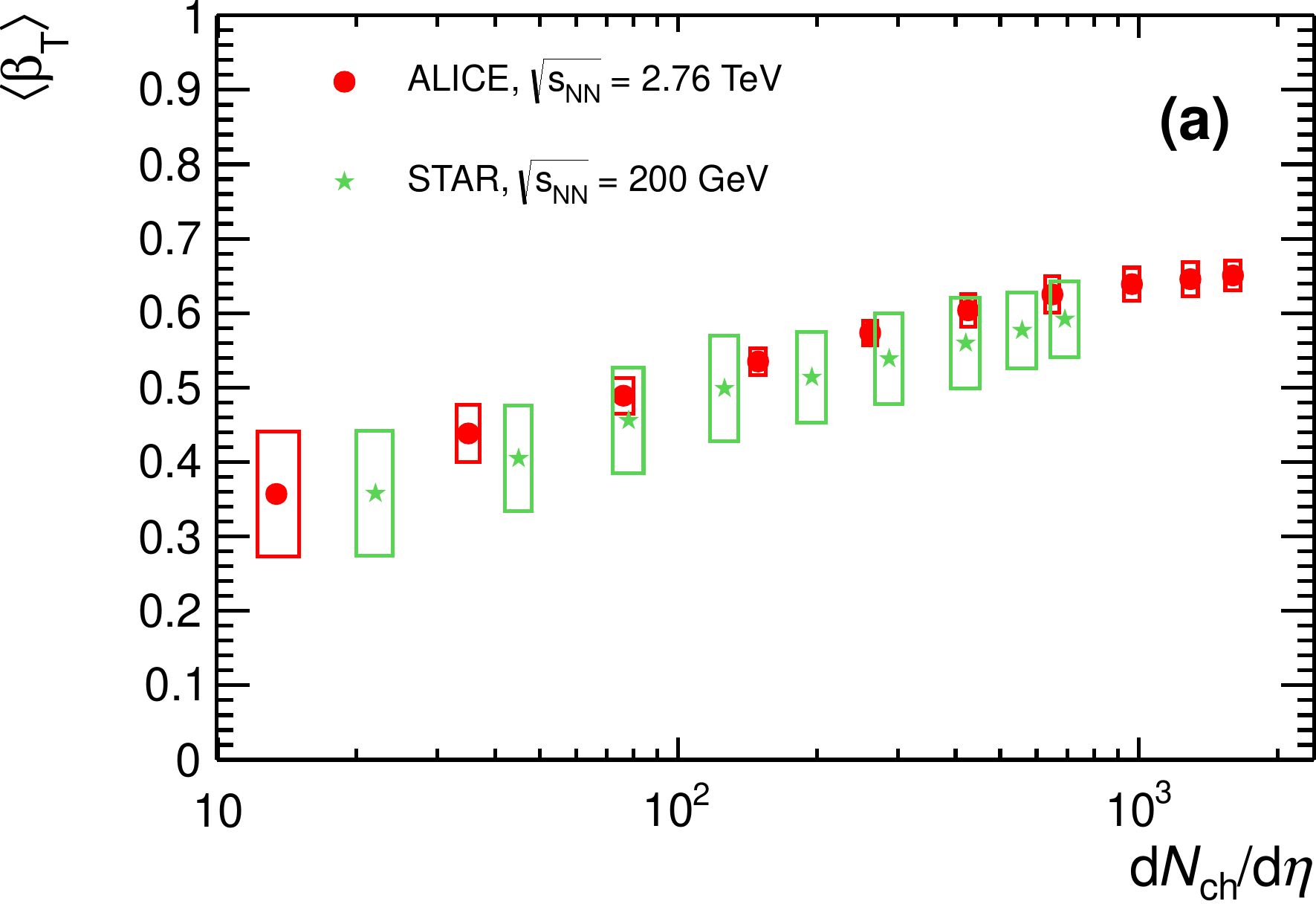}
    \quad
    \includegraphics[width=.53\textwidth]{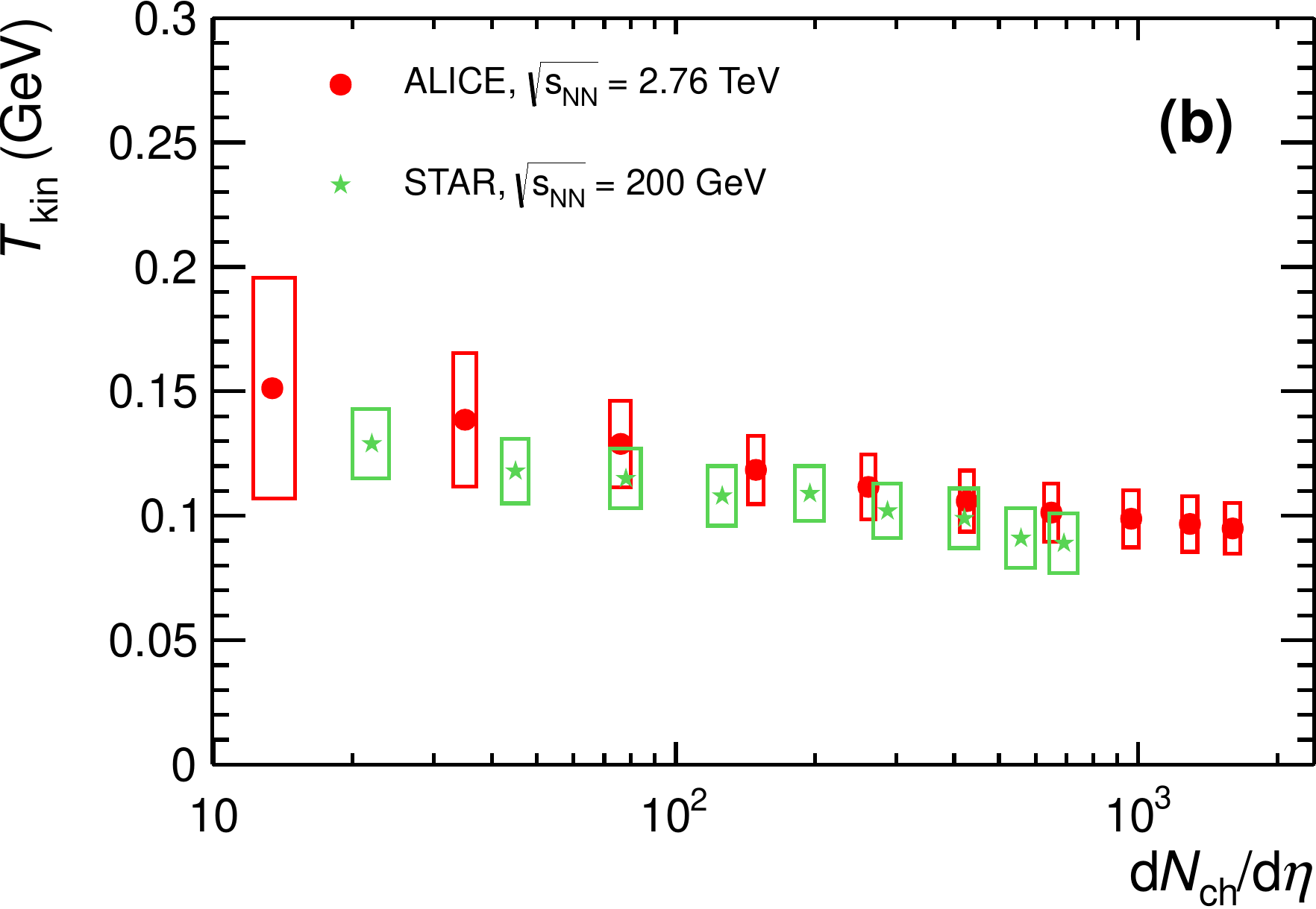}
  }
  \caption{
    Average transverse flow velocity $\avg{\beta_T}$ (left) and effective kinetic freeze-out temperature $T_\text{kin}$ (right) from blast-wave fits to transverse momentum spectra.
    Parameters from ALICE \cite{Abelev:2013vea} and STAR \cite{Adams:2005dq} are shown together as functions of midrapidity charged-particle production $d\Nch/d\eta$.
  }
  \label{fig:blast-wave}
\end{figure}

Figure \ref{fig:pT-spectra} shows typical transverse momentum distributions for several identified particle species measured by ALICE \cite{Abelev:2013vea,Abelev:2013xaa,Abelev:2013zaa,Abelev:2014uua}.
The distributions are approximately thermal in the hydrodynamic region, $p_T \lesssim 3$ GeV, with a peak at low $p_T$ and an exponential tail.
The height of each curve is proportional to the yield, which simply follows the mass hierarchy, while the slope relates to the kinetic freeze-out temperature and rate of transverse expansion.
Notably, the strange baryons ($\Lambda$, $\Xi$, $\Omega$) have shallower slopes and longer tails than the other species, indicating higher effective kinetic freeze-out temperatures;
they cease interacting earlier in the hadron gas expansion due to their smaller scattering cross sections.

The effective kinetic freeze-out temperature and transverse expansion velocity may be estimated by fitting spectra to the so-called ``blast-wave'' function \cite{Schnedermann:1993ws}, which incorporates thermal particle production and hydrodynamic flow.
As shown in figure \ref{fig:blast-wave}, the average transverse flow velocity $\avg{\beta_T}$ increases significantly with particle production, meaning that central collisions expand more explosively.
The kinetic freeze-out temperature $T_\text{kin}$ \emph{decreases} with centrality, presumably because as the system density increases, it must cool more before particles stop interacting.
In central collisions, $T_\text{kin} \sim 100$ MeV is well below the chemical freeze-out temperature $\sim$155 MeV, corroborating the picture that hadrons continue to scatter for some time after the chemical composition is fixed.
Compared to LHC, the RHIC flow velocities and temperatures are uniformly smaller given the same number of produced particles, reflecting the less explosive system created at lower beam energy.

\subsection{Collective behavior}

The observation of collective behavior is arguably the most compelling evidence that a strongly-interacting quark-gluon plasma is created in heavy-ion collisions.

Collectivity manifests as anisotropies in the azimuthal transverse momentum distribution \cite{Ollitrault:1992bk}, $dN/d\phi$, where $\phi = \text{arctan2}(p_y, p_x)$.
Why would such anisotropy occur?
Consider the diagram of a noncentral collision on the left:
\begin{figure}[h]
  \centering
  \medskip
  \includegraphics{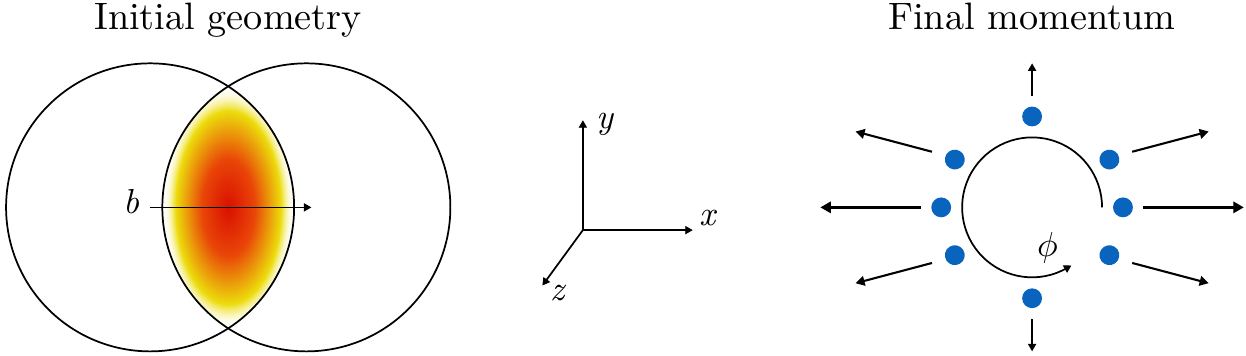}
  \caption{
    Left: Asymmetric overlap region created by a pair of nuclei (circles) colliding with impact parameter $b$.
    Right: The resulting anisotropic transverse particle emission.
  }
  \label{fig:anisotropy}
\end{figure} \\
The nuclei collide with impact parameter $b$ along the $x$-direction, creating an asymmetric almond-shaped overlap region where the hot and dense QGP medium forms.
This shape generates a steeper pressure gradient along the $x$-direction compared to $y$, since the same total pressure change---from the central pressure to surrounding vacuum---occurs over a shorter distance.
The pressure gradients then drive fluid dynamical expansion preferentially in the $x$-direction, and as the medium freezes into hadrons, it imparts that anisotropic momentum to the emitted particles, as shown on the right of the figure.
Ultimately, the observed transverse momentum distribution will have more particles near azimuth $\phi = 0$ and $\pi$.

\begin{wrapfigure}[38]{r}[.13\textwidth]{.43\textwidth}
  \vspace{-24pt}
  \includegraphics[height=.9\textheight]{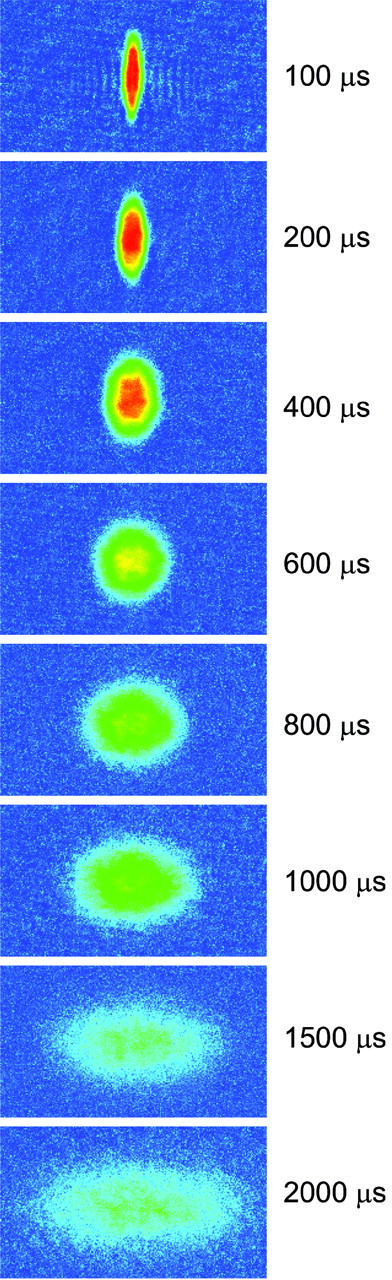}
  \caption{Anisotropic expansion of a strongly-interacting degenerate Fermi gas \cite{OHara:2002sci}.}
  \label{fig:fermi-gas}
\end{wrapfigure}

A similar phenomenon has been directly observed in a rather dissimilar strongly-interacting system:
an ultra-cold, degenerate gas of Fermionic lithium atoms.
In the experiment \cite{OHara:2002sci}, the gas is held by an asymmetric optical trap, then released and allowed to expand;
figure \ref{fig:fermi-gas} shows snapshots of the expanding gas from $t = 0.1$ to 2.0 milliseconds after release.
Beginning as a narrow ellipse with its short axis oriented horizontally, the Fermi gas expands almost exclusively in the horizontal direction, driven by the initial pressure gradients.
Although the images only show the spatial distribution, we can infer from the time evolution that the atoms are preferentially emitted in the horizontal direction.

The Fermi gas is many orders of magnitude cooler, larger, and longer-lived than the QGP created in heavy-ion collisions, yet it also behaves collectively.
Both systems are strongly coupled, where in this context, ``strong'' is more general than the strong nuclear force;
it means the quanta of the system have large cross sections, short mean free paths, and they interact frequently---so their motion is correlated.
In other words, such systems generally behave like fluids with low viscosity.
This is not terribly restrictive; the equations of fluid dynamics derive from universal conservation laws and treat the system as a continuous medium, ignoring the particulars of the underlying microscopic dynamics \cite{LandauLifshitz}.

It is widely accepted that the QGP medium behaves hydrodynamically, i.e.\ like a liquid.
Hydrodynamics explains the conversion of the initial geometric asymmetry to final-state momentum anisotropy, and viscous relativistic hydrodynamic models describe a diverse array of observables with exceptional accuracy, which I will highlight in the following subsections.

\subsubsection{Anisotropic flow coefficients}

\begin{figure}[h]
  \includegraphics{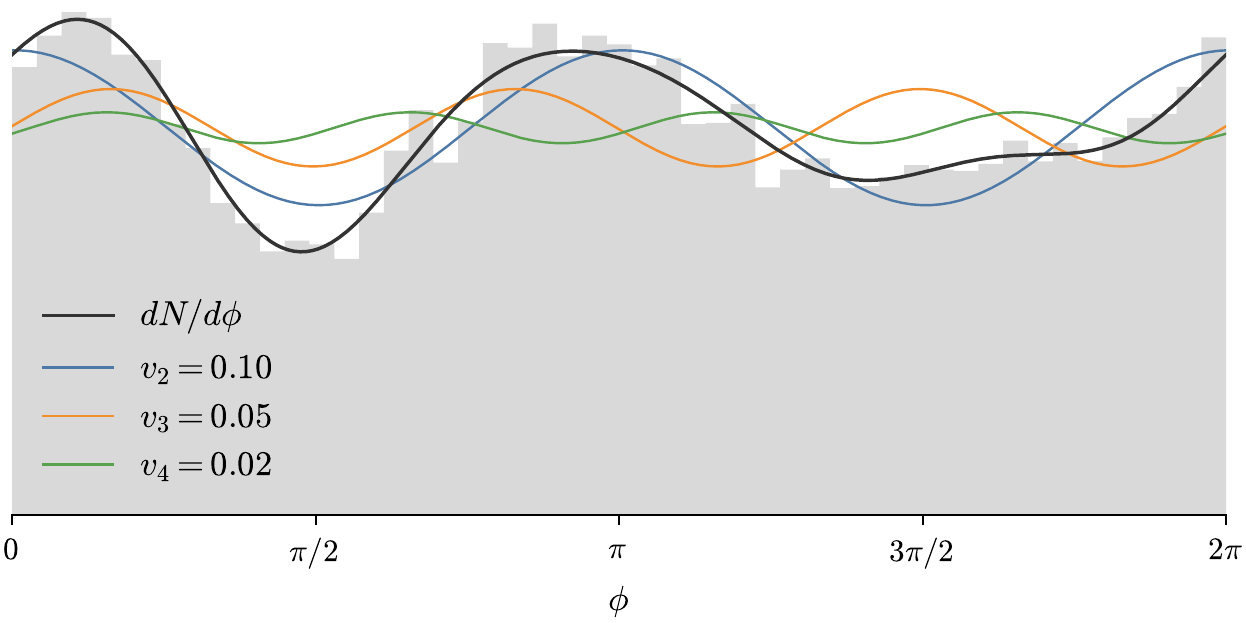}
  \caption{
    Fourier decomposition of an azimuthal particle distribution into flow harmonics $v_n$.
    The gray histogram is the ``observed'' distribution (randomly generated, not real experimental data), the colored lines are the Fourier components, and the black line is the total distribution.
  }
  \label{fig:flow-decomp}
\end{figure}

To quantify transverse momentum anisotropy, we expand the azimuthal distribution as a Fourier series \cite{Voloshin:1994mz,Poskanzer:1998yz}
\begin{equation}
  \frac{dN}{d\phi} \propto 1 + 2 \sum_{n=1}^\infty v_n \cos\bigl[ n(\phi - \Psi_n) \bigr],
\end{equation}
where the flow coefficient $v_n$ is the magnitude of $n$th-order anisotropy and the event-plane angle $\Psi_n$ is the corresponding phase;
figure \ref{fig:flow-decomp} shows a typical Fourier decomposition.
 \\
The flow coefficients, or harmonics, are given by
\begin{equation}
  v_n = \bigl\langle \cos\bigl[ n(\phi - \Psi_n) \bigr] \bigr\rangle,
  \label{eq:vn}
\end{equation}
where the average runs over particles (in a $p_T$ bin) and events (in a centrality bin).
In particular, $v_1$ is called directed flow, $v_2$ elliptic flow, and $v_3$ triangular flow.

\subsubsection{Fluctuations}

The simplified collision geometry shown in figure \ref{fig:anisotropy} explains only the existence of even-order anisotropy;
the almond shape would drive strong elliptic flow $v_2$ and contribute to higher-order even harmonics ($v_4$, $v_6$, $\ldots$), but cannot account for triangular flow $v_3$ or any other odd harmonics.
Triangular flow, universally observed at RHIC \cite{Adare:2011tg,Adamczyk:2013waa} and LHC \cite{ALICE:2011ab,Adam:2016izf,ATLAS:2012at,Chatrchyan:2013kba}, is thus attributed to event-by-event fluctuations in the collision geometry \cite{Alver:2010gr}.
\begin{figure}[h]
  \includegraphics{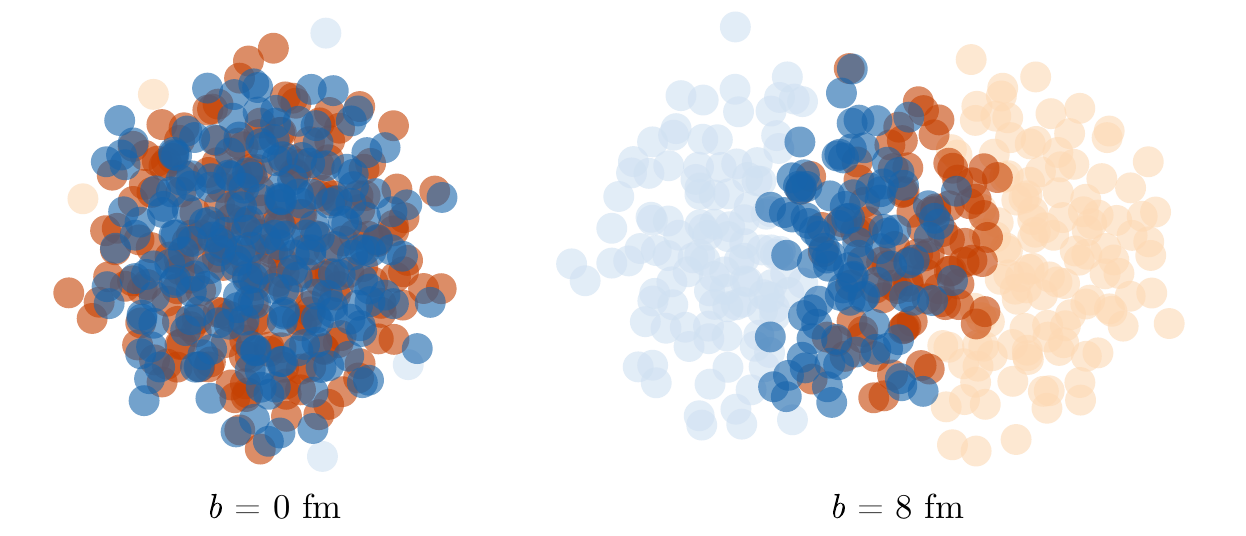}
  \caption{
    Initial collision geometry created by fluctuating nucleon positions.
    Blue and orange circles represent nucleons from each projectile nucleus;
    dark circles are participants, light are spectators.
    Left: ultra-central collision, right: intermediate centrality.
  }
  \label{fig:initial-conditions}
\end{figure} \\
Above, the impact of nucleon position fluctuations on overlap geometry;
the right side is a more realistic version of the perfect almond shape, while the left side shows that even perfectly central collisions may have spatial anisotropy.
These irregular overlap regions have nonzero ellipticity, triangularity, and higher-order deformations, which together drive all orders of anisotropic flow.

\subsubsection{Cumulants}

The definition of the flow coefficients \eqref{eq:vn} depends on the event-plane angles $\Psi_n$, characteristics of the initial collision geometry which are therefore not experimentally observable.
To circumvent this, flow coefficients are typically estimated via multiparticle azimuthal correlations, or cumulants \cite{Borghini:2000sa,Borghini:2001vi,Borghini:2001zr,Bilandzic:2010jr,Bilandzic:2012wva,Bilandzic:2013kga}.
Since collectivity induces particle correlations in momentum space, the flow can be extracted from measured correlation functions without knowledge of the event plane.

\begin{figure}[t]
  \makebox[\textwidth]{
    \includegraphics[width=1.14\textwidth]{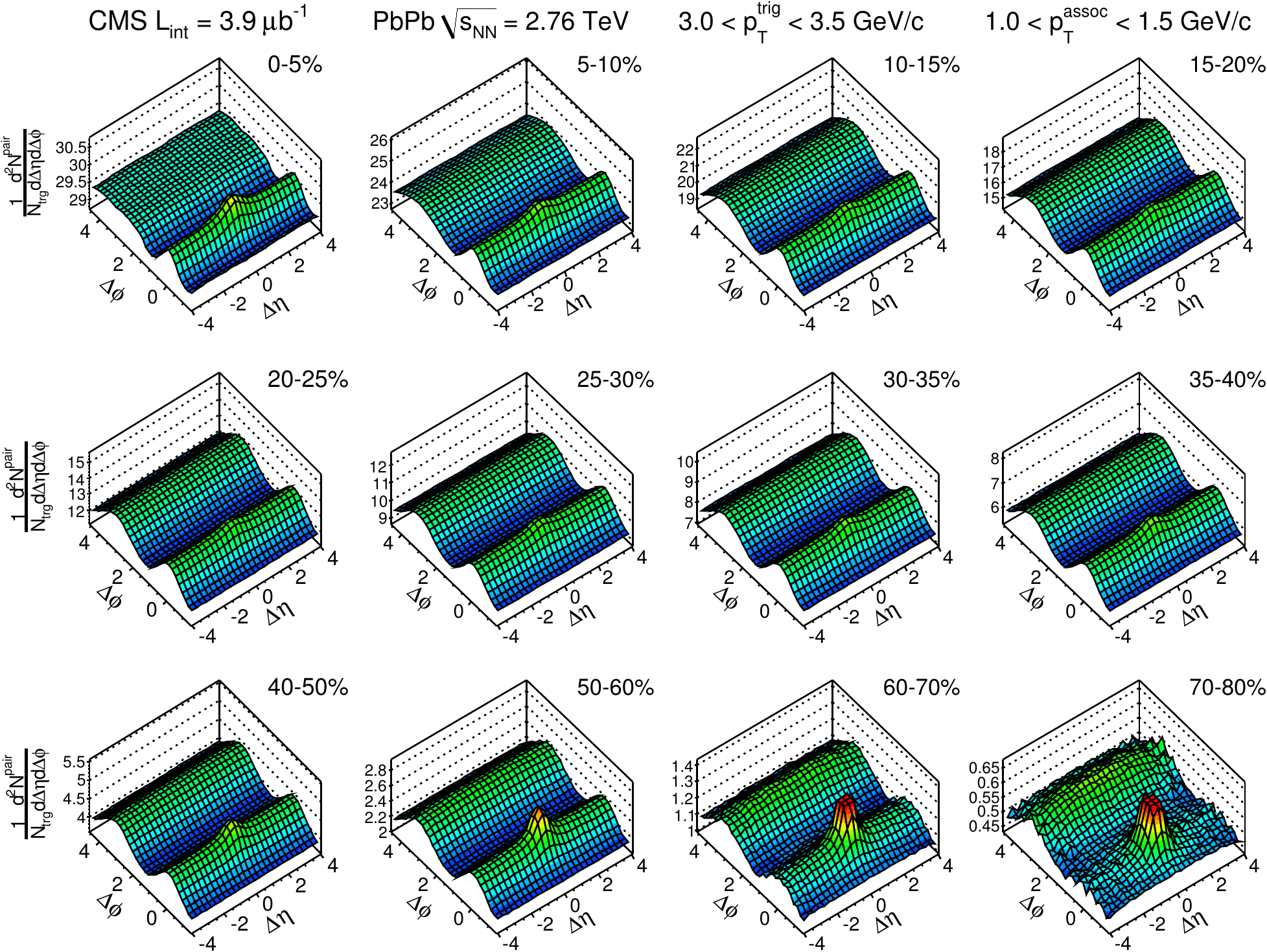}
  }
  \caption{
    Two-particle correlation functions in various centrality bins for Pb-Pb collisions at 2.76 TeV measured by CMS \cite{Chatrchyan:2012wg}.
    The height of each $(\Delta\phi, \Delta\eta)$ bin is proportional to the number of observed charged-particle pairs in the bin.
    Each pair consists of a trigger particle and an associated particle with transverse momenta in the annotated ranges for $p_T^\text{trig}$ and $p_T^\text{assoc}$, respectively.
  }
  \label{fig:corr-functions}
\end{figure}

Figure \ref{fig:corr-functions} shows typical two-particle correlation functions measured by the CMS experiment \cite{Chatrchyan:2012wg}.
They are histograms of $\Delta\phi$ and $\Delta\eta$, the differences in azimuthal angle and pseudorapidity between pairs of particles, where the height of each $(\Delta\phi, \Delta\eta)$ bin is proportional to the number of observed charged-particle pairs with those differences.
In all but the most peripheral collisions, there is a pronounced ridge structure at $\Delta\phi \sim 0$.
The fact that this ``near-side ridge'' extends to long range in $\Delta\eta$, and that it disappears in peripheral collisions, is taken as a signal of collective behavior.
A similar long-range ``away-side ridge'' forms at $\Delta\phi \sim \pi$ in mid-central collisions as a result of elliptic flow.
The peak at $\Delta\phi \sim \Delta\eta \sim 0$ is due to short-range correlations such as jets (collimated showers of particles that I discuss shortly, in subsection \ref{subsec:hard-processes}).

How can we extract flow coefficients from correlation functions?
This requires some formalism.
Let $\avg k$ denote the single-event $k$-particle azimuthal correlation function, then the two- and four-particle correlations are \cite{Bilandzic:2010jr}
\begin{equation}
  \begin{aligned}
    \avg 2 &= \bigl\langle e^{in(\phi_1 - \phi_2)} \bigr\rangle
      = \frac{1}{P_{M,2}} \sum_{i \neq j}^M e^{in(\phi_i - \phi_j)}, \\
    \avg 4 &= \bigl\langle e^{in(\phi_1 + \phi_2 - \phi_3 - \phi_4)} \bigr\rangle
      = \frac{1}{P_{M,4}} \sum_{i \neq j \neq k \neq l}^M e^{in(\phi_i + \phi_j - \phi_k - \phi_l)},
  \end{aligned}
  \label{eq:corr-functions}
\end{equation}
where $M$ is the event multiplicity and $P_{M,k} = M!/(M - k)!$ is the number of $k$-particle permutations, e.g.
\begin{equation}
  \begin{aligned}
    P_{M,2} &= M(M - 1), \\
    P_{M,4} &= M(M - 1)(M - 2)(M - 3).
  \end{aligned}
\end{equation}
The two-particle correlation function for a centrality bin is
\begin{equation}
  \davg 2 = {\bigl\langle\!\bigl\langle} e^{in(\phi_1 - \phi_2)} {\bigl\rangle\!\bigl\rangle}
    = \frac{\sum_i^{N_\text{events}} P_{M_i,2} \avg 2_i}{\sum_i^{N_\text{events}} P_{M_i,2}},
\end{equation}
where the outer average is performed over all events in the centrality bin, weighted by each event's number of permutations.
The definition of $\davg 4$ is analogous.

\label{loc:flow-corr-functions}

To see how the correlation functions relate to the flow coefficients, first add and subtract the event plane to the azimuthal angles:
\begin{equation}
  \davg 2 = {\bigl\langle\!\bigl\langle} e^{in[(\phi_1 - \psi_n) - (\phi_2 - \psi_n)]} {\bigl\rangle\!\bigl\rangle}.
\end{equation}
Now, as long as $\phi_1$ and $\phi_2$ are only correlated via the event plane, i.e.\ only due to collective flow, the inner average factorizes \cite{Bilandzic:2012wva}:
\begin{equation}
  \davg 2 \approx
    {\bigl\langle\!\bigl\langle} e^{in(\phi_1 - \psi_n)} \bigl\rangle
    \bigl\langle e^{-(\phi_2 - \psi_n)} {\bigl\rangle\!\bigl\rangle}
     = \avg{v_n^2}.
  \label{eq:pureflow}
\end{equation}
(The imaginary parts vanish by symmetry.)
Analogously, $\davg 4 \approx \avg{v_n^4}$, etc.

In reality, other physical processes besides collective flow, such as jets and resonance decays, can induce particle correlations, which is why the above relations are only approximate.
Certainly, some fraction of the away-side ridge is attributable to back-to-back jets.
When estimating flow via multiparticle correlations, it is crucial to remove as much of these ``nonflow'' effects as possible.
Using four-particle (or even higher-order) correlations is one way to suppress nonflow.

Continuing the derivation, let $c_n\{k\}$ be the $n$th-order cumulant from $k$-particle correlations, and specifically \cite{Borghini:2001vi}
\begin{equation}
  \begin{aligned}
    c_n\{2\} &= \davg 2, \\
    c_n\{4\} &= \davg 4 - 2\,\davg{2}^2.
  \end{aligned}
\end{equation}
Finally, defining $\vnk n k$ as the estimate of the flow coefficient $v_n$ from the cumulant $c_n\{k\}$:
\begin{equation}
  \begin{aligned}
    \vnk n 2 &= \sqrt{c_n\{2\}}, \\
    \vnk n 4 &= \sqrt[4]{-c_n\{4\}}.
  \end{aligned}
\end{equation}
Expressions for the six- and eight-particle cumulants $\vnk n 6$ and $\vnk n 8$ also exist but are rather lengthy, so I omit them here.
Each flow cumulant provides a different estimate of the underlying flow.
Notice that, in the absence of nonflow and statistical fluctuations, invoking equation \eqref{eq:pureflow} gives
\begin{equation}
  \begin{aligned}
    \vnk n 2 &\approx \sqrt{v_n^2} = v_n, \\
    \vnk n 4 &\approx \sqrt[4]{-[v_n^4 - 2(v_n^2)^2]} = v_n.
  \end{aligned}
\end{equation}
However, since these effects generally \emph{are} present, each flow cumulant will in general be different.

\label{loc:flow-cumulants-Qn}

Rather than evaluate the $k$-particle correlation functions via explicit nested loops over particle permutations---which may be feasible for two- or four-particle correlations, but quickly becomes unreasonable for six or eight---one typically uses $Q$-vectors, defined as \cite{Bilandzic:2010jr}
\begin{equation}
  Q_n = \sum_{i=1}^M e^{in\phi_i}.
  \label{eq:Qn}
\end{equation}
Each single-event correlation $\avg k$ can be analytically expressed in terms of $Q$-vectors, for example, the square of $Q_n$ is equivalent to a sum over pairs:
\begin{equation}
  |Q_n|^2 = \sum_{i,j=1}^M e^{in(\phi_i - \phi_j)} = M + \sum_{i \neq j}^M e^{in(\phi_i - \phi_j)},
\end{equation}
and comparing to equation \eqref{eq:corr-functions} immediately gives
\begin{equation}
  \avg 2 = \frac{|Q_n|^2 - M}{M(M - 1)}.
\end{equation}
A somewhat longer derivation yields \cite{Bilandzic:2010jr}
\begin{equation}
  \avg 4 = \frac{
    |Q_n|^4 + |Q_{2n}|^2 - 2\,\Re[Q_{2n}Q_n^*Q_n^*] - 4(M-2)\,|Q_n|^2 + 2M(M-3)
  }{
    M(M-1)(M-2)(M-3)
  }.
\end{equation}
Thus, all correlation functions can be evaluated with $\order M$ complexity instead of $\order{M^k}$.
The $Q$-vector method obviates the need to store lists of all particles for each event;
only the multiplicity $M$ and the $Q_n$ (a few complex numbers) are required.
It also provides several other benefits to experiments, such as dealing with nonuniform detector acceptance \cite{Bilandzic:2010jr,Bilandzic:2012wva,Bilandzic:2013kga}.
\begin{wrapfigure}[29]{l}[.20\textwidth]{.672\textwidth}
  \vspace{-5pt}
  \includegraphics[width=.672\textwidth]{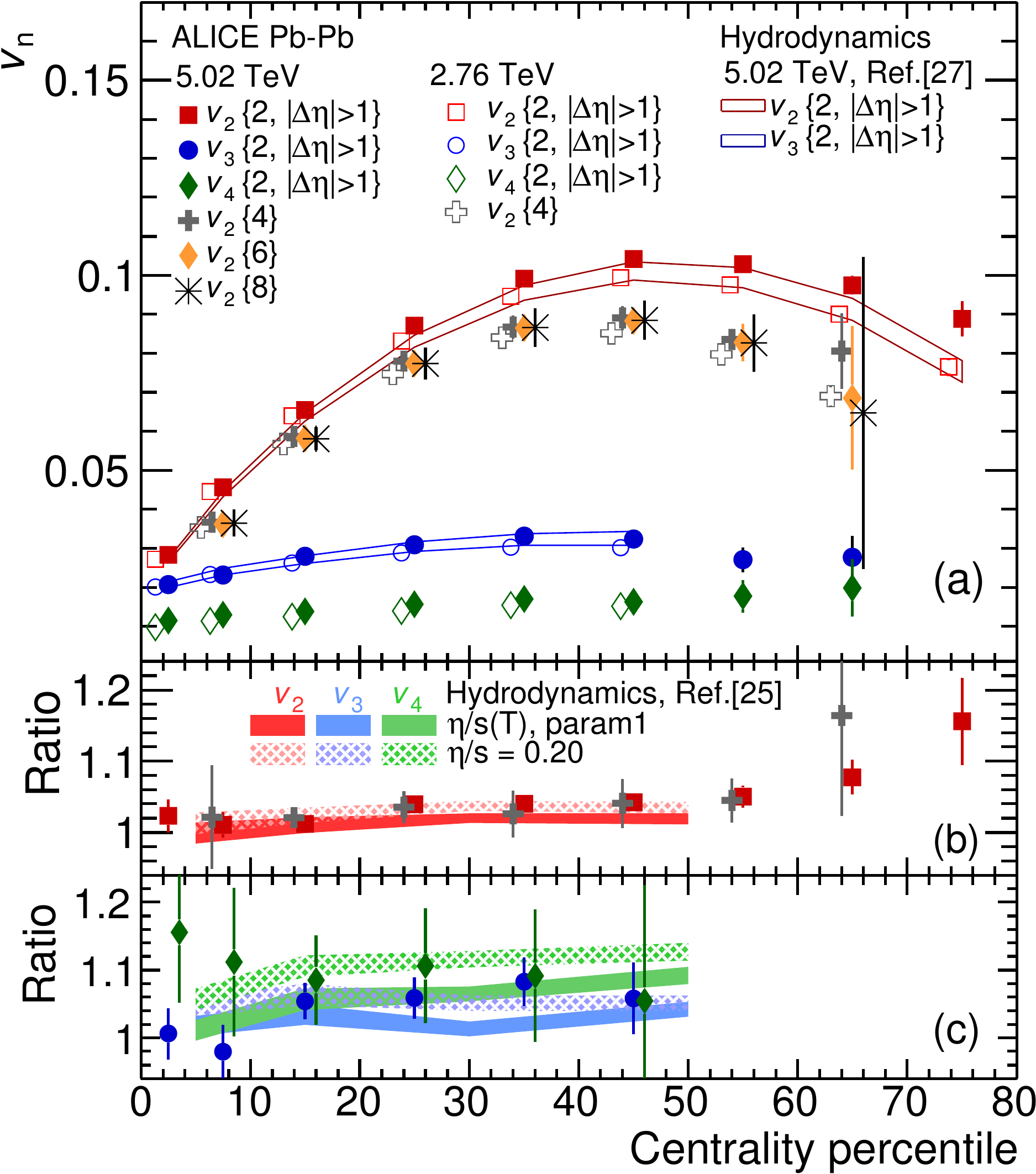}
  \caption{
    Integrated flow calculated with multiparticle cumulants as a function of centrality for Pb-Pb collisions at 2.76 and 5.02 TeV measured by ALICE \cite{ALICE:2011ab,Adam:2016izf}.
    Symbols are data as indicated in the legend; bands are predictions from a hydrodynamics model.
    The lower panels show the ratios of the two-particle cumulants $\vnk{n}{2,|\Delta\eta| > 1}$ between beam energies (symbols) and the corresponding model predictions of the ratios (bands) [$v_2$ in panel (b) and $v_3,v_4$ in panel (c)].
    References [25, 27] annotated in the figure are \cite{Niemi:2015voa,Noronha-Hostler:2015uye}.
  }
  \label{fig:vn-int}
\end{wrapfigure}

\subsubsection{Integrated flow}

The flow coefficients $v_n$ integrated over transverse momentum quantify the overall azimuthal anisotropy in a centrality class.

Figure \ref{fig:vn-int} shows the centrality dependence of integrated flow cumulants, calculated up to eight particles, for 2.76 and 5.02 collisions measured by ALICE \cite{ALICE:2011ab,Adam:2016izf}.
The notation $\vnk{n}{2, |\Delta\eta| > 1}$ means two-particle cumulants with a pseudorapidity gap, i.e.\ limited to pairs of particles separated by at least one unit of pseudorapidity.
This helps suppress nonflow, since azimuthal correlations caused by resonance decays, jets, etc tend to be short range in $\eta$.

Elliptic flow $v_2$ shows strong dependence on centrality due to the correlation with increasing impact parameter and initial-state anisotropy.
It increases until about 50\% centrality, above which it decreases, presumably because the QGP medium, while highly eccentric in these peripheral collisions, does not survive long enough for the flow to fully develop.
The hierarchy of the various cumulants is $\vnk22 > \vnk24 \approx \vnk26 \approx \vnk28$, implying that the two-particle cumulant contains some nonflow despite the $\eta$ gap, but the four-particle cumulant is sufficient to suppress this nonflow.

Meanwhile, triangular and quadrangular flow $v_3,v_4$ have much weaker centrality dependence since they are driven mostly by initial-state fluctuations.
In the most central bin, $v_2$ is much closer to $v_3,v_4$ since in this case, the impact parameter is small and the overlap roughly circular, so $v_2$ is also driven largely by fluctuations.

The bottom panels plot the ratios of the two-particle cumulants between 5.02 and 2.76 TeV (symbols).
In general, flow increases at the higher energy due to the hotter, longer-lived medium.
Elliptic flow increases slightly out to intermediate centrality and more significantly in peripheral bins.
The increase in $v_3$ is also slight, while $v_4$ is somewhat more pronounced (although the absolute increase is still small, but since the baseline is small the relative change is large).

\begin{wrapfigure}[23]{r}[.13\textwidth]{.57\textwidth}
  \vspace{-.5pt}
  \includegraphics[width=.57\textwidth]{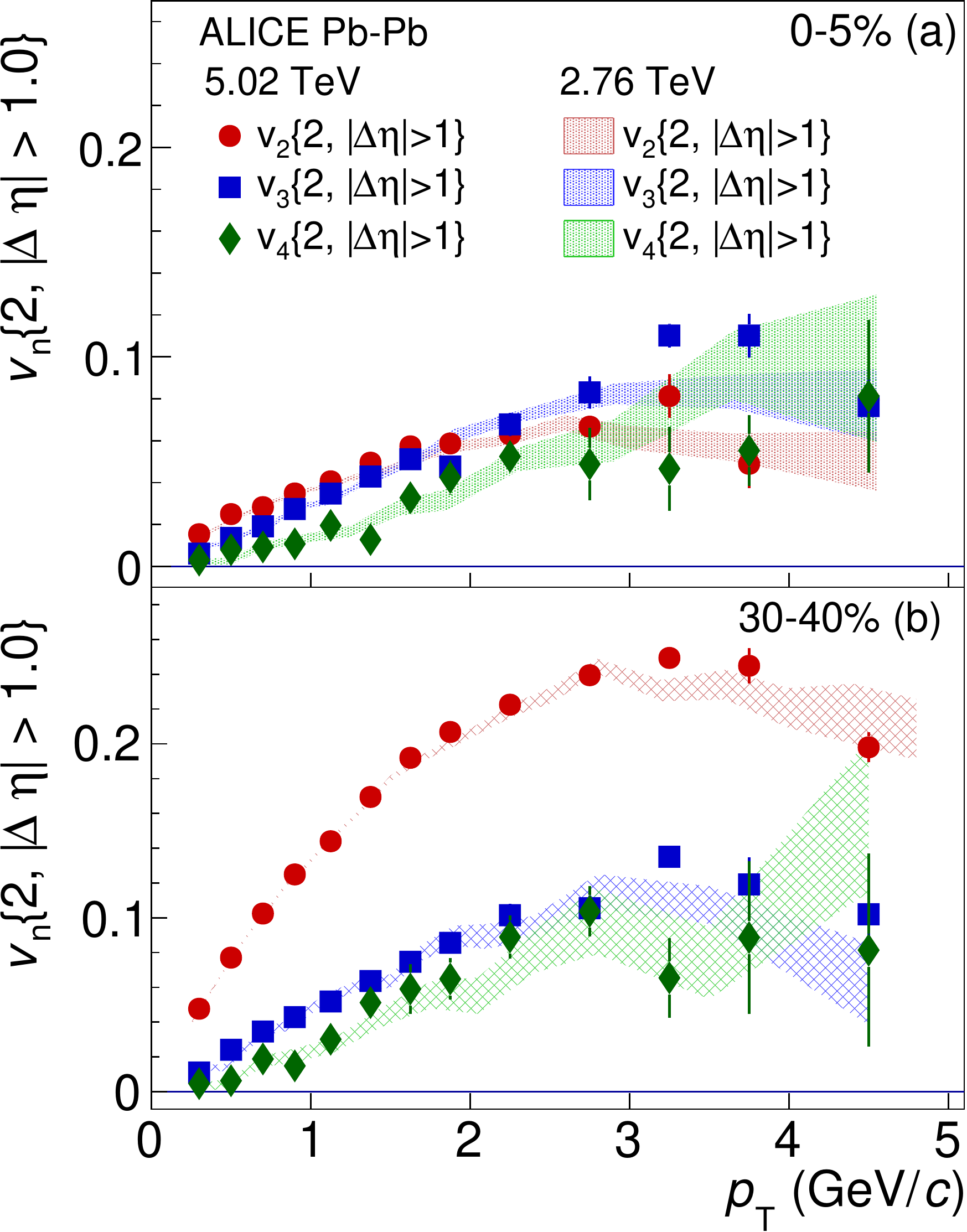}
  \caption{Differential two-particle flow cumulants in 0--5\% and 30--40\% centrality for Pb-Pb collisions at 2.76 and 5.02 TeV measured by ALICE \cite{Adam:2016izf}.}
  \label{fig:vn-diff}
\end{wrapfigure}

The figure includes hydrodynamic model predictions of the two-particle flow cumulants at 5.02 TeV and the ratios between beam energies \cite{Niemi:2015voa,Noronha-Hostler:2015uye}.
Overall, the model describes the data exceptionally well.

\subsubsection{Differential flow}

Flow coefficients may also be measured as a function of transverse momentum, $v_n(p_T)$, called differential flow.

Figure \ref{fig:vn-diff} shows differential flow cumulants for the two LHC beam energies \cite{Adam:2016izf}.
In central 0--5\% collisions, all measured harmonics have similar magnitude, with $v_3$ and $v_4$ becoming larger than $v_2$ at higher $p_T$.
However, integrated $v_2$ is still largest, since most particles reside in the low $p_T$ region where $v_2$ is slightly higher;
more precisely, the integrated flows are the integrals of these curves, weighted by the transverse momentum distribution.
In mid-central 30--40\% collisions, $v_2$ is much larger than $v_3$ and $v_4$ at all $p_T$.

There is little change in differential flow between the two beam energies, but as shown above, integrated flow increases slightly with energy.
This is because the mean transverse momentum is larger, so particles shift to higher regions of the differential flow curves.

\begin{figure}[h]
  \makebox[\textwidth]{
    \includegraphics[width=.6\textwidth]{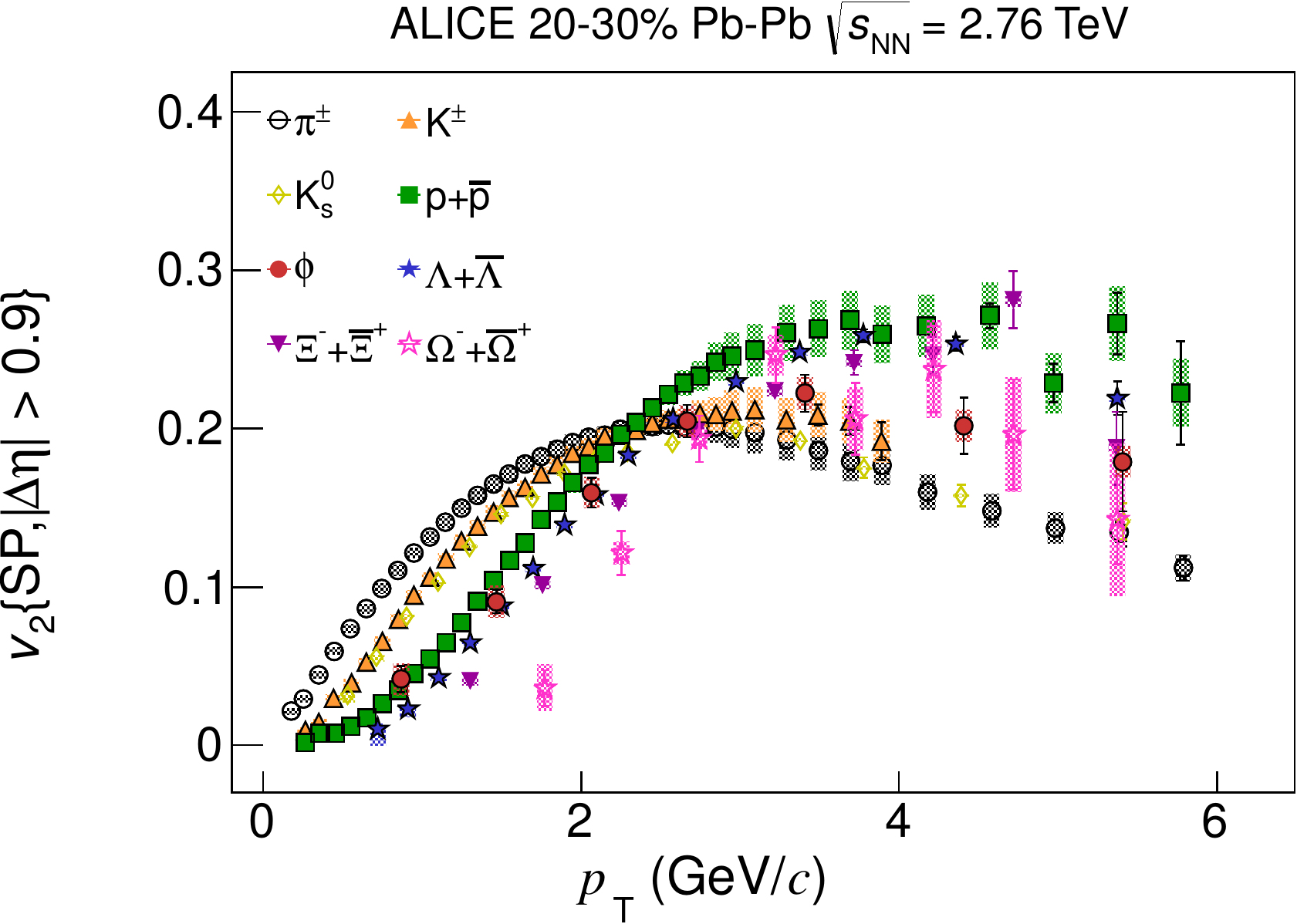}
    \quad
    \includegraphics[width=.6\textwidth]{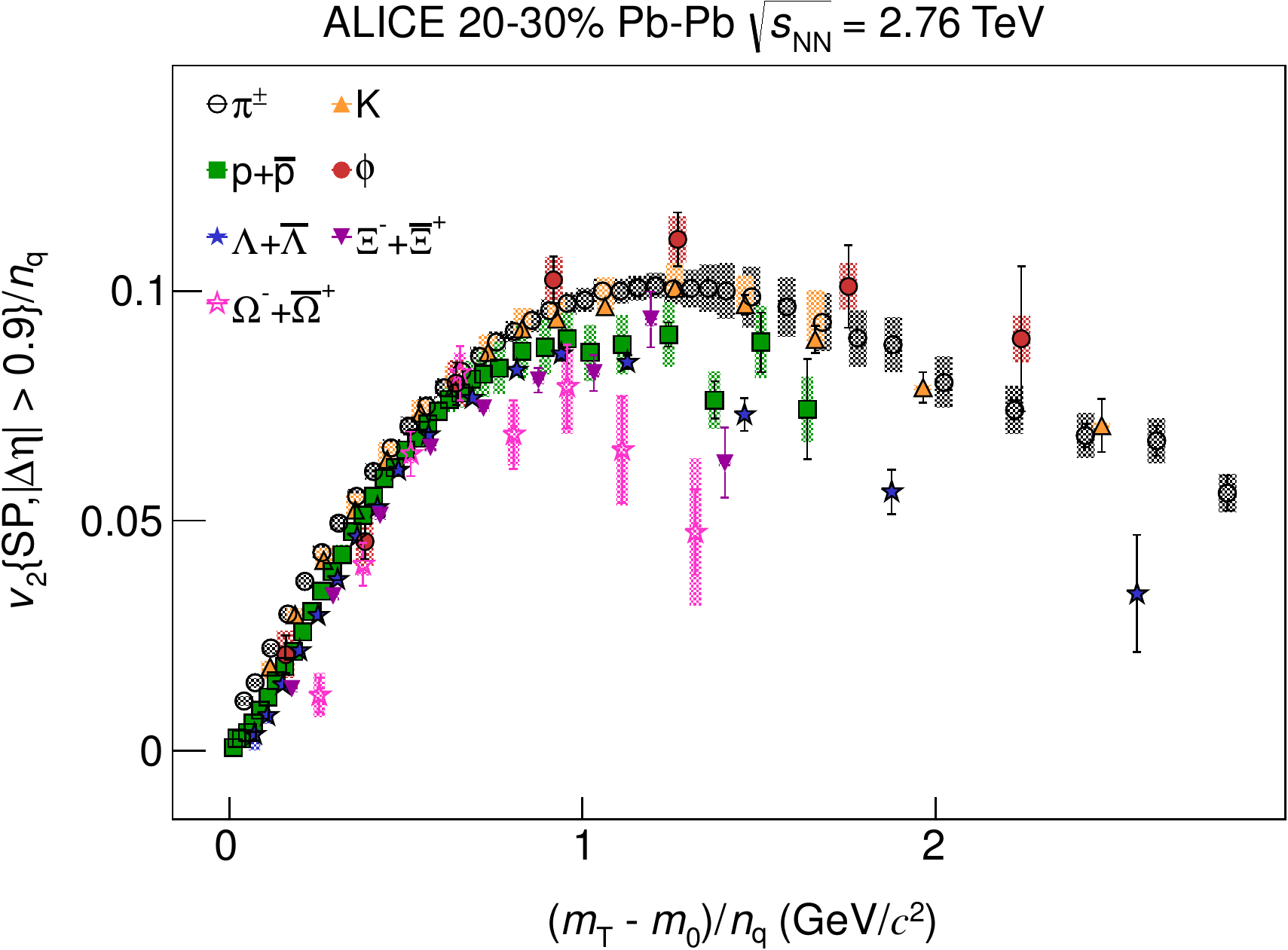}
  }
  \caption{
    Differential elliptic flow of identified hadrons in 20--30\% centrality for Pb-Pb collisions at 2.76 TeV measured by ALICE \cite{Abelev:2014pua}.
    Left: standard $p_T$-differential $v_2$.
    Right: $v_2$ as a function of the transverse kinetic energy $(m_T - m_0)$, both scaled by the number of constituent quarks $n_q$.
    (The underlying data are the same in both plots.)
  }
  \label{fig:vn-diff-id}
\end{figure}

The differential flow of identified particles \cite{Abelev:2014pua}, figure \ref{fig:vn-diff-id} left side, exhibits the characteristic ``mass splitting'':
Lighter particles (such as pions) have more flow at low $p_T$, while heavier particles (such as protons) have more flow at high $p_T$.
This occurs because all particles originate from the same expanding source and thus share a common average velocity, so heavier particles have higher $p_T$ (see figure \ref{fig:pT-spectra}), and consequently, the underlying anisotropic flow activates at higher $p_T$ for heavier species \cite{Huovinen:2001cy}.
The mass splitting also manifests in $v_3$ and $v_4$ \cite{Adam:2016nfo}.

The right-side plot is a test of quark deconfinement.
It shows the same data as on the left, but scaled by the number of constituent quarks, $n_q = 2$ for mesons and 3 for baryons, and as a function of the transverse kinetic energy per constituent, $(m_T  - m_0)/n_q$, where $m_T = \sqrt{m_0^2 + p_T^2}$ is the transverse mass.
The curves collapse much closer together at low $p_T$, signaling that collective flow develops partially when the medium consists of free quarks, which then coalesce into hadrons.
It is particularly compelling that the $\phi$ and proton, a meson and baryon with similar mass (1019 and 938 MeV), have comparable scaled flow.
Similar scaling behavior has been observed at RHIC \cite{Adams:2003am,Abelev:2007qg,Adler:2003kt,Adare:2006ti}, including some limitations \cite{Adare:2012vq}.
Clearly, the scaling is only approximate, but that it works at all is evidence of QGP formation.
There are numerous possible physical causes of the deviations, such as flow continuing to develop in the hadronic phase.

Although not shown here, hydrodynamic models do an overall excellent job of describing differential flow, including subtleties like the mass splitting \cite{Abelev:2014pua,Adam:2016nfo}.

\subsubsection{Other flow observables}

Besides the standard flow observables summarized here, a number of other flow-related quantities have been measured, including distributions of event-by-event flows \cite{Aad:2013xma}, correlations between flow harmonics \cite{ALICE:2016kpq,Acharya:2017gsw}, event-plane correlations \cite{Aad:2014fla}, the pseudorapidity dependence of flow \cite{Adam:2016ows}, and more.

\subsubsection{Small collision systems}

\begin{figure}[b!]
  \makebox[\textwidth]{
    \includegraphics[width=.55\textwidth]{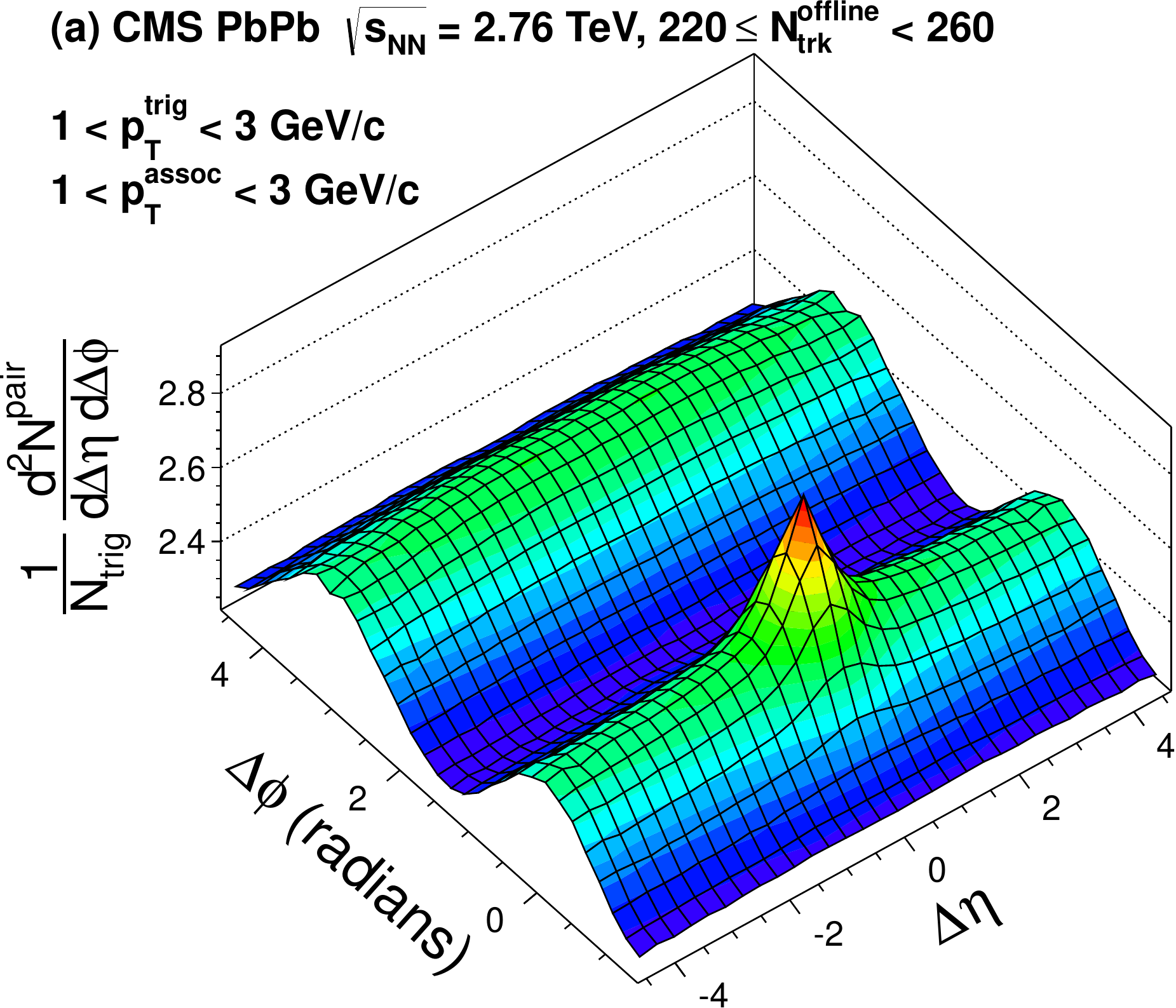}
    \enskip
    \includegraphics[width=.55\textwidth]{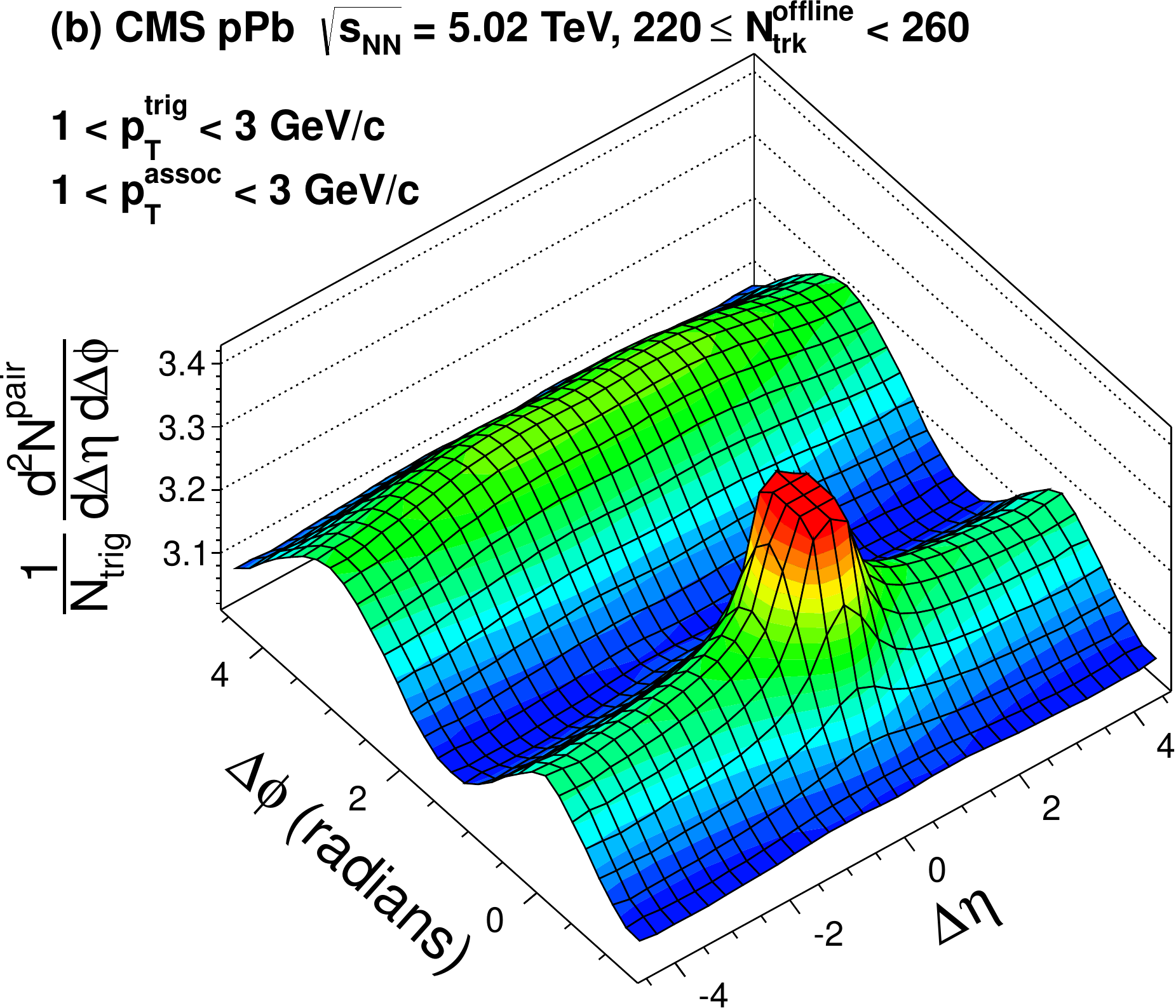}
  }
  \caption{Two-particle correlation functions for Pb-Pb (left) and p-Pb (right) collisions in the same multiplicity bin measured by CMS \cite{Chatrchyan:2013nka}.}
  \label{fig:corr-function-PbPb-pPb}
\end{figure}

Recent experimental results show unambiguous signatures of collective behavior in high-multiplicity events of small collision systems, such as proton-nucleus and even proton-proton.
Perhaps the clearest sign is the appearance of a long-range, near-side ridge in high-multiplicity bins.
Figure \ref{fig:corr-function-PbPb-pPb} compares two-particle correlation functions for Pb-Pb and p-Pb collisions in the same multiplicity bin, which corresponds to about 60\% centrality for Pb-Pb and ultra-central (at least top 1\%) for p-Pb \cite{Chatrchyan:2013nka}.
The two correlation functions are remarkably similar, with the near-side ridge clearly apparent (the ridge is not present in low-multiplicity p-Pb events \cite{CMS:2012qk}).
Other evidence includes nonzero cumulants $\vnk 2 k$ up to eight particles ($k = 8$) \cite{Khachatryan:2015waa,Khachatryan:2016txc} and similar $\vnk 3 2$ in p-Pb as Pb-Pb \cite{Chatrchyan:2013nka,Abelev:2014mda}.

It remains an open question whether the observed collective behavior originates from hydrodynamic flow, an initial state effect, or something else \cite{Dusling:2015gta,Schenke:2017bog,Nagle:2018nvi}.

\subsection{Hard processes}
\label{subsec:hard-processes}

\begin{figure}[b!]
  \graphicsandcaption{.72}{expt-data/RAA}{
    Nuclear modification factor $R_{AA}$ for charged particles in central (0--5\%) Pb-Pb collisions measured at $\sqrt s = 5.02$ TeV by CMS \cite{Khachatryan:2016odn} and at 2.76 TeV by CMS \cite{CMS:2012aa}, ALICE \cite{Abelev:2012hxa}, and ATLAS \cite{Aad:2015wga}.
  }
  \label{fig:RAA}
\end{figure}

The quantities discussed to this point are all bulk observables, meaning they describe the soft particles with $p_T \lesssim 3$ GeV which constitute the vast majority of particles produced in heavy-ion collisions.
However, some particles are produced with higher momentum, up to $\order{100\ \text{GeV}}$ (depending on the beam energy), by hard scatterings early in the collision evolution.
These high-$p_T$ particles then propagate through and interact with the hot and dense medium, thus, they serve as probes of the QGP.
Examples of hard probes include jets---collimated showers of high-$p_T$ hadrons---and heavy quarks (charm or bottom).

One of the simplest germane observables is the nuclear modification factor $R_{AA}$, which quantifies the modifications to transverse momentum distributions in nucleus-nucleus ($AA$) collisions relative to proton-proton ($pp$) collisions.
It is defined as
\begin{equation}
  R_{AA} = \frac{dN_{AA}/dp_T}{\avg{N_\text{coll}} \, dN_{pp}/dp_T},
\end{equation}
i.e.\ the ratio of the $AA$ spectrum to the $pp$ spectrum, scaled by the average number of binary nucleon-nucleon collisions $\avg{N_\text{coll}}$.
The denominator is a null hypothesis: the hypothetical spectrum if $AA$ collisions were simply a superposition of $pp$ collisions.
Thus, if $AA$ collisions did not produce a QGP medium, $R_{AA}$ would equal one.
As shown in figure \ref{fig:RAA}, $R_{AA}$ for charged particles is experimentally less than one out to very high $p_T$;
this ``suppression'' is taken as evidence of medium effects.

In this work, I focus on quantifying bulk properties of the QGP, so hard processes are not directly relevant.
But entire subfields of heavy-ion physics are devoted to theoretical and experimental study of various hard processes;
see, for example, recent reviews of jets \cite{Connors:2017ptx} and heavy quarks \cite{Andronic:2015wma}.

\section{Properties of hot and dense QCD matter}
\label{sec:properties}

Let us now turn our attention to the physical properties of hot and dense QCD matter---the quark-gluon plasma and the initial state that leads to its formation---the precise determination of which is a central goal of this work.
Many of the salient properties are defined in the context of viscous relativistic hydrodynamics, summarized below.

\subsection*{Viscous relativistic hydrodynamics}

The bulk dynamics of the QGP are well-described by viscous relativistic hydrodynamics, whose main equations of motion derive from conservation of energy and momentum:
\begin{equation}
  \partial_\mu T\mn = 0
  \label{eq:hydro}
\end{equation}
where
\begin{equation}
  T\mn = e \, u^\mu u^\nu - (P + \Pi)\Delta\mn + \pi\mn
\end{equation}
is the energy-momentum tensor;
$e$, $P$, and $u^\mu$ are the local energy density, pressure, and flow velocity, respectively, of the fluid, $\Delta\mn = g\mn - u^\mu u^\nu$ is the projector transverse to the flow velocity, $\pi\mn$ is the shear viscous pressure tensor, and $\Pi$ is the bulk viscous pressure.

An ideal (inviscid) fluid has five independent dynamical quantities:
the energy density, pressure, and three components of flow velocity.
Four of these are determined by the conservation equations \eqref{eq:hydro}, and the fifth by the equation of state $P(e)$.

The viscous pressures $\pi\mn$ and $\Pi$, which account for dissipative corrections to ideal hydrodynamics, introduce six additional independent quantities.
The shear tensor is traceless ($\pi^\mu_\mu = 0$) and orthogonal to the flow velocity ($\pi\mn u_\nu = 0$), so only five of its ten components are independent.
The bulk pressure, a scalar, effectively adds to the thermal pressure as $(P + \Pi)$ in $T\mn$.
In a simple relativistic generalization of Navier-Stokes theory, these terms connect to the fluid flow as \cite{Song:2007ux}
\begin{equation}
  \pi\mn = 2\eta\sigma\mn, \quad
  \Pi = -\zeta\theta,
\end{equation}
where $\eta$ and $\zeta$ are the shear and bulk viscosity, $\sigma\mn = \del^{\langle\mu} u^{\nu\rangle}$ is the velocity shear tensor, and $\theta = \del\cdot u$ is the expansion rate.
Notation: $\del^{\langle\mu} u^{\nu\rangle} = \frac{1}{2}(\del^\mu u^\nu + \del^\nu u^\mu) - \frac{1}{3}(\del\cdot u)\Delta\mn$, where $\del^\nu = \Delta\mn\partial_\nu$ is the gradient in the local rest frame.
However, the instantaneous connection between the fluid flow and viscous pressures leads to acausal signal propagation, so these equations do not suffice for relativistic hydrodynamics.
Israel and Stewart solved this problem \cite{Israel:1976tn,Israel:1979wp} with relaxation-type equations of the form
\begin{equation}
  \begin{aligned}
    \tau_\pi \dot\pi\mn + \pi\mn &= 2\eta\sigma\mn + \ldots, \\
    \tau_\Pi \dot\Pi + \Pi &= -\zeta\theta + \ldots,
  \end{aligned}
\end{equation}
where $\tau_\pi$ and $\tau_\Pi$ are timescales over which the viscous pressures relax to the Navier-Stokes limit.
A recent derivation from the relativistic Boltzmann equation yields \cite{Denicol:2010xn,Denicol:2012cn,Denicol:2014vaa}
\begin{equation}
  \begin{aligned}
    \tau_\pi\dot\pi^{\avg{\mu\nu}} + \pi\mn &=
      2\eta\sigma\mn
      + 2\pi_\alpha^{\langle\mu}\omega^{\nu\rangle\alpha}
      - \delta_{\pi\pi}\pi\mn\theta
      + \phi_7\pi_\alpha^{\langle\mu}\pi_{}^{\nu\rangle\alpha} \\
      &{}\qquad - \tau_{\pi\pi}\pi_\alpha^{\langle\mu}\sigma_{}^{\nu\rangle\alpha}
      + \lambda_{\pi\Pi}\Pi\sigma\mn
      + \phi_6\Pi\pi\mn, \\[1ex]
    \tau_\Pi \dot\Pi + \Pi &=
      -\zeta\theta
      - \delta_{\Pi\Pi}\Pi\theta
      + \phi_1\Pi^2
      + \lambda_{\Pi\pi}\pi\mn\sigma_{\mu\nu}
      + \phi_3\pi\mn\pi_{\mu\nu},
  \end{aligned}
  \label{eq:viscous}
\end{equation}
where $\pi^{\avg{\mu\nu}} = \Delta\mn_{\alpha\beta}\pi^{\alpha\beta}$, using the double projector $\Delta\mn_{\alpha\beta} = \frac{1}{2}(\Delta^\mu_\alpha\Delta^\nu_\beta + \Delta^\mu_\beta\Delta^\nu_\alpha - \frac{2}{3}\Delta\mn\Delta_{\alpha\beta})$, and $\omega^{\lambda\rho} = \frac{1}{2}(\del^\lambda u^\rho - \del^\rho u^\lambda)$ is the vorticity tensor.
These equations include all terms up to second order in the viscous pressures as well as coupling terms between shear and bulk.
The quantities multiplying each term are known as \emph{transport coefficients};
the shear and bulk viscosity, $\eta$ and $\zeta$, are the first-order transport coefficients.

\parbreakboldstart{Nothing here} is unique to the QGP---or any other fluid.
Within a hydrodynamic description, all that distinguish any given fluid are its transport coefficients and equation of state.
In the remainder of this section, I discuss these quantities and how they relate to the QGP.

\subsection{Transport coefficients}

Broadly, transport coefficients characterize the dynamical properties of a fluid, such as its response to external forces.
They are in general functions of temperature and, in the case of QGP, chemical potential.

\subsubsection{Viscosity}
\label{sec:viscosity}

Shear viscosity $\eta$ measures a fluid's resistance to shear strain.
A low-viscosity fluid is generally strongly-interacting, efficiently transmits shear strain through itself, and its constituents have a short mean free path;
on the other hand a nearly-ideal (weakly-interacting) gas has large viscosity because its constituents do not scatter enough to convey the information that a strain is being applied.

The ``quality'' of a fluid is quantified by its specific shear viscosity, the dimensionless%
\footnote{
  Dimensionless in natural units with $\hbar = k_B = 1$.
  To convert to SI units, multiply by $\hbar/k_B \simeq 7.6 \times 10^{-12}$~K~s.
}
ratio to the entropy density, $\eta/s$.
The entropy density $s$ is a proxy for the number density, so $\eta/s$ is in a sense the viscosity per unit (an intensive quantity).
The QGP specific shear viscosity is of particular interest since it is believed to be small---nearly zero---meaning that the QGP is nearly a ``perfect'' fluid.
The measurement of the temperature-dependent specific shear viscosity $(\eta/s)(T)$ is a primary goal of heavy-ion physics \cite{Bass:2012wp}.

It has been famously conjectured, based on a string theory calculation applicable to a wide range of strongly-interacting quantum field theories, that the minimum possible specific shear viscosity is $\eta/s \geq 1/4\pi \simeq 0.08$ \cite{Kovtun:2004de}.
Remarkably, a number of studies using viscous relativistic hydrodynamics, e.g.\ \cite{Romatschke:2007mq,Song:2010mg,Schenke:2010rr,Niemi:2015qia}, have found an approximate range for the QGP $\eta/s$ of $1/4\pi\text{--}2.5/4\pi \simeq 0.08\text{--}0.20$.
These results of course do not confirm the conjecture---the uncertainty is quite large, and even if the bound is correct, it would apply only to highly idealized systems, so the actual measured QGP value wouldn't necessary be exactly $1/4\pi$.
Nothing special happens to hydrodynamics when $\eta/s$ drops below $1/4\pi$.

\begin{figure}[t]
  \includegraphics{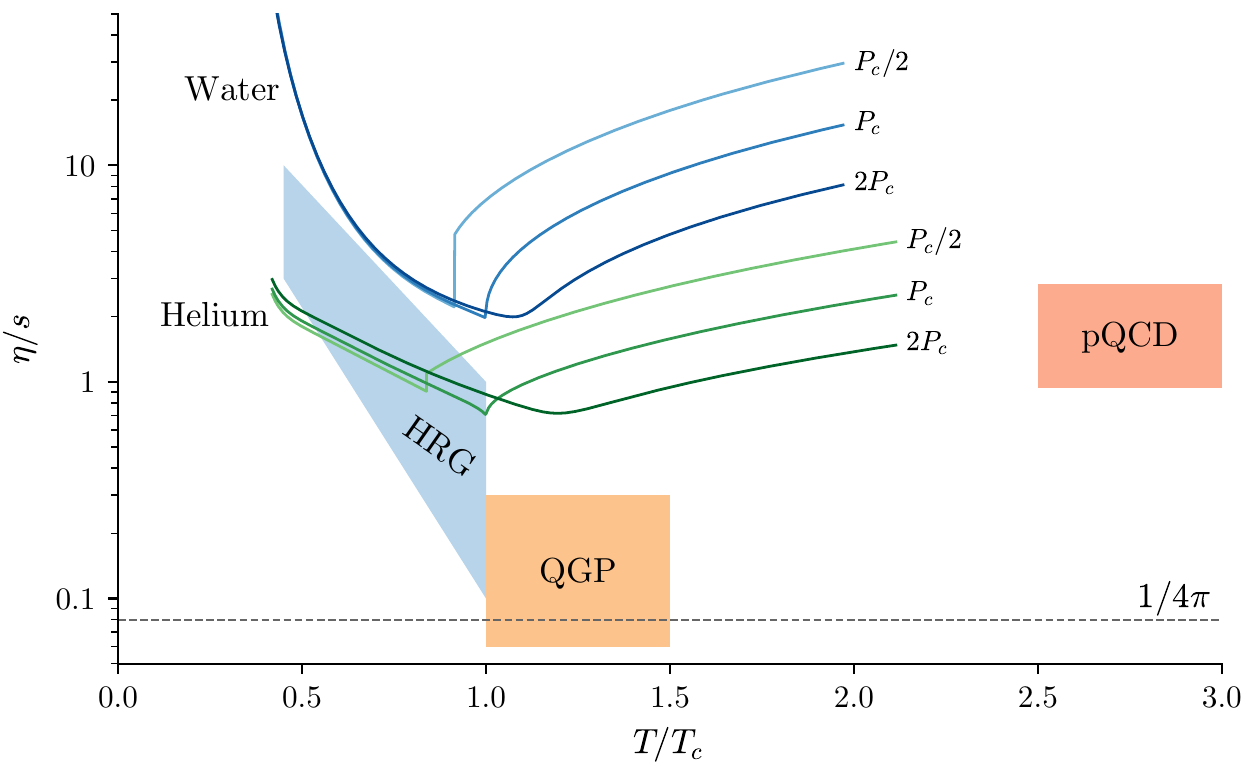}
  \caption{
    Specific shear viscosity $\eta/s$ of different fluids as a function of temperature relative to each fluid's critical temperature $T_c$.
    The colored lines represent common fluids, water and helium, at various pressures relative to their critical pressures $P_c$, as annotated.
    These curves were computed from NIST data \cite{NIST:fluid} with the entropy standardized so that it is zero at zero temperature, $S(T=0) = 0$, using standard-state thermochemistry data \cite{NIST:thermo}.
    The hadron resonance gas (HRG) area is based on a recent study \cite{Rose:2017bjz}, and the perturbative QCD (pQCD) area on a parametrization in the high-temperature limit \cite{Csernai:2006zz}.
    The QGP area is motivated by numerous studies, e.g.\ \cite{Romatschke:2007mq,Song:2010mg,Schenke:2010rr,Niemi:2015qia}, and results of this work that will be presented in chapter \ref{ch:quant-qcd-props}.
    The locations and shapes of all areas are approximate.
    The dashed line denotes the conjectured bound $1/4\pi$ \cite{Kovtun:2004de}.
  }
  \label{fig:specific-shear-viscosity}
\end{figure}

What is clear, however, is that the QGP is much closer to perfection than most ordinary fluids.
Figure \ref{fig:specific-shear-viscosity} compares our knowledge of the QGP $\eta/s$ to common fluids, whose properties have been measured and are tabulated by NIST \cite{NIST:fluid,NIST:thermo}.
The QGP $\eta/s$ is about an order of magnitude smaller than those of water and helium, which are $\order 1$ near their critical temperatures.
From the NIST data, we see that $\eta/s$ generally reaches a minimum near $T_c$, either as a continuous curve, a cusp, or a discontinuous jump, depending on whether the pressure is above, equal to, or below the critical pressure, respectively.
A similar functional form likely manifests for QCD matter:
Below the QCD transition temperature, various calculations with hadron resonance gas (HRG) models point to $\eta/s$ decreasing with temperature \cite{Rose:2017bjz};
in the high-temperature limit, where perturbative QCD (pQCD) is applicable, calculations show an increasing function of temperature \cite{Csernai:2006zz}.
Near $T_c$, neither the HRG model or pQCD is reliable, so we must rely on comparisons of hydrodynamic model calculations to data.

\begin{figure}[t]
  \makebox[\textwidth]{\includegraphics{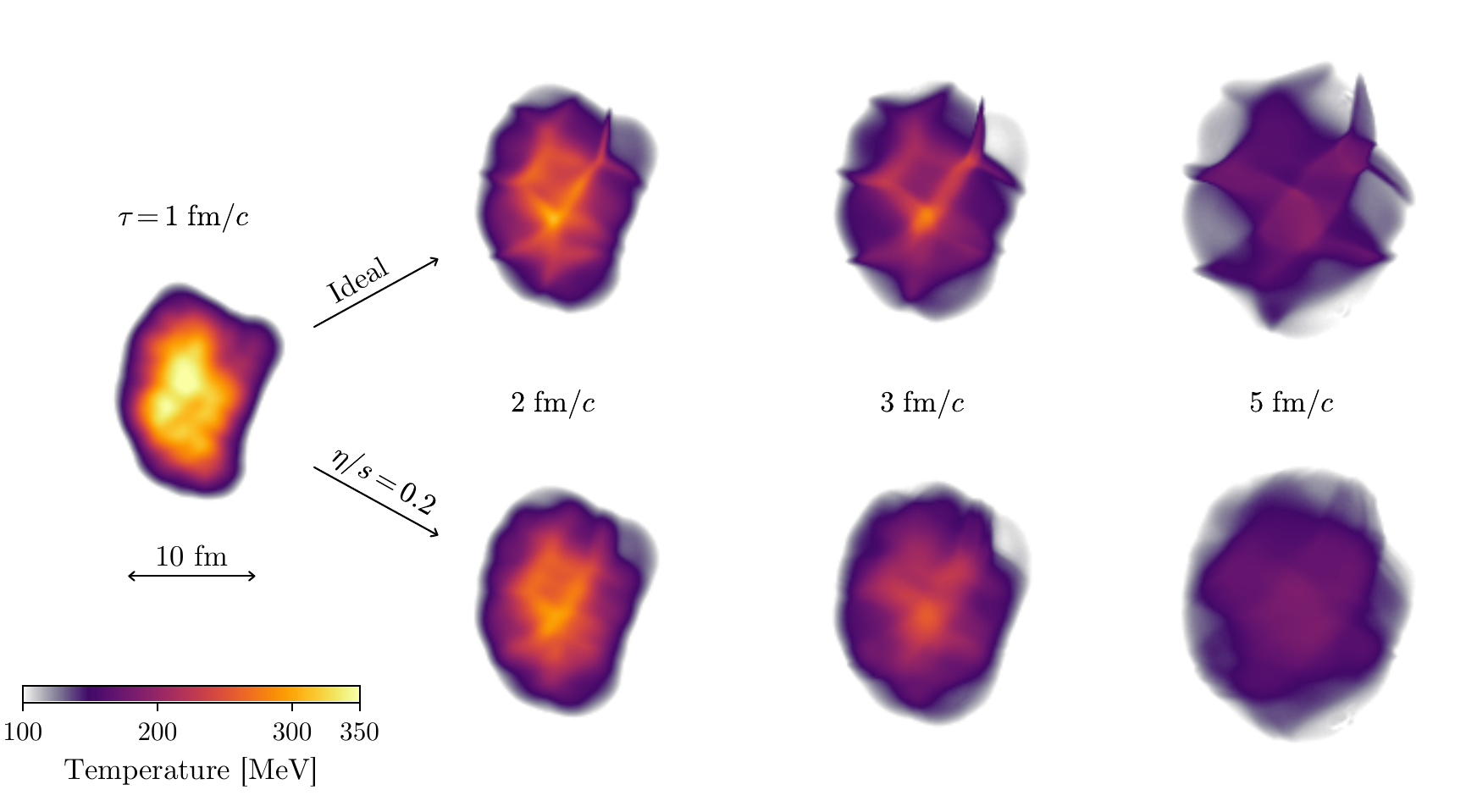}}
  \caption{
    Comparison of ideal and viscous hydrodynamics applied to a fluctuating intermediate centrality event.
    Shown is the time evolution, starting at proper time $\tau = 1\ \text{fm}/c$, of the temperature profile at midrapidity with $\eta/s = 0$ (top) and 0.2 (bottom).
  }
  \label{fig:hydro-evolution}
\end{figure}

How does shear viscosity impact hydrodynamics, and what is the connection to experimental observations of heavy-ion collisions?
Most apparent is the effect on collectivity: increasing $\eta/s$ reduces collective behavior and flow.
As shown in figure \ref{fig:hydro-evolution}, shear viscosity washes out small-scale structures and induces a more isotropic system, which would have smaller anisotropic flow coefficients $v_n$.
Thus, flow coefficients are the primary QGP viscometer.
The referenced studies \cite{Romatschke:2007mq,Song:2010mg,Schenke:2010rr,Niemi:2015qia} estimated $\eta/s$ by running hydrodynamic calculations with different values of $\eta/s$ and comparing the resulting $v_n$ to corresponding experimental data.

\parbreakboldstart{Bulk viscosity} $\zeta$ is related to the fluid expansion rate, as can be seen by the way it enters the evolution equation \eqref{eq:viscous}.
It influences QGP evolution primarily by suppressing the radial expansion rate, translating into a reduction in the transverse momentum of emitted particles (however, bulk viscosity makes no qualitative difference in the appearance of the hydrodynamic medium, which is why it isn't represented in figure \ref{fig:hydro-evolution}).
Recently, it was shown that a nonzero specific bulk viscosity $\zeta/s$ is necessary for hydrodynamic models to simultaneously describe mean transverse momentum and flow \cite{Ryu:2015vwa}, and other phenomenological studies have demonstrated that $\zeta/s$ modifies the transverse momentum spectra and, to a lesser extent, collective flow \cite{Bozek:2009dw,Song:2009rh,Dusling:2011fd,Noronha-Hostler:2013gga}.

As for the temperature dependence, $(\zeta/s)(T)$ is not well-known but is generally expected to peak near the QCD transition temperature and fall off on either side, based on calculations below $T_c$ \cite{NoronhaHostler:2008ju}, near \cite{Karsch:2007jc,Kharzeev:2007wb}, and above \cite{Arnold:2006fz}.
This picture is consistent with an approximate result from kinetic theory \cite{Dusling:2011fd}, $\zeta/\eta \approx 15(1/3 - c_s^2)^2$, where $c_s$ is the speed of sound.
In the high-temperature limit $c_s^2 \approx 1/3$, so $\zeta \ll \eta$, and near $T_c$ it is smaller, $c_s^2 \sim 0.15$--0.20, which gives $\eta/\zeta \sim 0.25$--0.50.
Most of the referenced phenomenological studies have used a peak value of $\zeta/s \sim 0.01$--0.05, the notable exception being \cite{Ryu:2015vwa}, whose parametrization peaked at $\sim$0.35.
Care must be taken with such large bulk viscosity, since $\zeta/s \gtrsim 0.1$ near $T_c$ can induce negative-pressure bubbles in the hydrodynamic medium (``cavitation'') \cite{Habich:2014tpa}.

\subsubsection{Other hydrodynamic coefficients}

Besides the first-order transport coefficients $\eta$ and $\zeta$, the viscous evolution equations \eqref{eq:viscous} contain several second-order coefficients, coupling coefficients, and relaxation times.
The second-order and coupling coefficients have been computed in the limit of small masses \cite{Denicol:2014vaa}, and it is reasonable to expect that varying them would not have a strong impact on hydrodynamic evolution.
The shear and bulk relaxation times $\tau_\pi$ and $\tau_\Pi$ are important for causal viscous relativistic hydrodynamics, but empirically, their specific values do not have much impact on physical observables \cite{Song:2008si,Bernhard:2015hxa} [and section \ref{sec:v1}].

\subsection{Equation of state}
\label{subsec:eos}

An equation of state (EoS) interrelates a system's various thermodynamic quantities: temperature, energy density, pressure, etc.
From a fluid dynamical perspective, an EoS $P = P(e)$ is required to close the system of conservation equations \eqref{eq:hydro}.

The QCD EoS has been computed numerically using modern lattice QCD techniques.
These calculations are complex and extremely computationally expensive (some of the largest NERSC computational allocations are given to lattice QCD groups), and lattice QCD is an entire field unto itself.
For a recent review of lattice techniques emphasizing applications to heavy-ion collisions see reference \cite{Soltz:2015ula}.

Lattice calculations begin by evaluating the QCD partition function $Z$ on a hypercubic spacetime lattice of size $N_\sigma^3 N_\tau$, where $N_\sigma$ and $N_\tau$ are the number of spatial and temporal steps.
Lattice sites are separated by lattice spacing $a$, which relates to the temperature and volume by
$T = 1/(a N_\tau)$ and $V = (a N_\sigma)^3$.
The calculation is repeated with different lattice sizes, usually with a fixed ratio of spatial and temporal steps $N_\sigma/N_\tau$, and the results are extrapolated to the continuum and thermodynamic  limits: $a \rightarrow 0$, $N_\tau \rightarrow \infty$, $V \rightarrow \infty$.

Constructing the EoS then hinges on the trace of the energy-momentum tensor, or trace anomaly,  $\Theta^{\mu\mu} = e - 3P$, from which all other thermodynamic quantities can be computed.
The trace anomaly is defined on the lattice by
\begin{equation}
  \Theta^{\mu\mu} = -\frac{T}{V}\frac{d \log Z}{d \log a}
\end{equation}
and related to the pressure as \cite{Cheng:2007jq}
\begin{equation}
  \frac{\Theta^{\mu\mu}}{T^4} = \frac{e - 3P}{T^4} = T \frac{d}{dT} \biggl( \frac{P}{T^4} \biggr),
\end{equation}
which, after integration, furnishes the pressure explicitly:\footnote{
  In natural units with $\hbar = c = 1$, pressure and energy density have the same units as temperature to the fourth power, so e.g.\ $P/T^4$ is dimensionless.
  To convert from units of fm$^{-4}$ to the more intuitive GeV/fm$^3$, multiply by $\hbar c \simeq 0.197$ GeV fm.
}
\begin{equation}
  \frac{P(T)}{T^4} = \frac{P_0}{T_0^4} + \int_{T_0}^T dT' \frac{\Theta^{\mu\mu}}{T'^5},
\end{equation}
where $P_0$ is the pressure at reference temperature $T_0$.
This reference point is usually computed using a hadron resonance gas (HRG) model, which sums the contributions of all hadrons and resonances as noninteracting particles, at a temperature below the QCD transition.
With the trace anomaly and pressure in hand, the energy density and entropy density $s = (e + P)/T$ follow immediately.

Below, thermodynamic quantities from the HotQCD Collaboration's recent lattice calculation of the EoS at zero net baryon density in ($2+1$)-flavor QCD, that is, with two light quarks (of equal mass) and a heavier strange quark \cite{Bazavov:2014pvz}.
\begin{figure}[h]
  \graphicsandcaption{.75}{etc/eos_hotqcd}{
    The QCD equation of state calculated by the HotQCD Collaboration \cite{Bazavov:2014pvz}.
    The colored bands show the normalized pressure, energy density, and entropy density as a function of temperature, where the width of the bands represents the uncertainty;
    the solid lines show the corresponding quantities from the HRG model.
    The vertical band denotes the crossover region, $T_c = 154 \pm 9$ MeV.
    The dashed horizontal line indicates the ideal gas (non-interacting) limit for the energy density, $e/T^4 = 95\pi^2/60$.
  }
\end{figure} \\
These calculations establish a crossover deconfinement transition region $T_c = 154 \pm 9$ MeV.
Unlike other classes of phase transitions, namely first- and second-order, a crossover transition lacks any discontinuities or divergences in thermodynamic quantities, their derivatives, or associated order parameters.
In the QCD EoS, there is a rapid rise in the specific heat, $C_V = \partial e/\partial T$, near $T_c$ but no peak or discontinuity.

The Wuppertal-Budapest Collaboration performed an earlier, independent lattice calculation of the EoS \cite{Borsanyi:2013bia}, with which current HotQCD results are consistent within systematic errors, despite some differences in the methodology.

Aside: The level of agreement between the HRG and lattice equations of state in the low-temperature region, given that they derive from entirely different physical considerations, is remarkable.

\begin{figure}[b!]
  \graphicsandcaption{.9}{etc/eos_prior_posterior}{
    Left: Randomly sampled, unconstrained equations of state represented by the squared speed of sound, $c_s^2 = \partial P/\partial e$, as a function of temperature.
    Right: Equations of state sampled from the posterior distribution, constrained by data.
    The thick red lines represent the range of lattice calculations and the green line is the HRG EoS, to which all samples connect at 165 MeV.
    Figure from \cite{Pratt:2015zsa}.
  }
  \label{fig:eos-prior-posterior}
\end{figure}

As properties of QCD matter go, the EoS stands out from most others:
While the aforementioned transport coefficients and upcoming initial state properties are highly uncertain and must be estimated through phenomenological models, the EoS is all but solved (at least at zero $\mu_B$) thanks to modern lattice techniques.
Indeed, I will not estimate the EoS in this work, but rather use the lattice results directly in a hydrodynamic model.
But it is reasonable to doubt that the lattice EoS, which is calculated in the infinite-time and infinite-volume limits, applies to the extremely transient and tiny QGP created in heavy-ion collisions.
A recent study \cite{Pratt:2015zsa} tested this as part of a Bayesian model-to-data comparison using a hydrodynamic model, RHIC and LHC data, and similar parameter estimation methods as in chapter \ref{ch:param-est}.
They simultaneously varied 14 parameters, including initial state properties, transport coefficients, and two which control the EoS, and found that the constrained EoS is indeed consistent with lattice, as shown in figure \ref{fig:eos-prior-posterior}.
Although the uncertainty on the constrained EoS is significant, some of that comes from folding in the uncertainty of all the other parameters, and the fact that the data prefer an EoS comparable to lattice is compelling evidence that the QGP truly has similar properties as the idealized QCD matter of lattice calculations.

\subsection{The initial state of heavy-ion collisions}
\label{subsec:properties-initial-state}

Hydrodynamics describes the spacetime evolution of the QGP medium, but not how it forms.
At the very least, some other physical process(es) must account for the initial energy density immediately after the collision, and, assuming the system takes a short time $\tau \sim 1\ \fmc$ to begin behaving hydrodynamically, another dynamical process must describe this early stage.
This also makes sense mathematically:
The hydrodynamic equations are differential equations and therefore require an initial condition.

But the initial state, as the earliest stage of heavy-ion collisions, is created under the most extreme conditions and is the furthest from what we observe.
Besides hindering the fundamental goals of characterizing and modeling the initial state itself, this has far reaching consequences for quantifying the transport properties of the QGP.
For example, one of the most well-known studies estimating $\eta/s$ \cite{Song:2010mg} found $(\eta/s)/(4\pi) \approx 1\text{--}2.5$, ascribing most of the uncertainty range to competing models of the initial state.

Why would the initial state affect estimates of medium properties like $\eta/s$?
One crucial way is through the anisotropy of the initial energy density.
Suppose model A tends to produce more anisotropic initial states than model B, then it will generally lead to larger anisotropic flow coefficients $v_n$, and if the goal is to match an experimental measurement of $v_n$, model A will require a larger $\eta/s$ to suppress its anisotropy (in fact, this is roughly what happened in the study just mentioned).
The precise initialization of the other dynamical quantities, namely the fluid flow velocity and viscous pressures, will also, in general, impact the QGP medium evolution and final state.

To rigorously estimate QGP properties, we must therefore also consider the variability and uncertainty in the initial state.
These degrees of freedom broadly fall into two categories:
characteristics of energy and/or entropy deposition immediately after the collision, and the subsequent dynamical properties of the system prior to QGP formation.

\subsubsection{Energy and entropy deposition}

Let us break this down even further:
How is nuclear density distributed in nuclei, and given a pair of colliding nuclei with known density distributions, what energy (or entropy) is deposited?

The problem can be factorized like this because ultra-relativistic collisions occur on such a short timescale.
As a rule of thumb, since the mass of a nucleon is about 1 GeV and the total energy of a massive particle is $E = \gamma m$, the Lorentz factor for a relativistic nucleon is approximately its kinetic energy in GeV.
At the LHC, where the beam energy is $\sqrt s = 2.76$ or 5.02 TeV per nucleon pair, this translates to Lorentz factors $\gamma > 1000$.
This means nuclei are longitudinally length-contracted so much as to be essentially flat, the collision occurs instantaneously, and the nuclear densities may be considered frozen for purposes of modeling the collision.

The radial density of a spherical heavy nucleus is typically parametrized by a Fermi (or Woods-Saxon) distribution
\begin{equation}
  \rho(r) \propto \frac{1}{1 + \exp\bigl( \frac{r-R}{a} \bigr)},
\end{equation}
which is basically a sphere with a blurry edge, where the nuclear radius $R$ and surface thickness (or skin depth) $a$ are measured for many nuclei, e.g.\ for $^{208}$Pb, $R = 6.62$ fm, $a = 0.546$ fm \cite{DeJager:1987qc}.
In Monte Carlo models, nucleon positions are randomly sampled, with the radii sampled from the full radial probability $P(r) \propto r^2 \rho(r)$ and the angles sampled isotropically (assuming a spherical nucleus).

This already raises a question:
How are the nucleon positions correlated?
It is reasonable to assume that nucleons cannot occupy the same spatial location, so one simple way to insert correlations is to impose a minimum distance between nucleons.
When ALICE estimates initial state properties for centrality bins, they vary the minimum distance from 0 to 0.8 fm as part of the systematic uncertainty \cite{Abelev:2013qoq}.
More realistic nucleon-nucleon correlations have also been implemented \cite{Alvioli:2009ab} and shown to influence anisotropic flow in ultra-central collisions \cite{Denicol:2014ywa}.

\begin{figure}[t]
  \includegraphics{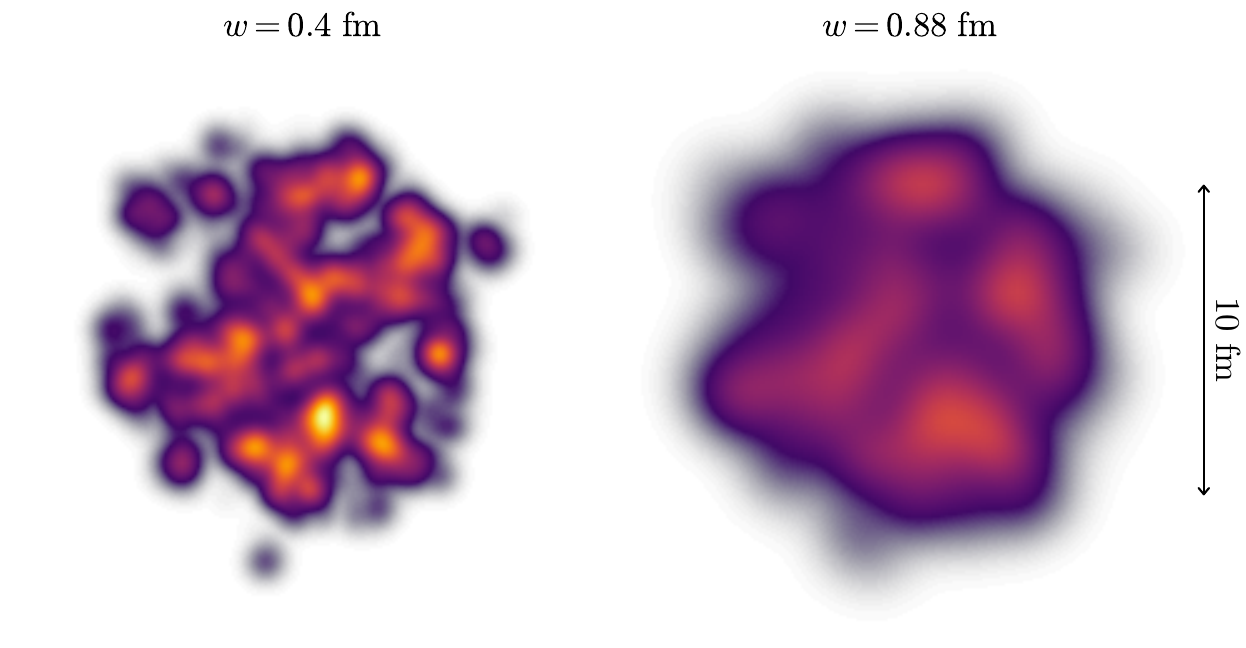}
  \caption{
    Fluctuating initial conditions with the Gaussian nucleon width $w$ set to the gluon radius (0.4 fm, left) and the proton charge radius (0.88 fm, right).
    The nucleon positions are the same in both cases; only the size is different.
  }
  \label{fig:nucleon-width}
\end{figure}

Another unknown degree of freedom is the effective size and shape of nucleons---``effective'' because although some physical properties of nucleons are independently measured, they don't necessarily directly connect to energy deposition in ultra-relativistic collisions.
For example the proton root-mean-square charge radius is 0.88 fm \cite{Mohr:2015ccw}, while the gluon radius is much smaller, approximately 0.4 fm \cite{Kowalski:2003hm}, and either (or neither) could be relevant in this context.
The impact of the nucleon size on the overall initial condition can hardly be overstated, as shown in figure \ref{fig:nucleon-width}.
Smaller nucleons create more compact structures with higher peak temperatures and steeper gradients, driving increased anisotropic flow and radial expansion, which has ramifications for estimating shear and bulk viscosity.
In general, initial state and medium properties are intrinsically entangled, and it is important to estimate them simultaneously while propagating all sources of uncertainty.

As for the nucleon shape, the transverse profile is often chosen to be a Gaussian:
\begin{equation}
  T_\text{nucleon}(x, y) = \int dz \, \rho_\text{nucleon}(x, y, z) = \frac{1}{2\pi w^2} \exp\biggl( -\frac{x^2 + y^2}{2w^2} \biggr),
  \label{eq:Tnucleon}
\end{equation}
where $T_\text{nucleon}$ is the beam-integrated density, or thickness, and $w$ is the effective nucleon width.
Although unproven, the Gaussian shape is computationally convenient and satisfies reasonable physical limits.
More empirical profiles are possible, for example one could use the charge distribution from the measured electric form factor \cite{Perdrisat:2006hj}, with the caveat that charge density may not directly translate to energy deposition in relativistic collisions.

Thickness is the relevant quantity at sufficiently high energy because of the aforementioned Lorentz contraction and instantaneity of the collision.
The full nuclear thickness of the projectile nuclei shall be denoted $T_A$ and $T_B$, where $T_A$ is the sum of $T_\text{nucleon}$ for all nucleons in nucleus $A$.

Now, restating the question posed earlier:
Given $T_A$ and $T_B$, what is the resulting transverse energy density $e(x, y)$ [or entropy density $s(x, y)$] at midrapidity?
(In terms of initializing hydrodynamics, either energy or entropy density is acceptable, since they can be interconverted via the equation of state.
For the remainder of this section I will only say energy for brevity.)

A simple model of energy deposition is the Glauber model \cite{Miller:2007ri,Loizides:2014vua}, which in its Monte Carlo formulation, deposits energy for each participant nucleon (sometimes called wounded nucleons) and each binary nucleon-nucleon collision, with the fraction of energy apportioned to binary collisions controlled by a phenomenological parameter $\alpha$.
Despite its basis on mostly geometrical arguments, the Glauber model has semi-quantitatively fit a variety of experimental measurements and is the de facto standard for unfolding centrality bins.

\begin{figure}[h]
  \medskip
  \makebox[\textwidth]{
    \small
    \begin{tabular}{ccc}
      MC-Glauber & MC-KLN & IP-Glasma \\
      \includegraphics[width=.45\textwidth]{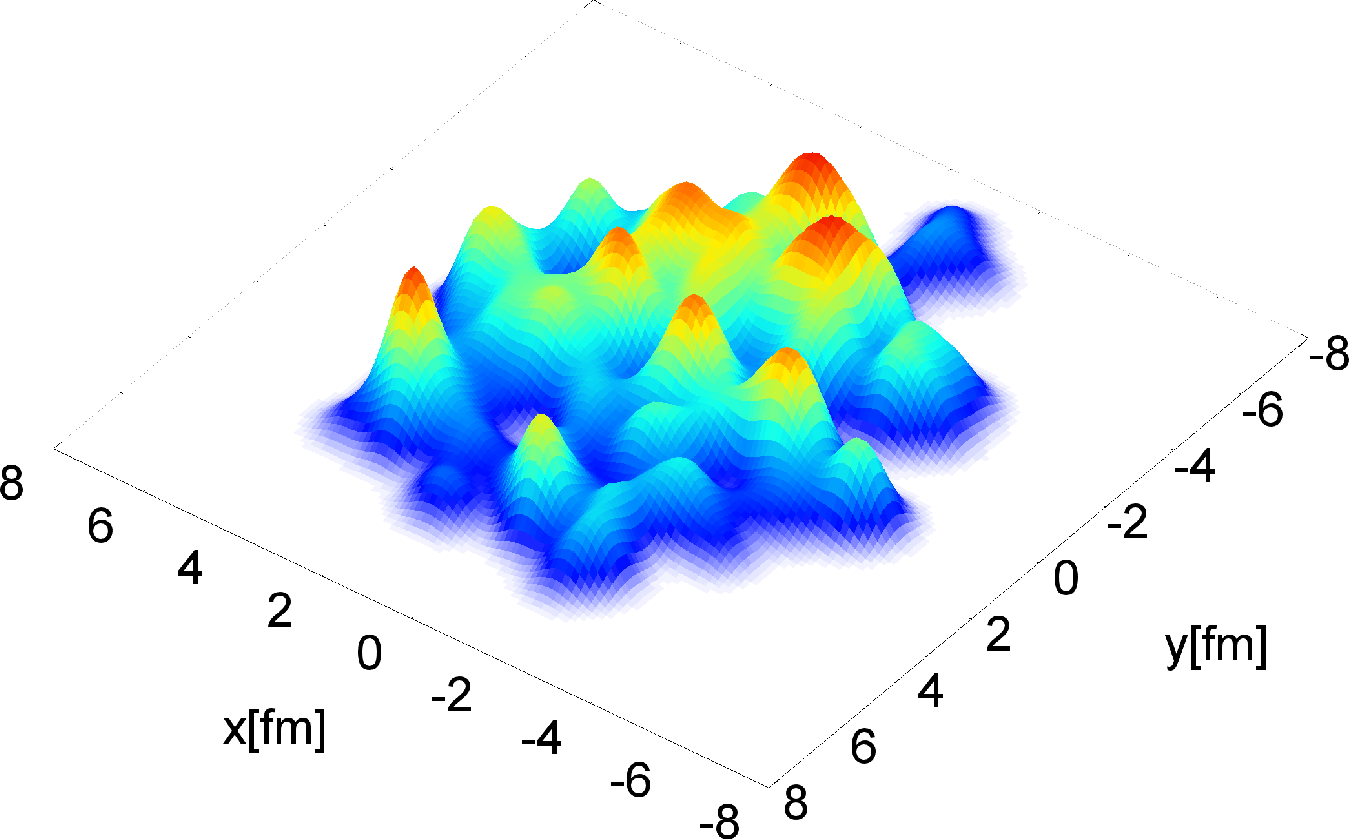} &
      \includegraphics[width=.45\textwidth]{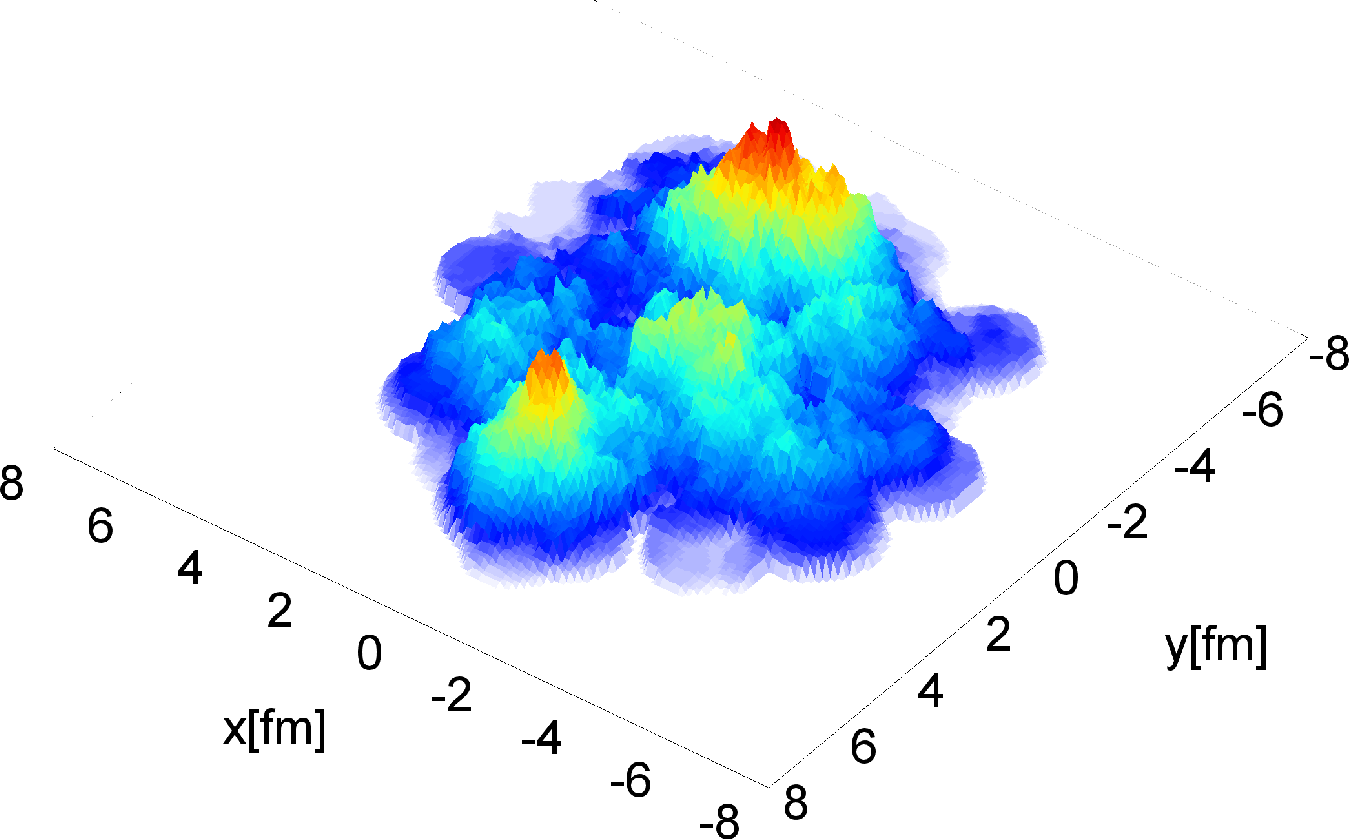} &
      \includegraphics[width=.45\textwidth]{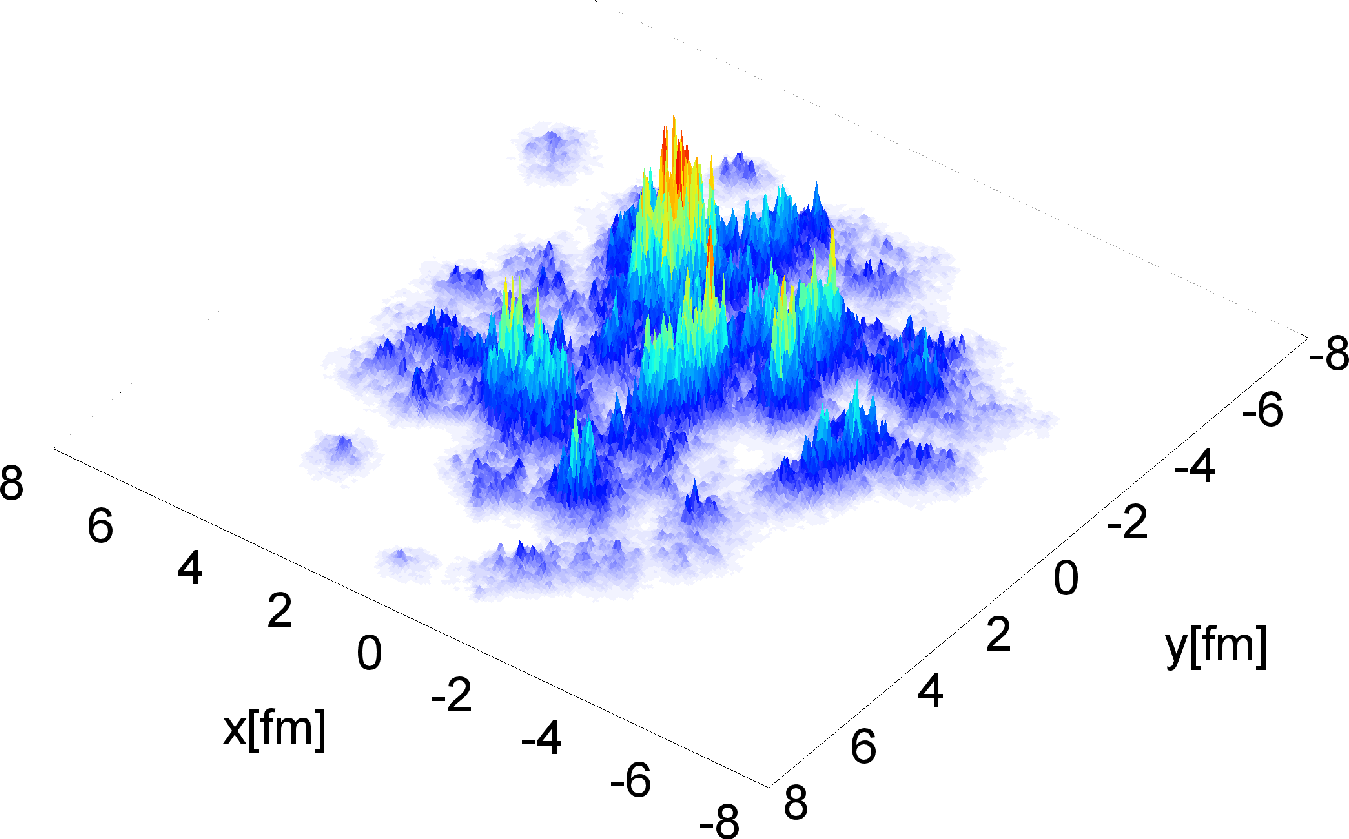}
    \end{tabular}
  }
  \caption{
    Transverse energy density generated by various initial condition models, as labeled.
    Figure from \cite{Schenke:2012wb}.
  }
\end{figure}

A more theoretically motivated initial state formalism is the color glass condensate (CGC) \cite{Gelis:2010nm}, an effective field theory based on gluon saturation at high energy.
The Monte Carlo Kharzeev-Levin-Nardi (MC-KLN) model \cite{Kharzeev:2001yq,Drescher:2006pi}, a CGC implementation, described the centrality dependence of yields and elliptic flow, but has fallen out of favor due to it overpredicting the difference between elliptic and triangular flow \cite{Qiu:2011hf,Bernhard:2015hxa} [and section \ref{sec:v1}].
Notably, the IP-Glasma model \cite{Schenke:2012wb,Schenke:2012fw}, which combines the CGC-based impact parameter dependent saturation model (IP-Sat) \cite{Bartels:2002cj,Kowalski:2003hm} with classical Yang-Mills dynamics of the produced glasma (gluon plasma) fields, has precisely described a wide array of observables, including integrated flow harmonics, differential flow, and event-by-event flow distributions \cite{Gale:2012rq}.

These models produce quite different initial energy density---even given the same nuclear thickness $T_A$ and $T_B$---and as with the nucleon width, this variability can be difficult or impossible to disentangle from QGP medium properties.
An alternative approach is to parametrize energy deposition as a function of thickness, retaining as much meaningful flexibility as possible, and constrain the parametrization simultaneously with medium properties.
This was part of the strategy in some previous Bayesian parameter estimation studies \cite{Novak:2013bqa,Pratt:2015zsa,Sangaline:2015isa}, and is the goal of the parametric initial condition model \trento\ \cite{Moreland:2014oya}, developed by myself and fellow Duke graduate student J.\ Scott Moreland.
See section \ref{sec:ic} for a detailed description of \trento.

\subsubsection{Dynamics and thermalization}

It is generally assumed that, for hydrodynamics to be valid, the system must be in (approximate) local thermodynamic equilibrium (although some recent work calls this into question \cite{Romatschke:2016hle}), and that the system is not born in equilibrium.
Since hydrodynamic evolution must begin early, by around $\tau \sim 1\ \fmc$, in order to leave sufficient time for the observed collective flow to build up before freeze-out, another dynamical process must rapidly drive the system to equilibrium.
This suggests a two-stage approach in which an energy deposition model describes the system immediately after the collision, at time $\tau = 0^+$, then a pre-equilibrium model handles the dynamics prior to QGP formation.

It should be noted that pre-equilibrium dynamics are not strictly required for computational models;
hydrodynamics can be initialized directly with the energy density and with zero initial flow and viscous pressures.
In this interpretation, the energy deposition model provides the energy at the hydrodynamic starting time, not at $\tau = 0^+$, skipping any pre-equilibrium stage.
Although not the most realistic, this scheme has been used in numerous studies and does not preclude a good description of the data, but it may force the initial conditions to somehow compensate for the lack of pre-equilibrium evolution.

The most rudimentary dynamical pre-equilibrium model is free streaming \cite{Broniowski:2008qk,Liu:2015nwa}, wherein the system is treated as an expanding, noninteracting gas of massless partons (quarks and/or gluons).
During free streaming, the energy density smooths out and radial flow increases, ultimately translating to larger mean $p_T$.
At a variable time $\tau_\text{fs}$, the system undergoes a sudden equilibration and hydrodynamic evolution begins;
this instantaneous transition from zero coupling to strong coupling cannot be the physical reality---it should be gradual---but the model can nonetheless help bracket the maximum pre-equilibrium time.
A multiparameter analysis of transverse momentum and flow data found $\tau_\text{fs} \approx 1\text{--}2.5\ \fmc$ and that increased free streaming time correlates with decreased $\eta/s$ \cite{Heinz:2015arc}.
See section \ref{sec:pre-eq} for a mathematical description of free streaming.

\begin{figure}[t]
  \makebox[\textwidth]{\includegraphics{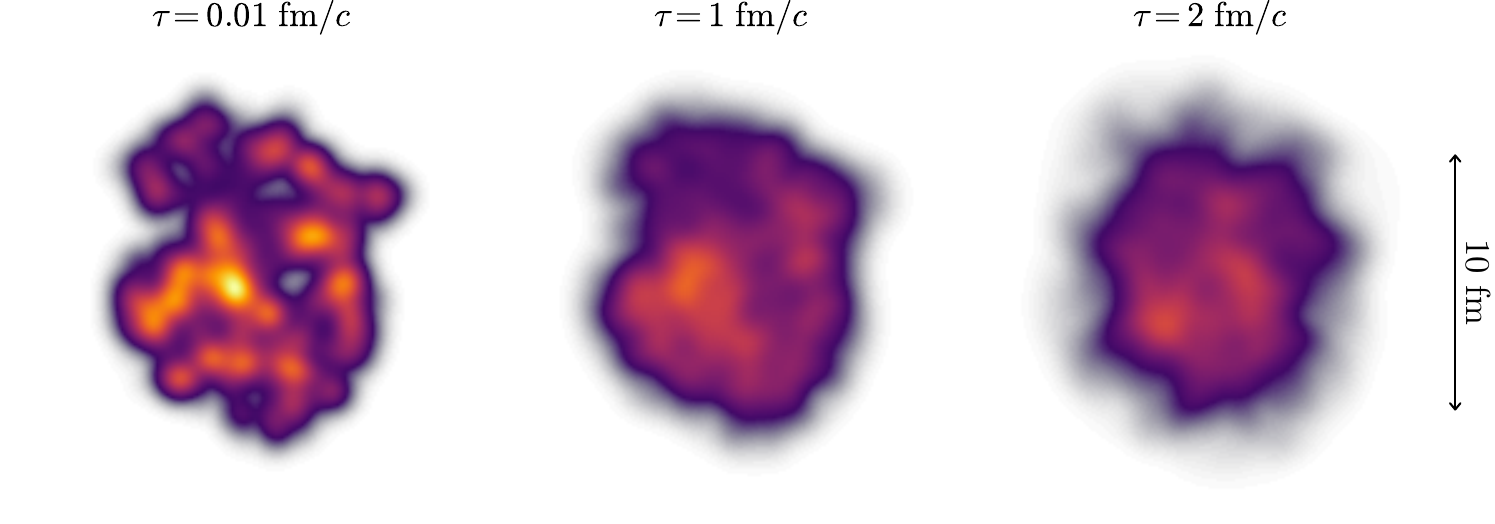}}
  \caption{
    Free streaming evolution of the transverse energy density for a typical fluctuating initial condition (nucleon width $w = 0.5$ fm).
    Each energy density profile is multiplied by its time $\tau$ to account for longitudinal expansion.
  }
\end{figure}

Other theories of pre-equilibrium dynamics include weakly-coupled effective kinetic theory \cite{Keegan:2016cpi}, which ought to be more realistic than free streaming (zero coupling) and may be a viable alternative for computational purposes;
the IP-Glasma model \cite{Schenke:2012wb,Schenke:2012fw}, in which the system is evolved by solving the classical Yang-Mills equations, another weakly-coupled theory;
and various strong coupling approaches, e.g.\ \cite{vanderSchee:2013pia}, although so far these have limited computational utility.

\chapter{Computational models of heavy-ion collisions}
\chaptermark{Computational models}
\label{ch:models}

\lettrine{T}{he} past two decades have seen considerable progress towards a ``standard model'' of the bulk dynamics of relativistic heavy-ion collisions \cite{Bass:2000ib,Teaney:2001av,Nonaka:2006yn,Petersen:2008dd,Song:2010aq,Song:2011hk,Song:2011qa,Shen:2014vra}.
Now well-established, the following multistage approach mirrors the presumptive \emph{true} collision spacetime evolution:
\begin{enumerate}
  \item Initial conditions: generates the energy density immediately after the collision.
  \item Pre-equilibrium: simulates the dynamics until QGP formation.
  \item Viscous relativistic hydrodynamics: calculates the evolution of the hot and dense QGP medium, including its collective expansion, cooling, and transition to a hadron gas.
  \item Particlization: converts the hydrodynamic system into a microscopic ensemble of hadrons.
  \item Boltzmann transport: computes hadronic scatterings and decays until freeze-out.
\end{enumerate}

\noindent
Over the next five sections, I assemble a set of these models tailored for use in Bayesian parameter estimation.
Of the various stages, I have developed new, original code for the initial conditions, pre-equilibrium, and particlization components.
I close the chapter with some details on performing large-scale model-to-data comparison.

Before proceeding, it's worth emphasizing that there are other viable approaches for modeling heavy-ion collision dynamics, such as using a parton cascade rather than hydrodynamics for the deconfined QGP phase, e.g.\ as in AMPT (A Multiphase Transport model) \cite{Lin:2004en} or BAMPS (Boltzmann Approach for Multiparton Scatterings) \cite{Xu:2004mz}.
Nevertheless, the stages listed above have together been broadly successful in describing a wide variety of observables and are the most natural option for quantitative analysis.
I shall not compare to alternative models in this work, but the parameter estimation method is certainly not specific to the present choices.

\section{Initial conditions}
\label{sec:ic}

Initial condition models are responsible for generating the energy or entropy density immediately after the collision.
In this work, I use \trento, a parametric initial condition model developed by myself and fellow Duke graduate student J.\ Scott Moreland.
The next few subsections introduce the model, demonstrate some of its capabilities, and illustrate why its flexibility makes it ideal for parameter estimation and uncertainty quantification;
text and figures have been adapted from our publication: \\[1ex]
\fullcite{Moreland:2014oya}. \\[1ex]
The model is publicly available at \url{https://github.com/Duke-QCD/trento}.

\subsection{The \texorpdfstring{\trento}{TRENTO} model}

\trento\ is an \emph{effective} model, intended to generate realistic Monte Carlo initial entropy profiles without assuming specific physical mechanisms for entropy production, pre-equilibrium dynamics, or thermalization.

Suppose a pair of projectiles labeled $A, B$ collide along beam axis $z$, and let $\rho^\text{part}_{A,B}$ be the density of nuclear matter that participates in inelastic collisions.
Each projectile may then be represented by its \emph{participant} thickness
\begin{equation}
  \T_{A,B}(x, y) = \int dz \, \rho^\text{part}_{A,B}(x, y, z).
  \label{eq:thickness}
\end{equation}
The construction of these thickness functions will be addressed shortly; first, we postulate the following:
\begin{enumerate}
  \item The eikonal approximation is valid:
    Entropy is produced if $\T_A$ and $\T_B$ eikonally overlap.
  \item There exists a scalar field $f(\T_A, \T_B)$ which converts projectile thicknesses into entropy
    deposition.
\end{enumerate}
The function $f$ is proportional to the entropy created at mid-rapidity and at the hydrodynamic thermalization
time:
\begin{equation}
  f \propto dS/dy \, |_{\tau = \tau_0}.
\end{equation}
It should provide an effective description of early collision dynamics:
it need not arise from a first-principles calculation, but it must obey basic physical constraints.

Perhaps the simplest such function is a sum, $f~\sim~\T_A~+~\T_B$, in fact this is equivalent to a wounded nucleon model since the present thickness functions \eqref{eq:thickness} only include participant matter.
The two-component Glauber ansatz adds a quadratic term to account for binary collisions, i.e.\ $f \sim (\T_A + \T_B) + \alpha \, \T_A \T_B$.

However, recent results from ultra-central uranium-uranium collisions at RHIC \cite{Pandit:2013uiv,Wang:2014qxa} show that particle production does not scale with the number of binary collisions, excluding the two-component Glauber ansatz \cite{Goldschmidt:2015qya}.
Therefore $N$ \mbox{one-on-one} nucleon collisions should produce the same amount of entropy as a single \mbox{$N$-on-$N$} collision, which is mathematically equivalent to the function $f$ being scale-invariant:
\begin{equation}
  f(c \, \T_A, c \, \T_B) = c \, f(\T_A, \T_B)
  \label{eq:scale-inv}
\end{equation}
for any nonzero constant $c$.
Note, this is clearly broken by the binary collision term $(\alpha \, \T_A \T_B)$.
We will justify this constraint later in the text; for the moment we take it as a postulate.

With these constraints in mind, we propose for $f$ the \emph{reduced thickness}
\begin{equation}
  f = \T_R(p; \T_A, \T_B) \equiv \biggl( \frac{\T_A^p + \T_B^p}{2} \biggr)^{1/p},
  \label{eq:tr}
\end{equation}
so named because it takes two thicknesses $\T_A, \T_B$ and ``reduces'' them to a third thickness, similar to a
reduced mass.
This functional form---known as the generalized mean---interpolates between the minimum and maximum of $\T_A, \T_B$ depending on the value of the dimensionless parameter $p$, and simplifies to the arithmetic, geometric, and harmonic means for certain values:
\begin{equation}
  \newlength{\extraspace}
  \setlength{\extraspace}{0.5ex}
  \T_R =
  \begin{cases}
    \max(\T_A, \T_B) & p \rightarrow +\infty, \\[\extraspace]
    (\T_A + \T_B)/2 & p = +1, \hfill \text{ (arithmetic)} \\[\extraspace]
    \sqrt{\T_A \T_B} & p = 0, \hfill \text{ (geometric)} \\[\extraspace]
    2\, \T_A \T_B/(\T_A + \T_B) & p = -1, \hfill \text{ (harmonic)} \\[\extraspace]
    \min(\T_A, \T_B) & p \rightarrow -\infty.
  \end{cases}
  \label{eq:means}
\end{equation}
Physically, $p$ interpolates among qualitatively different physical mechanisms for entropy production.
To see this, consider a pair of nucleon participants colliding with some nonzero impact parameter, as shown in figure \ref{fig:reduced-thickness}.
For $p = 1$, the reduced thickness is equivalent to a Monte Carlo wounded nucleon model and deposits a blob of entropy for each nucleon,
while for $p = 0$, the model deposits a single roughly symmetric blob at the midpoint of the collision,
and as $p$ becomes negative, it suppresses entropy deposition along the direction of the impact parameter.
Similar behavior was discussed in the context of small collision systems in \cite{Bzdak:2013zma}.
Note that the values $1, 0, -1$ are only special cases---$p$ is a continuous parameter---and the scale-invariant constraint \eqref{eq:scale-inv} is always satisfied.

\begin{figure}[t]
  \graphicsandcaption{.75}{trento/reduced_thickness}{
    Reduced thickness of a pair of nucleon participants.
    The nucleons collide with a nonzero impact parameter along the $x$-direction as shown in the upper right.
    The gray dashed lines are one-dimensional cross sections of the participant nucleon thickness functions $\T_A, \T_B$, and the colored lines are the reduced thickness $\T_R$ for $p = 1, 0, -1$ (green, blue, orange).
  }
  \label{fig:reduced-thickness}
\end{figure}

We now detail the construction of the thickness functions $\T_{A,B}(x, y)$, which combined with the definition
of the reduced thickness completes the specification of the model.  The procedure is constructed from the
ground up to handle a variety of collision systems; we begin with the simplest case.

Consider a collision of two protons $A, B$ with impact parameter $b$ along the $x$-direction and nuclear densities
\begin{equation}
  \rho_{A,B} = \rho_\text{proton}(x \pm b/2, y, z),
\end{equation}
and assume that the integral $\int dz \, \rho_\text{proton}$ either has a closed form or may be evaluated numerically, so that the proton thickness functions can be calculated.
The protons collide with probability \cite{dEnterria:2010xip}
\begin{equation}
  P_\text{coll} = 1 - \exp\biggl[ -\sigma_{gg} \int dx \, dy \int dz \, \rho_A \int dz \, \rho_B \biggr],
  \label{eq:pcoll}
\end{equation}
where the integral in the exponential is the overlap integral of the proton thickness functions and
$\sigma_{gg}$ is an effective parton-parton cross-section tuned so that the total proton-proton
cross-section equals the experimental inelastic nucleon-nucleon cross-section $\sigma_\text{NN}$.

The collision probability is sampled once to determine if the protons collide; assuming they do, we follow a procedure similar to \cite{Bozek:2013uha} and assign each proton a \emph{fluctuated} thickness
\begin{equation}
  \T_{A,B}(x, y) = w_{A,B} \int dz \, \rho_{A,B}(x, y, z),
\end{equation}
where $w_{A,B}$ are independent random weights sampled from a gamma distribution with unit mean,
\begin{equation}
  P_k(w) = \frac{k^k}{\Gamma(k)} w^{k-1} e^{-kw}.
  \label{eq:gamma}
\end{equation}
These gamma weights introduce additional multiplicity fluctuations in order to reproduce the large fluctuations observed in experimental proton-proton collisions.
The shape parameter $k$ may be tuned to optimally fit the data:
small values ($0 < k < 1$) correspond to large multiplicity fluctuations, while large values ($k \gg 1$) suppress fluctuations.

With the projectile thickness functions in hand, the reduced thickness is calculated to furnish the initial transverse entropy profile up to an overall normalization factor,
\begin{equation}
  dS/dy \, |_{\tau = \tau_0} \propto \T_R(p; \T_A, \T_B).
  \label{eq:dSdy}
\end{equation}

Composite collision systems such as proton-nucleus and nucleus-nucleus are essentially treated as
superpositions of proton-proton collisions.
A set of nucleon positions is chosen for each projectile, typically by sampling an uncorrelated Woods-Saxon distribution or from more realistic correlated nuclear configurations when available \cite{Alvioli:2009ab}.
The collision probability \eqref{eq:pcoll} is sampled for each pairwise interaction and those nucleons that collide with at least one partner are labeled ``participants'' while the rest are discarded.
The fluctuated thickness function of nucleus $A$ then reads
\begin{equation}
  \T_A = \sum_{i=1}^{N_\text{part}} w_i \int dz \, \rho_\text{proton}(x - x_i, y - y_i, z - z_i),
  \label{eq:nuc-thickness}
\end{equation}
where $w_i$ and $(x_i, y_i, z_i)$ are the weights and position, respectively, of participant $i$ in nucleus $A$.
$\T_B$ follows analogously.

This completes the construction of the model, \trento\ (Reduced Thickness Event-by-event Nuclear Topology).
In summary, the model deposits entropy proportional to the reduced thickness function \eqref{eq:tr}, defined as the generalized mean of fluctuated participant thickness functions \eqref{eq:nuc-thickness}, with each participant nucleon weighted by an independent gamma random number \eqref{eq:gamma}.

\subsection{Comparing to experimental data}

We now demonstrate \trento's ability to simultaneously describe a wide range of collision systems.
Note that the reduced thickness parameter $p$, gamma fluctuation parameter $k$, and nucleon profile $\rho_\text{proton}$ are not rigorously constrained here---see sections \ref{sec:v2} and \ref{sec:v3} for their quantitative estimates---so the following results do not necessarily represent the best-fit of the model to data.

We adopt a three-stage model for particle production similar to \cite{Bozek:2013uha}, in which the final multiplicity arises from a convolution of the initial entropy deposited by the collision, viscous entropy production during hydrodynamic evolution, and statistical hadronization at freeze-out.
The average charged-particle multiplicity $\avg\Nch$ after hydrodynamic evolution is to a good approximation proportional to the total initial entropy \cite{Song:2008si} and hence to the integrated reduced thickness via equation \eqref{eq:dSdy}:
\begin{equation}
  \avg\Nch \propto \int dx \, dy \, \T_R.
\end{equation}
Then, assuming independent particle emission at freeze-out, the final number of charged particles is Poisson distributed \cite{Kisiel:2005hn,Chojnacki:2011hb}, i.e.\ $P(\Nch) = \text{Poisson}(\avg\Nch)$.
The folding of the Poisson fluctuations with the gamma weights for each participant yields a negative binomial distribution \cite{Bozek:2013uha}, which has historically been used to fit proton-proton multiplicity fluctuations.

\begin{figure}[t]
  \makebox[\textwidth]{\includegraphics[width=1.5\textwidth]{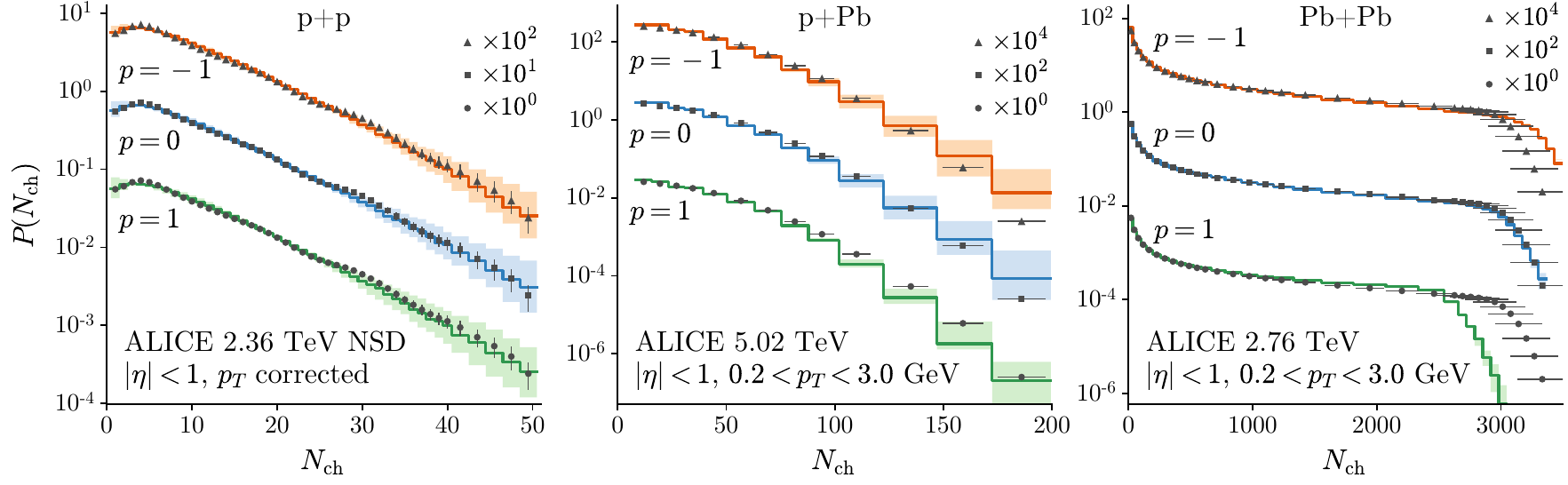}}
  \caption{
    Multiplicity distributions for proton-proton, proton-lead, and lead-lead collisions.
    The histograms are \protect\trento\ results for reduced thickness parameter $p = -1$ (top, orange), $p = 0$ (middle, blue), and $p = 1$ (bottom, green), with approximate best-fit fluctuation parameters $k$ and normalizations given in table~\ref{tab:nch}.
    The shaded bands show the sensitivity from varying $k$ by $\pm30\%$.
    Data points (triangles, squares, circles) are experimental distributions from ALICE \cite{Aamodt:2010ft,Abelev:2014mda} offset by powers of ten for comparison with the model.
  }
  \label{fig:nch}
\end{figure}

\begin{table}[b]
  \centering
  \captionsetup{width=.78\textwidth}
  \caption{
    Approximate best-fit fluctuation parameters $k$ and normalizations for each $p$ value and collision system in figure \ref{fig:nch}.
  }
  \label{tab:nch}
  \begin{tabular}{rcccc}
    \toprule
    $p\;$  & $k$ & p+p norm & p+Pb norm & Pb+Pb norm \\
    \midrule
    $+1$   & 0.8 & 9.7      & 7.0       & 13.        \\
    $ 0$   & 1.4 & 19.      & 17.       & 16.        \\
    $-1$   & 2.2 & 24.      & 26.       & 18.        \\
    \bottomrule
  \end{tabular}
\end{table}

To compare with experimental multiplicity distributions, we generate a large ensemble of minimum-bias events, integrate their $\T_R$ profiles, rescale by an overall normalization constant, and sample a Poisson number for the multiplicity of each event.
The left panel of figure \ref{fig:nch} shows the $\Nch$ distributions for proton-proton simulations with reduced thickness parameter $p = 1$, 0, $-1$, and Gaussian beam-integrated proton density
\begin{equation}
  \int dz \, \rho_\text{proton} = \frac{1}{2\pi B} \exp\biggr( -\frac{x^2 + y^2}{2B} \biggr)
  \label{eq:rhoproton}
\end{equation}
with effective area $B = (0.6\;\text{fm})^2$.
We tune the fluctuation parameter $k$ for each value of $p$ to qualitatively fit the experimental proton-proton distribution \cite{Aamodt:2010ft}, and additionally vary $k$ by $\pm30\%$ to explore the sensitivity of the model to the gamma participant weights.
For proton-lead and lead-lead collisions \cite{Abelev:2014mda} (middle and right panels), we use identical model parameters except for the overall normalization factor, which is allowed to vary independently across collision systems to account for differences in beam energy and kinematic cuts (annotated in the figure).
The $k$ values and normalizations are given in table~\ref{tab:nch}.

The model is able to reproduce the experimental proton-proton distribution for each value of $p$, provided $k$ is appropriately tuned.
Varying the best-fit $k$ value (by $\pm30\%$) has a noticeable effect on proton-proton and proton-lead systems, especially in the high-multiplicity tails, but is less important in lead-lead collisions, where the gamma weights are averaged over many participant nucleons.

Each $p$ value also yields a reasonable fit to the shapes of the proton-lead and lead-lead distributions, although lead-lead appears to favor $p \approx 0$.
Note that the normalizations for $p = 1$ (wounded nucleon model) in proton-lead and lead-lead collisions (table~\ref{tab:nch}) are not self-consistent, since proton-lead requires roughly half the normalization as lead-lead, even though the experimental data were measured at a higher beam energy.

\newcommand{\eccratio}{\sqrt{\langle \varepsilon_2^2 \rangle}/\sqrt{\langle \varepsilon_3^2 \rangle}^{\,0.6}}

\begin{figure}[t]
  \makebox[\textwidth]{\includegraphics[width=1.5\textwidth]{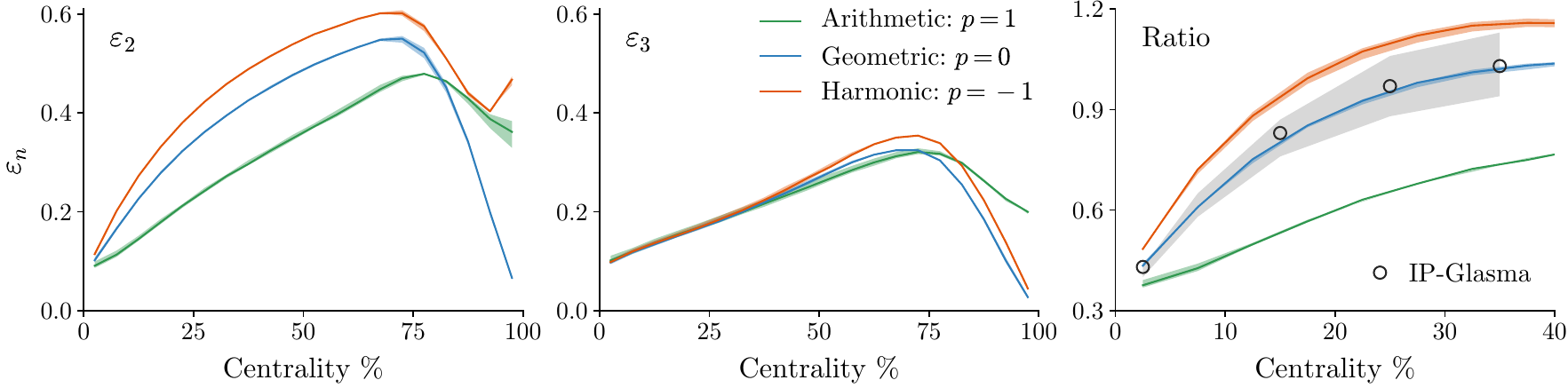}}
  \caption{
    Left and middle plots:
    Eccentricity harmonics $\varepsilon_2$ and $\varepsilon_3$ as a function of centrality for reduced thickness parameters $p = 1$, 0, $-1$ (green, blue, orange).
    The shaded bands show the sensitivity from varying $k$ by $\pm30\%$ from the values in table~\ref{tab:nch}.
    Right plot:
    Ratio of the rms eccentricities $\eccratio$ against the allowed region (gray band) and the ratio computed by IP-Glasma (circles) \cite{Retinskaya:2013gca}.
    Note that the axes have different ranges in the ratio plot.
  }
  \label{fig:eccentricity}
\end{figure}

Eccentricity harmonics $\varepsilon_n$ are calculated using the definition
\begin{equation}
  \varepsilon_n e^{i n\phi} = -\frac{\int dx \, dy\, r^n e^{i n \phi} \, \T_R}{\int dx \, dy \, r^n \, \T_R}.
\end{equation}
Figure~\ref{fig:eccentricity} shows ellipticity $\varepsilon_2$ and triangularity $\varepsilon_3$ as a function of centrality using the same lead-lead data as in figure \ref{fig:nch}.
There is a clear trend of increasing eccentricity (particularly $\varepsilon_2$) with decreasing $p$.
This is a larger-scale manifestation of the behavior in figure \ref{fig:reduced-thickness}:
as $p$ decreases, the generalized mean \eqref{eq:tr} attenuates entropy production in asymmetric regions of the collision, accentuating the elliptical overlap shape in non-central collisions and enhancing their eccentricity.
Meanwhile, varying the fluctuation parameter $k$ has limited effect.

In addition, we perform the test proposed by \cite{Retinskaya:2013gca}, which uses flow data and hydrodynamic calculations to determine an experimentally allowed band for the ratio of root-mean-square eccentricities $\eccratio$ as a function of centrality.
Among available initial condition models only IP-Glasma consistently falls within the allowed region.
As shown in the right panel of figure \ref{fig:eccentricity}, \trento\ with $p = 0$ (geometric mean) yields excellent agreement with the allowed band and is similar to IP-Glasma.

\label{loc:trento-uranium}

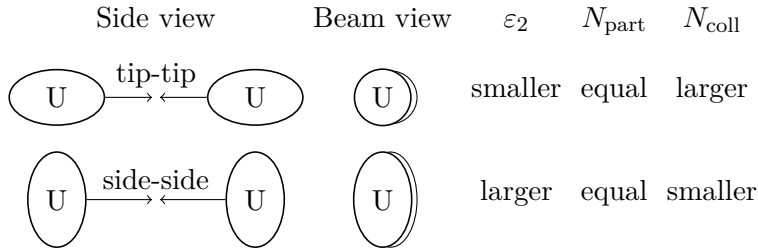
\begin{figure}[b!]
  \centering
  \begin{tikzpicture}[
    uranium/.style={draw, semithick, ellipse, anchor=center},
    small width/.style={minimum width=20},
    large width/.style={minimum width=36},
    small height/.style={minimum height=20},
    large height/.style={minimum height=36}
  ]
    \matrix (m) [matrix of nodes] {
      &[-1.5ex] Side view &[-1.5ex] & Beam view & $\varepsilon_2$ & $N_\text{part}$ & $N_\text{coll}$ \\[.7ex]
      |[uranium, large width, small height] (ttl)| U &
      |[above]| tip-tip &
      |[uranium, large width, small height] (ttr)| U &
      \node[draw, circle, small height, small width, xshift=3pt] {};
      \node[uranium, circle, small width, small height, fill=white] {U}; &
      smaller & equal & larger \\[2ex]
      |[uranium, large height, small width] (ssl)| U &
      |[above]| side-side &
      |[uranium, large height, small width] (ssr)| U &
      \node[draw, ellipse, large height, small width, xshift=3pt] {};
      \node[uranium, large height, small width, fill=white] {U}; &
      larger & equal & smaller \\
    };
    \begin{scope}[->]
      \draw (ttl) -- ($(ttl)!.48!(ttr)$);
      \draw (ttr) -- ($(ttr)!.48!(ttl)$);
      \draw (ssl) -- ($(ssl)!.48!(ssr)$);
      \draw (ssr) -- ($(ssr)!.48!(ssl)$);
    \end{scope}
  \end{tikzpicture}
  \captionsetup{width=.82\textwidth}
  \caption{
    Comparison of tip-tip and side-side uranium-uranium collisions.
    Schematics are shown from a side view and looking down the beam axis, and the following quantities are compared:
    ellipticity $\varepsilon_2$, number of participating nucleons $N_\text{part}$, and number of binary nucleon-nucleon collisions $N_\text{coll}$.
  }
  \label{fig:uu-schematic}
\end{figure}

As a final novel application, we return to the previously mentioned ultra-central uranium-uranium puzzle, where typical Glauber models are notably inconsistent with experimental data.
Unlike e.g.~gold and lead, uranium nuclei have a highly deformed prolate spheroidal shape, so uranium-uranium collisions may achieve maximal overlap via two distinct orientations:
``tip-tip'', in which the long axes of the spheroids are aligned with the beam axis and the overlap area is circular;
or ``side-side'', where the long axes are perpendicular to the beam axis and the overlap area is elliptical, as shown in figure \ref{fig:uu-schematic}.
Hence side-side collisions will in general have larger initial-state ellipticity $\varepsilon_2$ and final-state elliptic flow $v_2$ than tip-tip.

In the two-component Glauber model, tip-tip collisions produce more binary nucleon-nucleon collisions than side-side, so tip-tip collisions have larger charged-particle multiplicity $\Nch$.
Therefore, the most central uranium-uranium events are dominated by tip-tip collisions with maximal $\Nch$ and small $v_2$, while side-side collisions have a smaller $\Nch$ and somewhat larger $v_2$.
This predicted drop in elliptic flow as a function of $\Nch$ is known as the ``knee'' \cite{Voloshin:2010ut}.

\begin{figure}[t]
  \makebox[\textwidth]{\includegraphics[width=1.2\textwidth]{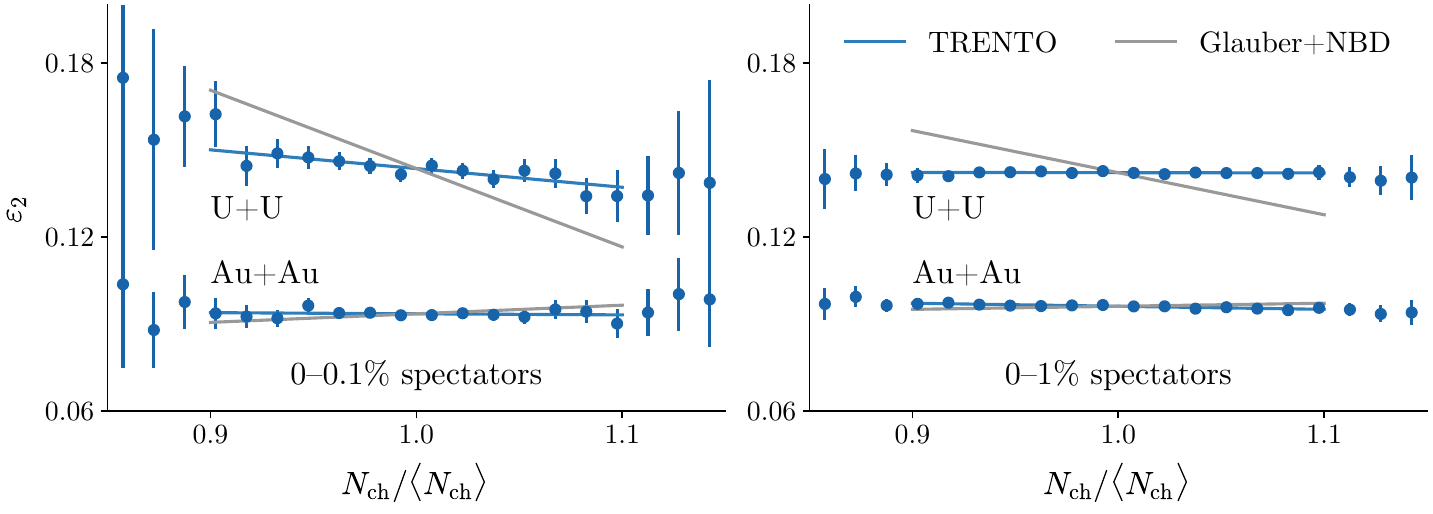}}
  \caption{
    Ellipticity $\varepsilon_2$ as a function of normalized charged-particle multiplicity $\Nch/\avg\Nch$ in ultra-central uranium-uranium and gold-gold collisions at RHIC.
    The left and right plots show the top 0.1\% and 1\% of collisions selected by number of spectators to mimic STAR's experimental ZDC selection \cite{Pandit:2013uiv}.
    Blue points with error bars are binned \protect\trento\ results with reduced thickness parameter $p = 0$ and best-fit fluctuation parameter $k = 1.4$.
    Blue lines are linear fits within $0.9~<~\Nch/\avg\Nch~<~1.1$.
    Gray lines represent the analogous Glauber+NBD slopes calculated in \cite{Pandit:2013uiv}.
  }
  \label{fig:uranium}
\end{figure}

Recent data by STAR on uranium-uranium collisions exhibits no evidence of a knee \cite{Pandit:2013uiv,Wang:2014qxa}, at odds with Glauber model predictions.
It has been proposed that fluctuations could wash out the knee \cite{Rybczynski:2012av}, but a recent flow analysis showed that it would still be visible \cite{Goldschmidt:2015qya}.

The data therefore imply that multiplicity is independent of the number of binary collisions, justifying the scale-invariant condition \eqref{eq:scale-inv} postulated during the construction of the reduced thickness ansatz \eqref{eq:tr}.
Consequently, \trento\ predicts roughly the same number of charged particles in tip-tip and side-side uranium-uranium collisions.
As shown in figure \ref{fig:uranium}, the slope of $\varepsilon_2$ as a function of $\Nch$ is approximately equal for uranium-uranium and gold-gold, in contrast to the Glauber model which predicts a much steeper slope for uranium.
Short of conducting a full hydrodynamic analysis, \trento\ appears to be more consistent with STAR data than the Glauber model, and behaves similarly to IP-Glasma \cite{Schenke:2014tga}.

\subsection{Reproducing existing models}
\label{subsec:reproducing-ic}

This subsection is adapted from: \\[1ex]
\fullcite{Bernhard:2016tnd}.

\medskip

\noindent \trento\ is constructed to achieve maximal flexibility using a minimal number of parameters and can mimic a wide range of existing initial condition models.
To demonstrate the efficacy of the generalized mean ansatz, equation \eqref{eq:tr}, we now show that the mapping can reproduce different theory calculations using suitable values of the parameter $p$.

Perhaps the simplest and oldest model of heavy-ion initial conditions is the so called participant or wounded nucleon model, which deposits entropy for each nucleon that engages in one or more inelastic collisions \cite{Bialas:1976ed}.
In its Monte Carlo formulation \cite{Shor:1988vk,Wang:1991hta,Alver:2008aq,Broniowski:2007nz}, the wounded nucleon model may be expressed in terms of participant thickness functions, equation \eqref{eq:nuc-thickness}, as
\begin{equation}
  s \propto \T_A + \T_B.
  \label{eq:wn}
\end{equation}
Comparing to equation \eqref{eq:means}, we see that the wounded nucleon model is equivalent to the generalized mean ansatz with $p=1$.

\newcommand{\Qs}[1]{Q_{s,\text{#1}}}

More sophisticated calculations of the mapping $f$ can be derived from color glass condensate effective field theory.
A common implementation of a CGC based saturation picture is the KLN model \cite{Kharzeev:2001yq,Kharzeev:2002ei,Kharzeev:2004if}, in which entropy deposition at the QGP thermalization time can be determined from the produced gluon density, $s \propto N_g$, where
\begin{equation}
  \frac{dN_g}{dy\,d^2r_\perp} \sim \Qs{min}^2 \biggl[
    2 + \log \biggl(\frac{\Qs{max}^2}{\Qs{min}^2} \biggr)
  \biggr],
  \label{eq:kln}
\end{equation}
and $\Qs{max}$ and $\Qs{min}$ denote the larger and smaller values of the two saturation scales in opposite nuclei at any fixed position in the transverse plane \cite{Drescher:2006ca}.
In the original formulation of the KLN model, the two saturation scales are proportional to the local participant nucleon density in each nucleus, $Q^2_{s,A} \propto \T_A$, and the gluon density can be recast as
\begin{equation}
  s \sim \T_\text{min} \bigl[ 2 + \log(\T_\text{max}/\T_\text{min}) \bigr]
\end{equation}
to put it in a form which can be directly compared with the wounded nucleon model.

Another saturation model which has attracted recent interest after it successfully described an extensive list of experimental particle multiplicity and flow observables \cite{Niemi:2015qia,Paatelainen:2013eea} is the EKRT model, which combines collinearly factorized pQCD minijet production with a simple conjecture for gluon saturation \cite{Eskola:1999fc,Eskola:2001bf}.
The energy density predicted by the model after a pre-thermal Bjorken free streaming stage is given by
\begin{equation}
  e(\tau_0, x, y) \sim \frac{K_\text{sat}}{\pi} p_\text{sat}^3(K_\text{sat}, \beta; T_A, T_B),
  \label{eq:ekrt_energy}
\end{equation}
where the saturation momentum $p_\text{sat}$ depends on nuclear thickness functions $T_A$ and $T_B$, as well as phenomenological model parameters $K_\text{sat}$ and $\beta$.
Calculating the saturation momentum in the EKRT formalism is computationally intensive, and hence---in its Monte Carlo implementation---the model parametrizes the saturation momentum $p_\text{sat}$ to facilitate efficient event sampling \cite{Niemi:2015qia}.
The energy density in equation \eqref{eq:ekrt_energy} can then be recast as an entropy density using the thermodynamic relation ${s \sim e^{3/4}}$ to compare it with the previous models.

\begin{figure}[t]
  \makebox[\textwidth]{\includegraphics[width=1.5\textwidth]{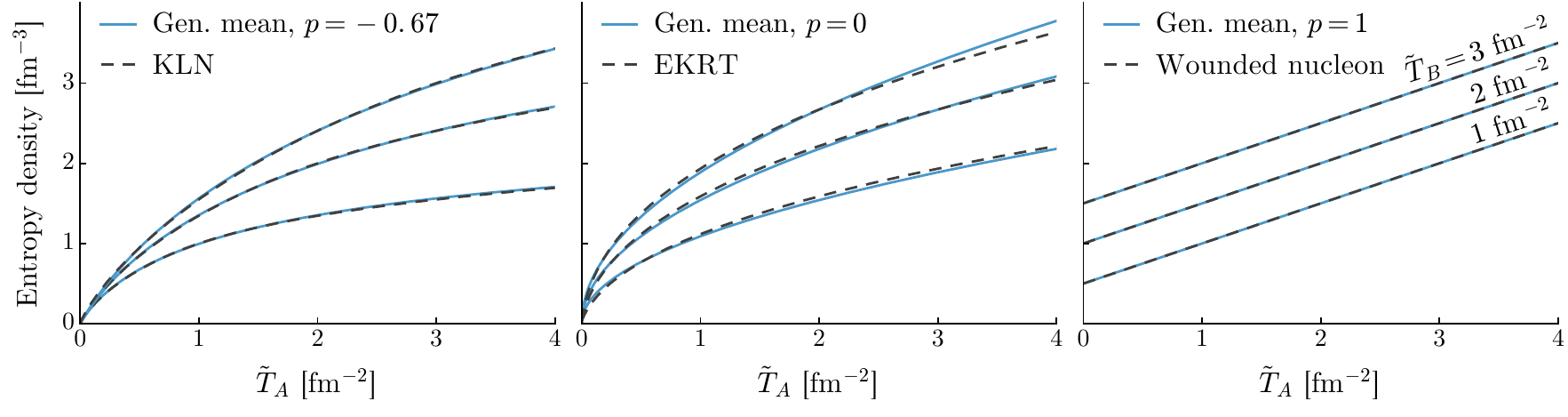}}
  \caption{
    Profiles of the initial thermal distribution predicted by the KLN (left), EKRT (middle), and wounded nucleon (right) models (dashed black lines) compared to a generalized mean with different values of the parameter $p$ (solid blue lines).
    Staggered lines show different slices of the initial entropy density $dS/(d^2r_\perp dy)$ as a function of the participant nucleon density $\T_A$ for several values of $\T_B = 1, 2, 3$ [fm$^{-2}$].
    The EKRT mapping is shown with model parameters $K=0.64$ and $\beta=0.8$ \cite{Niemi:2015qia}.
    Entropy normalization is arbitrary.
  }
  \label{fig:cgc-compare}
\end{figure}

Note that equation \eqref{eq:ekrt_energy} is expressed as a function of nuclear thickness $T$ which includes contributions from \emph{all} nucleons in the nucleus, as opposed to the participant thickness $\T$.
In order to express initial condition mappings as functions of a common variable one could, e.g.\ relate $\T$ and $T$ using an analytic wounded nucleon model.
The effect of this substitution on the EKRT model is small, as the mapping deposits zero entropy if nucleons are non-overlapping, effectively removing them from the participant thickness function.
We thus replace $T$ with $\T$ in the EKRT model and note that similar results are obtained by recasting the wounded nucleon, KLN, and \trento\ models as functions of $T$ using standard Glauber relations.

Figure~\ref{fig:cgc-compare} shows one-dimensional slices of the entropy deposition mapping predicted by the KLN, EKRT, and wounded nucleon models for typical values of the participant nucleon density sampled in Pb+Pb collisions at $\sqrt s = 2.76$~TeV.
The vertically staggered lines in each panel show the change in deposited entropy density as a function of $\T_A$ for several constant values of $\T_B$, where the dashed lines are the entropy density calculated using the various models and the solid lines show the generalized mean ansatz tuned to fit each model.
The figure illustrates that the ansatz reproduces different initial condition calculations and quantifies differences among them in terms of the generalized mean parameter $p$.
The KLN model, for example, is well-described by $p\sim-0.67$, the EKRT model corresponds to $p \sim 0$, and the wounded nucleon model is precisely $p=1$.
Smaller, more negative values of $p$ pull the generalized mean toward a minimum function and hence correspond to models with more extreme gluon saturation effects.

The three models considered in figure \ref{fig:cgc-compare} are by no means an exhaustive list of proposed initial condition models, see e.g.\ Refs.~\cite{Eremin:2003qn,Broniowski:2007nz,Pierog:2013ria,Drescher:2000ec,Chatterjee:2015aja,Zhang:1999bd}.
Notably absent, for instance, is the highly successful IP-Glasma model which combines IP-Sat CGC initial conditions with classical Yang-Mills dynamics to describe the full pre-equilibrium evolution of produced glasma fields \cite{Schenke:2012wb,Schenke:2012fw,Gale:2012rq}.
The IP-Glasma model lacks a simple analytic form for initial energy (or entropy) deposition at the QGP thermalization time and so cannot be directly compared to the generalized mean ansatz.
In lieu of such a comparison, we examined the geometric properties of IP-Glasma and \trento\ through their eccentricity harmonics $\varepsilon_n$.

\begin{figure}[t]
  \graphicsandcaption{.75}{trento/ipglasma}{
    Eccentricity harmonics $\varepsilon_2$ and $\varepsilon_3$ as a function of impact parameter $b$ for Pb+Pb collisions at ${\sqrt s = 2.76}$~TeV calculated from IP-Glasma and \protect\trento\ initial conditions.
    IP-Glasma events are evaluated after $\tau=0.4$~fm/$c$ classical Yang-Mills evolution \cite{Schenke:2012wb}; \protect\trento\ events after $\tau=0.4$~fm/$c$ free streaming \cite{Liu:2015nwa,Broniowski:2008qk} and using parameters $p = 0 \pm 0.1$, $k = 1.6$, and nucleon width $w=0.4$~fm to match IP-Glasma \cite{Schenke:2013dpa}.
  }
  \label{fig:ipglasma}
\end{figure}

We generated a large number of \trento\ events using entropy deposition parameter $p=0$, Gaussian nucleon width $w=0.4$~fm, and fluctuation parameter $k=1.6$, which were previously shown to reproduce the ratio of ellipticity and triangularity in IP-Glasma \cite{Moreland:2014oya}.
We then free streamed \cite{Liu:2015nwa,Broniowski:2008qk} the events for $\tau=0.4$~fm/$c$ to mimic the weakly coupled pre-equilibrium dynamics of IP-Glasma and match the evolution time of both models.
Finally, we calculated the eccentricity harmonics $\varepsilon_2$ and $\varepsilon_3$ weighted by energy density $e(x, y)$ according to the definition
\begin{equation}
    \varepsilon_n e^{i n \phi} = -\frac{\int dx\, dy\, r^n e^{i n \phi} e(x,y)}{\int dx\, dy\, e(x,y)},
\end{equation}
where the energy density is the time-time component of the stress-energy tensor after the free streaming phase, $T^{00}$.
The resulting eccentricities, pictured in figure \ref{fig:ipglasma}, are in good agreement for all but the most peripheral collisions, where sub-nucleonic structure becomes important.
This similarity suggests that \trento\ with $p \sim 0$ can effectively reproduce the scaling behavior of IP-Glasma, although a more detailed comparison would be necessary to establish the strength of correspondence illustrated in figure \ref{fig:cgc-compare}.

Additionally, a participant quark model has been proposed to describe the multiplicity and transverse-energy distributions of a variety of collision systems without a binary collision term \cite{Adler:2013aqf,Adare:2015bua}.
The model can be recast using an analytic Glauber formalism to construct an effective entropy deposition mapping $f$.
However, the resulting mapping cannot be encapsulated by a single value of the parameter $p$, so we do not attempt to support or exclude the participant quark model in the present analysis.

\subsection{Sampling nucleon positions with a minimum distance}
\label{subsec:dmin}

After publishing the preceding work, I implemented in \trento\ a minimum nucleon-nucleon distance parameter $\dmin$.
As discussed in subsection \ref{subsec:properties-initial-state}, this is an unknown degree of freedom that could impact initial geometry.

In the absence of a minimum distance, sampling uncorrelated nucleon positions is straightforward:
Independently sample the radial position of each nucleon from the Fermi distribution
\begin{equation}
  P(r) \propto r^2 \frac{1}{1 + \exp\bigl( \frac{r-R}{a} \bigr)},
\end{equation}
and sample the angles $\theta$ and $\phi$ isotropically.

Now, the na\"ive way to impose a minimum distance is, after sampling each nucleon's coordinates, check its distance to all previously sampled nucleons, and if it falls within $\dmin$ of any other nucleon, resample the coordinates.
However, since the spherical volume element is $dV = 4\pi r^2 dr$, there is less space available at small $r$, so positions sampled with small $r$ are more likely to need resampling.
This shifts density toward larger $r$, effectively modifying the target radial distribution, which is undesirable.

To avoid this, one should pre-sample the radii for all nucleons before attempting to place any of them.
Then, when choosing the full three-dimensional coordinates for each nucleon, sample only the angles, resampling as necessary to satisfy the minimum distance.
This way, the radial distribution is guaranteed not to change.

With this algorithm, if $\dmin$ is large it will occasionally be impossible to place a nucleon no matter how many times the angles are resampled.
To decrease the likelihood of this happening, one should sort the pre-sampled radii in increasing order;
this way, the nucleons with the smallest $r$, where space is at a premium, are placed first.

Using this method, $\dmin$ can be varied and estimated simultaneously with all other model parameters.

\section{Pre-equilibrium}
\label{sec:pre-eq}

Broadly speaking, pre-equilibrium models take the output of the initial condition model, compute the ensuing dynamics until the hydrodynamic starting time, and initialize the energy-momentum tensor $T\mn$.
The simplest model, and the present choice, is free streaming.

\subsection{Free streaming}

This scheme assumes that the system consists of noninteracting, massless partons which stream freely (zero coupling) for a tunable time $\tfs$ until a sudden equilibration and switch to hydrodynamics (strong coupling).
We therefore interpret the output of the initial condition model as the density of partons in the transverse plane, $n(x, y)$, at the initial time $\tau_0 = 0^+$.
This is different from the previous interpretation---that the initial condition provides the \emph{entropy} density directly at the hydrodynamic starting time---but not contradictory, since density has the same units as entropy density;
we are effectively asserting that each particle carries some number of entropy units.
And for a model like \trento, which stipulates that entropy deposition (or particle production) is purely eikonal, it arguably makes more sense to use its output immediately at $\tau_0$ rather than a later time.

Since the partons are massless and noninteracting, they propagate along straight trajectories at the speed of light;
at a later time $\tau > \tau_0$, the partons at transverse point $(x, y)$ were originally located on a ring of radius $c\Delta\tau$ centered at $(x, y)$, where $\Delta\tau = \tau - \tau_0$ is the elapsed time.
The energy-momentum tensor at position $(x, y)$ and time $\tau$ is therefore proportional to the integral of the density around the ring \cite{Broniowski:2008qk,Liu:2015nwa}:
\begin{equation}
  T\mn(x, y) = \frac{1}{\tau} \int d\phi \, \hat p^\mu \, \hat p^\nu \, n(x - \Delta\tau\cos\phi, y - \Delta\tau\sin\phi),
  \label{eq:fs-Tuv}
\end{equation}
where $\hat p^\mu = p^\mu/p_T$ is a transverse-momentum unit vector and the prefactor $1/\tau$ accounts for longitudinal expansion.
Assuming longitudinal boost invariance, we only need the tensor at midrapidity, in which case the unit vectors expand out to
\begin{equation}
  \hat p^\mu \, \hat p^\nu =
  \begin{pmatrix}
    1 & \cos\phi & \sin\phi \\
    \cos\phi & \cos^2\phi & \cos\phi\,\sin\phi \\
    \sin\phi & \cos\phi\,\sin\phi & \sin^2\phi
  \end{pmatrix}.
\end{equation}
The result for $T\mn$ may also be derived by analytically solving the collisionless Boltzmann equation, $p^\mu\partial_\mu f(x, p) = 0$ where $f$ is the parton distribution function, and noticing that the result is independent of the original transverse momentum distribution for massless particles \cite{Liu:2015nwa}.

At the switching time $\tfs$, we match the energy-momentum tensor to its hydrodynamic form
\begin{equation}
  T\mn = e \, u^\mu u^\nu - (P + \Pi)\Delta\mn + \pi\mn.
\end{equation}
The energy density and flow velocity are determined by the Landau matching condition
\begin{equation}
  T\mn u_\nu = e \, u^\mu,
  \label{eq:landau}
\end{equation}
which is an eigenvalue equation whose physical solution is the one with a timelike eigenvector $u^\mu$.
The equilibrium pressure $P = P(e)$ can then be obtained via the equation of state, and the bulk pressure $\Pi$ from the difference with the total effective pressure,
\begin{equation}
  P + \Pi = -\frac{1}{3} \Delta_{\mu\nu}T\mn.
\end{equation}
Lastly, the shear pressure tensor may be calculated as
\begin{equation}
  \pi\mn = T\mn - e \, u^\mu u^\nu + (P + \Pi)\Delta\mn,
\end{equation}
since everything on the right-hand side is now known.

\subsubsection{The corona}

At time $\tfs$, the system is assumed to equilibrate and begin evolving hydrodynamically.
The hydrodynamic calculation then runs until the system cools below a switching energy density $e_\text{switch}$ (usually parametrized as a temperature $\Tsw$, which can be converted to an energy density via the equation of state).
Presumably, most of the system has energy density $e > e_\text{switch}$ at time $\tfs$, but the periphery of the collision inevitably has $e < e_\text{switch}$.
This low-density region, known as the ``corona'', never enters the hydrodynamic calculation and is effectively neglected.
Longer free streaming times increase the relative size of the corona, since the energy density decreases as the system expands.

In the present scenario, for lead-lead collisions at LHC energies with $\tfs \sim 1\ \fmc$, the relative contribution of the corona is empirically negligible: less than 1\% of the total energy for all but the most peripheral collisions.
Thus, it is safe to neglect.
However, for smaller collision systems (such as proton-lead), lower beam energies, or longer free streaming times, the corona could become significant, and it may be necessary to consider its effects.

\subsubsection{Computational notes}

My implementation of free streaming, written in Python, is available at \url{https://github.com/Duke-QCD/freestream}.

Since the initial density is discretized onto a grid, it must be interpolated to obtain a continuous function $n(x, y)$ for integration in equation \eqref{eq:fs-Tuv}.
I use the bicubic interpolation provided by the Python class \href{https://docs.scipy.org/doc/scipy/reference/generated/scipy.interpolate.RectBivariateSpline.html}{scipy.interpolate.RectBivariateSpline}, which is a wrapper around the Fortran library \textsc{fitpack}.

Cubic interpolation faithfully captures the curvature of fluctuating initial conditions, but sometimes suffers from unphysical artifacts near where the density drops to zero, rapidly oscillating between small positive and negative values (clearly, density cannot be negative).
Linear interpolation, on the other hand, is unable to capture the curvature but does not have the same deficiencies.
One way to remove the artifacts from cubic interpolation is to interpolate the density grid with both bilinear and bicubic algorithms, then if both return values greater than zero, use the result of the cubic interpolation, otherwise use zero.
Letting $n_1$ and $n_3$ be the linear and cubic interpolating functions, this can be written
\begin{equation}
  n(x,y) =
  \begin{cases}
    n_3(x,y) & \text{if } n_3(x,y) > 0 \text{ and } n_1(x,y) > 0, \\
    0 & \text{if } n_3(x,y) \leq 0 \text{ or } n_1(x,y) = 0.
  \end{cases}
\end{equation}

The Landau matching condition \eqref{eq:landau} can be solved as $T^\mu_{\;\;\nu}u^\nu = e\,u^\mu$ by standard numerical eigensystem solvers.
In most cases, the resulting eigenvectors must be renormalized so that $u^\mu u_\mu = 1$.

\section{Viscous relativistic hydrodynamics}
\label{sec:hydro}

The workhorse of computational heavy-ion collision models, hydrodynamics is responsible for simulating the collective expansion of the hot and dense QGP medium and, via the equation of state, the transition to a hadron gas.
In the present work, the implementation of choice is the Ohio State University (2+1)-dimensional\footnote{
  Two spatial dimensions plus time, using boost invariance for the third (longitudinal) dimension.
}
viscous hydrodynamics code, originally published under the name {\small VISH2+1} \cite{Song:2007ux} and now updated to handle fluctuating initial conditions \cite{Shen:2014vra} and temperature-dependent shear and bulk viscosity \cite{Bernhard:2016tnd}.
My customized version is available at \url{https://github.com/jbernhard/osu-hydro}.

The boost-invariant approximation used in 2+1D hydrodynamics is a drawback for high-precision calculations, but the difference in midrapidity observables is small compared to full 3+1D calculations \cite{Vredevoogd:2012ui,Shen:2016zpp}, and 2+1D models run at least an order of magnitude faster.
Given a finite amount of computation time, many more events can be generated using a 2+1D model;
from the perspective of parameter estimation, this reduction in statistical uncertainty is more valuable than the marginal increase in accuracy gained by going full 3+1D.

Hydrodynamics models numerically solve the conservation equations (see section \ref{sec:properties})
\begin{equation}
  \partial_\mu T\mn = 0, \quad
  T\mn = e \, u^\mu u^\nu - (P + \Pi)\Delta\mn + \pi\mn,
\end{equation}
starting from initial profiles of the energy density $e$, flow velocity $u^\mu$, and viscous pressures $\pi\mn$ and $\Pi$ supplied by the initial condition and pre-equilibrium models;
the pressure $P$ relates to the energy density via the equation of state.
For the viscous pressures, the OSU code solves
\begin{equation}
  \begin{aligned}
    \tau_\pi\dot\pi^{\avg{\mu\nu}} + \pi\mn &=
      2\eta\sigma\mn
      - \delta_{\pi\pi}\pi\mn\theta
      + \phi_7\pi_\alpha^{\langle\mu}\pi_{}^{\nu\rangle\alpha} \\
      &{}\qquad - \tau_{\pi\pi}\pi_\alpha^{\langle\mu}\sigma_{}^{\nu\rangle\alpha}
      + \lambda_{\pi\Pi}\Pi\sigma\mn, \\[1ex]
    \tau_\Pi \dot\Pi + \Pi &=
      -\zeta\theta
      - \delta_{\Pi\Pi}\Pi\theta
      + \lambda_{\Pi\pi}\pi\mn\sigma_{\mu\nu},
  \end{aligned}
\end{equation}
which differ from equations \eqref{eq:viscous} by neglecting vorticity and some second-order terms, while retaining all first-order and shear-bulk coupling terms.
The viscosity coefficients $\eta$ and $\zeta$ are discussed in the next subsection;
the remaining transport coefficients and relaxation times are fixed to the values derived in the limit of small masses \cite{Denicol:2014vaa}.

\subsection{Parametrizations of temperature-dependent viscosity}

In order to estimate the temperature-dependent specific shear and bulk viscosity, $(\eta/s)(T)$ and $(\zeta/s)(T)$, we must parametrize them.

\begin{figure}[t]
  \makebox[\textwidth]{\includegraphics{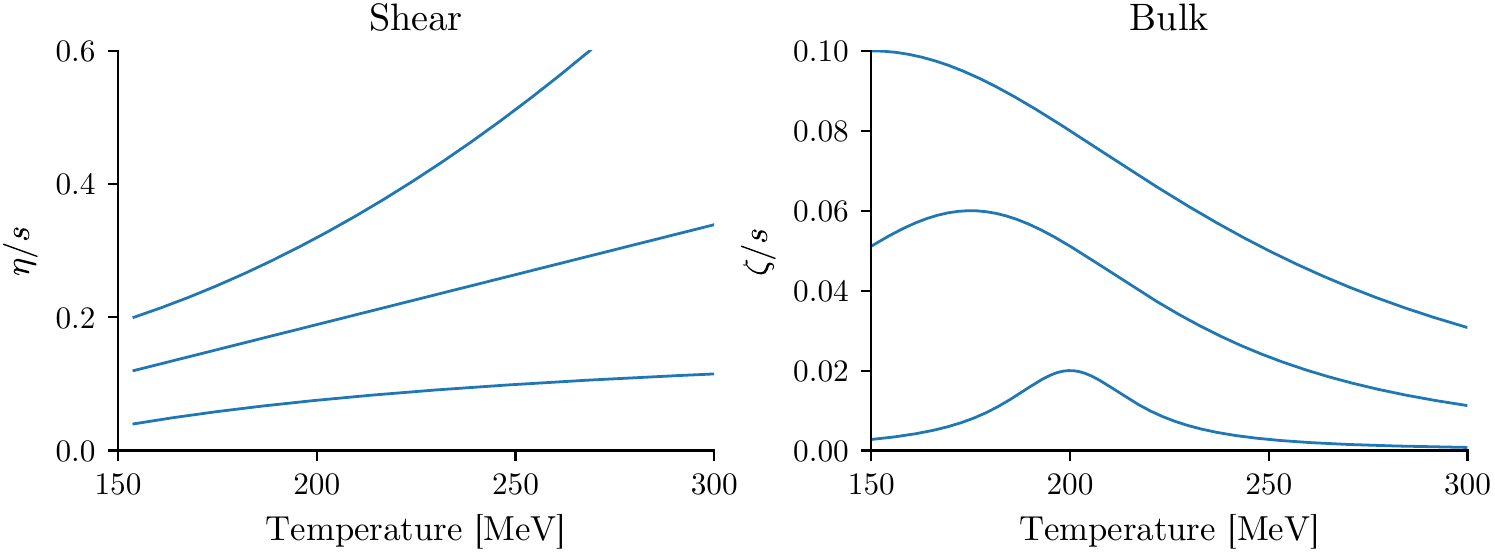}}
  \caption{
    Degrees of freedom of the temperature-dependent specific shear and bulk viscosity parametrizations, equations \eqref{eq:eta_s-T} and \eqref{eq:zeta_s-T}.
    The parameters for each curve are chosen for visual clarity and do not represent all possible variability, e.g.\ $\eta/s$ may have a large slope with negative curvature, or $\zeta/s$ may have a tall and narrow peak, neither of which are shown here.
  }
  \label{fig:viscosity-examples}
\end{figure}

The specific shear viscosity, as discussed in section \ref{sec:viscosity}, is expected to reach a minimum near the transition temperature $T_c$.
Above $T_c$, I use a modified linear ansatz
\begin{equation}
  (\eta/s)(T) = (\eta/s)_\text{min} + (\eta/s)_\text{slope} \cdot (T - T_c) \cdot (T/T_c)^{(\eta/s)_\text{crv}}
  \label{eq:eta_s-T}
\end{equation}
with three degrees of freedom:
a minimum value at $T_c$, a slope above $T_c$, and a curvature parameter (crv for short), which has intuitive meaning where zero curvature equates to a straight line and positive (negative) curvature to the function being concave up (down).
The left side of figure \ref{fig:viscosity-examples} shows these degrees of freedom.

In this parametrization, $\eta/s$ reaches its minimum value precisely at $T_c$, fixed to the HotQCD value 154 MeV \cite{Bazavov:2014pvz}.
But this may not exactly be the case;
consider that the minimum $\eta/s$ of other fluids can be located somewhat above or below the critical temperature (depending on the pressure), as shown in figure \ref{fig:specific-shear-viscosity}.
It would be reasonable to replace $T_c$ by a tunable parameter $T_0$ in \eqref{eq:eta_s-T}.

\label{loc:eta_s_hrg}

A fourth parameter $(\eta/s)_\text{hrg}$ sets a constant $\eta/s$ in the hadronic phase \emph{of the hydrodynamic model}, i.e.\ in the narrow temperature window below $T_c$ but before converting the medium to particles and switching to Boltzmann transport.
In practice, the value of this parameter matters little since it controls such a small fraction of the hydrodynamic evolution, and in any case, most flow develops at higher temperatures.
Note that $(\eta/s)_\text{hrg}$ is independent from $(\eta/s)_\text{min}$, so $\eta/s$ may be discontinuous at $T_c$, a feature observed in other fluids as shown in figure \ref{fig:specific-shear-viscosity}.

For the specific bulk viscosity, I use a three-parameter (unnormalized) Cauchy distribution
\begin{equation}
  (\zeta/s)(T) = \frac{(\zeta/s)_\text{max}}{1 + \biggl( \dfrac{T - (\zeta/s)_{T_0}}{(\zeta/s)_\text{width})} \biggr)^2},
  \label{eq:zeta_s-T}
\end{equation}
which is a symmetric peak with a tunable maximum, width, and location ($T_0$), shown on the right of figure \ref{fig:viscosity-examples}.
This form is qualitatively similar to the $(1/3 - c_s^2)^2$ dependence mentioned in section \ref{sec:viscosity}.

\subsection{Equation of state}

The hydrodynamic equation of state (EoS) shall consist of a lattice calculation for the high-temperature region and a hadron resonance gas (HRG) calculation at low temperatures.
Similar to previous work \cite{Huovinen:2009yb}, I construct this ``hybrid'' EoS by connecting the HRG and lattice trace anomalies in an intermediate temperature range;
the trace anomaly is the physical quantity computed directly on the lattice and may be integrated to obtain the pressure and other thermodynamic quantities.

I use the lattice EoS recently calculated by the HotQCD Collaboration \cite{Bazavov:2014pvz}, which they parametrize as
\begin{equation}
  \frac{P}{T^4} = \frac{1}{2} \bigl( 1 + \tanh[c_t(t - t_0)] \bigr)
  \biggl( \frac{p_{id} + a_n/t + b_n/t^2 + c_n/t^3 + d_n/t^4}{1 + a_d/t + b_d/t^2 + c_d/t^3 + d_d/t^4} \biggr),
\end{equation}
where $t = T/T_c$, $T_c = 154$ MeV, $p_{id} = 95\pi^2/180$ is the ideal gas value of $P/T^4$, and the fit coefficients are
\begin{align*}
  c_t &=  3.8706, &
  a_n &= -8.7704, &
  b_n &=  3.9200, &
  c_n &=       0, &
  d_n &=  0.3419, \\
  t_0 &=  0.9761, &
  a_d &= -1.2600, &
  b_d &=  0.8425, &
  c_d &=       0, &
  d_d &= -0.0475.
\end{align*}
This form is intended for differentiation; in particular, the trace anomaly is
\begin{equation}
  \frac{\Theta^{\mu\mu}}{T^4} = \frac{e - 3P}{T^4} = T \frac{d}{dT} \biggl( \frac{P}{T^4} \biggr).
  \label{eq:trace-deriv}
\end{equation}

There is some uncertainty in the lattice EoS, but the impact on actual observables is small:
A recent analysis of systematic uncertainties, using the HotQCD and Wuppertal-Budapest equations of state in hydrodynamic calculations, found $\sim$1\% differences in mean $p_T$ and $\sim$2--3\% in $v_2$ and $v_3$ \cite{Moreland:2015dvc}.

The HRG trace anomaly may be computed from the energy density and pressure
\begin{equation}
  e = \sumsp g \int \frac{d^3p}{(2\pi)^3} E \, f(p), \quad
  P = \sumsp g \int \frac{d^3p}{(2\pi)^3} \frac{p^2}{3E} f(p),
\end{equation}
where the sums run over all species in the hadron gas; $g$ and $f$ are the degeneracy and distribution function for each species.
See the next section for details on the hadron gas composition and particle distribution functions, which incorporate the effects of finite resonance width.

\begin{figure}[t]
  \makebox[\textwidth]{\includegraphics{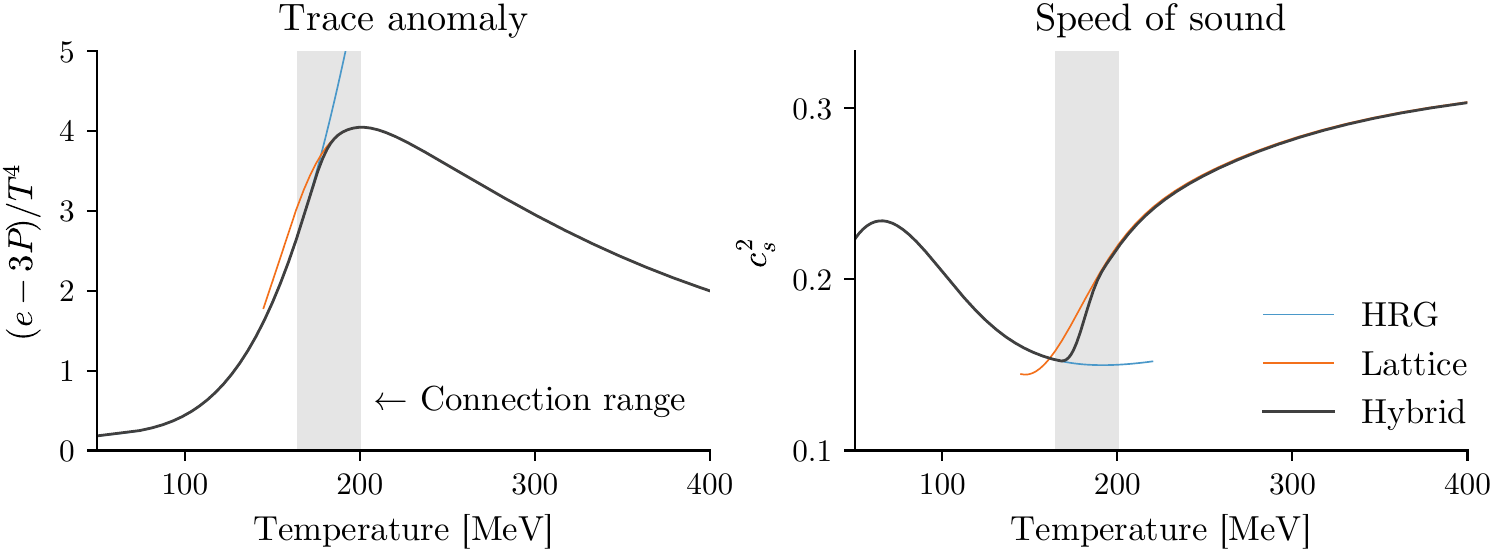}}
  \caption{
    Trace anomaly and speed of sound for the hybrid equation of state calculated using the HotQCD lattice EoS at high temperature and the HRG EoS at low temperature.
  }
  \label{fig:eos}
\end{figure}

\subsubsection{Procedure for constructing the hybrid EoS}

\begin{enumerate}
  \item Compute the HRG trace anomaly at an array of temperature points up to 165 MeV.
    This somewhat high temperature (above $T_c$) is necessary to ensure continuity of the EoS across the switch from hydrodynamics to Boltzmann transport---the EoS must exactly match the HRG calculation up to at least the maximum switching temperature, 165 MeV in the present work.
  \item Compute the lattice trace anomaly using equation \eqref{eq:trace-deriv} at an array of temperature points starting at 200 MeV.
  \item Connect the two curves between 165 and 200 MeV using a Krogh polynomial, which ensures continuity of the functions and their first several derivatives.
  \item Interpolate the trace anomaly across the full temperature range with a cubic spline.
  \item Numerically integrate the interpolating spline to obtain the pressure:
    \begin{equation}
      \frac{P(T)}{T^4} = \frac{P_0}{T_0^4} + \int_{T_0}^T dT' \frac{\Theta^{\mu\mu}}{T'^5},
    \end{equation}
    with reference pressure $P_0$ given by the HRG model at $T_0 = 50$ MeV.
  \item The energy density, entropy density $s = (e + P)/T$, and speed of sound $c_s^2 = \partial P/\partial e$ follow immediately.
\end{enumerate}
Figure \ref{fig:eos} shows the result.
An implementation of this procedure is included with my version of the OSU hydrodynamics code.

\section{Particlization}
\label{sec:particlization}

While hydrodynamics excels at modeling the high-temperature QGP, microscopic Boltzmann transport models are superior for the low-temperature hadron gas (I will justify this claim in the next section).
To switch to a microscopic model, the continuous hydrodynamic medium must be converted into an ensemble of discrete particles.
This process, ``particlization'', is a modeling artifact, distinct from the physical processes of hadronization and freeze-out, which is why such a neologism is necessary \cite{Huovinen:2012is}.
The physical system is the same before and after particlization; only the modeled representation changes.
In principle, there is a temperature window near the QCD crossover transition in which both hydrodynamics and microscopic transport are valid descriptions of the system, and it would be reasonable to particlize anywhere within this window.

\subsection{Cooper-Frye particle emission}

Particlization is performed on an isothermal spacetime hypersurface defined by a switching temperature $\Tsw$, a variable model parameter presumably close to $T_c$.
This four-dimensional surface encloses the spacetime region where $T > \Tsw$, which is modeled by hydrodynamics, and excludes the region where $T < \Tsw$, modeled by transport.
On the switching surface, particles are emitted with momentum distributions given by the Cooper-Frye formula \cite{Cooper:1974mv}
\begin{equation}
  E \frac{dN}{d^3p} = \frac{g}{(2\pi)^3} \int_\sigma f(p) \, p^\mu \, d^3\sigma_\mu,
\end{equation}
where the integral runs over the surface $\sigma$;
the integration element $d^3\sigma_\mu$ is a \emph{volume} element of the four-dimensional surface whose magnitude is its size and direction is normal to the surface.
In thermal equilibrium, the distribution function is a Bose-Einstein or Fermi-Dirac distribution
\begin{equation}
  f(p) = \frac{1}{\exp(p \cdot u/T) \mp 1},
\end{equation}
where $u$ is the velocity of the volume element; $p \cdot u$ is the energy, in the lab frame, of a particle with momentum $p$ in the frame of the volume element; and the upper sign corresponds to bosons, the lower to fermions.
Rearranging terms, it becomes apparent that the integrated yield is the total particle flux through the surface:
\begin{equation}
  N = \int_\sigma d^3\sigma_\mu \int g \, \frac{d^3p}{(2\pi)^3} \frac{f(p) \, p^\mu}{E},
\end{equation}
where the inner integral is effectively a particle four-current \cite{Huovinen:2012is}.
In the simple case of a single stationary volume element with zero normal vector, $d^3\sigma_\mu = (V, \mathbf 0)$, this reduces to something quite reasonable:
\begin{equation}
  N = V \int g \, \frac{d^3p}{(2\pi)^3} f(p) = V n,
\end{equation}
i.e.\ the product of the volume and the particle density.

In computational models, the Cooper-Frye integral is replaced by a sum over discrete volume elements,
\begin{equation}
  E \frac{dN}{d^3p} = \frac{g}{(2\pi)^3} \sum_\sigma f(p) \, p^\mu \, \Delta^3\sigma_\mu,
\end{equation}
with the elements and their normal vectors computed by a surface finding algorithm such as \textsc{Cornelius} \cite{Huovinen:2012is}.
To produce an ensemble of particles, momenta are randomly sampled from this function by treating it as a probability distribution.
In doing so, it is standard practice to discard particles with $p^\mu \, \Delta^3\sigma_\mu < 0$, meaning they are moving back into the hydrodynamic region; this is a physical effect but is difficult to model realistically.
Note also that the Cooper-Frye formula provides the \emph{average} number of emitted particles, which is in general not an integer.
A convenient way to convert the average to a discrete number of particles is to interpret it as the mean of a Poisson distribution.

I specify the complete sampling algorithm in subsection \ref{subsec:sampling-algorithm}, after addressing some other relevant aspects of particlization.

\subsection{Resonance width}

\newcommand{\Pm}{\mathcal P(m)}

Particlization models have traditionally neglected the width of resonances, instead assigning every sampled resonance its pole mass.
But it has been known for some time that accounting for finite width leads to increased pion production, especially at low $p_T$ \cite{Sollfrank:1991xm}, and a recent detailed study of the $\rho(770)$ resonance width confirmed the effect \cite{Huovinen:2016xxq}.

Why would this occur?
Consider that the density of a particle with mass $m_0$ is
\begin{equation}
  n = g \int \frac{d^3p}{(2\pi)^3} f(m_0, p), \quad
  f(m_0, p) = \frac{1}{e^{\sqrt{m_0^2 + p^2}/T} \pm 1},
\end{equation}
but if the particle is a resonance of finite width, then its mass probability distribution $\Pm$ must be integrated out of the distribution function:
\begin{equation}
  f(p) = \int dm \, \Pm \, f(m, p).
\end{equation}
Since mass is exponentially (rather than linearly) suppressed, the part of the distribution below $m_0$ is enhanced \emph{more} than the part above $m_0$ is suppressed.
The upshot is increased production of low-mass states relative to high-mass, and depending on the specific form of $\Pm$, a net change in overall production.

\begin{figure}[t]
  \makebox[\textwidth]{\includegraphics{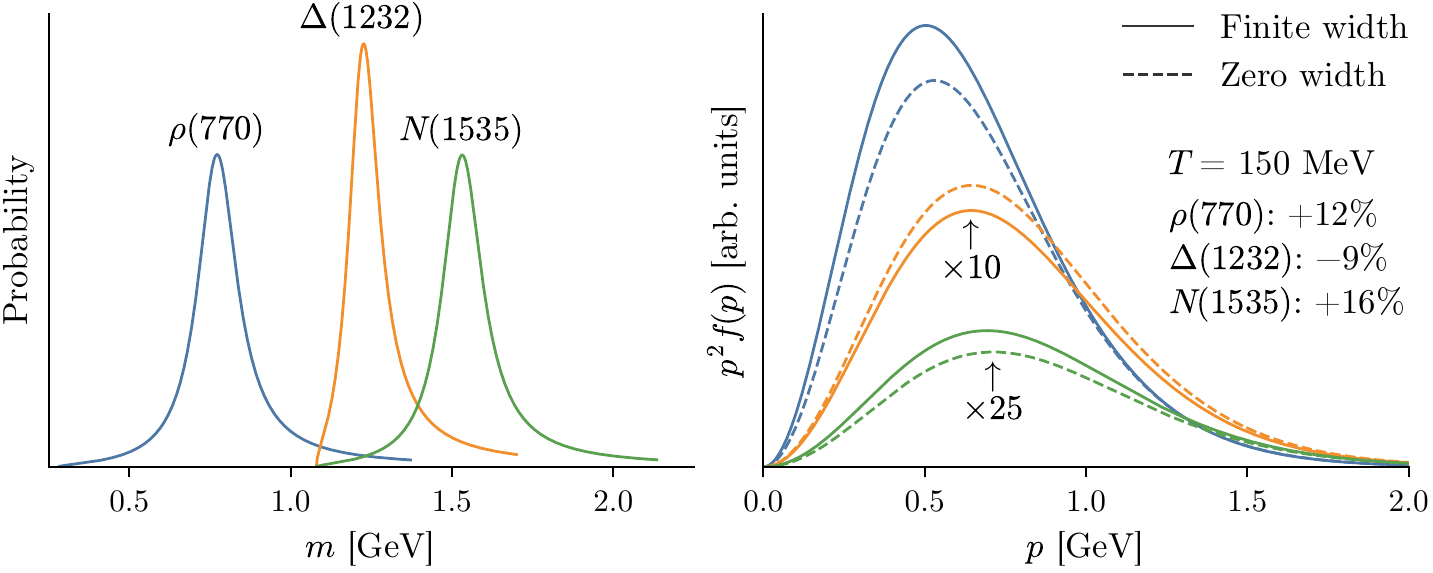}}
  \caption{
    Left: Mass distributions for the $\rho(770)$, $\Delta(1232)$, and $N(1535)$ resonances from equation \eqref{eq:mass-dist}.
    Right: Density distributions $p^2 f(p)$ including finite width (solid) and for zero width (dashed) at $T = 150$ MeV.
    Colors are the same as on the left.
    The $\Delta(1232)$ and $N(1535)$ distributions are scaled for visibility.
    Annotated are the relative changes in the particle densities from including finite width.
  }
  \label{fig:resonance-width}
\end{figure}

Given the precision goals of the present work, resonance width is too important to neglect; therefore, I randomly sample the masses of all (several hundred) resonances during particlization and allow the transport model to calculate their scatterings and decays as part of the full ensemble of particles.
I also account for finite width when calculating the HRG equation of state, as described in the previous section.
For the mass distribution, I assume a Breit-Wigner distribution with a mass-dependent width:
\begin{equation}
  \Pm \propto \frac{\Gamma(m)}{(m - m_0)^2 + \Gamma(m)^2/4}, \quad
  \Gamma(m) = \Gamma_0 \sqrt{\frac{m - m_\text{min}}{m_0 - m_\text{min}}},
  \label{eq:mass-dist}
\end{equation}
where $m_0$ and $\Gamma_0$ are the resonance's Breit-Wigner mass and width, the threshold mass $m_\text{min}$ is the total mass of the lightest decay products (e.g.\ $m_\text{min} = 2m_\pi$ for a resonance that can decay into a pair of pions), and the mass-dependent width $\Gamma(m)$ is designed to be physically reasonable and satisfy the constraints that $\Gamma(m_\text{min}) = 0$ and $\Gamma(m_0) = \Gamma_0$.
The distribution is normalized so that
\begin{equation}
  \int_{m_\text{min}}^{m_\text{max}} dm \, \Pm = 1, \quad
  m_\text{max} = m_0 + 4\Gamma_0.
\end{equation}
Figure \ref{fig:resonance-width} shows the mass distributions for several resonances and the impact on their densities.
The $\rho(770)$ and $N(1535)$ have roughly symmetric mass distributions and their densities significantly increase, especially at low momentum.
This is the general behavior of most species, but a minority of resonances, such as the $\Delta(1232)$, have their pole mass not far above their threshold mass, leading to asymmetric mass distributions with more weight above the pole mass, which reduces their total density.

The Breit-Wigner distribution \eqref{eq:mass-dist} is a simplifying assumption and is not accurate for all resonances.
But it is certainly closer to reality than assigning every resonance its pole mass, and the chosen mass-dependent width ensures that the distribution is physically reasonable.

It is difficult to predict the net effect of including finite resonance width on stable hadron yields, spectra, and other observables.
One likely consequence: Given the effects observed in figure \ref{fig:resonance-width}, and the fact that nearly all resonances decay to at least one pion, we can expect increased pion production relative to other species, especially at low $p_T$.

\subsubsection{The $f_0(500)$}

The $f_0(500)$ or $\sigma$ meson is an unusual resonance with a controversial history \cite{Pelaez:2015qba}.
It has an exceptionally small mass and large width, $m_0 = 475 \pm 75$ MeV and $\Gamma_0 = 550 \pm 150$ MeV in the 2017 Review of Particle Physics from the Particle Data Group (PDG) \cite{Patrignani:2016xqp}.
Since it is so light, it should be thermally produced in large quantities, and because it decays into pion pairs, should contribute significantly to the total pion multiplicity.

I include the $f_0(500)$ in the particlization routine, applying the same Breit-Wigner distribution with mass-dependent width as for all other resonances.
This is not formally correct---the $f_0(500)$ is known not to be a Breit-Wigner resonance---but it's preferable to neglecting the resonance or using only its pole mass.
Note that, with the chosen mass-dependent width and threshold mass $m_\text{min} = 2m_\pi \approx 280$ MeV, the mass distribution is not a typical Breit-Wigner peak, but a highly asymmetric distribution with a long high-mass tail (like the $\Delta(1232)$ distribution in figure \ref{fig:resonance-width} but even more extreme).

Another issue is that the $f_0(500)$ is unknown to many Boltzmann transport models, including the one used in this work.
To circumvent this, I decay each produced $f_0(500)$ into a pion pair before initializing the transport model.
This is physically justifiable since the resonance has such a short mean lifetime: about $10^{-24}$ seconds, or one-third \fmc, which is quick even on the timescale of heavy-ion collisions.

This scheme, while admittedly crude, should capture the basic physics of producing some pions that would otherwise be missing.

\subsection{Viscous corrections}

For the system to be physically self-consistent, the energy-momentum tensor $T\mn$ must be continuous across the transition from hydrodynamics to Boltzmann transport.
After particlization, kinetic theory gives (assuming a noninteracting hadron gas)
\begin{equation}
  T\mn = \sumsp g \int \frac{d^3p}{(2\pi)^3} \frac{p^\mu p^\nu}{E} f(p),
  \label{eq:Tuv-kinetic}
\end{equation}
where the sum runs over all species in the hadron gas; $g$ and $f$ are the degeneracy and distribution function of each species.
On the switching surface, the kinetic form must equal the hydrodynamic form
\begin{equation}
  T\mn = e \, u^\mu u^\nu - (P + \Pi)\Delta\mn + \pi\mn,
  \label{eq:Tuv-hydro}
\end{equation}
in particular, the sampled particles must have the same energy density, thermal pressure, and viscous pressures as the hydrodynamic medium.
Examining the kinetic form \eqref{eq:Tuv-kinetic}, it is clear that the only way to achieve continuity is to modify the distribution function $f(p)$;
if the hydrodynamic medium is out of thermal equilibrium, so should be the system of particles.

The standard modification to the distribution function is the addition of a small linear correction: $f = f_0 + \delta f$, where $f_0$ is the equilibrium Bose-Einstein or Fermi-Dirac distribution.
A simple form of the correction, derived from the Boltzmann equation using the relaxation time approximation (RTA) \cite{Dusling:2011fd,Teaney:2003kp,Bozek:2009dw}, is
\begin{equation}
  \delta f = f_0(1 \pm f_0) \frac{\tau}{E T} \biggl[
    \frac{1}{2\eta} p^i p^j \pi_{ij} +
    \frac{1}{\zeta} \biggl( \frac{p^2}{3} - c_s^2E^2 \biggr) \Pi
  \biggr],
  \label{eq:delta-f}
\end{equation}
where $\pi_{ij}$ is the shear tensor in the fluid rest frame\footnote{
  Per convention, Latin indices are purely spatial; vectors and tensors like $p^i$ and $\pi_{ij}$ represent only spatial components.
  The temporal components of $\pi_{\mu\nu}$ are zero in the rest frame, as required by orthogonality to the flow velocity, $\pi_{\mu\nu}u^\nu = 0$, which together with $u^\nu = (1, \mathbf 0)$ in the rest frame implies $\pi_{0\nu} = 0$.
}
and $\tau$ is a constant shear and bulk relaxation time for all species which gives the RTA its name.
But linear corrections break down for large viscous pressure and/or momentum;
eventually, $\delta f$ dominates the equilibrium distribution, invalidating the assumption of a ``small'' correction and sometimes causing unphysical negative densities ($f_0 + \delta_f < 0$).

An alternative method, which never causes negative densities, is to transform the momentum vector inside the distribution function as
\begin{equation}
  p_i \rightarrow p_i' = p_i + \sum_j \lambda_{ij} p_j, \quad
  \lambda_{ij} = (\lambda\shear)_{ij} + \lambda\bulk\delta_{ij},
  \label{eq:lambda}
\end{equation}
where $\lambda_{ij}$ is a linear transformation matrix, consisting of a traceless shear part and a bulk part proportional to the identity matrix, chosen to satisfy continuity of $T\mn$ \cite{Pratt:2010jt}.
This procedure lends itself naturally to computational particlization models: Simply sample momentum vectors from the equilibrium distribution and then apply the transformation.
I have adopted this general approach in this work.

\subsubsection{Shear corrections}

In the limit of small shear pressure, the shear transformation is \cite{Pratt:2010jt}
\begin{equation}
 (\lambda\shear)_{ij} = \frac{\tau}{2\eta} \pi_{ij},
  \label{eq:lambda-shear}
\end{equation}
where $\pi_{ij}$ is again the (spatial) shear pressure tensor in the local rest frame, and the ratio of the shear viscosity to the relaxation time in the noninteracting hadron gas model is
\begin{equation}
  \frac{\eta}{\tau} = \frac{1}{15T} \sumsp g \int \frac{d^3p}{(2\pi^3)} \frac{p^4}{E^2} f_0(1 \pm f_0).
\end{equation}
Inserting the transformed momentum vector $\mathbf p'$ into the equilibrium distribution and expanding for small $\lambda$ yields
\begin{equation}
  f_0(\mathbf p') - f_0(\mathbf p) \approx f_0(1 \pm f_0) \frac{\tau}{ET} \frac{1}{2\eta} p^i p^j \pi_{ij},
\end{equation}
which is the same as the shear part of $\delta f$ in equation \eqref{eq:delta-f} above,
hence, this ansatz is equivalent to the $\delta f$ correction for small shear pressure.

\begin{figure}[t]
  \makebox[\textwidth]{\includegraphics{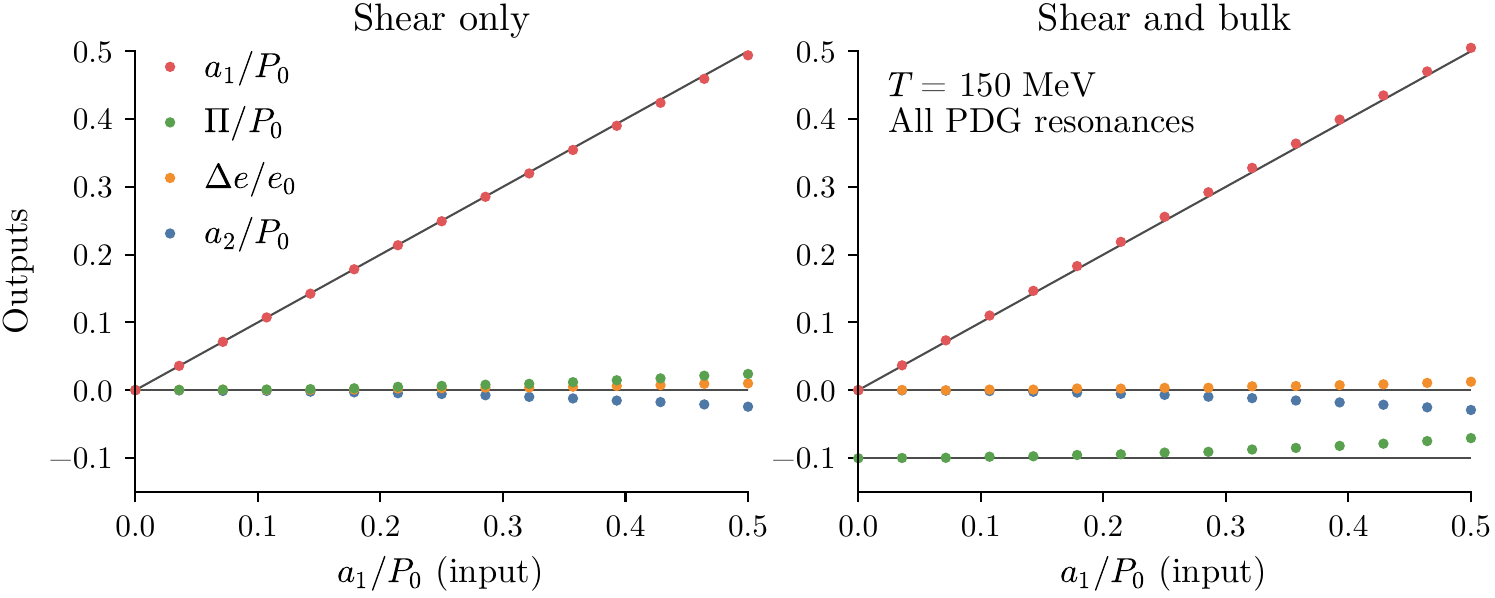}}
  \caption{
    Test of the viscous correction method.
    The input value of $a_1$, defined in equation \eqref{eq:a1}, is varied relative to the equilibrium pressure $P_0$ and the following output quantities are checked:
    $a_1/P_0$ itself, the bulk pressure $\Pi/P_0$, the change in the energy density $\Delta e/e_0$, and $a_2/P_0$ defined in equation \eqref{eq:a2}.
    Colored circles are the test calculations and lines are the targets.
    Left: zero bulk pressure, right: $\Pi = -0.1P_0$.
  }
  \label{fig:viscous-corrections}
\end{figure}

If this procedure works as intended, then given an input $\pi_{ij}$, the resulting sampled particles should actually have the specified $\pi_{ij}$.
In this vein, I have reproduced the test performed in \cite{Pratt:2010jt}, with the input $\pi_{ij}$ defined by
\begin{equation}
  a_1 = \pi_{xx} = -\pi_{yy} = \frac{T_{xx} - T_{yy}}{2},
  \label{eq:a1}
\end{equation}
and all other components set to zero.
For each value of $a_1$, I sample a large number of thermal particles, transform their momentum vectors by $(\lambda\shear)_{ij}$ as given in equation \eqref{eq:lambda-shear}, and compute the energy-momentum tensor
\begin{equation}
  T\mn = \frac{1}{V} \sum_\text{parts} \frac{p^\mu p^\nu}{E},
\end{equation}
where the sum runs over all sampled particles and $V$ is the volume of the thermal source.
From this, I calculate the output $a_1$ (which should equal the input), as well as the quantity
\begin{equation}
  a_2 = \frac{2T_{zz} - T_{xx} - T_{yy}}{\sqrt{12}},
  \label{eq:a2}
\end{equation}
which should be zero, and the energy density and pressure
\begin{equation}
  e = T_{tt} = \frac{1}{V} \sum_\text{parts} E, \quad
  P = \frac{T_{xx} + T_{yy} + T_{zz}}{3} = \frac{1}{V} \sum_\text{parts} \frac{p^2}{3E},
\end{equation}
which should not deviate from their equilibrium values $e_0$ and $P_0$.
In general, the pressure may deviate, the difference being the bulk pressure $\Pi$, but the bulk pressure is zero for this test.
The left panel of figure \ref{fig:viscous-corrections} shows the results;
$a_1$ is reproduced faithfully, with some small deviations in the other quantities for large $a_1$ relative to the equilibrium pressure.
This is expected since the procedure was derived in the limit of small shear pressure.
Note that the $\delta f$ correction would also induce deviations for large shear pressure because, as mentioned, it sometimes causes negative densities which are impossible to sample.
The right panel of the figure is the same test but with nonzero bulk pressure, which requires a separate correction that I will describe now.

\subsubsection{Bulk corrections}

The form of the bulk transformation in equation \eqref{eq:lambda} is $\lambda\bulk\delta_{ij}$, which translates to an overall scaling of the momentum: $p' = (1 + \lambda\bulk)p$.
As rationale, consider that the total effective pressure of the system is the sum of the thermal pressure $P$ and bulk pressure $\Pi$---as can be seen by how they enter the hydrodynamic energy-momentum tensor \eqref{eq:Tuv-hydro}---and the total kinetic pressure is
\begin{equation}
  P + \Pi = \sumsp g \int \frac{d^3p}{(2\pi)^3} \frac{p^2}{3E} f(p).
\end{equation}
This relation may be satisfied for a given bulk pressure by replacing $f(p) \rightarrow f(p') = f(p + \lambda\bulk p)$ and adjusting $\lambda\bulk$.
However, doing so would also change the energy density
\begin{equation}
  e = \sumsp g \int \frac{d^3p}{(2\pi)^3} E \, f(p),
\end{equation}
which should not deviate from its equilibrium value.
The energy density can be recovered by scaling the distribution function by a fugacity $z\bulk$, so that the complete replacement is $f(p) \rightarrow z\bulk f(p + \lambda\bulk p)$.
The parameters $\lambda\bulk$ and $z\bulk$ together account for bulk corrections and are uniquely determined by the requirement that the total pressure is reproduced without changing the energy density.

This parametric method, which I devised for use in computational particlization routines, is \emph{not} an approximation, but it does rely on some assumptions, namely:
It modifies the momentum distributions in a simple way, only by scaling the magnitude of momentum;
and it scales the density of all particle species by the same factor, maintaining their equilibrium ratios.
The physical interpretation is that bulk pressure implies a change in the momentum density of the system, and to conserve energy, the particle density must be adjusted accordingly.
Recall that the Navier-Stokes equation for bulk viscosity is $\Pi = -\zeta \, \del \cdot u$, where $\del \cdot u$ is the fluid expansion rate;
if the fluid is radially expanding (as is often the case in heavy-ion collisions), bulk viscosity acts as a kinetic ``brake'', reducing the momentum of emitted particles and converting that kinetic energy into increased production of low-momentum particles.

\begin{figure}[t]
  \makebox[\textwidth]{\includegraphics{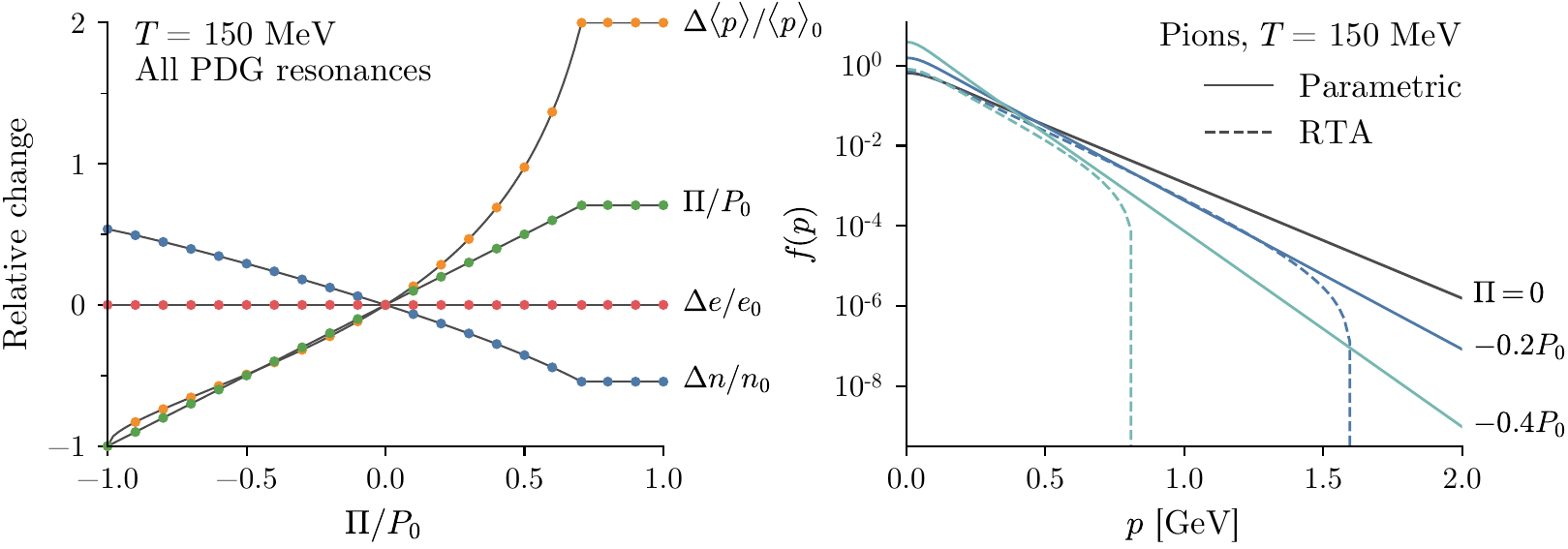}}
  \caption{
    Left: Effect of bulk pressure on thermodynamic quantities and verification of the parametric method.
    Shown are the changes in density, energy density, pressure, and mean momentum as a function of bulk pressure relative to the equilibrium pressure.
    Colored circles are test calculations from sampled particles and lines are the targets.
    Right: Effect of bulk pressure on the pion distribution function from the parametric method and the RTA.
  }
  \label{fig:bulk-parametric}
\end{figure}

The left panel of figure \ref{fig:bulk-parametric} verifies that the parametric method reproduces bulk pressure while preserving the energy density and shows the corresponding modifications to the particle density and mean momentum, which are closely related to the parameters $z\bulk$ and $\lambda\bulk$.
The method is accurate all the way down to $\Pi = -P_0$, meaning zero total pressure, at which point particles have zero momentum and all their energy is rest mass (this may not make much physical sense, but it works fine numerically).
For large positive bulk pressure, the mean momentum diverges and the corrections must be truncated, which is why everything becomes flat above $\Pi/P_0 \sim 0.7$.
This has negligible impact on heavy-ion collisions since very few volume elements have such large positive bulk pressure.

The right panel of the figure compares the modified distribution function from the parametric method and the RTA $\delta f$, equation \eqref{eq:delta-f}.
Calculating $\delta f$ requires the ratio of the bulk viscosity to the relaxation time
\begin{equation}
  \frac{\zeta}{\tau} = \frac{1}{3T} \sumsp g \int \frac{d^3p}{(2\pi^3)} \frac{m^2}{E^2} \biggl( c_s^2E^2 - \frac{p^2}{3} \biggr) f_0(1 \pm f_0)
\end{equation}
and the speed of sound
\begin{equation}
  c_s^2 = \frac{\partial P}{\partial e}
        = \frac{\partial P/\partial T}{\partial e/\partial T}
        = \frac{1}{3} \frac{
            \sumsp g \int d^3p \, p^2 f_0(1 \pm f_0)
          }{
            \sumsp g \int d^3p \, E^2 f_0(1 \pm f_0)
          }.
\end{equation}
Both methods generally decrease momentum with negative bulk pressure, but, importantly, the RTA distribution function goes (unphysically) negative for even moderately large bulk pressure and momentum.

Returning briefly to the previous figure \ref{fig:viscous-corrections}:
The right panel is a test of the complete viscous correction method with $\Pi = -0.1P_0$ and variable shear pressure.
As in the left panel, which has zero bulk pressure, there are some deviations for large $a_1$ input, but this is caused by the approximate shear correction method, not the parametric bulk method.

\subsection{Sampling algorithm}
\label{subsec:sampling-algorithm}

I have developed a new computational particlization model, available at \url{https://github.com/Duke-QCD/frzout}, with online code documentation including additional information and numerical tests.
The following summarizes the sampling algorithm.

\subsubsection{Preliminary steps}

\begin{enumerate}
  \item Choose a list of hadron species consistent with the Boltzmann transport model.
  \item Compute the density of each hadron species, including the effects of resonance width, using PDG data \cite{Patrignani:2016xqp}.
  \item Prepare for viscous corrections: For shear, compute $\eta/\tau$; for bulk, construct cubic interpolating splines that map $\Pi$ to the parameters $\lambda\bulk$ and $z\bulk$.
    These parameters are determined by the system of equations
    \begin{equation}
      \begin{aligned}
        P + \Pi &= z\bulk \sumsp g \int \frac{d^3p}{(2\pi)^3} \frac{p^2}{3E} f(p + \lambda\bulk p), \\
        e &= z\bulk \sumsp g \int \frac{d^3p}{(2\pi)^3} E \, f(p + \lambda\bulk p),
      \end{aligned}
    \end{equation}
    which can be inverted numerically, but it would be too slow to do so for every volume element, which would be necessary since each element in general has a different bulk pressure.
    Steps to create the interpolating splines:
    \begin{enumerate}
      \item Compute the equilibrium particle density $n_0$, energy density $e_0$, and pressure $P_0$.
      \item For an array of $\lambda\bulk$ values from $-1$ to $+2$, compute the resulting particle density $n$, energy density $e$, and pressure $P$.
        The $z\bulk$ necessary to preserve the equilibrium energy density is
        \begin{equation}
          z\bulk = \frac{\avg E_0}{\avg E} = \frac{e_0/n_0}{e/n},
        \end{equation}
        where $\avg E$ is the average energy per particle, and the resulting bulk pressure is
        \begin{equation}
          \Pi = P \frac{e_0}{e} - P_0.
        \end{equation}
      \item Interpolate the data points using $\Pi$ as the input variable and the bulk parameters as the outputs.
        The interpolating functions can then be evaluated quickly during the main sampling steps.
    \end{enumerate}
\end{enumerate}

\subsubsection{Main sampling steps}

Scott Pratt originally devised this algorithm \cite{Pratt:2010jt};
I wrote new code implementing it and made some minor modifications.

Rearranging the Cooper-Frye formula, the average number of particles emitted from a volume element $\Delta\sigma_\mu$ is
\begin{equation}
  \avg{dN} = \frac{p\cdot\Delta\sigma}{E} \frac{d^3p}{(2\pi)^3} \, g \, f(p)
           = \frac{p\cdot\Delta\sigma}{p \cdot u} \frac{d^3p'}{(2\pi)^3} \, g \, f(p'),
\end{equation}
where in the second form $p'$ is the momentum in the rest frame of the volume element.
Now multiplying and dividing by a volume $V$, this becomes
\begin{equation}
  \avg{dN} = w(p) \, V \frac{d^3p'}{(2\pi)^3} \, g \, f(p'), \quad
  w(p) = \frac{1}{V} \frac{p\cdot\Delta\sigma}{p \cdot u},
\end{equation}
where $w(p)$ is a particle emission probability.
The volume $V$ ensures $w(p) \le 1$; its optimal value is
\begin{equation}
  V = \max\biggl( \frac{p\cdot\Delta\sigma}{p \cdot u} \biggr)
    = u\cdot\Delta\sigma + \sqrt{(u\cdot\Delta\sigma)^2 - (\Delta\sigma)^2}.
\end{equation}
In view of these relations, the sampling algorithm is:
\begin{enumerate}
  \item Sample a particle four-momentum from a stationary thermal source of volume $V$.
    If the particle is a resonance, sample its mass in addition to the three-momentum.
  \item Apply the viscous correction transformation.
  \item Boost the momentum from the rest frame of the volume element, i.e.\ an inverse boost by four-velocity $u$.
  \item If $p\cdot\Delta\sigma < 0$, reject the particle, otherwise accept the particle with probability $w(p)$.
\end{enumerate}
This process should be repeated for each volume element and each species.
An efficient algorithm for achieving Poissonian particle production is:
\begin{enumerate}
  \item Initialize a variable $S$ with the negative of an exponential random number.
    Such a random number can be generated as $S = \log(U)$, where $U \in (0, 1]$ is a uniform random number.
  \item For each particle species in the hadron gas:
    \begin{enumerate}
      \item Add $V n$ to $S$, where $n$ is the density of the species, so $V n$ is the average number emitted from the volume.
        If the volume element has nonzero bulk pressure, determine the parameter $z\bulk$ and scale the density.
      \item If $S < 0$, continue to the next species, otherwise perform the above sampling algorithm and then subtract an exponential random number from $S$.
        Continue sampling particles and subtracting from $S$ until it again goes negative, then continue to the next species.
    \end{enumerate}
  \item Repeat for each volume element.
\end{enumerate}
This works because the time between Poisson events has an exponential distribution.

In boost-invariant hydrodynamics, the volume elements are $\Delta^3\sigma_\mu = \tau \, \Delta y \, \Delta^2\sigma_\mu$, where $\tau$ is the proper time of the element and $\Delta y$ is a rapidity range which must be chosen \emph{a priori}.
After accepting a particle in the above algorithm, its longitudinal momentum only determines the difference between the spacetime and momentum rapidity
\begin{equation}
  y - \eta_s = \frac{1}{2} \log\biggl( \frac{E + p_z}{E - p_z} \biggr),
\end{equation}
so some additional steps are required:
\begin{enumerate}
  \item Sample a momentum rapidity $y$ uniformly in the range $\Delta y$.
    Boost the particle's momentum vector longitudinally so that it has rapidity $y$.
  \item From $y$ and the difference $y - \eta_s$, calculate the spacetime rapidity $\eta_s$.
    Boost the particle's position vector longitudinally so that it has rapidity $\eta_s$, namely $t = \tau\cosh\eta_s$ and $z = \tau\sinh\eta_s$.
\end{enumerate}

There are many further subtleties which I omit here for brevity.
See the code documentation and comments for details.

\section{Boltzmann transport}
\label{sec:boltzmann}

After particlization, a Boltzmann transport model simulates the microscopic dynamics of the hadronic system, including scatterings and decays, until freeze-out.
As the name suggests, such models solve the Boltzmann equation
\begin{equation}
  \frac{df_i(x, p)}{dt} = \mathcal C_i(x, p),
\end{equation}
which stipulates that the time evolution of the distribution function $f_i$ for species $i$ is driven by the collision kernel, or source term, $\mathcal C_i$,
which accounts for collisions involving species $i$, including collisions with other species, so that the equations for each species are in general coupled.

The most widely used implementation of Boltzmann transport, and the present choice, is UrQMD (Ultra-relativistic Quantum Molecular Dynamics) \cite{Bass:1998ca,Bleicher:1999xi}.
UrQMD effectively solves the Boltzmann equation by propagating particles along classical (straight-line) trajectories, sampling their stochastic binary collisions, and calculating resonance formation and decays.
My version of UrQMD, tailored for use as a hadronic afterburner, i.e.\ as part of a multistage model following hydrodynamics and particlization, is available at \url{https://github.com/jbernhard/urqmd-afterburner}.

There are other Boltzmann transport implementations, but since the physics of hadronic scatterings and decays is well-understood, the priority is to use a stable, established code with a comprehensive set of hadronic resonances, which UrQMD satisfies.
A more recent model, SMASH (Simulating Many Accelerated Strongly-interacting Hadrons) \cite{Weil:2016zrk}, may ultimately replace UrQMD as the standard Boltzmann transport model for heavy-ion collisions, but at the time of this writing is not ready for production use.

\subsection{Advantages}

Microscopic transport models like UrQMD are ideal for modeling the late, hadronic stage of heavy-ion collisions.
There is no assumption of thermal equilibrium, the feed down of resonances to stable hadrons is calculated realistically, and the various stages of freeze-out arise naturally from the microscopic dynamics.
Chemical freeze-out may occur earlier and at a higher temperature than kinetic freeze-out, as is generally understood to happen in real collisions (see subsection \ref{subsec:particle-energy-production}).
Different species may kinetically freeze-out separately, for example because they have different scattering cross sections.

These models also innately account for hadronic transport properties, obviating the need to specify transport coefficients such as $\eta/s$ and $\zeta/s$.
In fact, the only free parameter is $\Tsw$, the particlization temperature.

\subsection{Limitations}

In the interest of computational tractability, the collision kernel usually includes only binary collisions and $2 \rightarrow n$ processes;
hence, microscopic transport is a valid description of the system provided it is sufficiently dilute that binary scatterings are the dominant process and higher-order scatterings are rare.
Hence, the system must have particle degrees of freedom and cannot be too hot and dense---this is why microscopic models are not suitable for the QGP phase.
As alluded to in previous section, in principle there is a temperature window near the QCD crossover transition in which the system is dense enough for hydrodynamics to apply, but not so dense as to invalidate the binary scattering picture.

\section{Comparing to experimental data}
\label{sec:comparing-expt}

The final step in the modeling workflow is to compute observables, such as multiplicities and anisotropic flow coefficients, for comparison with experimental observations.
I strive to replicate experimental data analysis methods as closely as possible.

\subsection{Centrality selection}

I run minimum-bias events (no centrality or impact parameter cuts), sort the events by charged-particle multiplicity $d\Nch/d\eta$ at midrapidity ($|\eta| < 0.5$), and apportion the events into the same centrality bins as the experimental data.
The definition of centrality by $d\Nch/d\eta$ is not exactly the same as most experiments, e.g.\ ALICE defines centrality by the energy deposited in its VZERO detectors \cite{Abelev:2013qoq}, which are not at midrapidity.
But this should not make much difference, since these measures of particle or energy production are strongly correlated.
In any case, since the present hydrodynamic model is boost-invariant, quantities away from midrapidity are fairly meaningless.

\subsection{Model observables}

After dividing the events into centrality bins, I compute observables from the particle data output by the final stage of the model (Boltzmann transport);
these virtual particles are analogous to their real counterparts recorded by an experimental detector.
I calculate quantities such as particle yields $d\Nch/d\eta$ and $dN/dy$, transverse energy $E_T$, and mean transverse momentum $\avg{p_T}$ by straightforward counting and averaging;
anisotropic flow cumulants $\vnk n k$ by the $Q$-cumulant method \cite{Bilandzic:2010jr} (see discussion on page \pageref{loc:flow-cumulants-Qn}).

It is always crucial to apply the same kinematic cuts as the experimental detector, for example ALICE measures flow cumulants using charged particles with $|\eta| < 0.8$ and $0.2 < p_T < 5.0$ GeV \cite{Adam:2016izf}.
Multiplicity and transverse momentum data are usually measured in the central rapidity unit, $|\eta| < 0.5$ or $|y| < 0.5$, and extrapolated to zero $p_T$ \cite{Abelev:2013vea}, so no $p_T$ cut is necessary when computing them from the model.

\subsection{Number of events}

How many events should one generate with the model?
It depends on the inherent statistical fluctuations in the desired observables:
Yields and mean $p_T$ converge quickly; flow cumulants are noisier and therefore require more events to stabilize.
I have found that ${\sim}2 \times 10^4$ minimum-bias events achieves acceptable statistical noise for two-particle flow cumulants in 10\% centrality bins.
Four-particle cumulants need more---at least $10^5$.

\subsection{Oversampling}

Since the hydrodynamic model usually takes much more time than the subsequent particlization and Boltzmann transport models, it is standard practice to run the particlization+transport combination multiple times per hydrodynamic evolution.
All particle data are then merged and used to compute low-noise observables.

This strategy, known as ``oversampling'' in reference to sampling the Cooper-Frye switching hypersurface, is advantageous because single events don't naturally produce enough particles to accurately measure their properties, and by sampling each event several times, more information can be extracted---without incurring much more computational cost.
Averaging over multiple samples certainly suppresses some event-by-event fluctuations, so one must take care that the observables of interest are not sensitive to these fluctuations.

To achieve a consistent statistical noise level across all events, I oversample each event until a target number of particles are emitted, which generally means more samples for peripheral events than central.
This is preferable to running a fixed number of samples, for then one would have to choose between wasting time running too many samples for central events, or having too few samples for peripheral events.

\subsection{Workflow for generating events}

I have developed a workflow for generating large quantities of heavy-ion collision events, available at
\url{https://github.com/Duke-QCD/hic-eventgen}.
It runs the five modules described in this chapter, implements the considerations just mentioned in this section, and provides utilities for running on high-performance computational systems, specifically the Open Science Grid (OSG) and the National Energy Research Scientific Computing Center (NERSC).
See the online documentation for details.

\chapter{Bayesian parameter estimation}
\label{ch:param-est}

\lettrine{N}{ot} coincidentally, the present situation conforms to the ``generic setup'' of the introduction (chapter \ref{ch:intro}):
We have assorted experimental observations of heavy-ion collisions (section \ref{sec:expt-obs}), some related properties of QCD matter that we wish to quantify (section \ref{sec:properties}), and a computational collision model which takes those properties as input parameters and produces simulated observables analogous to the experimental data (chapter \ref{ch:models}).

In order to rigorously quantify the model parameters---and further, to claim that they connect to genuine physical properties---the model must be a reasonable representation of real collisions, evidenced by a global fit to a wide variety of observables.
Complicating this endeavor is that each parameter is linked to multiple model observables, and vice versa;
for example, the specific shear viscosity $\eta/s$ affects the anisotropic flow coefficients $v_n$, but so too do the initial collision geometry and free-streaming time, which in turn also influence the transverse momentum distributions.
In general, it is safe to assume that all parameters affect every observable to some extent.
Undoubtedly, the only path to a global fit is a simultaneous treatment of all parameters and observables.

Putting aside how to achieve such a fit, it is essential to realize that parameters determined in this way are inherently uncertain.
Notable---and unavoidable---sources of uncertainty include measurement errors in the experimental data itself, the complex interplay among model parameters, and discrepancies between the model calculations and the data.
Thus, the objective is a quantitative \emph{estimate} of each parameter, including the associated uncertainties.

Bayesian statistics offers a natural framework for parameter estimation and uncertainty quantification, wherein the final result is a posterior probability distribution expressing the likely true values of the parameters.
The general approach is as follows:
Let the model parameters of interest be a vector $\xv = (x_1, x_2, \ldots, x_n)$ and denote the experimental data vector by $\yv$.
We then define the prior $P(\xv)$, a probability distribution encoding our initial knowledge of the parameters, and the likelihood $P(\yv|\xv)$, a conditional probability that quantifies the quality of the fit to data, accounting for all sources of uncertainty, given the parameters $\xv$.
Next, we apply Bayes' theorem to obtain the posterior distribution
\begin{equation}
  P(\xv|\yv) \propto P(\yv|\xv) \, P(\xv),
\end{equation}
which encapsulates all our knowledge of the parameters given the prior and the data.
Usually, we are interested in the marginal distributions for each parameter, calculated by marginalizing over (integrating out) all the rest, for example the marginal distribution for $x_1$ is
\begin{equation}
  P(x_1|\yv) = \int dx_2 \cdots dx_n \, P(\xv|\yv).
  \label{eq:marginal}
\end{equation}
From this, we can derive the desired estimate and uncertainty of $x_1$.

In a prominent application of this methodology, the Laser Interferometer Gravitational-Wave Observatory (LIGO) Scientific Collaboration has estimated properties of binary black hole and neutron star mergers from gravitational wave observations \cite{Aasi:2013jjl,Veitch:2014wba,TheLIGOScientific:2016wfe,TheLIGOScientific:2017qsa}.
Matching numerical relativity calculations to the observed gravitational waveforms, they extracted model parameters including the masses and spins of the progenitor objects and the final object.
Figure \ref{fig:ligo-posterior} shows the posterior distributions for the source-frame black hole masses in merger event GW150914 \cite{TheLIGOScientific:2016wfe,Abbott:2016blz}, from which they derived $m_1^\text{source}/M_\odot = 36^{+5}_{-4}$ and $m_2^\text{source}/M_\odot = 29^{+4}_{-4}$ ($M_\odot$ is the solar mass), where the reported values are the posterior medians and the uncertainties are 90\% credible intervals.
This means that, e.g., 90\% of the posterior density lies between $32 < m_1^\text{source}/M_\odot < 41$;
based on all the available information, there is a 90\% chance that the true value of $m_1^\text{source}$ lies within this range.

The figure also shows the joint probability distribution between the two masses, obtained from a marginalization integral similar to \eqref{eq:marginal}, but integrating out all but two parameters, instead of all but one.
From this visualization, we see that the estimates of the two masses are strongly correlated:
Large $m_1$ implies small $m_2$, and vice versa.
This suggests that the total mass is better constrained than the individual masses, and indeed, the reported total is $M^\text{source}/M_\odot = 65^{+5}_{-4}$, which has less relative uncertainty than $m_1$ and $m_2$.
In effect, the ambiguity in the mass apportionment contributes to the mutual uncertainty of both parameters.
If we later determined that $m_2$ is toward the lower end of its credible interval, we would then believe that $m_1$ is on the large side.
This mutual uncertainty is a typical characteristic of correlated parameter estimates.

\begin{figure}[t]
  \graphicsandcaption{.7}{etc/ligo_posterior}{
    Posterior distributions for the source-frame component masses of black hole merger GW150914 \cite{TheLIGOScientific:2016wfe,Abbott:2016blz}.
    The one-dimensional histograms are marginal distributions for the masses, where the colored lines correspond to different waveform models and the black line is the overall (average) result, and the dashed lines indicate the 90\% credible interval.
    The two-dimensional density plot is the joint distribution between the two masses with credible region contours (the sharp cut is due to the convention $m_2^\text{source} \le m_1^\text{source}$).
  }
  \label{fig:ligo-posterior}
\end{figure}

The one-dimensional marginal distributions in figure \ref{fig:ligo-posterior} are histograms, not smooth curves, because they were not actually obtained from direct calculation of marginalization integrals like equation \eqref{eq:marginal}.
In general, it is more computationally efficient and convenient to generate a large sample of the posterior distribution through Markov chain Monte Carlo (MCMC) sampling, after which marginalization is trivial:
Simply take the value of the desired parameter from each parameter vector in the sample.
If the sample of parameter vectors is $\{\xv_i\}$, with each $\xv_i = (x_{1i}, x_{2i}, \ldots, x_{ni})$, then the sample of $x_1$ is just $\{x_{1i}\}$.

Generating every parameter sample $\xv_i$ entails a model evaluation---a serious obstacle if the model is computationally expensive.
This is certainly the case for heavy-ion collisions:
Calculating the centrality dependence of bulk observables requires $\order{10^4}$ minimum-bias events, and at $\order{10^{-1}}$ hours per event, this works out to $\order{10^3}$ hours per parameter sample.
Assuming a statistically significant sample size $\order{10^6}$, the total computation time would be $\order{10^9}$ hours, which is out of the question---even the largest NERSC allocations provide ``only'' $\order{10^7}$ hours.

To circumvent this obstacle, we use a model emulator to predict the output of the full model in much less time than an explicit calculation.
The strategy, developed specifically for this type of scenario \cite{Kennedy:2001bc,OHagan:2006ba,Higdon:2008cmc,Higdon:2014tva}, proceeds by evaluating the model at a relatively small $\order{10^2}$ number of points in parameter space, training an emulator on the model input-output data, and then using the emulator as a fast surrogate to the full model during MCMC sampling.
This reduces the computation time requirement by several orders of magnitude, more than making up for the disparity.

The canonical choice for model emulators are Gaussian processes \cite{Rasmussen:2006gp}, statistical objects that can non-parametrically interpolate multidimensional functions.
When carefully constructed, Gaussian processes are sufficiently flexible to emulate a wide variety of models, and since they provide the uncertainty of their predictions, are ideal for parameter estimation with quantitative uncertainties.

Bayesian parameter estimation using Gaussian process emulators has been successfully deployed in heavy-ion physics \cite{Novak:2013bqa,Pratt:2015zsa,Sangaline:2015isa}, including my own previously published work \cite{Bernhard:2015hxa,Bernhard:2016tnd,Bernhard:2017vql,Bass:2017zyn}, and in numerous other fields, such as galaxy formation history \cite{Gomez:2012ak}.

\begin{figure}[t]
  \makebox[\textwidth]{
    \begin{tikzpicture}
      \foreach \nodename/\xy/\title/\text in {
          input/{4.5,6.6}/Input parameters/QGP properties,
          model/{8.5,4.6}/Model/Heavy-ion collision \\ spacetime evolution,
          gp/{1.3,4.6}/Gaussian process emulator/Surrogate model,
          cal/{3.8,2.2}/Bayesian calibration/Infer model parameters \\ from data,
          posterior/{8.5,0}/Posterior distribution/Quantitative estimates \\ of each parameter,
          exp/{0,0}/Experimental data/Heavy-ion collision \\ observables
        }
        \node[align=center, inner sep=1.6ex, fill=theblue!10, text=theblue!90!black]
          (\nodename) at (\xy) {\textbf{\title} \\[.3ex] \text};
      \begin{scope}[color=black!70, ->, semithick]
        \draw (input) -| (model);
        \draw (model) -- (gp);
        \draw (input) -| (gp);
        \draw
          let \p1 = (gp.south), \p2 = (cal.north), \p3 = ($(\p1)!.5!(\p2)$) in
          (\p1) -- (\x1, \y3)  -| (\p2);
        \draw (exp) |- (cal);
        \draw (cal) -| (posterior);
      \end{scope}
    \end{tikzpicture}
  }
  \caption{
    Overview of the parameter estimation process.
  }
  \label{fig:param-est-flowchart}
\end{figure}
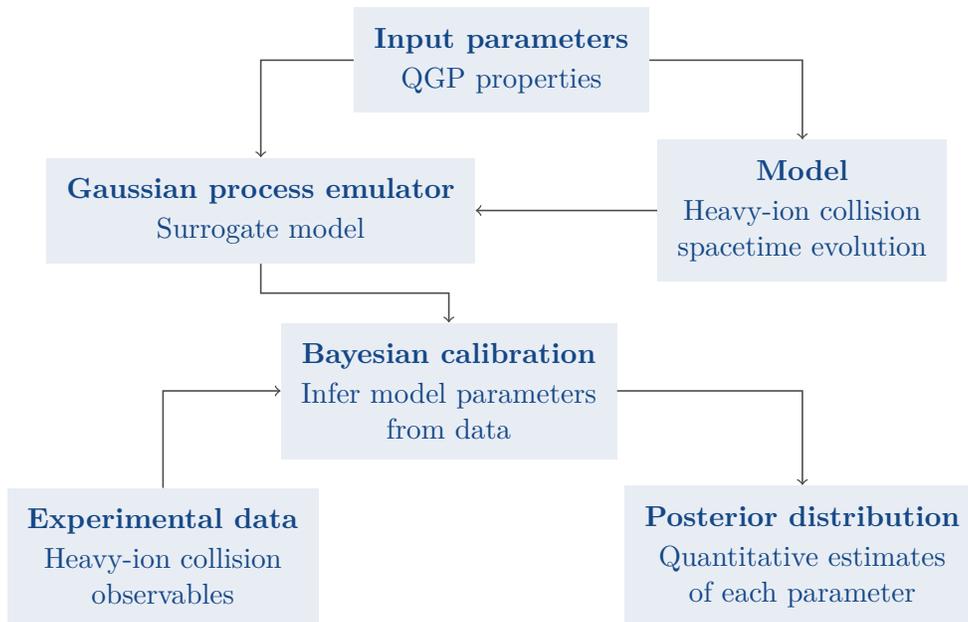

In this chapter, I fully develop the parameter estimation procedure, represented graphically in figure \ref{fig:param-est-flowchart}.
Sections \ref{sec:design} and \ref{sec:postprocess} address the model inputs and outputs, respectively;
I elaborate on the choice of input parameters, their distribution in parameter space, and postprocessing of model calculations for emulation.
In section \ref{sec:gp}, I discuss the theory of Gaussian processes and the practical details of building model emulators.
In section \ref{sec:calibration}, I expand upon model calibration, including MCMC sampling, construction of the posterior distribution, and uncertainty quantification.
Lastly, in section \ref{sec:comp-impl}, I point out my computer code implementing Bayesian parameter estimation for heavy-ion collisions.

\section{Parameter design}
\label{sec:design}

The goal of this section is to choose $n$ model input parameters for estimation, $\xv = (x_1, x_2, \ldots, x_n)$, and $d$ points\footnote{
  I use $d$ for the number of design points because $m$ shall be the number of model outputs.
  Mnemonic: $d \rightarrow$ \textbf{d}esign points, $m \rightarrow$ \textbf{m}odel outputs.
}
in parameter space at which to evaluate the full model, arranged into a $d \times n$ design matrix $X = (\xv_1, \xv_2, \ldots, \xv_d)$.
These choices will have important downstream consequences for uncertainty quantification and the performance of the Gaussian process emulator.

\subsection{Choice of parameters}

The adage ``as much as necessary, as little as possible'' is sometimes invoked regarding antibiotics, meaning that they should be used to treat bacterial illness, but avoided to prevent antibiotic resistance.
Similar considerations apply here, although the potential repercussions are, fortunately, much less dire.

Any parameter that \emph{might} have a \emph{meaningful} impact on the model calculation should be included in the design.
Physical properties certainly satisfy this criterion, but all parameters need not have a direct physical connection.
It is important to vary parameters that could change the behavior of the model, even if we don't care about their optimal values, in order to propagate their uncertainty to the parameters we \emph{do} care about.
We saw in the LIGO example, figure \ref{fig:ligo-posterior}, how parameters can contribute to their mutual uncertainties through marginalization.
Fixing a parameter to a nominal value---even a model-dependent nuisance parameter---can artificially bias the results for other parameters.

At the same time, we should not get carried away introducing frivolous parameters.
That is why I say a \emph{meaningful} impact, though what is meaningful is of course subjective.

Sometimes, we may not know whether a parameter will affect the model.
When in doubt, it is usually better to include such parameters in the design rather than risk bias.
The primary drawback of adding parameters is that, as we shall see in the next subsection, more parameters require more design points, which means more computation time.

In summary: As many parameters as necessary, as few as possible.

\subsection{Distribution in parameter space}

Having chosen a set of parameters to estimate, we must now decide the number of design points and their locations in parameter space.
The guiding motive is to create an efficient scaffolding of parameter space for emulation using as few design points as possible.

First, we specify ranges, i.e.\ minimum and maximum values, for each parameter.
Effectively, this imposes a prior distribution which is \emph{zero} outside the design range, a very strong assumption.
The ranges should therefore enclose any possibly reasonable values, erring on the side of generosity rather than risking truncation.

Sometimes, choosing the ranges is a somewhat paradoxical problem, where part of the reason for performing parameter estimation is to determine reasonable ranges.
One strategy I have used in this case is first performing a coarse-grained ``pilot study'', that is, running a wide design range with few design points and low statistics.
Based on the resulting low-precision posterior distribution, adjust the parameter ranges as necessary and re-run with normal precision.

\begin{figure}[t]
  \makebox[\textwidth]{\includegraphics{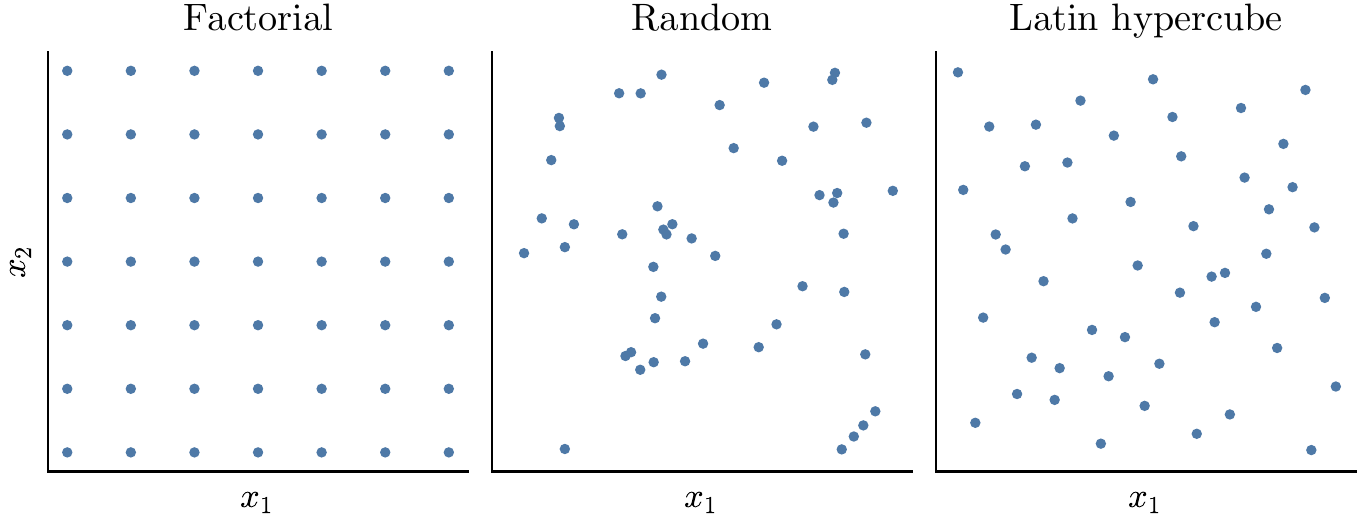}}
  \caption{
    Examples of factorial, random, and maximin Latin hypercube designs in two dimensions.
    Each has $7^2 = 49$ points.
  }
  \label{fig:design}
\end{figure}

Now, how many design points, and where?
Figure \ref{fig:design} shows three possible strategies.
Factorial design, in which points are placed on a uniform lattice, is an obvious choice in low dimensions, but fails in high dimensions.
A factorial design of $k$ points in each of $n$ dimensions has $k^n$ total points---far too many even for a modest $k = 10$ and $n > 2$ or 3.

Another simple design is purely random points.
This is certainly more reasonable than factorial design in high dimensions, but still suboptimal, because there's no guarantee that random points fill the space.
Often, purely random samples leave large regions with no points, which will preclude accurate emulation in those empty regions.

A common method for generating semi-random, space-filling designs is Latin hypercube sampling \cite{Tang:1993lh,Morris:1995lh}, in which points are generated in an $n$-dimensional unit hypercube, $[0, 1]^n$ (which can then be scaled to the desired ranges), such that if each dimension is divided into equal subintervals, there is exactly one point in each subinterval (like a Sudoku grid).
For example, four points in two dimensions could be distributed like this:
\begin{center}
  \begin{tikzpicture}[scale=.7]
    \draw (0, 0) grid (4, 4);
    \foreach \xy in {(.5, 1.5), (1.5, 2.5), (2.5, .5), (3.5, 3.5)}
      \fill \xy  circle [radius=.08];
  \end{tikzpicture}
\end{center}
Such designs provide the desired efficient scaffolding because they uniformly fill the space with relatively few points;
the required number of points grows only \emph{linearly} with the number of dimensions.
As a rule of thumb, 10 points per dimension yields acceptable emulation accuracy \cite{Loeppky:2009ss}, although more is always better if computation time permits.
I usually aim for at least 20 per dimension.

In using a sparse, space-filling design, we implicitly assume that the model is well-behaved and smoothly-varying, which is indeed almost always the case for physical models.
This implies that nearby parameter points produce similar model output, so running multiple nearby points would be redundant.
To improve computational efficiency, we impose a ``maximin'' criterion to the Latin hypercube sample, which \emph{maximizes} the \emph{minimum} distance between points.

A numerical library for generating Latin hypercube samples is publicly available \cite{Carnell:2018lhs}.
I have used it in my work, but I have not contributed to it.

Comparing the example designs in figure \ref{fig:design}, we see that the maximin Latin hypercube fills the space much more uniformly than the random design, with smaller gaps and no points on top of each other.
Note that, in two dimensions, the factorial design may actually be the best choice, but as explained above, it is not a viable option in higher dimensions.

One final note:
It is sometimes desirable to nonlinearly transform a parameter so that it affects the model smoothly across its range.
Like a shower hot water knob, we want linear behavior as we turn the virtual knob of each parameter.
In particular, this will facilitate training the Gaussian process emulator.

\subsection{Design for the present study}
\label{subsec:design-present-study}

I use a 500 point maximin Latin hypercube design, repeated for Pb-Pb collisions at 2.76 and 5.02 TeV, for 1000 total design points.

\subsubsection{Initial condition}

\begin{enumerate}
  \item Normalization factor for the initial density profile (different normalization for each beam energy).

  \item \trento\ entropy deposition parameter $p$ defined in equation \eqref{eq:tr}.
    With a free-streaming stage, the initial condition provides the transverse density of partons, parametrized as
    \begin{equation}
      n = \text{Norm} \times \biggl( \frac{\T_A^p + \T_B^p}{2} \biggr)^{1/p},
    \end{equation}
    where $\T$ is a participant thickness function.

  \item Gaussian nucleon width $w$ of the nucleon thickness function
    \begin{equation}
      T_p(x, y) = \frac{1}{2\pi w^2} \exp\biggl( -\frac{x^2 + y^2}{2w^2} \biggr).
    \end{equation}
    See also equations \eqref{eq:Tnucleon} and \eqref{eq:rhoproton}.

  \item Standard deviation of nucleon multiplicity fluctuations $\sigma_\text{fluct} = 1/\sqrt k$, where $k$ is the shape parameter of the gamma distribution, equation \eqref{eq:gamma}, reproduced here:
    \begin{equation}
      P_k(u) = \frac{k^k}{\Gamma(k)} u^{k-1} e^{-ku}.
    \end{equation}
    The fluctuated participant thickness functions are
    \begin{equation}
      \T_A = \sum_{i=1}^{N_\text{part,A}} u_i \, T_p(x - x_i, y - y_i),
    \end{equation}
    where $(x_i, y_i)$ is the transverse position of nucleon participant $i$ in nucleus $A$, and the $u_i$ are sampled from the gamma distribution.
    I use the standard deviation $\sigma_\text{fluct}$, instead of $k$ itself, because it is more intuitive and allows setting $k$ to very large values ($\sigma_\text{fluct} \rightarrow 0$, $k \rightarrow \infty$), which effectively disables fluctuations.

  \item Minimum distance between nucleons $\dmin$ (subsection \ref{subsec:dmin}), transformed to the volume $\dmin^3$.
\end{enumerate}

\subsubsection{Pre-equilibrium}

\begin{enumerate}[resume]
  \item Free-streaming time $\tfs$ (section \ref{sec:pre-eq}).
\end{enumerate}

\subsubsection{QGP medium}

\begin{enumerate}[resume]
  \stepcounter{enumi}
  \setcounter{enumisave}{\value{enumi}}
  \addtocounter{enumi}{2}
\item[\theenumisave--\theenumi.] $\eta/s$ min, slope, and curvature, which set the temperature dependence of the QGP specific shear viscosity in equation \eqref{eq:eta_s-T}, reproduced here:
    \begin{equation}
      (\eta/s)(T) = (\eta/s)_\text{min} + (\eta/s)_\text{slope} \cdot (T - T_c) \cdot (T/T_c)^{(\eta/s)_\text{crv}}
    \end{equation}

  \item Constant value of $\eta/s$ in the hadronic phase \emph{of the hydrodynamic model} (see discussion on page \pageref{loc:eta_s_hrg}).

  \stepcounter{enumi}
  \setcounter{enumisave}{\value{enumi}}
  \addtocounter{enumi}{2}
\item[\theenumisave--\theenumi.] $\zeta/s$ max, width, and location ($T_0$), which set the temperature dependence of the QGP specific bulk viscosity in equation \eqref{eq:zeta_s-T}, reproduced here:
    \begin{equation}
      (\zeta/s)(T) = \frac{(\zeta/s)_\text{max}}{1 + \biggl( \dfrac{T - (\zeta/s)_{T_0}}{(\zeta/s)_\text{width})} \biggr)^2}.
    \end{equation}

  \item Particlization temperature $\Tsw$ (section \ref{sec:particlization}).
\end{enumerate}

\section{Postprocessing model output}
\label{sec:postprocess}

Generically, the computational model takes a vector of $n$ input parameters $\xv = (x_1, x_2, \ldots, x_n)$ and produces a vector of $m$ outputs $\yv = (y_1, y_2, \ldots, y_m)$.
For a heavy-ion collision model, each of the outputs $y_i$ is an observable in a particular centrality class or kinematic bin ($p_T$ or $\eta$).
If the outputs include the centrality and/or kinematic dependence of several observables, the total number $m$ quickly becomes quite large.

As we shall see in the next section, Gaussian processes (which we will use to interpolate the model output) are scalar functions, i.e.\ they map a vector input to a single output.
The na\"ive way to handle all the model outputs is to use $m$ independent Gaussian processes, but this could be computationally expensive, and it ignores correlations among the outputs.
Since physical models generally produce many highly-correlated outputs, this is unsatisfactory.

Instead, we transform the model outputs into a smaller number of uncorrelated variables using principal component analysis (PCA), then treat each new variable independently.

\subsection{Principal component analysis}

\begin{figure}[t]
  \makebox[\textwidth]{\includegraphics{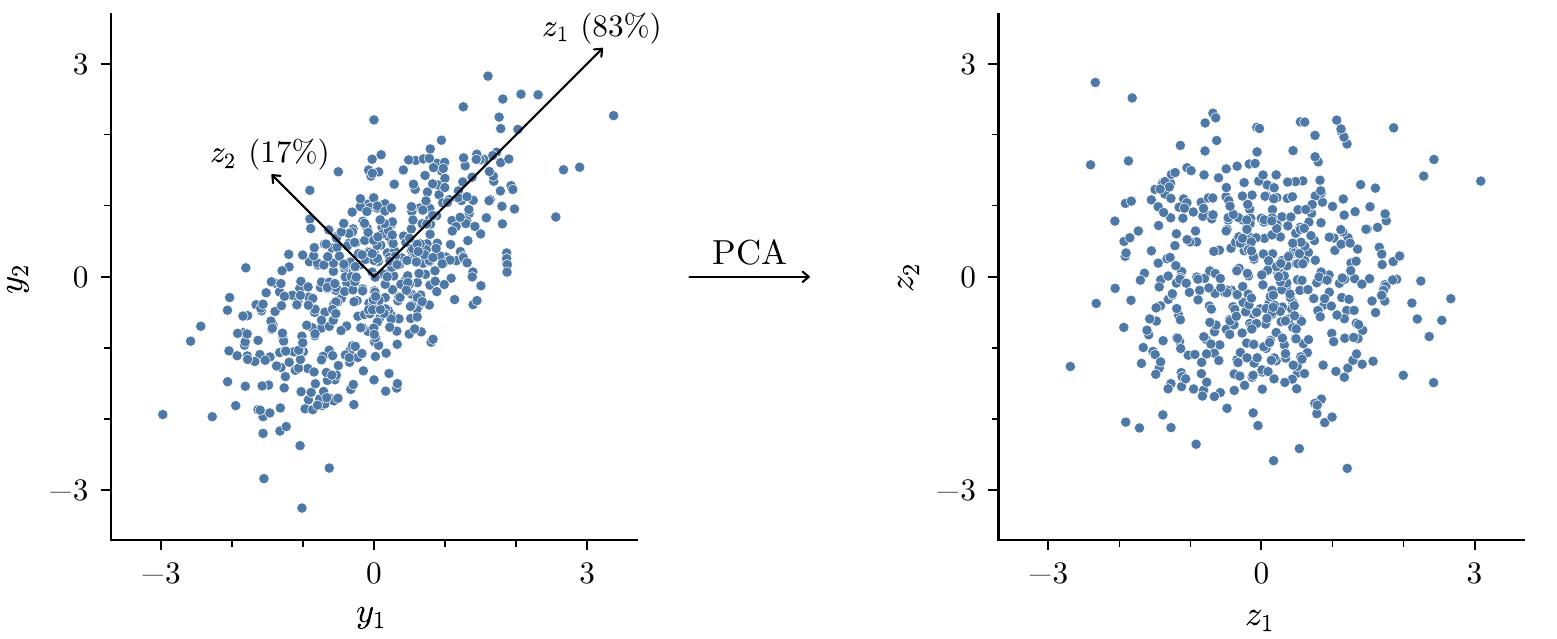}}
  \caption{
    Principal component analysis (PCA) transformation of two correlated variables $(y_1, y_2)$ into linearly uncorrelated variables $(z_1, z_2)$.
    On the left, arrows represent the principal component vectors, with labels including the fraction of explained variance.
  }
  \label{fig:pca}
\end{figure}

PCA \cite{Tipping:1999pca} is a general procedure that defines an orthogonal linear transformation from a set of correlated variables to a new set of linearly uncorrelated variables, aptly called principal components (PCs), which explain the maximum possible variance of the original data.
Figure \ref{fig:pca} shows a typical PCA transformation of two (randomly generated) correlated variables $(y_1, y_2)$;
it is effectively a rotation around the empirical mean into a different orthonormal basis.
The first PC, $z_1 \approx (y_1 + y_2)/\sqrt2$, explains over 80\% of the original variance---in other words, based on the observed correlation of $y_1$ and $y_2$, their sum contains most of the information about the individual variables.
Meanwhile, the second PC, $z_2 \approx (y_1 - y_2)/\sqrt2$, is orthogonal to the first and accounts for the remaining variance (i.e.\ information).

In the present situation, the original variables are the $m$ model outputs $\yv = (y_1, y_2, \ldots, y_m)$, to be transformed into the principal components $\zv = (z_1, z_2, \ldots, z_m)$, where each $z_i$ is a linear combination of the $y_i$.
To construct the PCA transformation, we first concatenate all the model outputs into an $d \times m$ matrix $Y = (\yv_1, \yv_2, \ldots, \yv_d)$ whose rows correspond to design points and columns to model outputs.
Each vector $\yv_i = (y_{1i},y _{2i}, \ldots, y_{mi})$ contains the $m$ model outputs at the $i$th design point, $y_{ji}$ being the $j$th model output at design point $i$.
We then standardize the data by centering and scaling each column of $Y$ to zero mean and unit variance.
Zero mean is required since PCA is a rotation around the empirical mean;
unit variance is not explicitly required, but the columns must all have the same units and similar magnitude, and scaling to unit variance is a convenient way to achieve this.

The transformation is now determined by the standardized data $Y$, such that the first principal component has the maximum possible variance (explaining as much variance of $Y$ as possible), the second component has maximal variance while being orthogonal to the first, and so forth.
This results in an orthonormal $m \times m$ matrix $V$ which transforms the (standardized) model data as
\begin{equation}
  Z = Y V,
\end{equation}
where $Z = (\zv_1, \zv_2, \ldots, \zv_d)$ is another $d \times m$ matrix whose rows correspond to design points and columns to principal components, with the columns sorted from greatest variance to least---in contrast to $Y$, whose columns all have unit variance.
Analogous to the above notation, each vector $\zv_i = (z_{1i}, z_{2i}, \ldots, z_{mi})$, where $z_{ji}$ is the value of the $j$th PC at design point $i$.
The columns of $V$ are orthonormal vectors $\vv_j$, i.e.\ satisfying $\vv_j \cdot \vv_k = \delta_{jk}$, each containing the linear combination coefficients for PC $j$, so that
\begin{equation}
  z_{ji} = \vv_j \cdot \tilde\yv_i,
\end{equation}
where the tilde denotes \emph{standardized} model output.

In numerical implementations \cite{scikit-learn}, the PCA transformation is computed efficiently via the singular value decomposition (SVD) of the data matrix $Y$.
The SVD, a generalization of the eigendecomposition for non-square matrices, is the factorization
\begin{equation}
  Y = U \Sigma V\tran,
  \label{eq:svd}
\end{equation}
where $U$ and $V$ are orthogonal matrices containing the left and right singular vectors and $\Sigma$ is diagonal containing the singular values.
The matrix $V$ of the right singular vectors is the PCA transformation matrix.
Using the SVD, we can also see that PCA is related to the eigendecomposition of the sample covariance matrix as
\begin{equation}
  Y\tran Y = (U \Sigma V\tran)\tran U \Sigma V\tran = V \Sigma^2 V\tran.
\end{equation}
Hence, $V$ contains the eigenvectors of the sample covariance matrix ($Y\tran Y$) and $\Sigma^2$ has the eigenvalues on the diagonal.

\begin{figure}[t]
  \makebox[\textwidth]{\includegraphics{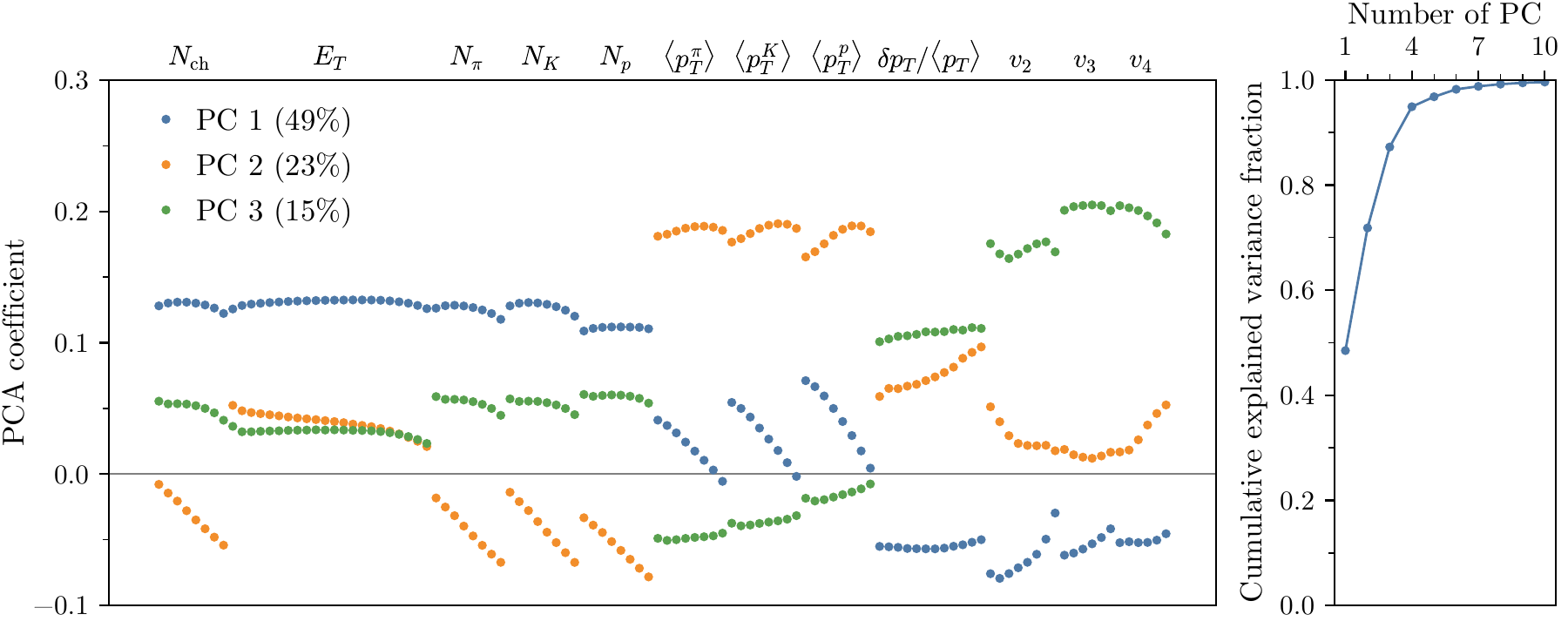}}
  \caption{
    Application of PCA to heavy-ion collision model output for Pb-Pb collisions at 2.76 TeV.
    Left: Linear combination coefficients for the observables labeled along the top:
    charged-particle yield, transverse energy production, identified particle yields, identified particle mean $p_T$, mean $p_T$ fluctuations, and flow coefficients (two-particle cumulants).
    Each point represents a centrality bin.
    The legend entries include the explained variance of each component.
    Right: Cumulative explained variance fraction for up to 10 components.
  }
  \label{fig:pca-vectors-variance}
\end{figure}

Figure \ref{fig:pca-vectors-variance} shows some properties of a realistic application of PCA to the present heavy-ion collision model, using output from the design specified in subsection \ref{subsec:design-present-study}.
The main left panel shows the linear combination coefficients for the first three components, i.e.\ the values of $\vv_j$, $j = 1, 2, 3$.
The first component, which by itself explains about half of the model's variance, accounts for the mutual correlation of all the particle and energy production data, and their anti-correlation with the flow data.
In this context, correlations refer to correlations across the parameter design space, for example changing a parameter that increases the charged-particle yield is likely to also increase energy production and identified particle yields.
The remaining components, of which only the second and third are shown here, are orthogonal to the first (and to each other) and describe various other correlations and anti-correlations among the observables.

The side right panel shows the convergence of the explained variance.
Despite there being over 100 original observables, the first four principal components explain about 95\% of the total variance;
the 99\% threshold is attained with eight components.
This trend is valuable for dimensionality reduction.

\subsubsection{Dimensionality reduction}

If the original data are strongly correlated, which is often the case for physical models, the first few principal components will usually explain most of the original variance.
Thus, we can use a smaller number of principal components $k$ than the number of original model outputs $m$, sacrificing a small amount of information in the process.
Since the columns of the transformation matrix $V$ are sorted by their explained variance, we simply take the first $k < m$ columns and transform the data as
\begin{equation}
  Z_k = Y V_k,
\end{equation}
where $V_k$ is $d \times k$ containing the first $k$ principal components.
The inverse transformation is
\begin{equation}
  Y \simeq Z_k V_k\tran,
\end{equation}
where the equality is only approximate since we have discarded some information.

PCA dimensionality reduction is particularly effective for data containing statistical noise.
Since noise is, by definition, uncorrelated with the true variability of the model, PCA will naturally separate the true variability from the noise, and assuming the noise is small, it will be relegated to the unimportant PCs, which we discard.
In the first example, figure \ref{fig:pca}, it could be that $y_1 = y_2$ in reality, with the differences caused entirely by statistical fluctuations.
Thus if we modeled only the first PC, we would account for all the true variability.

\subsubsection{Caveats}

PCA works best if the original data have a joint multivariate-normal distribution.
It is not necessary for every model output to have a perfect normal distribution, but they should roughly have a peak with tails.
In fact, this usually happens automatically when several parameters are varied, due to the central limit theorem.
Non-normal distributions can sometimes be made more normal by applying a nonlinear transformation such as a Box-Cox power transformation, but since this would later complicate propagation of uncertainty, it should be avoided unless necessary.

More important is that the model outputs are only \emph{linearly} correlated;
as a linear transformation, PCA can only remove linear correlations.
Practically speaking, this means that scatterplots of $y_i$ vs.\ $y_j$ should look approximately elliptical, with no curved ``S'' or ``C'' shapes.

Outliers will have an undue influence on the principal component directions.
But if outlier points are determined to be true model behavior, it may be desirable to still include them in the analysis.
They should nonetheless be excluded from the data matrix $Y$ when computing the SVD, then, after determining the PCA transformation, the outlier points can be transformed as usual.

One should always check these considerations before applying PCA to a dataset.

\subsection{Postprocessing steps}

PCA is implemented in \textsc{scikit-learn} \cite{scikit-learn}, a Python machine learning library.

\begin{enumerate}
  \item Check model outputs for approximate normality, linear correlations, and outliers.
  \item Concatenate into the matrix $Y = (\yv_1, \yv_2, \ldots, \yv_d)$ and standardize the columns to zero mean and unit variance.
  \item Compute the PCA transformation via the SVD \eqref{eq:svd}.
  \item Choose the desired number of principal components, e.g.\ to satisfy a minimum explained variance threshold.
  \item Apply the PCA transformation and dimensionality reduction.
\end{enumerate}

\section{Gaussian processes}
\label{sec:gp}

Having evaluated the model at each of the parameter design points, the time has come to construct an emulator to serve as a fast surrogate to the full model, that is, to quickly predict the model output at any point in parameter space.
Gaussian processes are ideal for this purpose since they operate in arbitrarily high-dimensional space, require only minimal assumptions about the model, and naturally quantify the uncertainty of their predictions.
They are not the \emph{only} valid emulation scheme, but exploring the alternatives is beyond the scope of this work, and Gaussian processes are the de facto choice for parameter estimation with computationally expensive models.

In the following subsections, I summarize the theory of Gaussian processes and discuss relevant practicalities of building model emulators.
For a complete treatment, see the seminal book \citetitle{Rasmussen:2006gp} by \citeauthor{Rasmussen:2006gp} \cite{Rasmussen:2006gp}, especially chapters 2, 4, and 5.

\subsection{Interpolation, regression, and emulation}

The essential ingredients of a model emulator are:
\begin{itemize}
  \item A set of training points $X_t = (\xv_1, \xv_2, \ldots, \xv_d)$, where each $\xv_i$ is an $n$-dimensional input vector.
  \item A corresponding set of model outputs $\yv_t = (y_1, y_2, \ldots, y_d)$, where each $y_i$ is the result of evaluating the model at $\xv_i$.
\end{itemize}
Given these training data, the emulator shall predict model outputs $\yv_p$ at new points $X_p$.
Gaussian process (GP) emulators achieve this provided another key ingredient:
\begin{itemize}
  \item A covariance function, which dictates the similarity between pairs of outputs $(y_i, y_j)$.
\end{itemize}
Given such a function, a GP predicts new model outputs $\yv_p$ by exploiting their covariance, i.e.\ similarity, with the training outputs $\yv_t$.
In fact, the predictions are probability distributions---specifically normal (Gaussian) distributions---for the likely values of $\yv_p$, from which we can extract mean values and associated uncertainties.

A brief word on notation:
The subscripts $t$ and $p$ mean ``training'' and ``predictive'', respectively, and I shall use them consistently throughout this section.
However, I am somewhat overloading the vector notation:
Here, the vectors $\yv_t$ and $\yv_p$ contain multiple values of a single model output, while in other contexts, $\yv$ (without a subscript) is the vector of all the model output variables, and $\yv_i$ (with an index subscript) is a single observation of all the model outputs at design point $i$.
The meaning of vector symbols will hopefully be clear from context.

\begin{figure}[t]
  \makebox[\textwidth]{\includegraphics{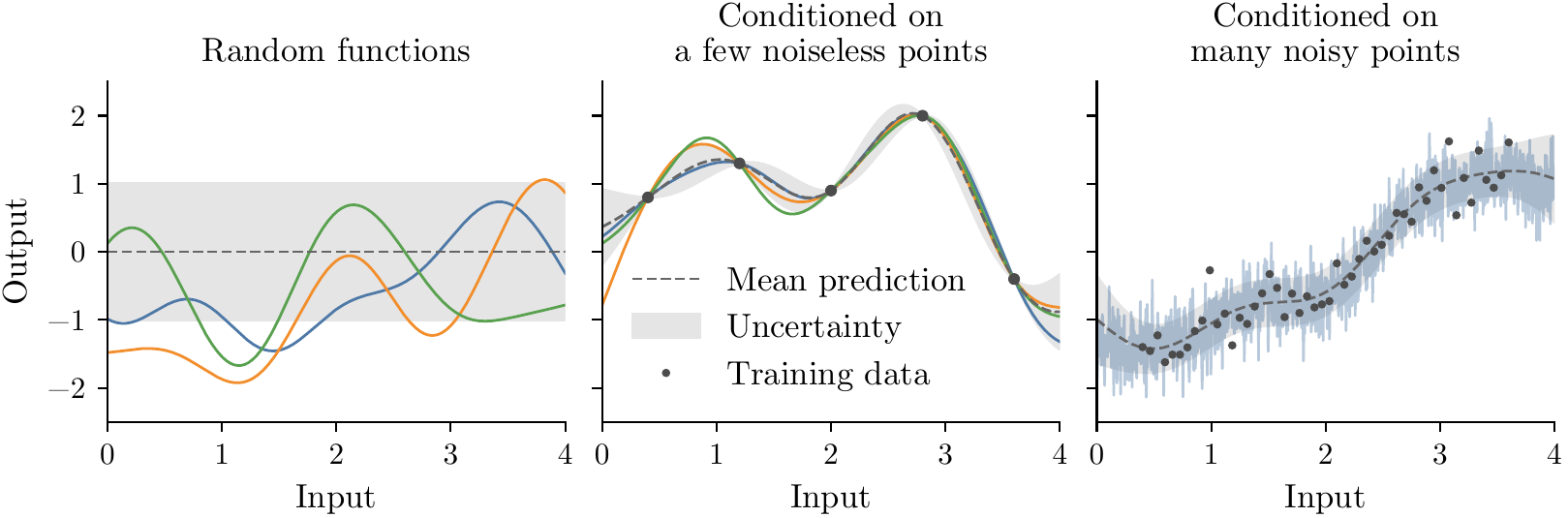}}
  \caption{
    Left: Random functions drawn from a Gaussian process.
    Center: Functions drawn from a GP conditioned on a few noiseless training points.
    Right: A function drawn from a GP conditioned on many noisy training points (only one semitransparent line for visual clarity).
    In all plots, the gray dashed line and band are the GP predictive mean and uncertainty (one standard deviation), respectively.
    All GPs have a squared exponential covariance function \eqref{eq:cov-se-simple} with length scale $\ell = 0.6$.
    On the right, the covariance function \eqref{eq:cov-se-noise} also has a noise term with variance $\sigma_n^2 = 0.1$.
  }
  \label{fig:gp}
\end{figure}

Before formalizing Gaussian process predictions, let us take a step back.
A GP is, in one interpretation, a distribution over \emph{functions}.
Analogously to sampling random numbers from a probability distribution, we can draw random functions from a GP, demonstrated in the left panel of figure \ref{fig:gp} (I will explain precisely how to do this shortly).
The colored lines are the random functions, the dashed line is the GP mean (zero), and the band is the standard deviation (one).

After \emph{conditioning} the GP on some training data, its mean and standard deviation become functions of the input, as in the center panel.
The mean interpolates the training points, while the standard deviation is large in the gaps between points and small near the points, reflecting the uncertainty of the interpolation.
Functions drawn from the GP pass through the training points.

Gaussian processes can also be used for regression with noisy training data, as in the right panel.
Now, the mean traces the center of the cloud, not passing exactly through all the points, and the standard deviation accounts for the underlying noise.
The randomly drawn function (only one shown for visual clarity) is also noisy.

Formally, a GP is a type of stochastic process---a collection of random variables.
A random walk is a classic example of a stochastic process:
The random variables are the positions in space, and a realization of a random walk is a particular path through space.
For a GP, the random variables are the function outputs, and the realizations are randomly sampled functions.
A GP is defined by the property that any finite collection of its random variables have a multivariate normal (joint Gaussian) distribution, written as
\begin{equation}
  \yv \sim \N(\muv, \Sigma)
\end{equation}
for mean vector $\muv$ and covariance matrix $\Sigma$.
This property also means that any single variable has a (univariate) normal distribution
\begin{equation}
  y \sim \N(\mu, \sigma^2),
\end{equation}
for mean $\mu$ and variance $\sigma^2$.

Before drawing functions from a GP, we must specify a covariance function, or kernel, $k(\xv, \xv')$.
A standard choice is the squared exponential (SE) covariance function
\begin{equation}
  \cov(y_i, y_j) = k(\xv_i, \xv_j) =
    \exp\biggl( -\frac{|\xv_i - \xv_j|^2}{2\ell^2} \biggr),
  \label{eq:cov-se-simple}
\end{equation}
which we shall to refer to as the SE function, even though it's obviously a Gaussian function, to distinguish it from the ``Gaussian'' in ``Gaussian process'';
it's also known as the radial basis function (RBF).
Notice that the covariance function describes the similarity between pairs of \emph{outputs}, but is a function of the \emph{inputs}.
With this particular covariance function, outputs from nearby input points are strongly correlated, while distant points become uncorrelated over a characteristic length scale $\ell$.

Now, to sample functions:
We choose some input points\footnote{
  The input points are denoted by $X_p$ because they are technically predictive points, even though they are not actually predicting anything in this case.
  Soon, $X_t$ and $X_p$ will appear together and it will be important to distinguish between them.
}
$X_p$ and construct the covariance matrix $K_{pp}$, where this notation means a matrix from applying the covariance function to each pair of points in $X_p$:
\begin{equation}
  K_{pp} =
    \begin{pmatrix}
      k(\xv_{p1}, \xv_{p1}) & k(\xv_{p1}, \xv_{p2}) & \cdots \\
      k(\xv_{p2}, \xv_{p1}) & k(\xv_{p2}, \xv_{p2}) & \cdots \\
      \vdots                & \vdots                & \ddots \\
    \end{pmatrix}.
  \label{eq:Kcov}
\end{equation}
We then launch our favorite statistical software, generate random vectors from the multivariate normal distribution
\begin{equation}
  \yv_p \sim \N(\zerov, K_{pp}),
  \label{eq:yprior}
\end{equation}
and plot the resulting vectors as smooth curves.
The left panel of figure \ref{fig:gp} is the result of following this procedure, setting $X_p$ to an array of one-dimensional points from 0 to 4 and using the SE covariance function \eqref{eq:cov-se-simple} with length scale $\ell = 0.6$.

As foreshadowed at the beginning of this subsection, a GP represents an infinitely large family of functions $f(\xv)$ with a specified covariance structure $k(\xv, \xv')$.
The curves on the left of figure \ref{fig:gp} are samples from the family of functions, defined by the chosen SE covariance function \eqref{eq:cov-se-simple}, which vary smoothly over the chosen characteristic length scale.

To use a GP as a model emulator, we assume that the training data $(X_t, \yv_t)$ are the inputs and outputs of a function $f(\xv)$ from a GP.
This is quite general, tantamount to assuming that there exists a covariance function $k(\xv, \xv')$ that describes the relationships between model outputs.
The SE covariance function used to this point is in fact appropriate for many physical models, which tend to be well-behaved and smoothly varying.
Other functions allow control over the degree of smoothness.
More sophisticated covariance functions can be constructed by adding together different kernels as $k = k_1 + k_2 + \cdots$, for example the sum of two SE functions with different length scales implies a covariance structure with both small- and large-scale trends.
If the model is periodic, a periodic covariance function would be suitable.

After designating a covariance function, we \emph{condition} a GP on the training data $(X_t, \yv_t)$, furnishing the predictive distribution for new model outputs $\yv_p$ at input points $X_p$,
\begin{equation}
  \begin{aligned}
    \yv_p &\sim \N(\muv, \Sigma), \\
    \muv &= K_{pt} K_{tt}^{-1} \yv_t, \\
    \Sigma &= K_{pp} - K_{pt} K_{tt}^{-1} K_{tp},
  \end{aligned}
  \label{eq:ypost}
\end{equation}
with $K_{pp}$ defined in \eqref{eq:Kcov} and the other $K_{**}$ matrices following analogously, for example $K_{pt}$ is the covariance matrix from applying the covariance function to each pair of predictive and training points, i.e.\ its $ij$ element is $(K_{pt})_{ij} = k(\xv_{pi}, \xv_{tj})$.
The center panel of figure \ref{fig:gp} shows the effects of conditioning a GP on the plotted training points using the same SE covariance function as before;
the dashed line is the predictive mean $\muv$ plotted as a smooth curve, the gray band is the mean plus or minus one predictive standard deviation, and the colored lines are sampled functions.
In general, conditioning a GP restricts its function space to functions that are consistent with the training data;
the plotted curves in the figure are several possible functions that could have given rise to the training points.

If the model calculations are non-deterministic---perhaps due to averaging over a finite sample---the training data will contain statistical noise as $y = f(\xv) + \epsilon$, where $\epsilon$ is a fluctuating noise term.
We may account for this by adding a noise kernel $k(\xv_i, \xv_j) = \sigma_n^2 \,  \delta_{ij}$ to the covariance function, which describes uncorrelated (independent for each training point) Gaussian noise of variance $\sigma_n^2$.
Combining the noise kernel with the SE function, for example, gives the total covariance function
\begin{equation}
  k(\xv_i, \xv_j) =
    \exp\biggl( -\frac{|\xv_i - \xv_j|^2}{2\ell^2} \biggr) + \sigma_n^2\,\delta_{ij}.
  \label{eq:cov-se-noise}
\end{equation}
The right panel of figure \ref{fig:gp} shows the result of conditioning a GP on noisy training data using this covariance function.
Since the data are noisy, the predictive mean does not pass through every point exactly, but rather behaves more like a regression line, and the predictive standard deviation accounts for the noise.
Functions drawn from the GP have random fluctuations with variance $\sigma_n^2$.

We can gain more intuition for how GP emulators work by examining the conditional (predictive) distribution for a single output $y_p$.
Writing $\kv_p = (k(\xv_p, \xv_1), k(\xv_p, \xv_2), \ldots, k(\xv_p, \xv_d))$ for the vector of covariances between the predictive point and the training points, equation \eqref{eq:ypost} reduces to
\begin{equation}
  \begin{aligned}
    y_p &\sim \N(\mu, \sigma^2), \\
    \mu &= \kv_p\tran K_{tt}^{-1} \yv_p, \\
    \sigma^2 &= k(\xv_p, \xv_p) - \kv_p\tran K_{tt}^{-1} \kv_p.
  \end{aligned}
\end{equation}
From this, we see that the mean $\mu$ is a linear combination of \emph{all} the training points, with the relative contributions depending on the covariance function.
The variance $\sigma^2$ consists of the variance at $\xv_p$ from the covariance function, minus a second (positive) term, which reduces the total variance by assimilating additional information from the training points.

The conditioning process is a Bayesian update, in which the prior (unconditioned) GP is updated with the training data to form a posterior GP.
Equation \eqref{eq:yprior} is the prior distribution for new model outputs $\yv_p$, determined by the covariance function;
\eqref{eq:ypost} is the posterior distribution, determined by the covariance function in combination with the training data.
In figure \ref{fig:gp}, the left panel shows a prior GP with the SE covariance function \eqref{eq:cov-se-simple} and the center shows a posterior GP after conditioning the prior on the training data.
The right panel shows another posterior GP, conditioned on noisy training data using the covariance function \eqref{eq:cov-se-noise}.

\subsection{Multivariate output}
\label{subsec:multivariate-output}

Fundamentally, Gaussian processes map vector inputs to scalar outputs, but computational models often have many outputs.
As detailed in the previous section \ref{sec:postprocess}, we deal with this by transforming the model outputs using principal component analysis (PCA) and building an independent GP emulator for each principal component.
Given an input point $\xv$, we compute the predictive distributions for each PC, collect the mean predictions into a vector $\zv$, and transform it into the desired model outputs
\begin{equation}
  \yv = V \zv,
\end{equation}
where $V$ is the PCA transformation matrix.

Calculating the uncertainty on $\yv$ is straightforward since it is related to $\zv$ by a linear transformation;
writing $\Sigma_z$ for the predictive covariance matrix of $\zv$, the covariance of $\yv$ is
\begin{equation}
  \Sigma_y = V \Sigma_z V\tran.
\end{equation}
The principal components are uncorrelated by construction, so the covariance matrix is diagonal,
\begin{equation}
  \Sigma_z = \diag(\sigma_1^2, \ldots, \sigma_m^2),
  \label{eq:Sigma-z}
\end{equation}
where each $\sigma_k^2$ is the predictive variance of the $k$th principal component, i.e.\ of the $k$th element of $\zv$.
Hence, the covariance of $\yv$ reduces to
\begin{equation}
  (\Sigma_y)_{ij} = \sum_k V_{ik} \sigma_k^2 V_{jk}.
\end{equation}
Note that $\Sigma_y$ is in general not diagonal, meaning that the uncertainties on the model outputs are correlated.
These correlations manifest because we are predicting principal components, which are linear combinations of the model outputs, so any uncertainty on a principal component translates into correlated uncertainty on the actual outputs.

Also mentioned in the previous section:
We use PCA dimensionality reduction and emulate only the first $k < m$ principal components of the largest variance.
But the remaining components do contain a small amount of information, so neglecting them contributes some uncertainty, which we must take into account.
Consider that neglecting a PC is equivalent to treating it as an unconditioned (prior) GP with zero mean and constant variance equal to the sample variance, therefore, we should take each neglected component's sample variance as its predictive variance.
In the diagonal covariance matrix $\Sigma_z$ \eqref{eq:Sigma-z}, we set the first $k$ variances $(\sigma_1^2, \ldots, \sigma_k^2)$ according to the GP emulators, and set the remaining variances $(\sigma_{k+1}^2, \ldots, \sigma_m^2)$ to the sample variance of the respective principal component.

\subsection{Training}
\label{subsec:training}

To this point I have glossed over the free parameters that are often present in covariance functions, such as the characteristic length scale $\ell$ in the squared exponential function.
These \emph{hyperparameters} are usually not known \emph{a priori} and must be estimated from the data.
The selection of hyperparameters, known as training, is typically accomplished by maximizing the likelihood
\begin{equation}
  \mathcal L(\thetav) = \frac{1}{\sqrt{(2\pi)^d \det K_{tt}}} \exp\biggl( -\frac{1}{2} \yv_t\tran K_{tt}^{-1} \yv_t \biggr),
\end{equation}
where $\thetav$ is the vector of hyperparameters and $K_{tt} = K_{tt}(\thetav)$ is the covariance matrix from applying the covariance function---which depends on $\thetav$---to the training points.
The form of $\mathcal L(\thetav)$ is nothing but a multivariate normal probability density, whose logarithm
\begin{equation}
  \log\mathcal L(\thetav) = -\frac{1}{2} \yv_t\tran K_{tt}^{-1} \yv_t - \frac{1}{2} \log(\det K_{tt}) - \frac{d}{2}\log 2\pi
  \label{eq:ll-hyperpars}
\end{equation}
is preferable for numerical optimization, and more clearly separates into meaningful components:
The first term is a fit to the data, the second is a complexity penalty which prevents overfitting, and the third is a normalization constant.

\begin{figure}[t]
  \makebox[\textwidth]{\includegraphics{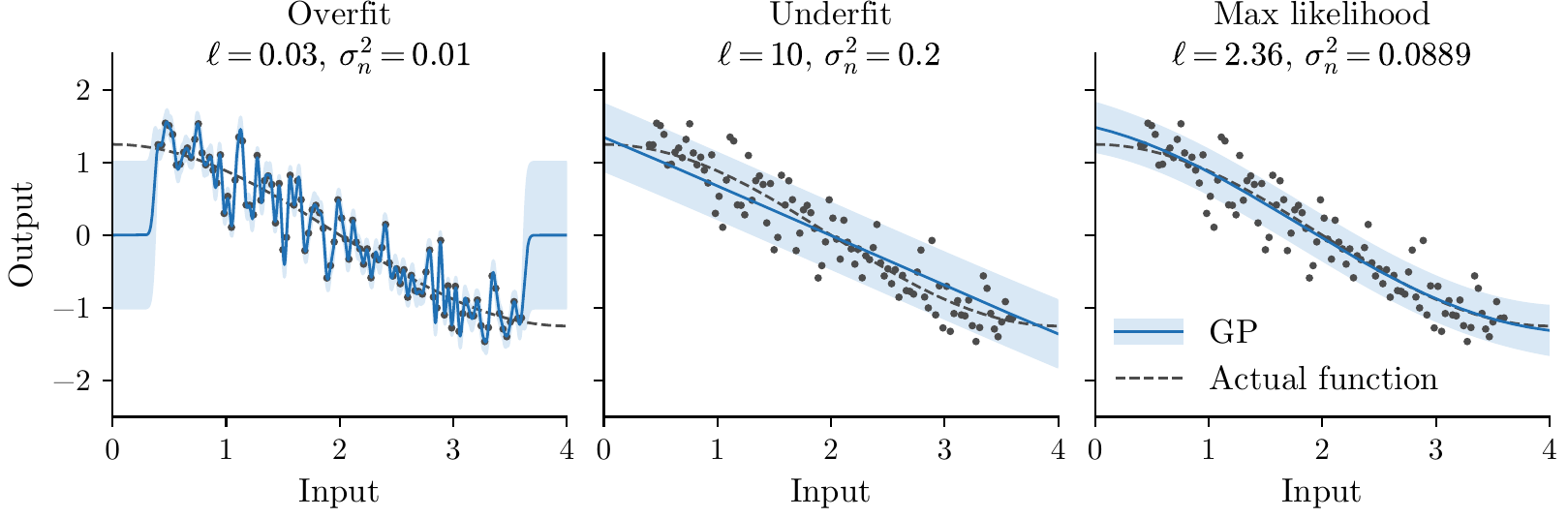}}
  \caption{
    Example of training a Gaussian process.
    The data points, which are the same in all plots, were generated by evaluating the function plotted as the dashed line and adding random Gaussian noise of variance $\sigma^2 = 0.09$.
    Each plot shows a GP conditioned on the noisy data using the SE covariance function \eqref{eq:cov-se-noise} with variable length scale $\ell$ and noise term $\sigma_n^2$.
    On the left and center, the hyperparameters were set manually;
    on the right, they were determined by numerically maximizing the likelihood \eqref{eq:ll-hyperpars}.
  }
  \label{fig:gp-train}
\end{figure}

To see how this works, let us consider an instructive example:
Figure \ref{fig:gp-train} shows three GPs conditioned on the same noisy data, all using the SE covariance function with a noise term \eqref{eq:cov-se-noise}, but with different values of the hyperparameters $\thetav = (\ell, \sigma_n^2)$.
On the left, a too-short length scale and too-small noise variance lead to an ``overfit'' model that passes through every point, mistakenly treating the noise as true variability, and thus offering no predictive value.
These hyperparameter values have a low likelihood due to a high complexity penalty.
In the center, the opposite:
The model is ``underfit'' with a long length scale and large noise term, ascribing too much of the true variability to noise.
These values also have a low likelihood, this time because of a poor fit to the data.
On the right, the maximum likelihood hyperparameters strike a compromise, accurately capturing the actual underlying function and noise.

In the present work, I use an anisotropic squared exponential covariance function
\begin{equation}
  k(\xv_i, \xv_j) = \sigma_f^2 \exp\biggl[
    -\frac{1}{2} \sum_k \biggl( \frac{x_{ki} - x_{kj}}{\ell_k} \biggr)^2
  \biggr] + \sigma_n^2\delta_{ij},
\end{equation}
whose hyperparameters are the independent length scales $\ell_k$ for each input dimension (hence, anisotropic), overall variance scale $\sigma_f^2$, and noise variance $\sigma_n^2$.
Using this covariance function essentially amounts to assuming that the model is well-behaved and smoothly varying, with no discontinuities, divergences, or other anomalous features.
The noise variance allows for some statistical fluctuations in the training data.

One possible criticism of the SE function is that it's \emph{too} smooth---a GP with this covariance function is infinitely differentiable, which may not be the case for some physical models.
The Mat\'ern class of covariance functions attempts to resolve this by introducing a smoothness parameter while otherwise being similar to the SE function.
I trained GPs to the model data using once- and twice-differentiable Mat\'ern covariance functions, which are both somewhat less smooth than SE, but found no difference in practical performance.
Thus, in the interest of simplicity, I use the SE covariance function.

I determine the hyperparameters $\thetav = (\sigma_f^2, \ell_1, \ldots, \ell_n, \sigma_n^2)$ by maximizing the likelihood using a numerical optimization algorithm.
To help prevent over or underfitting, I constrain the length scales to within an order of magnitude of the corresponding parameter's design range, i.e.\ if $\Delta_k = \max(x_k) - \min(x_k)$ is the design range of parameter $x_k$, then the constraint is $0.1 < \ell_k/\Delta_k < 10$.
As previously discussed, I train an independent GP on each principal component;
the optimal hyperparameters are in general different for each.

Numerical optimizers sometimes converge to a local rather than global maximum.
To ensure this is not the case, we can repeat the hyperparameter optimization several times starting from different initial values of $\thetav$, then take the best result.
However, if we do find several competing local maxima, it may be a sign that we do not have enough information to uniquely determine the hyperparameters, i.e.\ there are too few training points.
In my experience, using Latin hypercube designs with at least 20 points per dimension, the optimization algorithm converges to the same result almost every time, regardless of the initial values.
This lends confidence that the hyperparameters are well-determined by the data.

If the hyperparameters were not well-determined, we could account for that uncertainty by sampling them during the MCMC calibration phase of the analysis (next section, \ref{sec:calibration}).
Formally, we should always do this, since the hyperparameters are not known exactly, but it incurs significant computational cost, as the likelihood requires calculating the inverse covariance matrix, an $\order{n^3}$ operation.
In any case, provided a sufficient number of training points, the actual emulator predictions will not depend strongly on the hyperparameters, as long as they are not egregiously over or underfit.

\begin{figure}[t!]
  \makebox[\textwidth]{\includegraphics{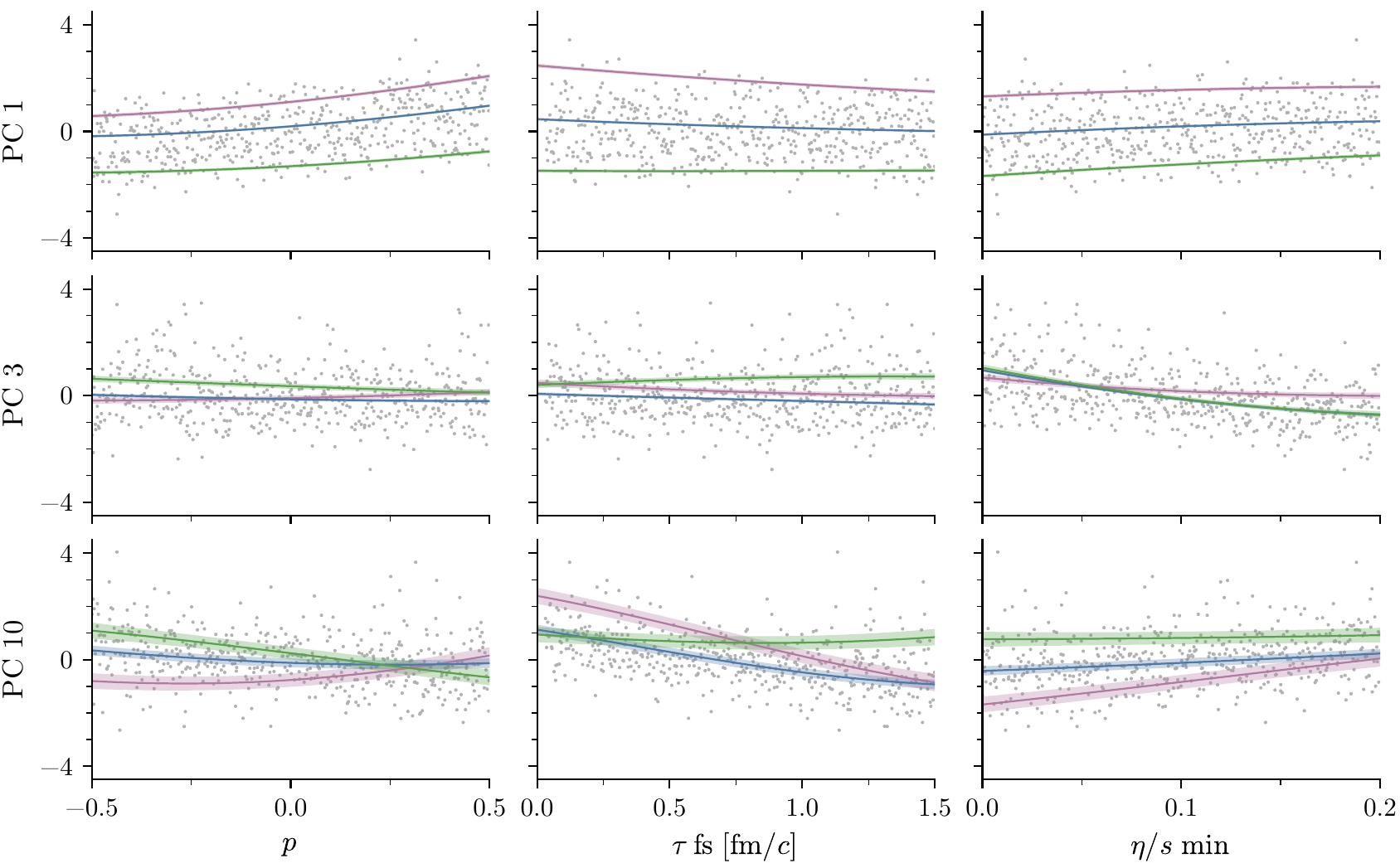}}
  \caption{
    Emulator diagnostic visualization.
    Each subplot shows the dependence of a principal component on a model parameter, as labeled on the axes.
    The dots are the training data.
    The lines with bands are GP emulator predictions, with uncertainty, as a function of the given parameter over its full design range, holding all other parameters fixed.
    The blue lines are with all other parameters fixed to the midpoint of their design range (50\%), purple is 20\%, and green is 80\%.
  }
  \label{fig:diag-emu-partial}
\end{figure}

Still, it is reasonable to doubt whether a GP emulator with the maximum likelihood hyperparameters truly captures the underlying model behavior.
The ultimate test of emulator performance is whether it accurately predicts new model calculations, which I will address in the next subsection.
First, we can perform some simple checks that the emulator is behaving reasonably, for example plotting the dependence of model outputs on input parameters and verifying that the relationships align with expectations.
Another diagnostic visualization that I have found quite useful is shown in figure \ref{fig:diag-emu-partial}.
Without repeating the information in the caption, here are some characteristics we can check:
\begin{itemize}
  \item Are the emulator predictions smooth and sensible?
    Changing a single parameter can only affect the model so much;
    there should not be any rapid oscillations or extreme behavior, which could be a sign of overfitting.
    But there should be some variability---the predictions should not all be flat.
  \item Are the predictions consistent with the training data?
    The 50\% curve should probably track through the middle of the cloud, while the 20\% and 80\% curves should be distinct (the values 20 and 80 are not special, the point is to probe closer to the corners of the design space).
  \item The uncertainties should be much smaller than the spread of the training data, which is due to varying \emph{all} parameters simultaneously.
    In other words, the predictive variance should be smaller than the total variance of the model.
    Equivalently, check that $\sigma_n^2 \ll \sigma_f^2$.
  \item The uncertainty should usually increase for the higher order principal components, since they describe more noise.
    This is why I have shown components 1, 3, and 10, to emphasize the increase of the uncertainty.
    Equivalently, check that $\sigma_n^2$ generally trends upward.
\end{itemize}
The subplots in the figure are only a small subset of all the possible input-output combinations;
I chose these representative instances to keep the figure a reasonable size.

\subsection{Validation}
\label{subsec:validation}

The most important test of emulator performance is if it faithfully predicts model calculations, that is, given an arbitrary input point $\xv$, the predicted model output $\yv_\text{pred}(\xv)$ should be close to the result of a full model calculation $\yv_\text{calc}(\xv)$.
We should check a large sample of validation points to ensure statistical significance.

In the present work, I have a sample of model calculations from an earlier version of the design that I will use for validation.
However, it sometimes may be too computationally expensive to run a separate validation sample.
An alternative is cross-validation, a general technique in which the training data is split into two sets, one for training and the other for validation.
In $k$-fold cross-validation, the training data is partitioned into $k$ equally sized subsets, then one subset is used for validation and the other ${k-1}$ for training.
This is repeated for each of the $k$ subsets, so that eventually all training points have been used for validation.

\begin{figure}[t]
  \makebox[\textwidth]{\includegraphics{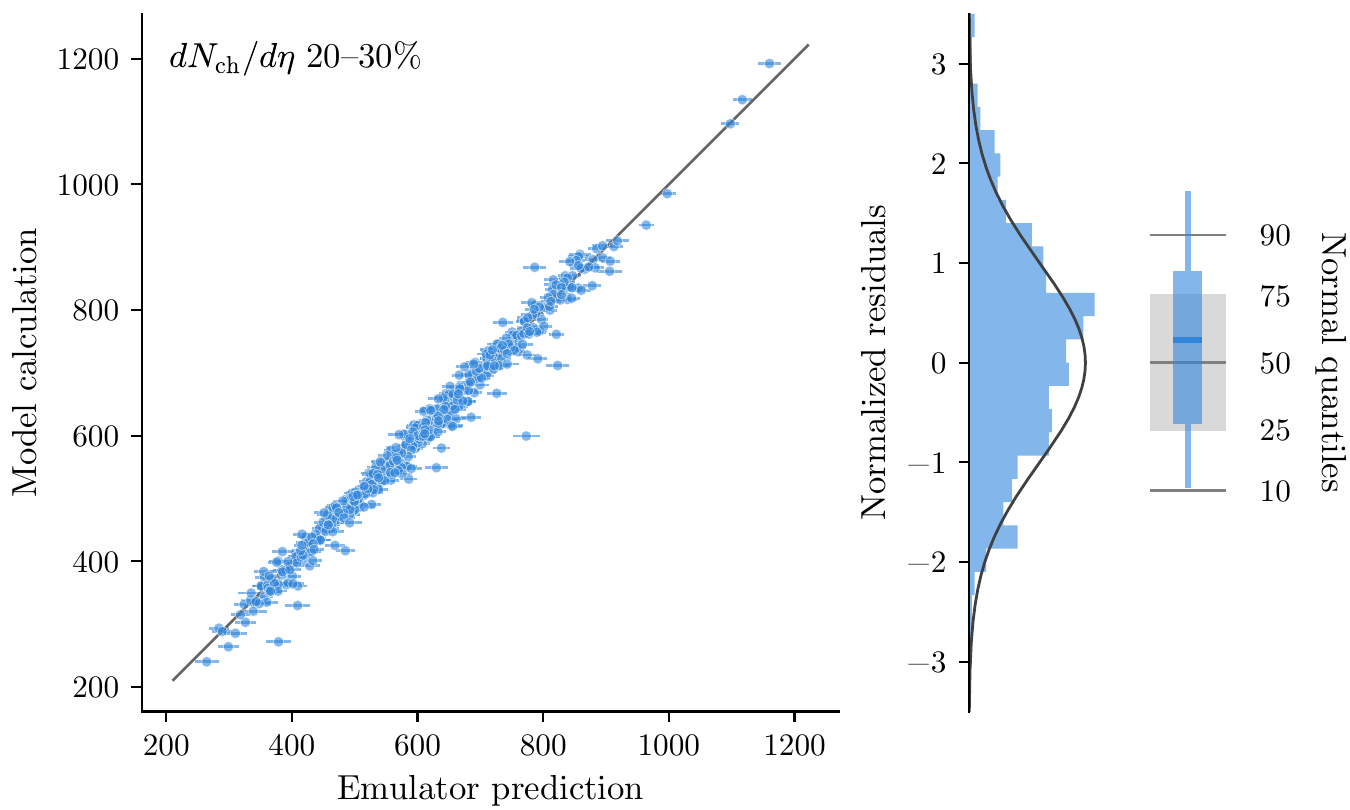}}
  \caption{
    Validation of emulator predictions of a single model output.
    Left: Scatterplot of model calculations vs.\ emulator predictions of $d\Nch/d\eta$ in 20--30\% centrality.
    The horizontal error bars are the standard deviation of the predictive uncertainty, and the diagonal line is a reference for calculation = prediction.
    Center: Histogram of the normalized residuals, overlaid with a standard normal distribution $\N(0, 1)$ probability density.
    Right: Box plot of the normalized residuals compared to normal distribution quantiles.
  }
  \label{fig:validation-example}
\end{figure}

The simplest validation test is a scatterplot of a calculated vs.\ predicted model output, for example in the main (left) panel of figure \ref{fig:validation-example}.
It appears that the emulator is performing reasonably well, although it is difficult to say precisely how well from this plot alone.
To quantify this, consider that since Gaussian processes predict probability distributions, they need not predict every validation output exactly, but rather should predict the \emph{distribution} of outputs.
Specifically, GP predictions are normal distributions, therefore, it should be the case that for every output $y$,
\begin{equation}
  \frac{y_\text{pred} - y_\text{calc}}{\sigma_\text{pred}} \sim \N(0, 1).
\end{equation}
The left-hand side is a normalized residual:
The difference of the predictive mean and the actual calculation, divided by the predictive uncertainty.
If the emulator is performing perfectly, these normalized residuals would have a standard zero-mean, unit-variance normal distribution.
The center panel of figure \ref{fig:validation-example} compares a histogram of the normalized residuals to the $\N(0, 1)$ probability density, revealing that the distribution of the residuals is indeed close to the ideal.

We can use a box plot to validate the quantiles (or percentiles) of the normalized residuals, as in the right the figure.
The horizontal blue line marks the median of the distribution, the box extends from the 25th to 75th quantile (the interquartile range), and the tails extend to the 10th and 90th quantiles.
The gray reference lines and box indicate where these elements would be located in the ideal case.
From this, we see that the median is somewhat high, meaning that more validation points were overpredicted than underpredicted, which we can also see qualitatively on the scatterplot.
In general, the predictions are skewed high, but not unreasonably so.

\begin{figure}[t]
  \makebox[\textwidth]{\includegraphics{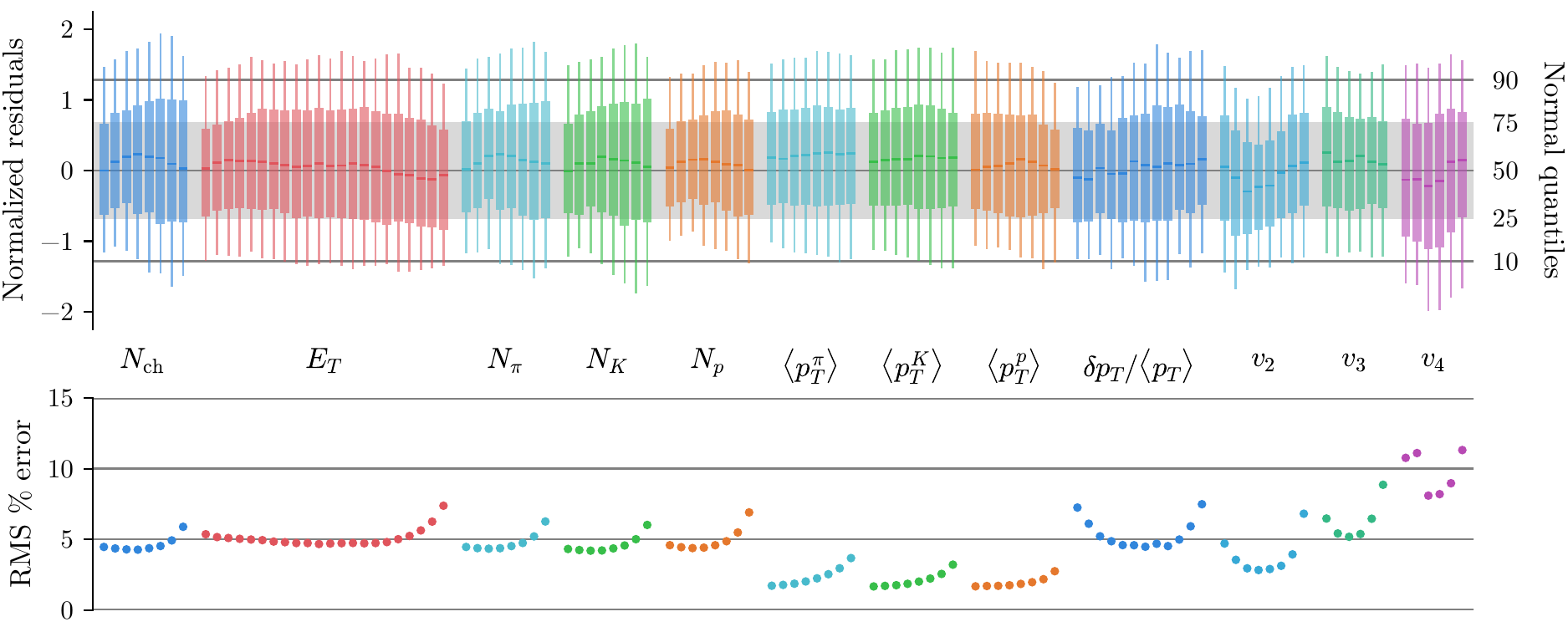}}
  \caption{
    Validation of emulator predictions of all model outputs.
    The outputs (observables) are grouped by type, as labeled, with each box plot or dot corresponding to a centrality bin (most central on the left to most peripheral on the right).
    Top: Box plots of normalized residuals compared to normal distribution quantiles.
    Bottom: Root mean square (RMS) percentage predictive error for each observable.
  }
  \label{fig:validation-all}
\end{figure}

Comparing quantiles like this is a sensitive test, and the box plots are quite compact.
Taking advantage of this, the top of figure \ref{fig:validation-all} shows box plots for \emph{all} the model observables.
To reiterate, in the ideal case, the median line would coincide with the reference line at zero, the box would match the range indicated by the gray band, and the tails would extend to the positive and negative reference lines.
We observe overall very good performance, although many of the box tails extend a little too far, suggesting that the uncertainty may be slightly underpredicted.
This is likely because the validation data have fewer events per point than the training data and therefore have more noise.

The quantile test is a stringent assessment of emulator performance, but does not immediately convey the relative predictive error of the physical observables.
To this end, the bottom of figure \ref{fig:validation-all} plots the root mean square relative predictive error of each model output, $(y_\text{pred} - y_\text{calc})/y_\text{calc}$.
Most observables are predicted to roughly 5\% precision, which is quite good considering that there are only 500 training points in a 14-dimensional space, so some interpolation uncertainty is inevitable.
Some observables are also intrinsically noisier than others, which is why the relative error increases for more peripheral bins, and why the error of $v_4$ is greater than that of $v_3$, which is greater than that of $v_2$.
The error of the mean $p_T$ is smaller than the rest simply because it does not vary as much across the design space (imagine if an observable did not change at all across the design space, then it would be easy to predict).

The salient point to take away from this validation:
The emulator accurately quantifies its own uncertainty, so it is safe to use for parameter estimation as long as we take that uncertainty into account.

\section{Calibration}
\label{sec:calibration}

We are now prepared to calibrate the model to experimental data, thereby inferring quantitative estimates of the model parameters, including uncertainties.
This is an inverse problem---we wish to learn about unknown model inputs using data we have collected about the outputs---for which Bayesian inference offers a natural solution.
In this framework, we extract parameter estimates from the posterior distribution for the model parameters
\begin{equation}
  P(\xv|\D) \propto P(\D|\xv) \, P(\xv),
\end{equation}
where $\xv$ are the parameters and $\D$ represents all the collected data, from both experiments and model calculations.
In this relation, now familiar as Bayes' theorem, the left-hand side is the posterior distribution:
the conditional probability of the parameters given the data.
Written like so, as a proportionality, the posterior distribution is unnormalized, which is acceptable for the present purposes since we are only concerned with relative probabilities.
On the right, $P(\D|\xv)$ is the likelihood, the probability of observing the data conditional on some assumed parameter values, and $P(\xv)$ is the prior distribution, which embodies our initial knowledge of the parameters.

The following subsections discuss the prior and likelihood and describe the MCMC sampling of the posterior distribution, as outlined at the start of this chapter.
Reminder: The procedure for Bayesian calibration of computationally expensive models is based on established statistical methods \cite{Kennedy:2001bc,OHagan:2006ba,Higdon:2008cmc,Higdon:2014tva}.

\subsection{Choice of priors}

The prior distribution $P(\xv)$ expresses any information we have about the parameters before observing the data.
If we know little about the parameters, a uniform prior would be appropriate, $P(\xv) = \text{constant}$.
In the present method, we have designated a finite design range for each parameter, and the Gaussian process emulator can only make predictions within those ranges, thus, we may choose a prior which is constant inside the hyperrectangular design region and zero outside,
\begin{equation}
  P(\xv) \propto
  \begin{cases}
    1 & \text{if } \min(x_i) \le x_i \le \max(x_i) \text{ for all } i, \\
    0 & \text{else}.
  \end{cases}
  \label{eq:uniform-prior}
\end{equation}
The prior being zero outside the design region is a very strong assumption:
It means we believe it is \emph{impossible} for the true value of any parameter to be outside its design range.
To ensure that plausible parameter combinations are not excluded \emph{a priori}, we ought to err on the side of too-wide design ranges.

We should not be fooled into thinking that a uniform prior is \emph{uninformative} or \emph{unbiased};
it does not amount to the absence of a prior.
A uniform prior encodes a specific assumption:
that any equally-sized volume of parameter space is equally probable, regardless of location.
For example, if we place a uniform prior on parameter $x_1$ from zero to one, we are asserting a belief that the true value of $x_1$ is equally likely to fall within $[0, 1/2]$ as $[1/2, 1]$ (or any other pair of equally-sized ranges).
This may be reasonable, but we should not take it as a given.

Further, uniform priors become nonuniform if the parameter is nonlinearly transformed.
Continuing the above example, suppose that $x_1$ enters the model only as its square, then we might instead place a uniform prior on $x_1^2$, but that would encode a different assumption:
that the true value of $x_1$ is equally likely to fall within $[0, 1/\sqrt2]$ as $[1/\sqrt2, 1]$.
This is clearly different from above;
which is preferable depends on the specific nature of the parameter and any additional information we might have.

In some cases, it may be advisable to place a joint prior on multiple parameters to discourage unreasonable combinations.
For instance, if both $x_1$ and $x_2$ have natural ranges of zero to one, but, based on physical considerations, it's unlikely that \emph{both} parameters are close to one, we could choose a prior that decreases when, e.g., $x_1^2 + x_2^2 > 1$.

Having said all this, a uniform prior is a satisfactory default in the absence of more informative knowledge.
And in any case, a strongly-peaked likelihood will ultimately overcome any nonzero prior---if the parameters are well-constrained by the data, the posterior distribution will be essentially independent of the prior.

\subsection{Likelihood and uncertainty quantification}
\label{subsec:likelihood-uq}

The likelihood $P(\D|\xv)$ is the probability of observing the data given the parameters;
it quantifies the compatibility of the model calculations, at a particular parameter point $\xv$, with the experimental data.
Here, the symbol $\D$ is shorthand for all the collected data, including the experimental observations, model calculations, and associated uncertainties.

Before specifying the likelihood function, we define some terms.
Let $\yv_e$ be the vector of experimental data, which is the result of observing the hypothetical ``true'' values $\yv_e^\text{true}$ with some measurement error $\epsv_e$.
We write this as
\begin{equation}
  \yv_e = \yv_e^\text{true} + \epsv_e, \quad
  \epsv_e \sim \N(\zerov, \Sigma_e),
  \label{eq:ye}
\end{equation}
where the second relation means that the error is distributed as a multivariate normal distribution with mean zero and covariance matrix $\Sigma_e$, which accounts for all sources of experimental uncertainty, namely statistical and systematic.
Similarly, the model outputs $\yv_m$ for input parameters $\xv$ are
\begin{equation}
  \yv_m(\xv) = \yv_m^\text{ideal}(\xv) + \epsv_m, \quad
  \epsv_m \sim \N(\zerov, \Sigma_m),
  \label{eq:ym}
\end{equation}
where the ``ideal'' model outputs represent the hypothetical calculations of a perfect physical model to unlimited precision.
Since we are using a model emulator, $\yv_m(\xv)$ is an emulator prediction, and the model covariance matrix $\Sigma_m$ accounts for predictive uncertainty, model statistical uncertainty (e.g.\ from averaging over a finite sample), and model systematic uncertainty (e.g.\ from discretizing a continuous system onto a grid).

Now, we assume that there exists some true values of the parameters $\xv_\star$ at which the ideal model calculations would match the true experimental data: $\yv_e^\text{true} = \yv_m^\text{ideal}(\xv_\star)$.
Combining this with \eqref{eq:ye} and \eqref{eq:ym} gives
\begin{equation}
  \yv_e = \yv_m(\xv_\star) + \epsv, \quad
  \epsv \sim \N(\zerov, \Sigma), \quad
  \Sigma = \Sigma_e + \Sigma_m,
  \label{eq:y-expt-model}
\end{equation}
where $\Sigma$ is the total covariance matrix, which subsumes all sources of uncertainty.
This relation between the model and experimental data implies that the likelihood is the multivariate normal distribution
\begin{equation}
  P(\D|\xv) = \frac{1}{\sqrt{(2\pi)^m \det \Sigma}} \exp\biggl\{
    -\frac{1}{2} [\yv_m(\xv) - \yv_e]\tran \Sigma^{-1} [\yv_m(\xv) - \yv_e]
  \biggr\}.
  \label{eq:likelihood}
\end{equation}

\begin{figure}[t]
  \makebox[\textwidth]{
    \includegraphics[width=.9\paperwidth]{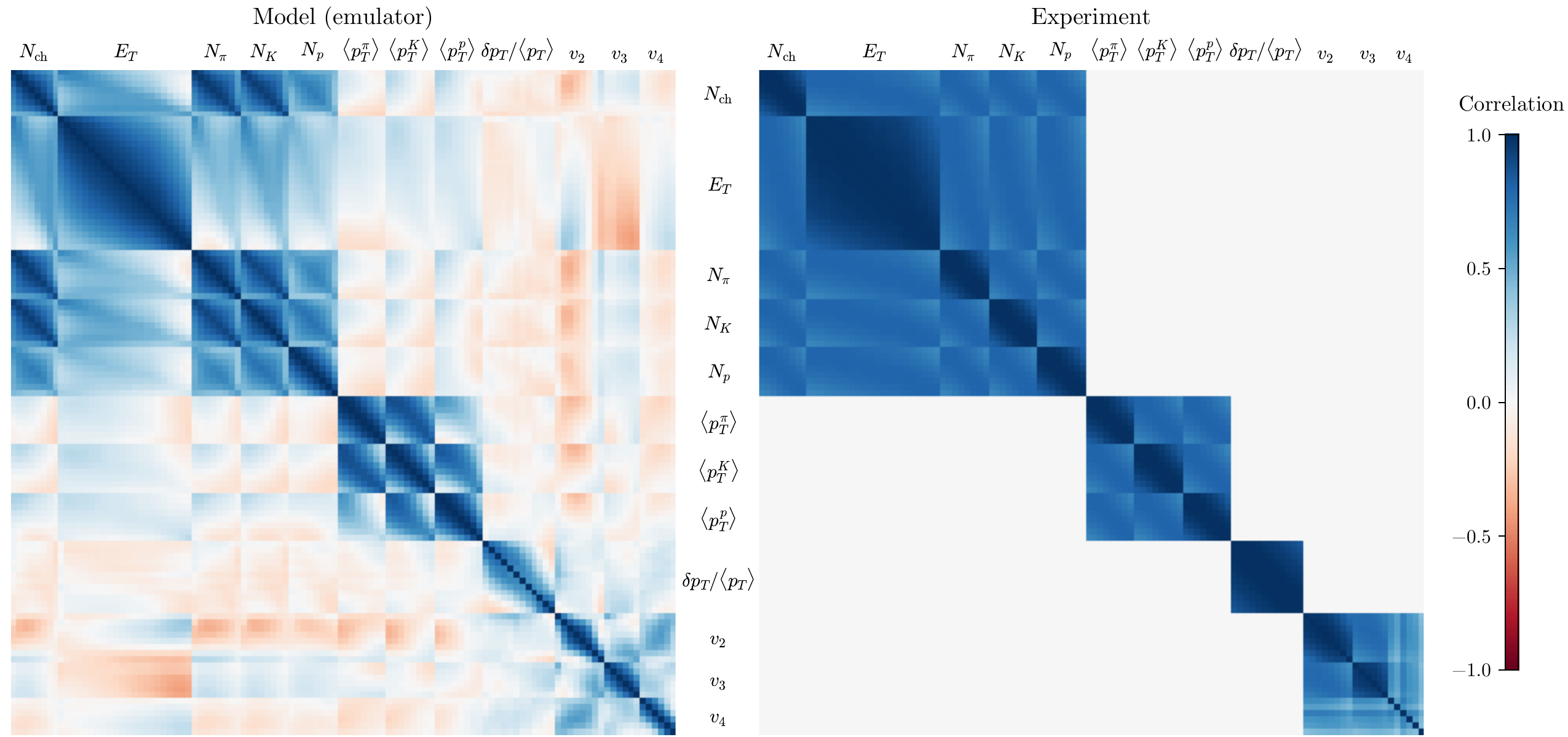}
  }
  \caption{
    Visualizations of the model (emulator) and experimental correlation matrices, whose elements are $\operatorname{corr}(y_i, y_j) = \cov(y_i, y_j)/(\sigma_i\sigma_j)$.
    Observables are grouped by type, as labeled on the axes, where each cell represents a centrality bin.
  }
  \label{fig:correlation-matrices}
\end{figure}

All that remains is to calculate the covariance matrix.
We further break down the experimental part into statistical and systematic components,
\begin{equation}
  \Sigma_e = \Sigma_e\stat + \Sigma_e\sys.
\end{equation}
Statistical uncertainties are uncorrelated by definition, so the statistical covariance matrix is diagonal,
\begin{equation}
  \Sigma_e^\text{stat} = \diag\Bigl[
    (\sigma_1\stat)^2,
    (\sigma_2\stat)^2,
    \ldots,
    (\sigma_m\stat)^2
  \Bigr],
\end{equation}
where $\sigma_i^\text{stat}$ is the statistical uncertainty of experimental observable $y_i$ (the $i$th element of $\yv_e$).
Systematic uncertainties are in general correlated, so $\Sigma_e\sys$ is not diagonal.
However, while experimental collaborations typically report separate statistical and systematic uncertainties, they usually do not report the systematic correlation structure, so we shall assume something reasonable.
Quite generally, we can express the covariance between observables $(y_i, y_j)$ as
\begin{equation}
  \Sigma_{ij} = \cov(y_i, y_j) = \rho_{ij} \sigma_i \sigma_j,
  \label{eq:cov-ij}
\end{equation}
where $\sigma_i$ is the uncertainty of $y_i$ and $\rho_{ij}$ is a correlation coefficient satisfying
\begin{equation}
  |\rho_{ij}| \le 1, \quad \rho_{ii} = 1\;\;\text{(not a sum)},
\end{equation}
with the following meaning
\begin{equation}
  \begin{cases}
    \rho_{ij} = 1 & (y_i, y_j)\ \text{are fully correlated}, \\
    0 < \rho_{ij} < 1 & \text{partially correlated}, \\
    \rho_{ij} = 0 & \text{uncorrelated}, \\
    -1 \le \rho_{ij} < 0 & \text{anticorrelated}.
  \end{cases}
\end{equation}
Indeed, $\Sigma_e^\text{stat}$ can be cast in the form \eqref{eq:cov-ij} with $\rho_{ij}\stat = \delta_{ij}$.
For systematic uncertainty, I assume that observables within a centrality dataset (e.g.\ $d\Nch/d\eta$ as a function of centrality) have correlation coefficients
\begin{equation}
  \rho_{ij}^\text{sys} = \exp\biggl[ -\frac{1}{2} \biggl(\frac{c_i - c_j}{\ell}\biggr)^2 \biggr],
  \label{eq:rho-ij-sys}
\end{equation}
where $c_i$ is the midpoint of the centrality bin for observable $y_i$ and $\ell$ is a correlation length, which I set to $\ell = 1$.
I reduce the correlation by 20\%, i.e.\ multiply $\rho_{ij}$ by 0.8, for pairs of observables in different datasets but of the same type, e.g.\ pion and kaon yield;
and I assume that observables of different types are uncorrelated.
All of this results in the block diagonal correlation matrix visualized in figure \ref{fig:correlation-matrices}.
Clearly, these are assumptions, but the behavior is qualitatively correct and certainly preferable to neglecting systematic error correlations.

Model uncertainty consists of emulator predictive uncertainty, statistical fluctuations, and systematic uncertainty:
\begin{equation}
  \Sigma_m = \Sigma_m^\text{pred} + \Sigma_m\stat + \Sigma_m\sys.
\end{equation}
In fact, the Gaussian process emulator accounts for both predictive and statistical uncertainty since the GPs have estimated noise terms (subsection \ref{subsec:training}), thus
\begin{equation}
  \Sigma_m = \Sigma_m^\text{GP} + \Sigma_m\sys.
\end{equation}
The GP covariance matrix, derived in subsection \ref{subsec:multivariate-output}, is
\begin{equation}
  \Sigma_m^\text{GP} = V \Sigma_{m,z}^\text{GP} V\tran,
\end{equation}
where $\Sigma_{m,z}^\text{GP}$ is the (diagonal) predictive covariance in principal component space and $V$ is the PCA transformation matrix.
The GP matrix depends on the position $\xv$ in parameter space, but not strongly;
figure \ref{fig:correlation-matrices} shows a representative correlation matrix from a random point in parameter space.
Although it shares some qualitative features with the (assumed) experimental correlation matrix, the emulator correlation structure is \emph{not} assumed, it's a direct consequence of the empirical correlations in the model output data.

Model \emph{systematic} uncertainty arises from non-random imperfections in the computational model, such as grid discretization effects, uncertainty in the hydrodynamic equation of state, and negative contributions to Cooper-Frye.
It would be futile to attempt to enumerate every source of uncertainty and compute a covariance matrix for each;
instead, I define a simple parameter $\sigma_m\sys$ which is added in quadrature to the diagonal of $\Sigma_{m,z}^\text{GP}$ in principal component space, so that the complete model covariance matrix is
\begin{equation}
  \Sigma_m = V \Bigl[ \Sigma_{m,z}^\text{GP} + (\sigma_m\sys)^2I \Bigr] V\tran
  = \Sigma_m^\text{GP} + (\sigma_m\sys)^2 V\tran V.
  \label{eq:sigma-m-sys}
\end{equation}
The natural range of this parameter is zero to one, relative to the overall variance of the model:
$\sigma_m\sys = 0$ means no systematic uncertainty, $\sigma_m\sys = 1$ means that \emph{all} the model's variability is due to systematic uncertainty (which is obviously not the case).
Since we do not know the ``true'' value of $\sigma_m\sys$, I leave it as a free parameter with a gamma distribution prior,
\begin{equation}
  P(\sigma) \propto \sigma^2 e^{-\sigma/s}, \quad
  s = 0.05,
\end{equation}
which encodes that $\sigma_m\sys$ is greater than zero but less than about 0.4.
I will eventually marginalize over the posterior distribution for $\sigma_m\sys$, thereby accounting for our uncertainty in the uncertainty (not a typo).
This treatment, while rudimentary, is preferable to neglecting model systematic uncertainty.

\subsubsection{Multiple collision systems}

We may calibrate the model to data from multiple collision systems (or beam energies) by calculating an independent likelihood for each system and then the joint likelihood as the product
\begin{equation}
  P(\D|\xv) = \prod_s P(\D_s|\xv_s),
\end{equation}
where $s$ is an index over systems;
$\D_s$ and $\xv_s$ are the data and parameters for system $s$.
This factorized likelihood implicitly assumes that the uncertainties on observables from different collision systems are uncorrelated.

The parameters $\xv_s$ for each system are in general different:
There could be parameters which are specific to a particular system, or parameters with potentially different values for each system.
Example: Suppose we are calibrating to data from $n_s$ collision systems, and parameter $x_1$ depends on the system, but the other parameters $(x_2, \ldots, x_n)$ are common to all systems.
Writing $x_{1,s}$ for the value of $x_1$ for system $s$, the parameter vectors are
\begin{equation}
  \begin{aligned}
    \xv &= (x_{1,1}, x_{1,2}, \ldots, x_{1,n_s}, x_2, \ldots, x_n), \\
    \xv_s &= (x_{1,s}, x_2, \ldots, x_n),
  \end{aligned}
\end{equation}
so that $\xv_s$ contains the parameters for system $s$ and $\xv$ contains the union of all the parameters.
In such cases, we calibrate all the parameters $\xv$, distributing them to the appropriate system-specific likelihood functions.
This entails constructing an independent Gaussian process emulator for each system, each taking the system's particular parameters $\xv_s$ and predicting its outputs $\yv_{m,s}$.

\subsection{MCMC sampling}

Markov chain Monte Carlo (MCMC) sampling is the key to computational Bayesian inference.
A general class of algorithms for sampling probability distributions, MCMC methods produce a representative sample of the posterior distribution by generating a random walk through parameter space weighted by the posterior probability.
The sample (also called the chain) can then be used to calculate marginal distributions, derive parameter estimates, and create visualizations.

A simple, widely-used MCMC method is the Metropolis-Hastings algorithm \cite{Metropolis:1953,Hastings:1970}, which proceeds iteratively as follows:
Given a position $\xv_i$, randomly choose a new proposal position $\xv'$, then accept or reject $\xv'$ with probability based on the ratio of the posterior probabilities at $\xv_i$ and $\xv'$.
If accepted, set the next position to the proposal, $\xv_{i+1} = \xv'$, otherwise repeat the current position, $\xv_{i+1} = \xv_i$.
After repeating this many times, the distribution of the resulting positions $\{\xv_1, \xv_2, \ldots, \xv_n\}$ approximates the posterior distribution.

In this work, I use the affine-invariant ensemble sampler \cite{Goodman:2010}, an MCMC algorithm that uses a large ensemble of interdependent walkers.
Ensemble sampling tends to performs well in most contexts and converges to the posterior distribution faster than Metropolis-Hastings sampling.
Additionally, the walkers can be updated in parallel, affording a significant computational speed-up.
A stable, well-tested implementation of ensemble sampling is available in the Python library \textsc{emcee} \cite{ForemanMackey:2012ig}.

Since I have not personally developed the MCMC algorithm, I will not describe it in detail, but instead comment on some relevant practicalities.

\subsubsection{Computing the posterior probability}

Since the posterior typically varies over many orders of magnitude, it is numerically preferable to operate on its logarithm,
\begin{equation}
  \log P(\xv|\D) = \log P(\D|\xv) + \log P(\xv) + \text{const},
\end{equation}
where the additive constant is irrelevant in this context because only the ratio of probabilities, i.e.\ the difference of the logs, enters MCMC sampling.
The logs of the uniform prior \eqref{eq:uniform-prior} and likelihood \eqref{eq:likelihood} are
\begin{align}
  \log P(\xv) &=
  \begin{cases}
    0 & \text{if } \min(x_i) \le x_i \le \max(x_i) \text{ for all } i, \\
    -\infty & \text{else},
  \end{cases} \\[1ex]
  \log P(\D|\xv) &= -\frac{1}{2} \dv\tran \Sigma^{-1} \dv - \frac{1}{2}\log(\det\Sigma), \quad
  \dv = \yv_m(\xv) - \yv_e,
\end{align}
where I have dropped normalization constants.
Note that since the emulator predictive covariance is in general a function of $\xv$, the determinant of the covariance matrix $\Sigma$ is not constant and must be computed (if the covariance matrix were constant, we could safely neglect this term).

The likelihood contains the inverse and determinant of the covariance matrix, which are both $\order{n^3}$ operations.
Rather than evaluate the likelihood as written,  it is numerically faster and more stable to use the Cholesky decomposition of the covariance matrix,
\begin{equation}
  \Sigma = LL\tran,
\end{equation}
where $L$ is a lower triangular matrix.
This factorization is also an $\order{n^3}$ operation, but allows us to avoid computing the inverse or determinant explicitly.
Given a Cholesky decomposition, numerical linear algebra libraries can efficiently solve the linear equation
\begin{equation}
  LL\tran\alphav = \dv \quad \text{for} \quad \alphav = \Sigma^{-1}\dv.
\end{equation}
Since $L$ is a triangular matrix, its determinant is simply the product of its diagonal entries, so
\begin{equation}
  \det\Sigma = \det(LL\tran) = \det(L)^2 = \prod_i L_{ii}^2.
\end{equation}
Inserting these intermediate results, the log likelihood reduces to
\begin{equation}
  \log P(\D|\xv) = -\frac{1}{2} \dv\cdot\alphav - \sum_i \log L_{ii}.
\end{equation}

\subsubsection{Burn-in}

It takes a number of MCMC steps for the chain to converge to the posterior distribution, so it is almost always necessary to discard the first part of the chain.
This is called ``burn-in''.
After the burn-in phase, the chain (in principle) no longer depends on the starting position.
The necessary number of burn-in steps depends strongly on the specific problem and MCMC algorithm, but is usually hundreds or thousands.

\subsubsection{Number of walkers and steps}

In ensemble sampling, a large number of walkers is usually necessary for sampling high-dimensional distributions.
I use 1000 walkers as a default number, although that is likely overkill;
a few hundred would probably suffice in most cases.

I initialize the walkers at random positions in parameter space and run several hundred burn-in steps, perhaps up to 1000.
Sometimes, walkers that were initialized in very low-probability regions may become stuck and take a very long time to burn-in.
To accelerate this process, we can perform a two-stage burn-in:
Randomly initialize the walkers and run some burn-in steps, then resample the walker positions around the most probable positions sampled so far, and finally run some more burn-in steps.

After burn-in, I run $\order{10^3\text{--}10^4}$ steps to generate the posterior sample.
This is enough to create smooth histogram visualizations but is overkill for most other purposes, such as calculating medians or other summary statistics.
Keep in mind that the total number of samples is the number of walkers times the number of steps.

The fraction of accepted proposal points is an important MCMC performance metric:
If the acceptance fraction is very small (close to zero), that indicates that the walkers are stuck;
if the acceptance fraction is too large (close to one), that means the parameter space is being sampled completely randomly.
In both cases, the MCMC sample will not be representative of the posterior distribution.
I have typically observed acceptance fractions around 15--40\%, with higher-dimensional distributions usually having lower fractions.

\subsubsection{Marginal distributions}

A marginal distribution is a posterior distribution for a subset of the parameters, obtained by marginalizing over (integrating out) all the rest;
for example the marginal distribution for $x_1$ is
\begin{equation}
  P(x_1|\D) = \int dx_2 \cdots dx_n \, P(\xv|\D),
\end{equation}
and the joint marginal distribution for $(x_1, x_2)$ would be
\begin{equation}
  P(x_1,x_2|\D) = \int dx_3 \cdots dx_n \, P(\xv|\D).
\end{equation}
Given an MCMC sample $\{\xv_i\}$ of the posterior distribution, marginalization is trivial:
\begin{itemize}
  \item The values of $x_j$ from the MCMC sample, $\{x_{ji}\}$, is a sample of the marginal distribution $P(x_j|\D)$.
  \item The values $\{(x_{ji}, x_{ki})\}$ is a sample of $P(x_j,x_k|\D)$.
\end{itemize}
And so forth.

\subsubsection{Parameter uncertainties: credible intervals}

We quantify the uncertainty on a parameter by a credible interval---a range containing a certain fraction of the marginal distribution.
For example, a 90\% credible interval contains 90\% of the posterior density, and means that the true value of the parameter is expected to fall within the interval 90\% of the time (recall, the gravitational wave posterior distribution figure \ref{fig:ligo-posterior} showed 90\% credible intervals).
For a generic parameter $x$, let $x_l$ and $x_h$ be the lower and upper endpoints of a credible interval containing a fraction $0 < c < 1$, then assuming the marginal distribution of $x$ is unimodal, we can write
\begin{equation}
  \int_{x_l}^{x_h} dx \, P(x|\D) = c \int_{x_\text{min}}^{x_\text{max}} dx \, P(x|\D).
\end{equation}
More practically, we can extract credible intervals from an MCMC sample via its percentiles, for example 0--90\%, 1--91\%, \ldots, 10--100\% are all 90\% credible intervals.
The narrowest interval containing the desired fraction is called the highest posterior density (HPD) interval.
Given some samples $\{x_1, x_2, \ldots, x_n\}$, and again assuming a unimodal distribution, we can find the HPD interval as follows:
\begin{enumerate}
  \item Set $m = \operatorname{int}(c \times n)$, the number of samples contained in an interval of fraction $c$.
  \item Sort the samples in ascending order.
  \item Compute the widths of all $n - m$ intervals containing the fraction $c$, $\{x_m - x_1, x_{m + 1} - x_2, \ldots, x_n - x_{n - m}\}$.
  \item Choose the smallest interval.
\end{enumerate}
However, this algorithm is inefficient since it sorts all the samples, when we only need the upper and lower ends to be sorted.
To avoid this inefficiency, we can partition the samples on indices $m - n$ and $m$, then sort only the samples up to index $m - n$ and after index $m$.
The procedure is otherwise identical.

A credible \emph{region} is a generalization to multiple dimensions, e.g.\ an area enclosing some fraction of a two-dimensional joint posterior distribution between a pair of parameters (see again figure \ref{fig:ligo-posterior}).

\subsubsection{Visualizations}

The standard visualization of a posterior distribution is a triangle (or corner) plot:
A triangular grid of subplots with the marginal distributions for each parameter on the diagonal subplots and the joint distributions between each pair of parameters on the off-diagonal subplots.
Such visualizations compactly display the probability densities for all parameters and reveal correlations between parameters.
Operationally, the marginal distributions on the diagonal are histograms of MCMC samples, and the off-diagonal joint distributions are two-dimensional histograms (density plots).
In the next chapter, figures \ref{fig:v1-posterior-glb}, \ref{fig:v1-posterior-kln}, \ref{fig:v2-posterior}, and \ref{fig:v3-posterior} are triangle plots of actual posterior distributions for heavy-ion collision parameters.

Besides the distributions for the parameters themselves, it's also useful to visualize the model calculations compared to the experimental data.
In particular, we can plot the model calculations from each design point overlaid on the data points, then after calibration, make a similar plot showing emulator predictions of the model output from random draws of the posterior MCMC sample.
The first version, which effectively represents the prior on the model parameters, generally exhibits a wide spread around the data points, since there are several parameters varying across wide ranges.
In the second version---the posterior---the emulator predictions should be tightly clustered around the data, with the remaining spread arising from the finite width of the posterior distribution.
Figures of this type: \ref{fig:v1-observables}, \ref{fig:v2-observables-samples}, \ref{fig:v3-observables-design}, \ref{fig:v3-observables-posterior}.

\section{Computational implementation}
\label{sec:comp-impl}

I have developed a complete parameter estimation code implementing the methods and strategies detailed in this chapter;
it is the basis for my latest analysis, the results of which are presented in section \ref{sec:v3}.
The source code is publicly available at \url{https://github.com/jbernhard/hic-param-est} with documentation at \url{http://qcd.phy.duke.edu/hic-param-est}.
I encourage interested readers to peruse the code and documentation, since in many cases it is not obvious how to translate theoretical concepts into functioning code.

The code makes use of several open-source Python libraries:
\textsc{NumPy} \cite{numpy} and \textsc{SciPy} \cite{scipy} for general scientific computing,
\textsc{scikit-learn} \cite{scikit-learn} for principal component analysis and Gaussian processes,
\textsc{emcee} \cite{ForemanMackey:2012ig} for MCMC sampling,
\textsc{h5py} \cite{h5py} for data storage,
\textsc{matplotlib} \cite{matplotlib} for generating plots.

\chapter{Quantifying properties of hot and dense QCD matter}
\chaptermark{Quantifying QCD properties}
\label{ch:quant-qcd-props}

\lettrine{O}{ver} the past several years, I have conducted a series of case studies applying Bayesian parameter estimation to relativistic heavy-ion collisions, each time improving the analysis and advancing toward the ultimate goal:
to quantitatively determine the properties of the quark-gluon plasma.
This chapter is an exhibit of these case studies.

The first two studies, which are published \cite{Bernhard:2015hxa,Bernhard:2016tnd}, are somewhat limited---primarily by earlier and less sophisticated versions of both the computational model of chapter \ref{ch:models} and the parameter estimation method of chapter \ref{ch:param-est}---but they nonetheless represent significant steps forward.
After explaining the meaningful differences in these earlier iterations, I will defer to the discussion I previously wrote.

The third and final study is state of the art:
It eliminates many of the shortcomings in the first two and realizes the most precise estimates of QGP properties to date.
These are new results, as of yet unpublished.

\bigskip

\noindent The trio:
\begin{enumerate}[label=\Roman*.]
  \item \nameref{sec:v1}
  \item \nameref{sec:v2}
  \item \nameref{sec:v3}
\end{enumerate}

\section{A proof of concept}
\label{sec:v1}

The purpose of this first study \cite{Bernhard:2015hxa} is to begin developing the parameter estimation method and validate that it is a viable strategy in heavy-ion physics.
It succeeds in doing so, with the results \emph{quantitatively} confirming prior \emph{qualitative} knowledge about the model parameters while revealing some previously unknown details.
On a philosophical level, this is a positive outcome.

\subsection{Differences from the present work}
\label{subsec:v1-diffs}

\paragraph{Initial conditions}

This study precedes \trento, instead using two existing initial condition models:
the Monte Carlo Glauber model \cite{Miller:2007ri}, a widely-used geometric model, and the Monte Carlo KLN model \cite{Drescher:2006pi}, an implementation of color-glass condensate (CGC) effective field theory.
The parameter estimation process is carried out separately for the two models.
There is no pre-equilibrium free-streaming stage.

\paragraph{Hydrodynamics and particlization}

The hydrodynamic model has constant shear viscosity $\eta/s$ (no temperature dependence) and lacks bulk viscosity.
The particlization routine was contributed by the Ohio State University group \cite{Shen:2014vra}.

\paragraph{Parameters and observables}

There is a modest set of five calibration parameters:
\begin{enumerate}
  \item Initial condition normalization factor.
  \item A parameter specific to the initial condition model.
    Glauber: The binary collision fraction $\alpha$, which controls how entropy is distributed to wounded nucleons and binary collisions.
    KLN: The saturation scale exponent $\lambda$, a CGC parameter.
  \item QGP thermalization time (and hydrodynamic starting time) $\tau_0$.
  \item Constant QGP specific shear viscosity $\eta/s$.
  \item Shear relaxation time $\tau_\pi$, controlled via the coefficient $k_\pi$ in the relation $\tau_\pi = 5k_\pi\eta/(sT)$.
\end{enumerate}
Table \ref{tab:v1-design} summarizes the parameters and their design ranges.

\begin{table}[t]
  \centering
  \captionsetup{width=.79\textwidth}
  \caption{
    Input parameter ranges for the Glauber and KLN initial condition models and for the hydrodynamic model.
  }
  \label{tab:v1-design}
  \small
  \begin{tabular}{lll}
    \toprule
    Parameter & Description & Range \\
    \midrule
    Glauber Norm & Overall normalization & 20--60 \\
    Glauber $\alpha$ & Binary collision fraction & 0.05--0.30 \\
    KLN Norm & Overall normalization & 5--15 \\
    KLN $\lambda$ & Saturation scale exponent & 0.1--0.3 \\
    $\tau_0$ & Thermalization time & 0.2--1.0 fm \\
    $\eta/s$ & Specific shear viscosity & 0--0.3 \\
    $k_\pi$ & Shear relaxation time coefficient & 0.2--1.1 \\
    \bottomrule
  \end{tabular}
\end{table}

The observables are the centrality dependence of the average charged-particle multiplicity $\avg{\Nch}$ and the flow cumulants $\vnk 2 2$, $\vnk 2 3$, with experimental data from the ALICE experiment, Pb-Pb collisions at $\sqrt s = 2.76$ TeV \cite{Abelev:2014mda}.

\paragraph{Parameter estimation method}

Most aspects of the parameter estimation method are similar or identical to chapter \ref{ch:param-est}, including the Latin-hypercube parameter design, principal component analysis of the model output, Gaussian process emulator, and MCMC algorithm.
The primary difference is much less sophisticated uncertainty quantification than in subsection \ref{subsec:likelihood-uq}.
The likelihood is evaluated in principal component space as
\begin{equation}
  P(\D|\xv) \propto \exp\biggl\{
    -\frac{1}{2} [\zv_m(\xv) - \zv_e]\tran \Sigma_z^{-1} [\zv_m(\xv) - \zv_e]
  \biggr\},
\end{equation}
where $\zv_e$ is the PCA transformation of the experimental data $\yv_e$ and $\zv_m(\xv)$ contains the values of the principal components, predicted by the Gaussian processes, at parameter point $\xv$.
The covariance matrix is diagonal in principal component space with a simple fractional uncertainty:
\begin{equation}
  \Sigma_z = \diag(\sigma_z^2\,\zv_e), \quad
  \sigma_z = 0.06.
\end{equation}
This assumption precludes rigorous quantitative uncertainties on the model parameters, but does not invalidate the overall results.
(Editorial comment: This was a stopgap.
As I wrote in the original publication,
``The primary goal of this study is to develop and test a model-to-data comparison framework;  details such as the precise treatment of uncertainties can be improved later.'')

\subsection{Results and discussion}

This subsection is adapted from: \\[1ex]
\fullcite{Bernhard:2015hxa}.

\medskip

\begin{figure}[p]
  \makebox[\textwidth]{
    \includegraphics[width=1.25\textwidth]{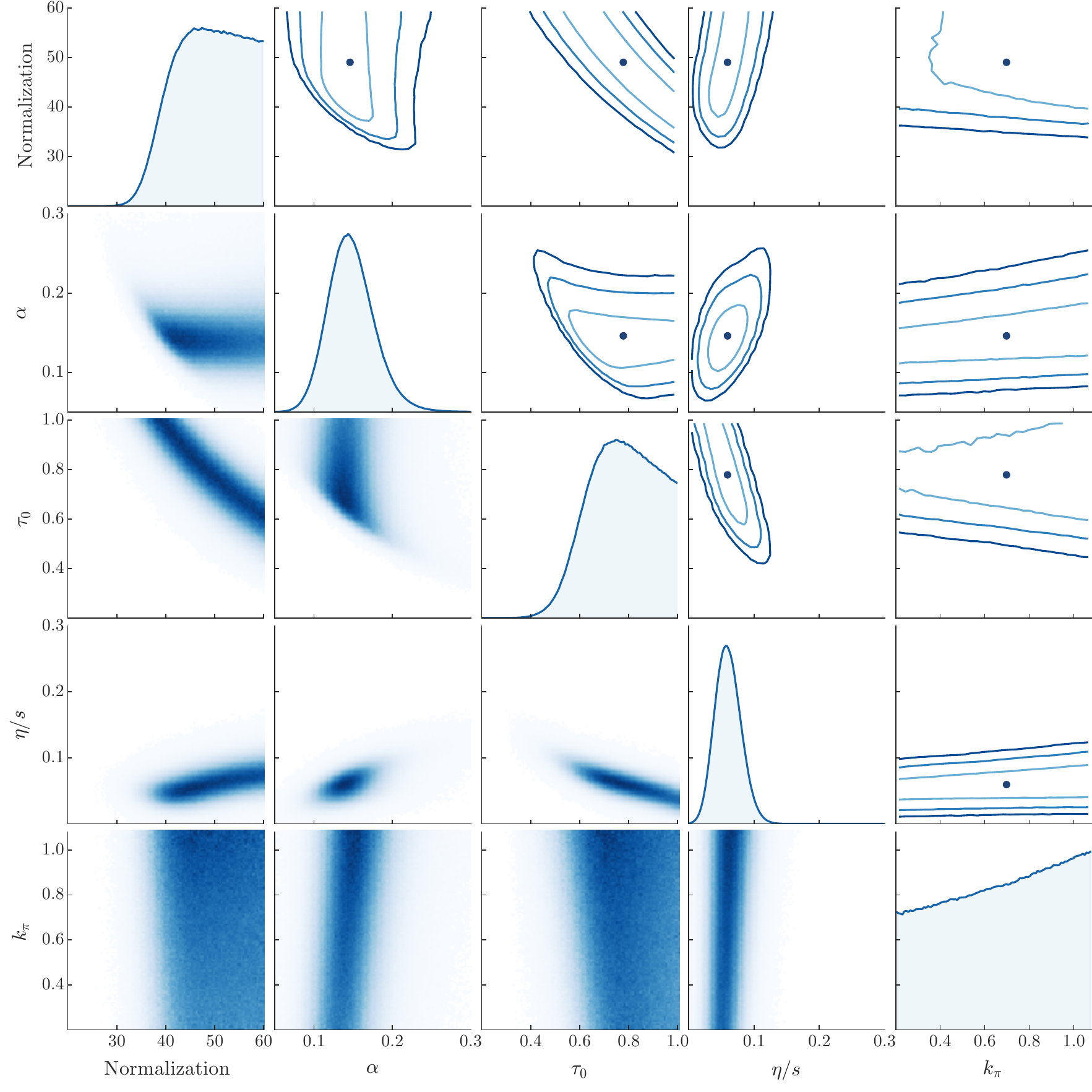}
  }
  \caption{
    Posterior marginal and joint distributions of the calibration parameters for the Glauber model.
    On the diagonal are histograms of MCMC samples for the respective parameters,
    on the lower triangle are two-dimensional scatter histograms of MCMC samples showing the correlation between pairs of parameters,
    and on the upper triangle are approximate contours for 68\%, 95\%, and 99\% credible regions along with a dot indicating the median.
  }
  \label{fig:v1-posterior-glb}
\end{figure}

\begin{figure}[p]
  \makebox[\textwidth]{
    \includegraphics[width=1.25\textwidth]{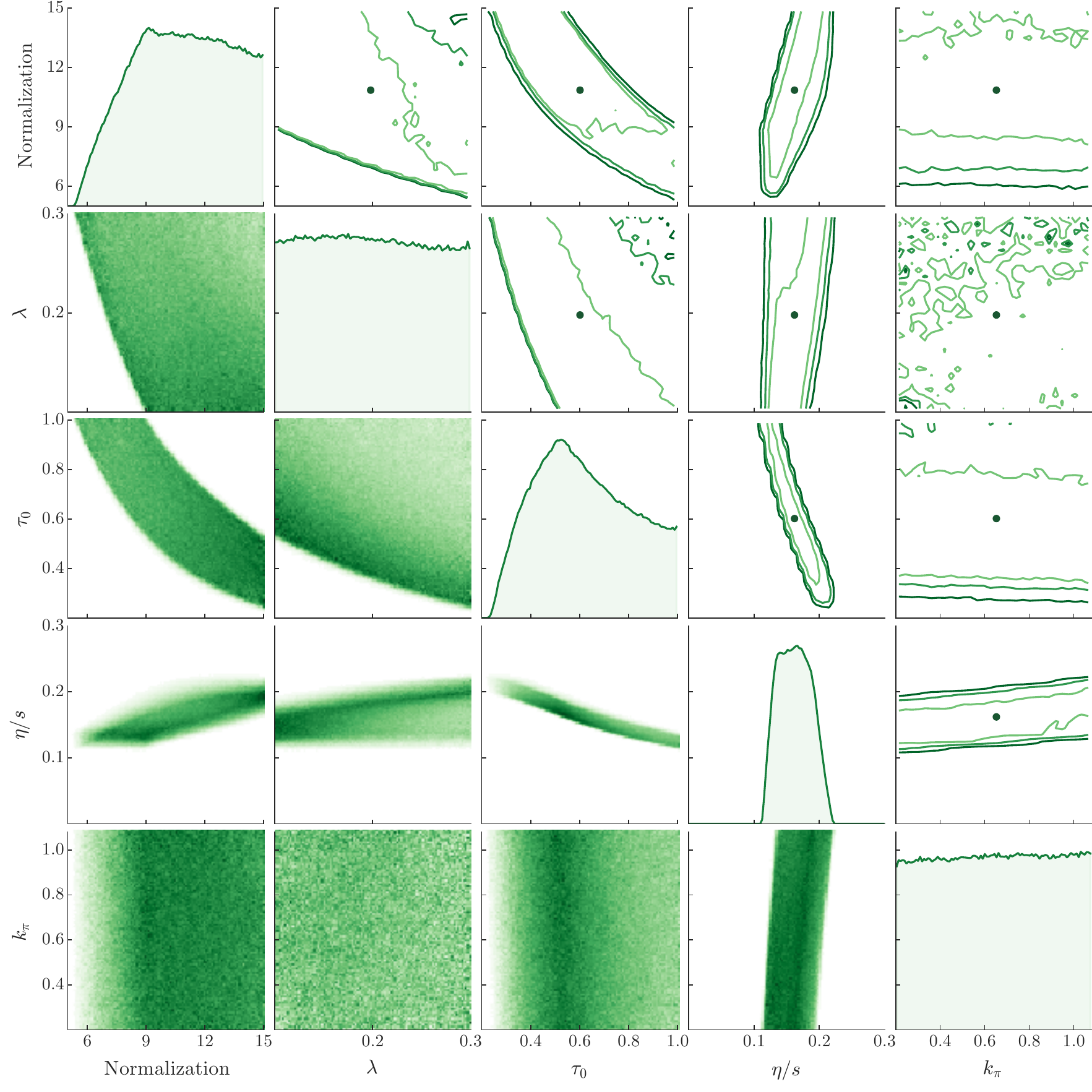}
  }
  \caption{Same as figure \ref{fig:v1-posterior-glb} for the KLN model.}
  \label{fig:v1-posterior-kln}
\end{figure}

\noindent The primary MCMC calibration results are presented in figures \ref{fig:v1-posterior-glb} and \ref{fig:v1-posterior-kln} for the Glauber and KLN models, respectively.
These are visualizations of the posterior probability distributions of the true parameters, including the distribution of each individual parameter and all correlations.
The diagonal histograms show the marginal distributions for each parameter (all other parameters integrated out);
the lower-triangle plots are two-dimensional scatter histograms of joint distributions between pairs of parameters, where darker color denotes higher probability density;
and the upper triangle has contour plots of the same joint distributions, where the contour lines enclose the 68\%, 95\%, and 99\% credible regions.

A wealth of information may be gained from these posterior visualizations; the following highlights some important features.

Focusing on the Glauber results in figure \ref{fig:v1-posterior-glb}, we see the shear viscosity $\eta/s$ (fourth diagonal plot) has a narrow approximately normal distribution located near the commonly quoted value 0.08.
As expected, $\eta/s$ is tightly constrained by experimental flow data.
Going across the fourth row, we observe nontrivial correlations among $\eta/s$ and other parameters, for example, $\eta/s$ and the hydrodynamic thermalization time $\tau_0$ are negatively correlated (fourth row, third column).
As $\tau_0$ increases, the medium expands as a fluid for less time, so less flow develops, and viscosity must decrease to compensate.

Both $\tau_0$ and normalization (third and first diagonals) have broad distributions without strong peaks, and they are strongly-correlated (third row, first column).
This is because the hydrodynamic model is boost-invariant and lacks any pre-equilibrium dynamics, so $\tau_0$ is effectively an inverse normalization factor.
The joint distribution shows a narrow acceptable band whose shape is governed by the inverse relationship.

The wounded nucleon / binary collision parameter $\alpha$ (second diagonal) has a roughly-normal distribution located near the typical value 0.12.
It is mainly related to the slope of multiplicity vs.\ centrality and hence has a nontrivial correlation with normalization and $\tau_0$, e.g.\ we can decrease the normalization to the lower end of its distribution provided we also increase $\alpha$ to compensate.

Meanwhile, the shear stress relaxation time coefficient $k_\pi$ (fifth diagonal) has an almost flat distribution and its joint distributions show no correlations.
Evidently, this parameter does not influence flow coefficients or multiplicity.

The KLN results in figure \ref{fig:v1-posterior-kln} generally exhibit wider, less normal distributions than Glauber.
This could indicate an inferior fit to the data and suggests that KLN is somewhat less flexible than Glauber, i.e.\ its overall behavior is relatively insensitive to the specific values of input parameters.

\begin{table}[t]
  \caption{
    Quantitative summary of posterior distributions.
    For each parameter, the previous estimate \cite{Shen:2011zc,Heinz:2011kt}, mean, median, and credible intervals are given.
    Credible intervals are computed from central percentiles, e.g.\ the 68\% interval is 16--84\%.
  }
  \label{tab:v1-posterior}
  \small
  \makebox[\textwidth]{
    \begin{tabular}{lllllccc}
      \toprule
      &           &             &      &        & \multicolumn{3}{c}{Credible intervals} \\
      \cmidrule(l){6-8}
      & Parameter & Prev.\ est. & Mean & Median & 68\% & 95\% & 99\% \\
      \midrule
      \multirow{5}{*}{\rotatebox{90}{Glauber\ \ }}
      & Norm. & 57 & 48.9 & 49.0 & 41.6--56.4 & 36.5--59.4 & 33.9--59.9 \\
      & $\alpha$ & 0.12 & 0.148 & 0.146 & 0.119--0.176 & 0.0954--0.212 & 0.0808--0.242 \\
      & $\tau_0$ & 0.6 & 0.776 & 0.778 & 0.638--0.922 & 0.527--0.987 & 0.461--0.997 \\
      & $\eta/s$ & 0.08 & 0.0604 & 0.0595 & 0.0407--0.0801 & 0.0244--0.101 & 0.0149--0.116 \\
      & $k_\pi$ & 0.5 & 0.682 & 0.698 & 0.373--0.978 & 0.228--1.08 & 0.206--1.09 \\
      \midrule
      \multirow{5}{*}{\rotatebox{90}{KLN\ \ }}
      & Norm. & 9.9 & 10.8 & 10.9 & 8.15--13.6 & 6.40--14.8 & 5.82--15.0 \\
      & $\lambda$ & 0.14 & 0.199 & 0.198 & 0.132--0.267 & 0.105--0.295 & 0.101--0.299 \\
      & $\tau_0$ & 0.6 & 0.620 & 0.602 & 0.415--0.846 & 0.302--0.975 & 0.265--0.995 \\
      & $\eta/s$ & 0.20 & 0.163 & 0.162 & 0.135--0.190 & 0.121--0.208 & 0.116--0.215 \\
      & $k_\pi$ & 0.5 & 0.651 & 0.653 & 0.347--0.955 & 0.223--1.07 & 0.205--1.09 \\
      \bottomrule
    \end{tabular}
  }
\end{table}

\begin{figure}[p]
  \centering
  \color[HTML]{262626}
  \hspace{1.6em} Prior \\[1ex]
  \makebox[\textwidth]{
    \includegraphics[width=1.25\textwidth]{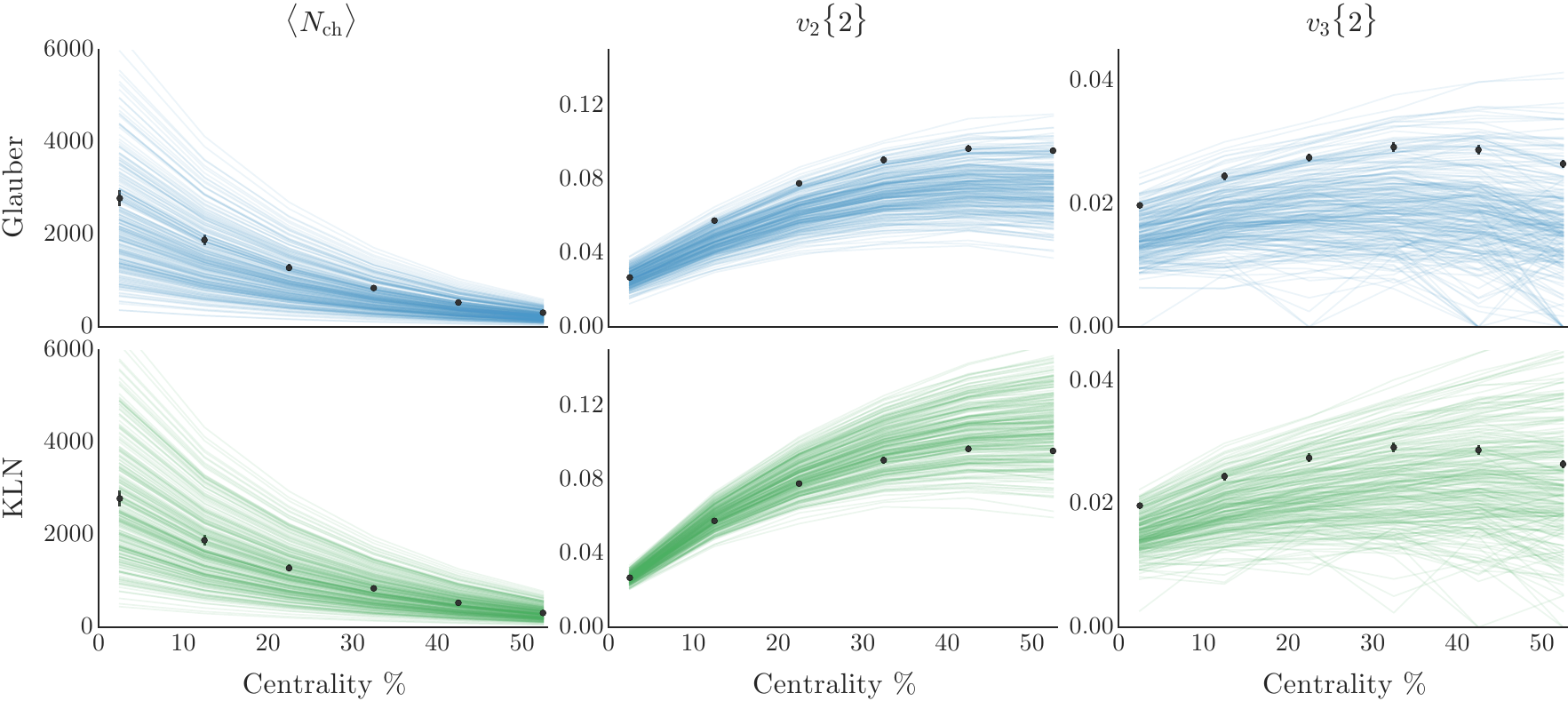}
  } \\[1.5ex]
  \hspace{1.6em} Posterior \\[1ex]
  \makebox[\textwidth]{
    \includegraphics[width=1.25\textwidth]{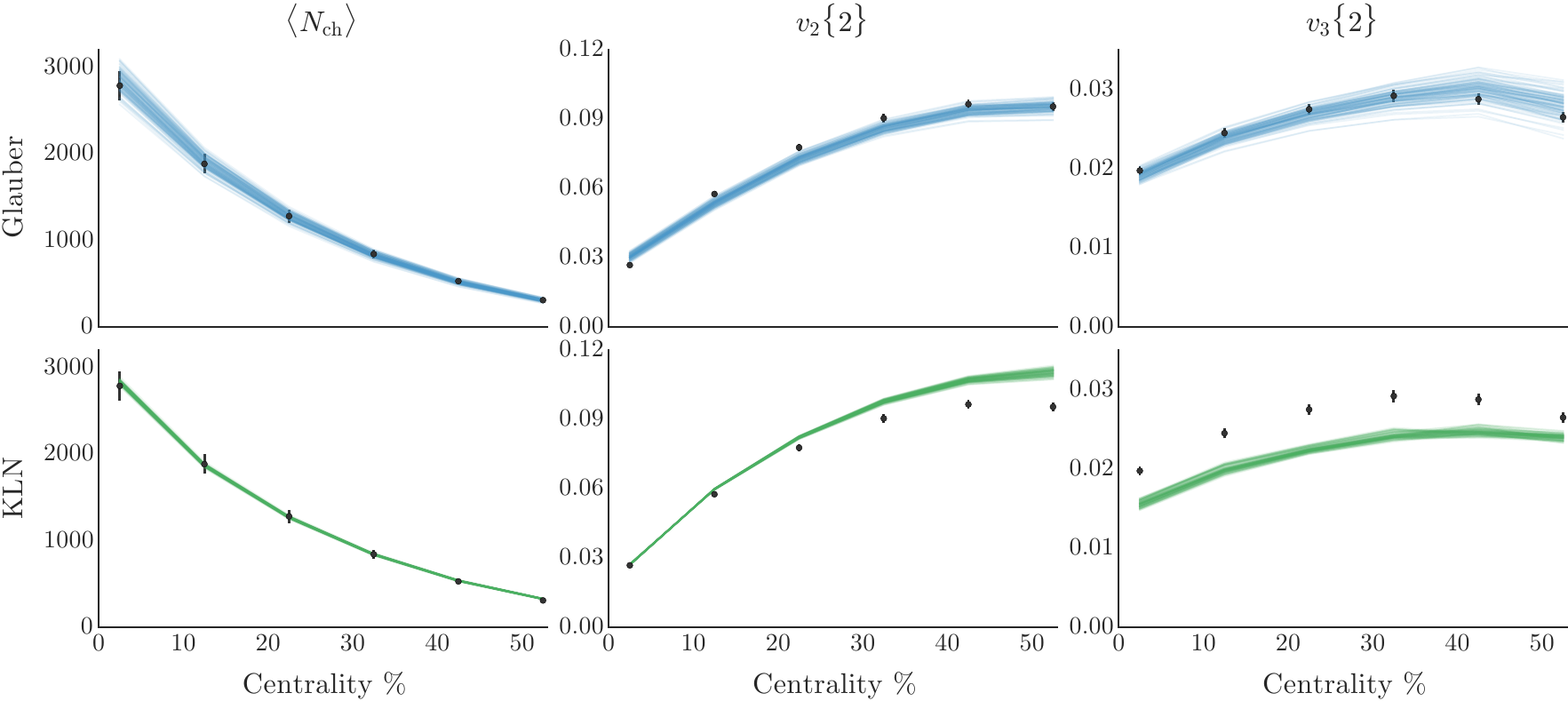}
  }
  \caption{
    Top two rows (prior): Model calculations from Glauber (blue) and KLN (green) initial conditions at each design point.
    Bottom two rows (posterior): Random samples of the calibrated posterior distributions for Glauber and KLN.
    From left to right:
    average charged-particle multiplicity $\avg\Nch$,
    elliptic flow two-particle cumulant $\vnk 2 2$,
    and triangular flow two-particle cumulant $\vnk 3 2$.
    Data points are experimental measurements from ALICE \cite{Abelev:2014mda}.
  }
  \label{fig:v1-observables}
\end{figure}

The shear viscosity $\eta/s$ has a narrow, irregular distribution covering the common value 0.20.
As with Glauber, $\eta/s$ has a negative correlation with $\tau_0$, there is a strong inverse relationship between normalization and $\tau_0$, and $k_\pi$ has no effect.
The KLN parameter $\lambda$ has a flat marginal distribution, but there are strongly excluded regions in the joint distributions with normalization and $\tau_0$.
This appears to be the same effect as observed with Glauber $\alpha$, except the dependence on $\lambda$ is significantly weaker.

The posteriors may be validated by drawing samples from the calibrated distributions and visualizing the corresponding emulator predictions:
if the model is correct and properly calibrated, the posterior samples will be close to experimental measurements.
Figure \ref{fig:v1-observables} confirms---for the most part---that the posteriors are indeed tightly clustered around the data points.
Visualizations such as this will always have some uncertainty since samples are drawn from the full posterior, however, the posterior samples in the bottom of the figure are markedly narrower than the prior calculations in the top, in which the input parameters varied across their full ranges and were not tuned to match experiment.

\begin{figure}[t]
  \graphicsandcaption{.8}{quant-qcd-props/v1/posterior_compare}{
    Comparison of posterior distributions of $\eta/s$ for Glauber (blue) and KLN (green).
    These are the same histograms as in figures \ref{fig:v1-posterior-glb} and \ref{fig:v1-posterior-kln}, expanded and placed on the same axis.
    The vertical grey lines indicate the common values 0.08 for Glauber and 0.20 for KLN \cite{Shen:2011zc,Heinz:2011kt}.
  }
  \label{fig:v1-posterior-compare}
\end{figure}

As shown in the posterior samples of figure \ref{fig:v1-observables}, the Glauber model nearly fits the centrality dependence of all the present observables ($\avg\Nch$, $\vnk 2 2$, $\vnk 3 2$).
The $v_3$ samples have a somewhat larger variance than the others, in part due to the underlying noise in the model calculations and also because $v_3$ is explicitly given a lower weight (recall that $\avg\Nch$~:~$\vnk 2 2$~:~$\vnk 2 3$ are weighted 1.2~:~1.0~:~0.6).

The KLN results in the bottom row tell a somewhat different story, as they cannot fit all observables simultaneously.
While the fit to $\avg\Nch$ is excellent, the ratio of $v_2$ to $v_3$ is simply too large and the model has no choice but to compromise between the two, similar to previous KLN results \cite{Qiu:2011hf}.
The posterior biases more towards $v_2$ than $v_3$ due to the explicit higher weight on $v_2$.

Figure \ref{fig:v1-posterior-compare} shows an expanded view of the $\eta/s$ marginal distributions for Glauber and KLN.
The Glauber distribution is approximately normal with mean ${\sim}$0.06 and 95\% credible interval ${\sim}$0.02--0.10, consistent with but mostly below 0.08.
This is unsurprising and easily within the uncertainty of existing results.
KLN has a wider plateau-like distribution with mean ${\sim}$0.16 and 95\% credible interval ${\sim}$0.12--0.21.
While the common estimate 0.20 was derived primarily from comparisons to $v_2$, the additional constraint from $v_3$ shifts the distribution to somewhat smaller values and causes the plateau shape:
Rather than a strong peak, there is a range of values which all fit the data roughly equally well.

Table \ref{tab:v1-posterior} quantitatively summarizes the posterior distributions for each parameter including basic statistics, credible intervals, and comparisons to previous estimates from earlier work with the same models \cite{Shen:2011zc,Heinz:2011kt}.
All previous estimates fall within 95\% credible intervals, and most within 68\%.

\section{A more flexible approach}
\label{sec:v2}

With a markedly improved computational model, more parameters, and increased constraining power from additional observables, this analysis \cite{Bernhard:2016tnd} delivers new insights on the initial state of heavy-ion collisions and on QGP medium properties, especially the temperature dependence of shear and bulk viscosity.

Compared to \emph{\nameref{sec:v1}}, the present model is much more flexible, owing in large part to the parametric initial condition model \trento\ (section \ref{sec:ic}).
This adaptability is of paramount importance to ensure faithful uncertainty quantification.
Consider, for example, the posterior distributions for $\eta/s$ in figure \ref{fig:v1-posterior-compare}, obtained using the Glauber and KLN initial condition models;
the two distributions are almost entirely incompatible, despite the hydrodynamic model and the rest of the analysis being identical.
This happened, in short, because the KLN model tends to produce more elliptic initial geometry than Glauber, so requires a larger $\eta/s$ to describe elliptic flow $v_2$.

More generally, the choice of initial condition model can strongly affect the estimates of $\eta/s$ and other QGP medium properties.
Since we do not know the precise nature of the initial state, we should incorporate that uncertainty into our estimates of other model parameters.
The \trento\ model enables this by parametrically interpolating among a family of physically reasonable initial condition models, so that when we marginalize over its parameters, we propagate any remaining uncertainty into all other parameter estimates.
Thus, by employing a flexible model, we can \emph{simultaneously} characterize the initial state and QGP medium.

Another way to view this:
Choosing a specific initial condition model is a strong prior, equivalent to asserting that particular model \emph{is} the true initial condition.
The posterior distribution will then reflect this prior.
So perhaps the $\eta/s$ posterior distributions for Glauber and KLN are compatible after all---they are simply the consequences of different priors.
On the other hand, \trento\ is effectively a weak prior on the initial condition;
as demonstrated in section \ref{sec:ic}, it can mimic the behavior of---and continuously interpolate among---various particular initial condition models, including KLN, IP-Glasma, EKRT, and wounded nucleon.

\subsection{Differences from the present work}

\paragraph{Initial conditions}

The \trento\ model is identical to the description in section \ref{sec:ic}, except there is no minimum nucleon distance parameter.
There is no pre-equilibrium free-streaming stage.

\paragraph{Hydrodynamics and particlization}

The hydrodynamic model has tem\-perature-dependent shear and bulk viscosity, although the parametrizations are somewhat different from section \ref{sec:hydro} (see below).
The particlization model lacks bulk viscous corrections (but does implement shear corrections).
As stated in the original publication,
``This precludes any quantitative conclusions on bulk viscosity, since we are only allowing bulk viscosity to affect the hydrodynamic evolution, not particlization.
We will, however, be able to determine whether $\zeta/s$ is nonzero.''

\paragraph{Parameters}

There are nine model parameters for estimation, summarized with their ranges in table \ref{tab:v2-design}.
Four control the parametric initial state:
\begin{enumerate}
  \item Initial condition normalization factor.
  \item \trento\ entropy deposition parameter $p$ in the generalized mean ansatz
    \begin{equation}
      s \propto \left( \frac{\T_A^p + \T_B^p}{2} \right)^{1/p},
      \label{eq:genmean}
    \end{equation}
    where $\T$ is a fluctuated participant thickness function
    \begin{equation}
      \T(x, y) = \sum\limits_{i=1}^{N_\text{part}} u_i\, T_p(x - x_i, y - y_i),
    \end{equation}
    with $u_i$ (a random fluctuation factor) and $T_p$ (the nucleon thickness function) defined below.
  \item Multiplicity fluctuation parameter $k$.
    Nucleon fluctuation factors $u_i$ are sampled from a gamma distribution with unit mean and variance $1/k$, whose probability density is
    \begin{equation}
      P_k(u) = \frac{k^k}{\Gamma(k)} u^{k-1} e^{-ku}.
    \end{equation}
  \item Gaussian nucleon width $w$, which determines initial-state granularity through the nucleon thickness function
    \begin{equation}
      T_p(x, y) = \frac{1}{2\pi w^2} \exp\bigg(\!-\frac{x^2 + y^2}{2 w^2}\bigg).
    \end{equation}
\end{enumerate}
\begin{table}[t]
  \centering
  \captionsetup{width=.75\textwidth}
  \caption{
    Input parameter ranges for the initial condition and hydrodynamic models.
  }
  \label{tab:v2-design}
  \small
  \begin{tabular}{lll}
    \toprule
    Parameter         & Description                        & Range           \\
    \midrule
    Norm              & Overall normalization              & 100--250        \\
    $p$               & Entropy deposition parameter       & $-1$ to $+1$    \\
    $k$               & Multiplicity fluct.\ shape         & 0.8--2.2        \\
    $w$               & Gaussian nucleon width             & 0.4--1.0 fm     \\
    $\eta/s$ hrg      & Const.\ shear viscosity, $T < T_c$ & 0.3--1.0        \\
    $\eta/s$ min      & Shear viscosity at $T_c$           & 0--0.3          \\
    $\eta/s$ slope    & Slope above $T_c$                  & 0--2 GeV$^{-1}$ \\
    $\zeta/s$ norm    & Prefactor for $(\zeta/s)(T)$       & 0--2            \\
    $T_\text{switch}$ & Particlization temperature         & 135--165 MeV    \\
    \bottomrule
  \end{tabular}
\end{table}
The remaining five parameters are related to the QGP medium:
\begin{enumerate}[itemsep=0pt]
  \item[5--7.] The three parameters ($\eta/s$ hrg, min, and slope) that set the temperature dependence of the specific shear viscosity in the piecewise linear parametrization
    \begin{equation}
      (\eta/s)(T) =
      \begin{cases}
        (\eta/s)_\text{min} + (\eta/s)_\text{slope} (T - T_c) & T > T_c \\
        (\eta/s)_\text{hrg}                                   & T \le T_c
      \end{cases}.
      \label{eq:v2-etas}
    \end{equation}
  \setcounter{enumi}{7}
  \item Normalization prefactor $(\zeta/s)_\text{norm}$ for the temperature dependence of bulk viscosity, parametrized as \cite{Denicol:2009am,Ryu:2015vwa}
    \begin{equation}
      (\zeta/s)(T) = (\zeta/s)_\text{norm}
      \begin{cases}
        \begin{aligned}
          C_1 &+ \lambda_1 \exp [(x-1)/\sigma_1]  \\ &+ \lambda_2 \exp [ (x-1)/\sigma_2]
        \end{aligned}
        &T < T_a \\[3ex]
        A_0 + A_1 x + A_2 x^2 &T_a \le T \le T_b \\[2ex]
        \begin{aligned}
          C_2 &+ \lambda_3 \exp [-(x-1)/\sigma_3]  \\ &+ \lambda_4 \exp [-(x-1)/\sigma_4]
        \end{aligned}
        &T > T_b
      \end{cases},
      \label{eq:v2-zetas}
    \end{equation}
    with $x = T/T_0$ and coefficients
    \begin{align*}
      &C_1=0.03,\quad C_2=0.001, \\
      &A_0=-13.45,\quad A_1=27.55,\quad A_2=-13.77, \\
      &\sigma_1=0.0025,\quad \sigma_2=0.022,\quad \sigma_3=0.025,\quad \sigma_4=0.13, \\
      &\lambda_1=0.9,\quad \lambda_2=0.22,\quad \lambda_3=0.9,\quad \lambda_4=0.25, \\
      &T_0 = 0.18 \text{ GeV},\quad T_a = 0.995\, T_0,\quad T_b = 1.05\, T_0.
    \end{align*}
    Qualitatively, this form peaks near $T_0 = 180$~MeV and falls off exponentially on either side.
  \item Particlization temperature $T_\text{switch}$.
\end{enumerate}

\paragraph{Observables}

Centrality dependence of identified particle yields $dN/dy$ and mean transverse momenta $\avg{p_T}$, for charged pions, kaons, and protons, as well as two-particle anisotropic flow coefficients $\vnk n 2$ for $n = 2$, 3, 4.
Table \ref{tab:v2-observables} summarizes the observables including kinematic cuts, centrality classes, and experimental data, which are all from the ALICE experiment, Pb-Pb collisions at $\sqrt{s_{NN}} = 2.76$ TeV \cite{Abelev:2013vea,ALICE:2011ab}.

\begin{table}[h]
  \caption{
    Experimental data to be compared with model calculations.
  }
  \label{tab:v2-observables}
  \small
  \makebox[\textwidth]{
    \begin{tabular}{lcccc}
      \toprule
      Observable & Particle species & Kinematic cuts & Centrality classes & Ref. \\
      \midrule
      Yields $dN/dy$                       & $\pi^\pm$, $K^\pm$, $p\bar p$ &
      $|y| < 0.5$ & 0--5, 5--10, 10--20, \ldots, 60--70 & \cite{Abelev:2013vea} \\
      \noalign{\smallskip}
      Mean transverse momentum $\avg{p_T}$ & $\pi^\pm$, $K^\pm$, $p\bar p$ &
      $|y| < 0.5$ & 0--5, 5--10, 10--20, \ldots, 60--70 & \cite{Abelev:2013vea} \\
      \noalign{\smallskip}
      Two-particle flow cumulants $\vnk n 2$ & \multirow{2}{*}{all charged} &
      $|\eta| < 1$ & 0--5, 5--10, 10--20, \ldots, 40--50 &
      \multirow{2}{*}{\cite{ALICE:2011ab}} \\
      $n = 2$, 3, 4 & & $0.2 < p_T < 5.0$ GeV & $n = 2$ only: 50--60, 60--70 & \\
      \bottomrule
    \end{tabular}
  }
\end{table}

\paragraph{Parameter estimation method}

Nearly the same as in \emph{\nameref{sec:v1}} (subsection \ref{subsec:v1-diffs}), namely, the likelihood is
\begin{equation}
  P(\D|\xv) \propto \exp\biggl\{
    -\frac{1}{2} [\zv_m(\xv) - \zv_e]\tran \Sigma_z^{-1} [\zv_m(\xv) - \zv_e]
  \biggr\},
\end{equation}
with covariance matrix
\begin{equation}
  \Sigma_z = \diag(\sigma_z^2\,\zv_e), \quad
  \sigma_z = 0.10.
  \label{eq:v2-cov}
\end{equation}
The sole difference is the uncertainty fraction: a more conservative 10\% compared to 6\% previously.

\subsection{Results and discussion}

This subsection is adapted from: \\[1ex]
\fullcite{Bernhard:2016tnd}.

\medskip

\begin{figure}[p]
  \makebox[\textwidth]{
    \includegraphics[width=1.3\textwidth]{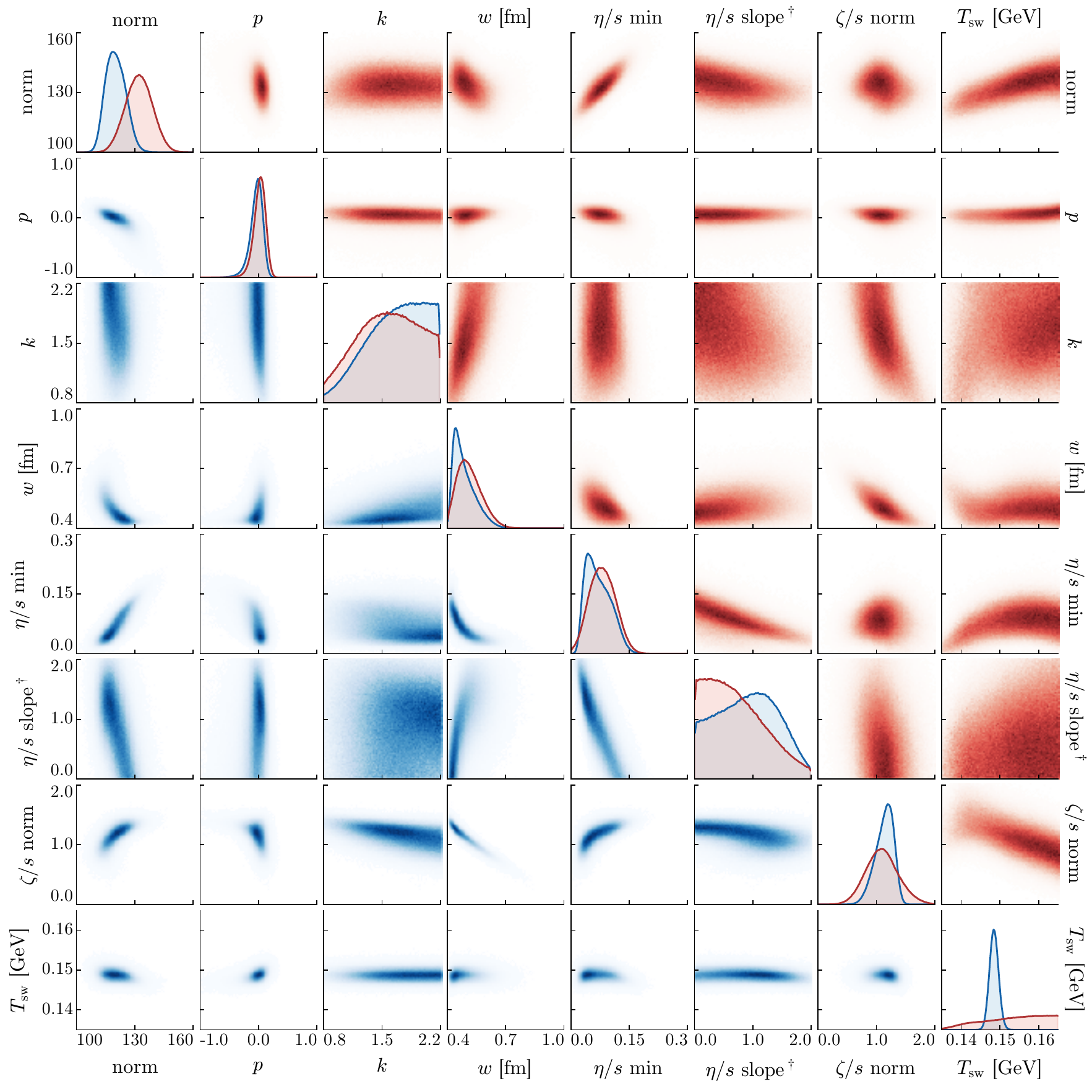}
  }
  \caption{
    Posterior distributions for the model parameters from calibrating to identified particles yields (blue, lower triangle) and charged particles yields (red, upper triangle).
    The diagonal has marginal distributions for each parameter, while the off-diagonal contains joint distributions showing correlations among pairs of parameters.
    $^\dagger$The units for $\eta/s$ slope are [GeV$^{-1}$].
  }
  \label{fig:v2-posterior}
\end{figure}

\noindent The primary result of this study is the posterior distribution for the model parameters, figure \ref{fig:v2-posterior}.
In fact, this figure contains two posterior distributions:
one from calibrating to identified particle yields $dN/dy$ (blue, lower triangle),
and the other from calibrating to charged particle yields $d\Nch/d\eta$ (red, upper triangle).
We performed the alternate calibration to charged particles because the model could not simultaneously describe all identified particle yields for \emph{any} parameter values, as will be demonstrated shortly.

\begin{figure}[t]
  \makebox[\textwidth]{
    \includegraphics[width=1.25\textwidth]{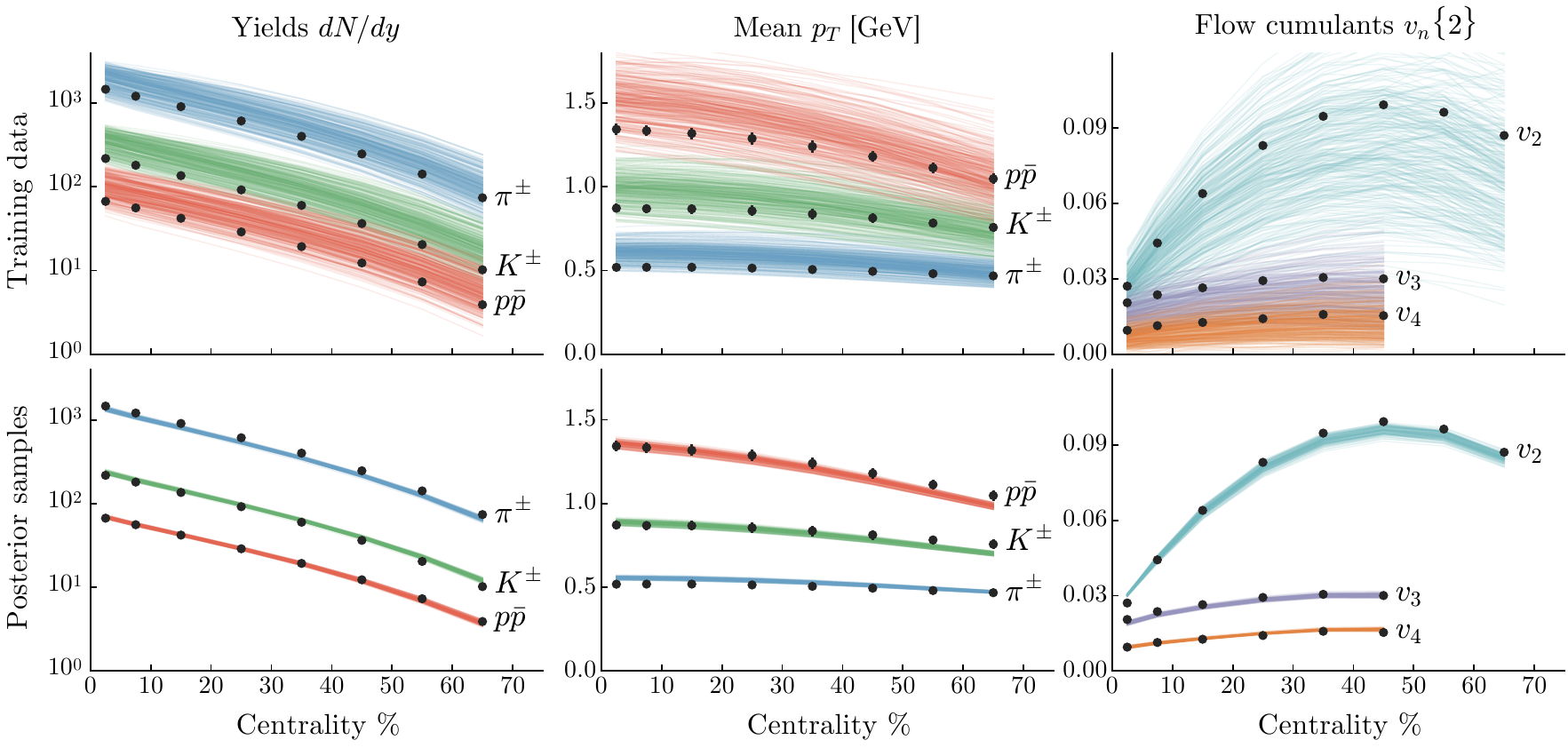}
  }
  \caption{
    Simulated observables compared to experimental data from the ALICE experiment \cite{Abelev:2013vea,ALICE:2011ab}.
    Top row: explicit model calculations for each of the 300 design points,
    bottom: emulator predictions of 100 random samples drawn from the posterior distribution.
    Left column: identified particle yields $dN/dy$,
    middle: mean transverse momenta $\avg{p_T}$,
    right: flow cumulants $v_n\{2\}$.
  }
  \label{fig:v2-observables-samples}
\end{figure}

In figure \ref{fig:v2-posterior}, the diagonal plots are marginal distributions for each model parameter (all other parameters integrated out) from the calibrations to identified (blue) and charged (red) particles, while the off-diagonals are joint distributions showing correlations among pairs of parameters from the calibrations to identified (blue, lower triangle) and charged (red, upper triangle) particles.
Operationally, these are all histograms of MCMC samples.

We discuss the posterior distributions in detail in the following subsections.
First, let us introduce several ancillary results.

Table \ref{tab:v2-posterior} contains quantitative estimates of each parameter extracted from the posterior distributions.
The reported values are the medians of each parameter's distribution, and the uncertainties are highest posterior density (HPD) 90\% credible intervals.
Note that some estimates are influenced by limited prior ranges, e.g.\ the lower bound of the nucleon width $w$.

Figure \ref{fig:v2-observables-samples} compares simulated observables (see table \ref{tab:v2-observables}) to experimental data.
The top row has explicit model calculations at each of the 300 design points;
recall that all model parameters vary across their full ranges, leading to the large spread in computed observables.
The bottom row shows emulator predictions of 100 random samples from the identified particle posterior distribution (these are visually indistinguishable for the charged particle posterior).
Here, the model has been calibrated to experiment, so its calculations are clustered tightly around the data---although some uncertainty remains since the samples are drawn from a posterior distribution of finite width.
Overall, the calibrated model provides an excellent simultaneous fit to all observables except the pion/kaon yield ratio, which (although it is difficult to see on a log scale) deviates by roughly 10--30\%.
We address this deficiency in the following subsections.

\subsubsection{Initial condition parameters}

\begin{figure}[t]
  \graphicsandcaption{.8}{quant-qcd-props/v2/posterior_p_arrows}{
    Posterior distribution of the \protect\trento\ entropy deposition parameter $p$ introduced in equation \eqref{eq:genmean}.
    Approximate $p$-values are annotated for the KLN ($p \approx -0.67 \pm 0.01$), EKRT ($p \approx 0.0 \pm 0.1$), and wounded nucleon ($p = 1$) models.
  }
  \label{fig:v2-posterior-p-arrows}
\end{figure}

The first four parameters are related to the initial condition model.
Proceeding in order:

The normalization factor is not a physical parameter but nonetheless must be tuned to fit overall particle production.
Both calibrations produced narrow posterior distributions, with the identified particle result located slightly lower to compromise between pion and kaon yields.
There are some mild correlations between the normalization and other parameters that affect particle production.

The \trento\ entropy deposition parameter $p$ introduced in equation \eqref{eq:genmean} has a remarkably narrow distribution, with the two calibrations in excellent agreement.
The estimated value is essentially zero with approximate 90\% uncertainty $\pm0.2$, meaning that initial state entropy deposition is roughly proportional to the geometric mean of participant nuclear thickness functions, $s \sim \sqrt{\T_A\T_B}$.
This confirms previous analysis of the \trento\ model which demonstrated that $p \approx 0$ simultaneously produces the correct ratio between initial state ellipticity and triangularity and fits multiplicity distributions for a variety of collision systems \cite{Moreland:2014oya}.
We observe little correlation between $p$ and any other parameters, suggesting that its optimal value is mostly factorized from the rest of the model.

Further, recall that the $p$ parameter smoothly interpolates among different classes of initial condition models;
figure \ref{fig:v2-posterior-p-arrows} shows an expanded view of the posterior distribution along with the approximate $p$-values for the other models in figure \ref{fig:cgc-compare}.
The EKRT model (and presumably IP-Glasma as well) lie squarely in the peak---this helps explain their success---while the KLN and wounded nucleon models are considerably outside.

The distributions for the multiplicity fluctuation parameter $k$ are quite broad, indicating that it's relatively unimportant for the present model and observables.
Indeed, these fluctuations are overwhelmed by nucleon position fluctuations in large collision systems such as Pb+Pb.

The Gaussian nucleon width $w$ has fairly narrow distributions mostly within 0.4--0.6 fm.
It appears we did not extend the initial range low enough and so the posteriors are truncated;
however we still resolve peaks at ${\sim}0.43$ and ${\sim}0.49$ fm for the identified and charged particle calibrations, respectively.
Since the distributions are asymmetric, the median values are somewhat higher than the modes.
The quantitative estimates and uncertainties are in good agreement with the gluonic widths extracted from deep inelastic scattering data at HERA \cite{Chekanov:2004mw,Kowalski:2006hc,Rezaeian:2012ji} and support the values used in EKRT and IP-Glasma studies \cite{Niemi:2015qia,Schenke:2012wb}.
We also observe striking correlations between the nucleon width and QGP viscosities---this is because decreasing the width leads to smaller scale structures and steeper gradients in the initial state.
So e.g.\ as the nucleon width decreases, average transverse momentum increases, and bulk viscosity must increase to compensate.
This explains the strong anti-correlation between $w$ and $\zeta/s$ norm.

\subsubsection{QGP medium parameters}

The shear viscosity parameters $(\eta/s)_\text{min,slope}$ set the temperature dependence of $\eta/s$ according to the linear ansatz
\begin{equation}
  (\eta/s)(T) = (\eta/s)_\text{min} + (\eta/s)_\text{slope} (T - T_c)
  \label{eq:etas2}
\end{equation}
for $T > T_c$.
The full parametrization, equation \eqref{eq:v2-etas}, also includes a constant $(\eta/s)_\text{hrg}$ for $T < T_c$; this parameter was included in the calibration but yielded an essentially flat posterior distribution, implying that it has little to no effect.
This is not surprising, since hadronic viscosity is largely handled by UrQMD, not the hydrodynamic model.
Therefore, we omit $(\eta/s)_\text{hrg}$ from the posterior distribution visualizations and tables.

\begin{table}[t]
  \centering
  \captionsetup{width=.83\textwidth}
  \caption{
    Estimated parameter values (medians) and uncertainties (90\% credible intervals) from the posterior distributions calibrated to identified and charged particle yields (middle and right columns, respectively).
    The distribution for $\Tsw$ based on charged particles is essentially flat, so we do not report a quantitative estimate.
  }
  \label{tab:v2-posterior}
  \small
  \begin{tabular}{lll}
    \toprule
    & \multicolumn{2}{c}{Calibrated to:} \\
    \cmidrule(l){2-3}
    Parameter & \multicolumn{1}{c}{Identified} & \multicolumn{1}{c}{Charged} \\
    \midrule
    Normalization & \parbox{3.5em}{\hfill $120.$}$_{-8.}^{+8.}$ & \parbox{3.5em}{\hfill $132.$}$_{-11.}^{+11.}$ \\ \noalign{\smallskip}
    $p$ & \parbox{3.5em}{\hfill $-0.02$}$_{-0.18}^{+0.16}$ & \parbox{3.5em}{\hfill $0.03$}$_{-0.17}^{+0.16}$ \\ \noalign{\smallskip}
    $k$ & \parbox{3.5em}{\hfill $1.7$}$_{-0.5}^{+0.5}$ & \parbox{3.5em}{\hfill $1.6$}$_{-0.5}^{+0.6}$ \\ \noalign{\smallskip}
    $w$ [fm] & \parbox{3.5em}{\hfill $0.48$}$_{-0.07}^{+0.10}$ & \parbox{3.5em}{\hfill $0.51$}$_{-0.09}^{+0.10}$ \\ \noalign{\smallskip}
    $\eta/s$ min & \parbox{3.5em}{\hfill $0.07$}$_{-0.04}^{+0.05}$ & \parbox{3.5em}{\hfill $0.08$}$_{-0.05}^{+0.05}$ \\ \noalign{\smallskip}
    $\eta/s$ slope [GeV$^{-1}$] & \parbox{3.5em}{\hfill $0.93$}$_{-0.92}^{+0.65}$ & \parbox{3.5em}{\hfill $0.65$}$_{-0.65}^{+0.77}$ \\ \noalign{\smallskip}
    $\zeta/s$ norm & \parbox{3.5em}{\hfill $1.2$}$_{-0.3}^{+0.2}$ & \parbox{3.5em}{\hfill $1.1$}$_{-0.5}^{+0.5}$ \\ \noalign{\smallskip}
    $\Tsw$ [GeV] & \parbox{3.5em}{\hfill $0.148$}$_{-0.002}^{+0.002}$ & \hspace{3em}--- \\ \noalign{\smallskip}
    \bottomrule
  \end{tabular}
\end{table}

Examining the marginal distributions for $\eta/s$ min and slope, we see a clear preference for $(\eta/s)_\text{min} \lesssim 0.15$ and a slight disfavor of steep slopes;
however, the marginal distributions do not paint a complete picture.
The joint distribution shows a salient correlation between the two parameters, hence, while neither $\eta/s$ min nor slope are strongly constrained independently, a linear combination is quite strongly constrained.
Figure \ref{fig:v2-etas-estimate} visualizes the complete estimate of the temperature dependence of $\eta/s$ via the median min and slope from the posterior (for identified particles) and a 90\% credible region.
This visualization corroborates that the posterior for $(\eta/s)(T)$ is markedly narrower than the prior and further reveals that the uncertainty is smallest at intermediate temperatures, $T \sim {}$200--225 MeV.
We hypothesize that this is the most important temperature range for the present observables at $\sqrt{s_{NN}} = 2.76$~TeV---perhaps it is where the system spends most of its time and hence where most anisotropic flow develops, for instance---and thus the data provide a ``handle'' for $\eta/s$ around 200 MeV.
Data at other beam energies and other, more sensitive observables could provide additional handles at different temperatures, enabling a more precise estimate of the temperature dependence of $\eta/s$.

\begin{figure}[t]
  \graphicsandcaption{.75}{quant-qcd-props/v2/etas_estimate}{
    Estimated temperature dependence of the shear viscosity $(\eta/s)(T)$ for $T > T_c = 0.154$ GeV.
    The gray shaded region indicates the prior range for the linear $(\eta/s)(T)$ parametrization equation \eqref{eq:etas2},
    the blue line is the median from the posterior distribution,
    and the blue band is a 90\% credible region.
    The horizontal gray line indicates the KSS bound $\eta/s \geq 1/4\pi$ \cite{Danielewicz:1984ww,Policastro:2001yc,Kovtun:2004de}.
  }
  \label{fig:v2-etas-estimate}
\end{figure}

This result for $(\eta/s)(T)$ supports several recent findings using other models:
a detailed study using the EKRT model \cite{Niemi:2015qia} showed that a combination of RHIC and LHC data prefer a flat or shallow high-temperature slope, while an analysis using a three-dimensional constituent quark model \cite{Denicol:2015nhu} demonstrated that a similar flat or shallow slope best describes the rapidity dependence of elliptic flow at RHIC.
In addition, the estimated temperature-averaged shear viscosity is consistent with the (constant) $\eta/s = 0.095$ reported \cite{Ryu:2015vwa} using the IP-Glasma model and the same bulk viscosity parametrization, equation \eqref{eq:v2-zetas}.
Finally, the present result remains compatible (within uncertainty) with the KSS bound $\eta/s \geq 1/4\pi$ \cite{Danielewicz:1984ww,Policastro:2001yc,Kovtun:2004de}.

One should interpret the estimate of $(\eta/s)(T)$ depicted in figure \ref{fig:v2-etas-estimate} with care.
We asserted a somewhat restricted linear parametrization reaching a minimum at a fixed temperature, and evidently may not have extended the prior range for the slope high enough to bracket the posterior distribution;
these assumptions, along with the flat 10\% uncertainty [see equation \eqref{eq:v2-cov}], surely affect the precise result.
And in general, a credible region is not a strict constraint---the true function may lie partially or completely (however improbably) outside the estimated region.
Yet the overarching message holds: we find the least uncertainty in $\eta/s$ at intermediate temperatures, and estimate that its temperature dependence has at most a shallow positive slope.

For the $\zeta/s$ norm [the prefactor for the parametrization equation \eqref{eq:v2-zetas}], the calibrations yielded clearly peaked posterior distributions located slight\-ly above one.
Hence, the estimate is comfortably consistent with leaving the parametrization unscaled, as in \cite{Ryu:2015vwa}.
As noted in the previous subsection, there is a strong anti-correlation between $\zeta/s$ norm and the nucleon width.
We also observe a positive correlation with $\eta/s$ min, which initially seems counterintuitive.
This dependence arises via the nucleon width:
increasing bulk viscosity requires decreasing the nucleon width, which in turn necessitates increasing shear viscosity to damp out the excess anisotropy.
Given the previously mentioned shortcomings in the current treatment of bulk viscosity (neglecting bulk corrections at particlization, lack of a dynamical pre-equilibrium phase), we refrain from making any quantitative statements.
What is clear, however, is that a nonzero bulk viscosity is necessary to simultaneously describe transverse momentum and flow data.

The distributions for the particlization temperature $T_\text{switch}$ have by far the most dramatic difference between the two calibrations.
The posterior from identified particle yields shows a sharp peak centered at $T \approx 148$~MeV, just below $T_c = 154$~MeV;
but with charged particle yields, the distribution is nearly flat.
This is because the final particle ratios---while somewhat modified by scatterings and decays in the hadronic phase---are largely determined by the thermal ratios at the particlization temperature.
So, when we require the model to describe identified particle yields, $T_\text{switch}$ is tightly constrained;
on the other hand, lacking these data there is little else to determine an optimal switching temperature.
This reinforces the original hybrid model postulate---that both hydro and Boltzmann transport models predict the same medium evolution within a temperature window \cite{Bass:2000ib,Nonaka:2006yn,Petersen:2008dd}.

Note that, while we do see a narrow peak for $T_\text{switch}$, the model cannot simultaneously fit pion, kaon, and proton yields;
in particular, the pion/kaon ratio is 10--30\% low.
The peak thus arises from a compromise between pions and kaons---not an ideal fit---so we do not consider the quantitative value of the peak to be particularly meaningful.
This is a long-standing issue in hybrid models \cite{Song:2013qma} and therefore likely indicates a more fundamental problem with the particle production scheme rather than one with this specific model.

\subsubsection{Verification of high-probability parameters}

\begin{table}[b!]
  \centering
  \captionsetup{width=.79\textwidth}
  \caption{
    High-probability parameters chosen based on the posterior distributions and used to generate figure \ref{fig:v2-mode-observables}.
    Pairs of values separated by slashes are based on identified / charged particle yields, respectively.
    Single values are the same for both cases.
  }
  \label{tab:v2-mode-params}
  \small
  \begin{tabular}{lll@{\hspace{2em}}ll}
    \toprule
    \multicolumn{2}{c}{Initial condition} & & \multicolumn{2}{c}{QGP medium} \\
    \midrule
    norm & 120. / 129.    & & $\eta/s$ min      & 0.08  \\
    $p$  & 0.0            & & $\eta/s$ slope    & 0.85 / 0.75 GeV$^{-1}$  \\
    $k$  & 1.5  / 1.6     & & $\zeta/s$ norm    & 1.25 / 1.10 \\
    $w$  & 0.43 / 0.49 fm & & $T_\text{switch}$ & 0.148 GeV \\
    \bottomrule
  \end{tabular}
\end{table}

\begin{figure}[t]
  \makebox[\textwidth]{
    \includegraphics[width=1.25\textwidth]{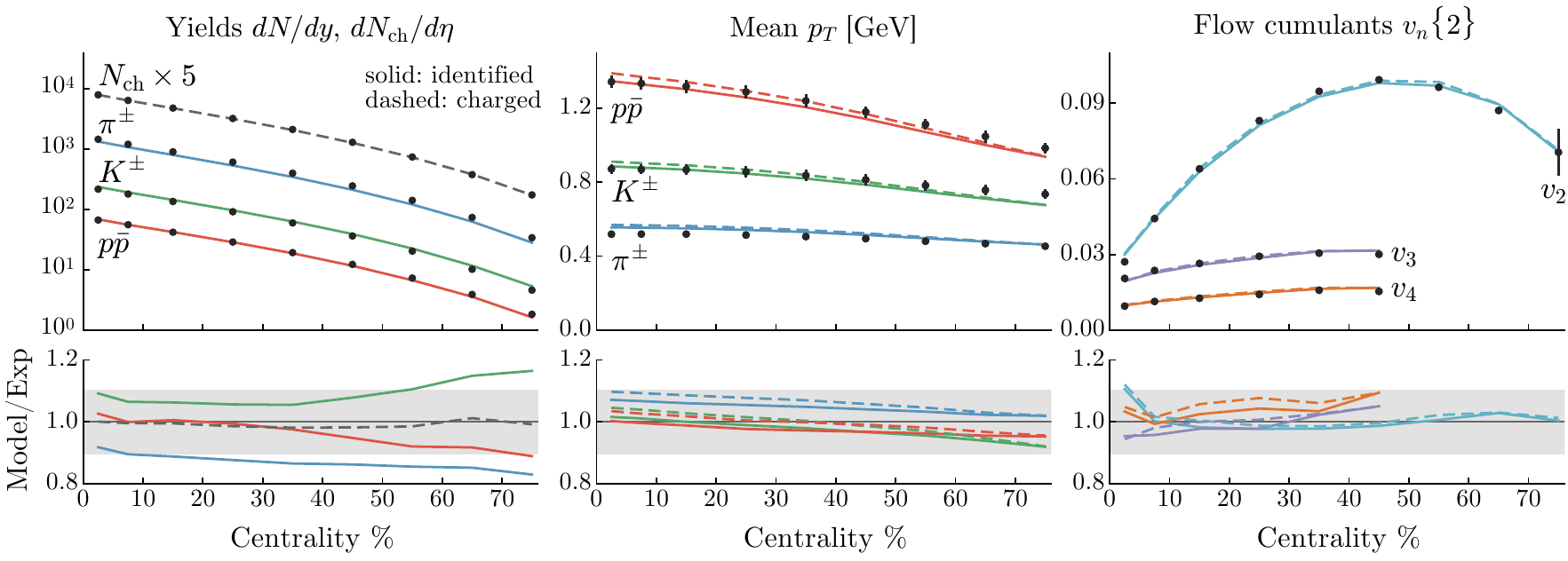}
  }
  \caption{
    Model calculations using the high-probability parameters listed in table \ref{tab:v2-mode-params}.
    Solid lines are calculations using parameters based on the identified particle posterior,
    dashed lines are based on the charged particle posterior,
    and points are data from the ALICE experiment \cite{Abelev:2013vea,ALICE:2011ab}.
    Top row: calculations of identified or charged particle yields $dN/dy$ or $d\Nch/d\eta$ (left), mean transverse momenta $\avg{p_T}$ (middle), and flow cumulants $\vnk n 2$ (right) compared to data.
    Bottom: ratio of model calculations to data, where the gray band indicates $\pm10$\%.
  }
  \label{fig:v2-mode-observables}
\end{figure}

As a final verification of emulator predictions and the model's accuracy, we calculated a large number of events using high-probability parameters and compared the resulting observables to experiment.
We chose two sets of parameters based on the peaks of the posterior distributions, listed in table \ref{tab:v2-mode-params}.
These values approximate the ``most probable'' parameters and the corresponding model calculations should optimally fit the data.

We evaluated $\order{10^5}$ minimum-bias events (no emulator) for each set of parameters and computed observables, shown along with experimental data in figure \ref{fig:v2-mode-observables}.
Solid lines represent calculations using parameters based on the identified particle posterior while dashed lines are based on the charged particle posterior.
Note that these calculations include a peripheral centrality bin (70--80\%) that was not used in parameter estimation.

We observe an excellent overall fit; most calculations are within 10\% of experimental data, the notable exceptions being the pion/kaon ratio (discussed in the previous subsection) and central elliptic flow, both of which are general problems within this class of models.
Total charged particle production is nearly perfect---within 2\% of experiment out to 80\% centrality---indicating that the issues with identified particle ratios arise in the particlization and/or hadronic phases, not in initial entropy production.
The $v_2$ mismatch in the most central bin is a manifestation of the experimental observation that elliptic and triangular flow converge to nearly the same value in ultra-central collisions \cite{ALICE:2011ab,CMS:2013bza}, a phenomenon that hydrodynamic models have yet to explain \cite{Denicol:2014ywa,Shen:2015qta}.

\section{A precision extraction}
\label{sec:v3}

This final act represents the culmination of this work.
Leveraging an advanced computational model, Bayesian parameter estimation with rigorous uncertainty quantification, and diverse experimental data from two beam energies, it lives up to its title.

Building upon \emph{\nameref{sec:v2}}, the model is now more physically realistic and has several additional degrees of freedom, including a pre-equilibrium free streaming stage of variable duration and an improved treatment of bulk viscosity.
Calibrating simultaneously to experimental data from both LHC Pb-Pb collision beam energies, $\sqrt s = 2.76$ and 5.02 TeV, reduces the uncertainties on important model parameters, such as the temperature-dependent shear and bulk viscosity.
Those uncertainties are the first in this series with true quantitative meaning, as they now account for all experimental and model errors.

Before proceeding to the details and results, I should note that I presented a preliminary version of this analysis at the Quark Matter 2017 conference \cite{Bernhard:2017vql,Bass:2017zyn}.
But with only minor differences compared to this final version, it does not warrant a separate discussion.

\subsection{Computational model}

The model is exactly as described in chapter \ref{ch:models}, with the five stages:
\begin{enumerate}
  \item \trento\ parametric initial conditions.
  \item Pre-equilibrium free streaming.
  \item Viscous relativistic 2+1D hydrodynamics, implemented by the Ohio State University group.
  \item Particlization, performed by new implementation \textsc{frzout}.
  \item UrQMD for the hadronic phase.
\end{enumerate}
The model is identical at the two beam energies except for the inelastic nucleon cross section and the initial condition normalization factor.
The parameters to be estimated are listed below and summarized in table \ref{tab:v3-design}.

\subsubsection{Initial condition parameters}

\begin{enumerate}
  \addtocounter{enumi}{2}
  \item[1--2.] Normalization factor for the initial density profile (independent values at each beam energy).

  \item \trento\ entropy deposition parameter $p$.
    The initial density of partons is
    \begin{equation}
      n = \text{Norm} \times \biggl( \frac{\T_A^p + \T_B^p}{2} \biggr)^{1/p},
    \end{equation}
    where $\T$ is a participant thickness function
    \begin{equation}
      \T(x, y) = \sum_{i=1}^{N_\text{part}} u_i \, T_p(x - x_i, y - y_i),
    \end{equation}
    with the fluctuation factors $u_i$ and nucleon thickness function $T_p$ defined below.

  \item Gaussian nucleon width $w$ of the nucleon thickness function
    \begin{equation}
      T_p(x, y) = \frac{1}{2\pi w^2} \exp\biggl( -\frac{x^2 + y^2}{2w^2} \biggr).
    \end{equation}

  \item Standard deviation of nucleon multiplicity fluctuations $\sigma_\text{fluct} = 1/\sqrt k$, where $k$ is the shape parameter of the unit-mean gamma distribution
    \begin{equation}
      P_k(u) = \frac{k^k}{\Gamma(k)} u^{k-1} e^{-ku}.
    \end{equation}

  \item Minimum distance between nucleons $\dmin$ (subsection \ref{subsec:dmin}), transformed to the volume $\dmin^3$.
\end{enumerate}

\subsubsection{Pre-equilibrium parameter}

\begin{enumerate}[resume]
  \item Free-streaming time $\tfs$.
\end{enumerate}

\subsubsection{QGP medium parameters}

\begin{enumerate}[resume]
  \stepcounter{enumi}
  \setcounter{enumisave}{\value{enumi}}
  \addtocounter{enumi}{2}
\item[\theenumisave--\theenumi.] $\eta/s$ min, slope, and curvature (crv), which set the temperature dependence of the QGP specific shear viscosity for $T > T_c$ as the modified linear ansatz
    \begin{equation}
      (\eta/s)(T) = (\eta/s)_\text{min} + (\eta/s)_\text{slope} \cdot (T - T_c) \cdot (T/T_c)^{(\eta/s)_\text{crv}}.
      \label{v3:eta_s-T}
    \end{equation}

  \item Constant value of $\eta/s$ in the hadronic phase ($T < T_c$) \emph{of the hydrodynamic model} (see discussion on page \pageref{loc:eta_s_hrg}).

  \stepcounter{enumi}
  \setcounter{enumisave}{\value{enumi}}
  \addtocounter{enumi}{2}
\item[\theenumisave--\theenumi.] $\zeta/s$ max, width, and location ($T_0$), which set the temperature dependence of the QGP specific bulk viscosity as the three-parameter (unnormalized) Cauchy distribution
    \begin{equation}
      (\zeta/s)(T) = \frac{(\zeta/s)_\text{max}}{1 + \biggl( \dfrac{T - (\zeta/s)_{T_0}}{(\zeta/s)_\text{width})} \biggr)^2}.
      \label{v3:zeta_s-T}
    \end{equation}

  \item Particlization temperature $\Tsw$.
\end{enumerate}

\begin{table}[t]
  \centering
  \caption{Model parameters to be estimated and their design ranges.}
  \label{tab:v3-design}
  \small
  \begin{tabular}{lll}
    \toprule
    Parameter             & Description                           & Range             \\
    \midrule
    \multirow{2}{*}{Norm} & \multirow{2}{*}{Normalization factor} & 8--20 (2.76 TeV)  \\
                          &                                       & 10--25 (5.02 TeV) \\
    $p$                   & Entropy deposition parameter          & $-1/2$ to $+1/2$  \\
    $\sigma_\text{fluct}$ & Multiplicity fluct.\ std.\ dev.       & 0--2              \\
    $w$                   & Gaussian nucleon width                & 0.4--1.0 fm       \\
    $\dmin^3$             & Minimum nucleon volume                & 0--1.7 fm$^3$     \\
    $\tfs$                & Free streaming time                   & 0--1.5 \fmc       \\
    $\eta/s$ hrg          & Const.\ shear viscosity, $T < T_c$    & 0.1--0.5          \\
    $\eta/s$ min          & Shear viscosity at $T_c$              & 0--0.2            \\
    $\eta/s$ slope        & Slope above $T_c$                     & 0--8 GeV$^{-1}$   \\
    $\eta/s$ crv          & Curvature above $T_c$                 & $-1$ to $+1$      \\
    $\zeta/s$ max         & Maximum bulk viscosity                & 0--0.1            \\
    $\zeta/s$ width       & Peak width                            & 0--0.1 GeV        \\
    $\zeta/s$ $T_0$       & Peak location                         & 150--200 MeV      \\
    $\Tsw$                & Particlization temperature            & 135--165 MeV      \\
    \bottomrule
  \end{tabular}
\end{table}

\subsection{Calibration observables}
\label{subsec:v3-observables}

All experimental data are from ALICE, Pb-Pb collisions at $\sqrt s = 2.76$ and 5.02 TeV.
At the time of this writing, some datasets are not available at 5.02 TeV, as noted below.
The calibration observables are the centrality dependence of:
\begin{itemize}
  \item Charged-particle multiplicity $d\Nch/d\eta$ at midrapidity ($|\eta| < 0.5$) \cite{Aamodt:2010cz,Adam:2015ptt}.
  \item Identified particle yields $dN/dy$ of pions, kaons, and protons at midrapidity ($|y| < 0.5$) (2.76 TeV only) \cite{Abelev:2013vea}.
  \item Transverse energy production $dE_T/d\eta$ at midrapidity ($|\eta| < 0.6$) (2.76 TeV only) \cite{Adam:2016thv}.
  \item Identified particle mean $p_T$ of pions, kaons, and protons at midrapidity ($|y| < 0.5$) (2.76 TeV only) \cite{Abelev:2013vea}.
  \item Mean transverse momentum fluctuations $\delta p_T/\avg{p_T}$ (charged particles, $|\eta| < 0.8$, $0.15 < p_T < 2.0$ GeV) (2.76 TeV only) \cite{Abelev:2014ckr} (see below).
  \item Anisotropic flow cumulants $\vnk n 2$ from two-particle correlations, $n = 2$, 3, 4 (charged particles, $|\eta| < 0.8$, $0.2 < p_T < 5.0$ GeV) \cite{ALICE:2011ab,Adam:2016izf}.
\end{itemize}
Figure \ref{fig:v3-observables-design} shows the experimental data along with model calculations at each design point.

\subsubsection{Mean transverse momentum fluctuations}

The event-by-event fluctuations of the mean transverse momentum is a new calibration observable in this analysis, included to provide more information on the $p_T$ distributions (beyond simply the mean).
The dynamical fluctuations of mean $p_T$ (as opposed to random statistical fluctuations) are quantified by the two-particle correlator \cite{Abelev:2014ckr}
\begin{equation}
  (\delta p_T)^2 = \Bigl\langle\!\Bigl\langle \bigl( p_{T,i} - \avg{p_T} \bigr)\bigl( p_{T,j} - \avg{p_T} \bigr) \Bigr\rangle\!\Bigr\rangle,
  \label{eq:pTfluct}
\end{equation}
where the outer double average runs over pairs of particles $i,j$ in the same event and over events in a centrality class, and $\avg{p_T}$ is the usual mean transverse momentum of the centrality class.
This is typically normalized by the mean $p_T$ to form the dimensionless ratio $\delta p_T/\avg{p_T}$, i.e.\ the relative dynamical fluctuations.

The expression \eqref{eq:pTfluct} is numerically inconvenient since it involves a sum over pairs of particles.
To recast it in a more favorable form, we first write out the sums as
\begin{equation}
  (\delta p_T)^2 = \frac{1}{\sum_k^{n_\text{ev}} N_k^\text{pairs}} \sum_k^{n_\text{ev}} \sum_{i,j>i}^{N_k} \bigl( p_{T,i} - \avg{p_T} \bigr)\bigl( p_{T,j} - \avg{p_T} \bigr),
  \label{eq:pTfluct-sums}
\end{equation}
where index $k$ runs over all $n_\text{ev}$ events in the centrality class, $N_k$ is the number of particles that satisfy the kinematic cuts in event $k$, and indices $i,j$ run over all $N_k^\text{pairs} = N_k(N_k-1)/2$ pairs of particles in event $k$.
Now, in general, a sum over pairs can be expanded as
\begin{equation}
  \sum_{i,j>i}^N a_i a_j = \frac{1}{2} \Biggl[ \biggl(\sum_i^N a_i\biggr)^2 - \sum_i^N a_i^2 \Biggr].
\end{equation}
Applying this to \eqref{eq:pTfluct-sums} and collecting terms gives
\begin{align}
  (\delta p_T)^2 = \frac{1}{\sum_k^{n_\text{ev}} N_k^\text{pairs}} \sum_k^{n_\text{ev}}
  \Biggl[
    &\frac{1}{2} \biggl(\sum_i^{N_k} p_{T,i}\biggr)^2 - \frac{1}{2} \sum_i^{N_k} p_{T,i}^2  \\
    &\quad {} + \avg{p_T}(N_k-1)\sum_i^{N_k} p_{T,i} + \avg{p_T}^2N_k^\text{pairs}
  \Biggr], \nonumber
\end{align}
which involves only sums of $p_T$ and $p_T^2$.

\subsection{Parameter estimation method}

The method is exactly as described in chapter \ref{ch:param-est}.
A few specifics:
\begin{itemize}
  \item The parameter design is a 500 point Latin hypercube sample, repeated at the two beam energies for 1000 total design points.
  \item Model outputs at each beam energy are postprocessed separately, i.e.\ with independent PCA transformations.
    The first 10 principal components are used, accounting for about 99.6\% of the total variance.
  \item Independent Gaussian process emulators predict the model outputs for each beam energy.
    Subsection \ref{subsec:validation} validates their performance.
  \item The likelihood is computed as described in subsection \ref{subsec:likelihood-uq}, with a joint likelihood
    \begin{equation}
      P(\D|\xv) = P(\D_{2.76}|\xv_{2.76}) \, P(\D_{5.02}|\xv_{5.02}).
    \end{equation}
    Here, the only difference between each energy's parameter vector is the normalization factor;
    all other parameters are the same.
\end{itemize}

\subsection{Posterior parameter estimates}

\begin{figure}[p]
  \makebox[\textwidth]{
    \includegraphics[width=.95\paperwidth]{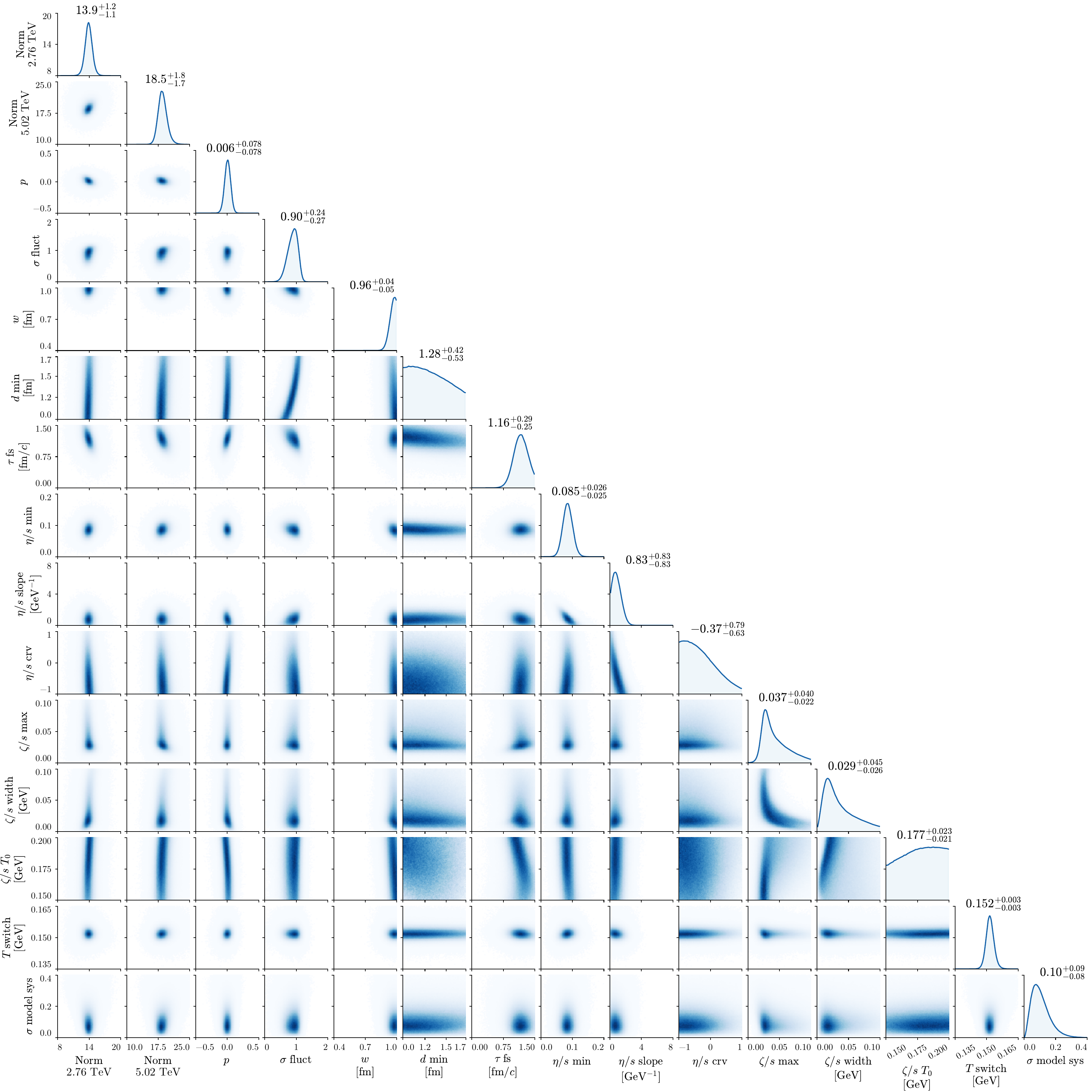}
  }
  \captionsetup{
    margin={.2\textwidth,0pt},
    justification=raggedleft,
    skip=-.95\textheight
  }
  \caption{
    Posterior distribution for the model parameters.
    Diagonal: marginal distributions for each parameter; \\
    off-diagonal: joint marginal distributions between \\ pairs of parameters.
    The annotated estimates \\ are the posterior medians with 90\% \\ HPD credible intervals.%
    \vspace{.7\textheight}  
  }
  \label{fig:v3-posterior}
\end{figure}

Figure \ref{fig:v3-posterior} shows the now-familiar triangular visualization of the posterior distribution, where the diagonal subplots are marginal distributions for each parameter, and the off-diagonals are joint marginal distributions showing correlations between pairs of parameters.
The annotations along the diagonal, also listed in table \ref{tab:v3-posterior}, are quantitative parameter estimates, consisting of the posterior median and highest posterior density (HPD) 90\% credible interval for each parameter.
Note that for asymmetric distributions, the median does not coincide with the mode (peak value).

\begin{table}[t]
  \centering
  \captionsetup{width=.82\textwidth}
  \caption{
    Posterior parameter estimates.
    Reported values are the posterior medians;
    uncertainties are the 90\% HPD credible intervals.
  }
  \label{tab:v3-posterior}
  \small
  \newlength{\cellwidth}
  \settowidth{\cellwidth}{$-0.00$}
  \newcommand{\est}[3]{\parbox{\cellwidth}{\hfill$#1$}$_{-#2}^{+#3}$}
  \begin{tabular}{llll}
    \toprule
    \multicolumn{2}{c}{Initial condition / Pre-eq}          & \multicolumn{2}{c}{QGP medium}                      \\
    \cmidrule(r){1-2}                                         \cmidrule(l){3-4}
    \addlinespace[.4ex]
    \multirow{2}{*}{Norm} & \est{13.9}{1.1}{1.2} (2.76 TeV) & $\eta/s$ min    & \est{0.085}{0.025}{0.026}         \\[1.1ex]
                          & \est{18.5}{1.7}{1.8} (5.02 TeV) & $\eta/s$ slope  & \est{0.83}{0.83}{0.83} GeV$^{-1}$ \\[1.1ex]
    $p$                   & \est{0.006}{0.078}{0.078}       & $\eta/s$ crv    & \est{-0.37}{0.63}{0.79}           \\[1.1ex]
    $\sigma_\text{fluct}$ & \est{0.90}{0.27}{0.24}          & $\zeta/s$ max   & \est{0.037}{0.022}{0.040}         \\[1.1ex]
    $w$                   & \est{0.96}{0.05}{0.04} fm       & $\zeta/s$ width & \est{0.029}{0.026}{0.045} GeV     \\[1.1ex]
    $\dmin$               & \est{1.28}{0.53}{0.42} fm       & $\zeta/s$ $T_0$ & \est{0.177}{0.021}{0.023} GeV     \\[1.1ex]
    $\tfs$                & \est{1.16}{0.25}{0.29} \fmc     & $\Tsw$          & \est{0.152}{0.003}{0.003} GeV     \\
    \addlinespace[.4ex]
    \bottomrule
  \end{tabular}
\end{table}

\begin{figure}[p]
  \makebox[\textwidth]{
    \includegraphics[width=1.1\textwidth]{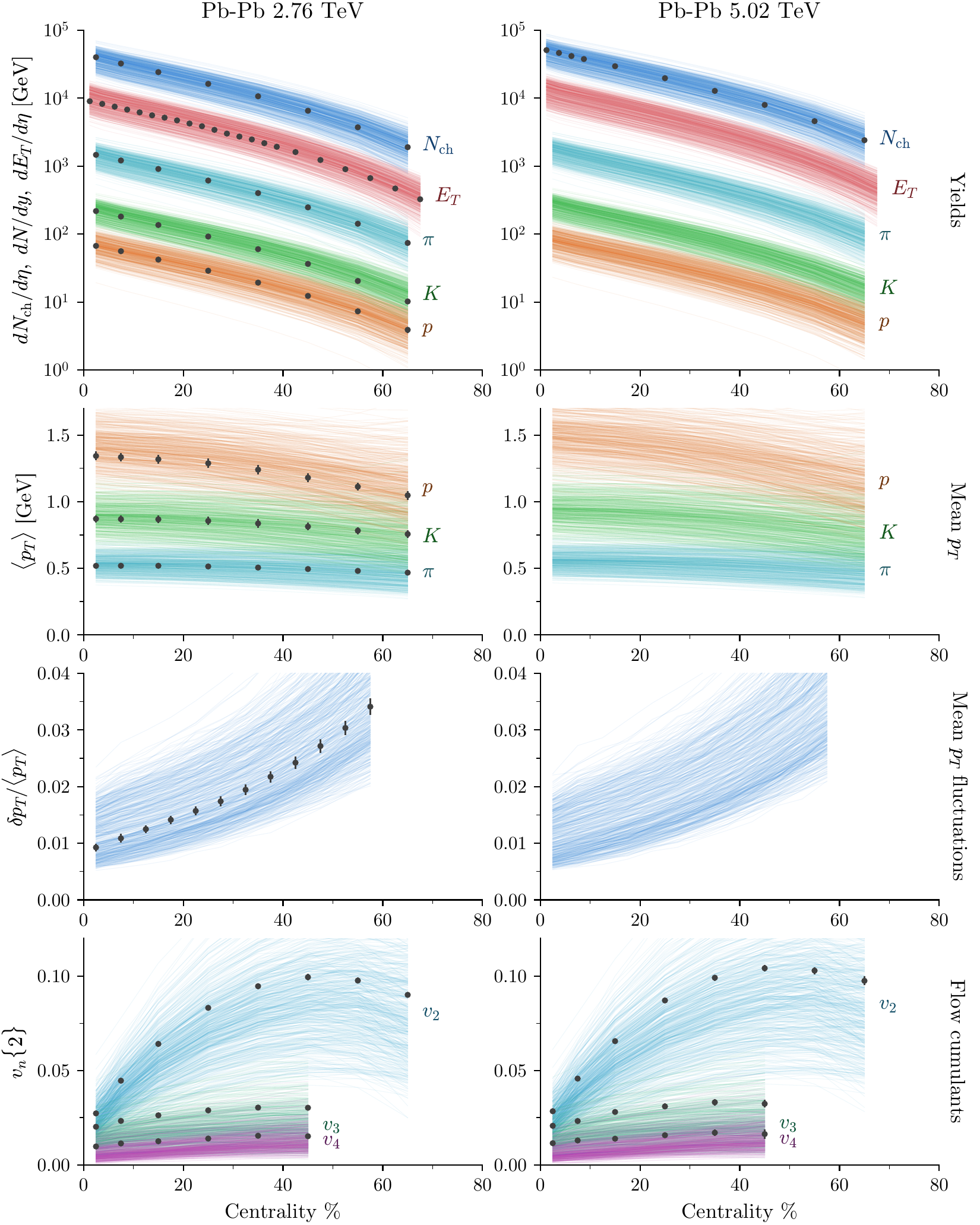}
  }
  \caption{
    Model calculations at each design point.
    Experimental data are from ALICE, Pb-Pb collisions at $\sqrt s = 2.76$ TeV (left column) \cite{Aamodt:2010cz,Adam:2016thv,Abelev:2013vea,Abelev:2014ckr,ALICE:2011ab} and 5.02 TeV (right) \cite{Adam:2015ptt,Adam:2016izf}.
    Some datasets are not available at 5.02 TeV.
  }
  \label{fig:v3-observables-design}
\end{figure}

\begin{figure}[p]
  \makebox[\textwidth]{
    \includegraphics[width=1.1\textwidth]{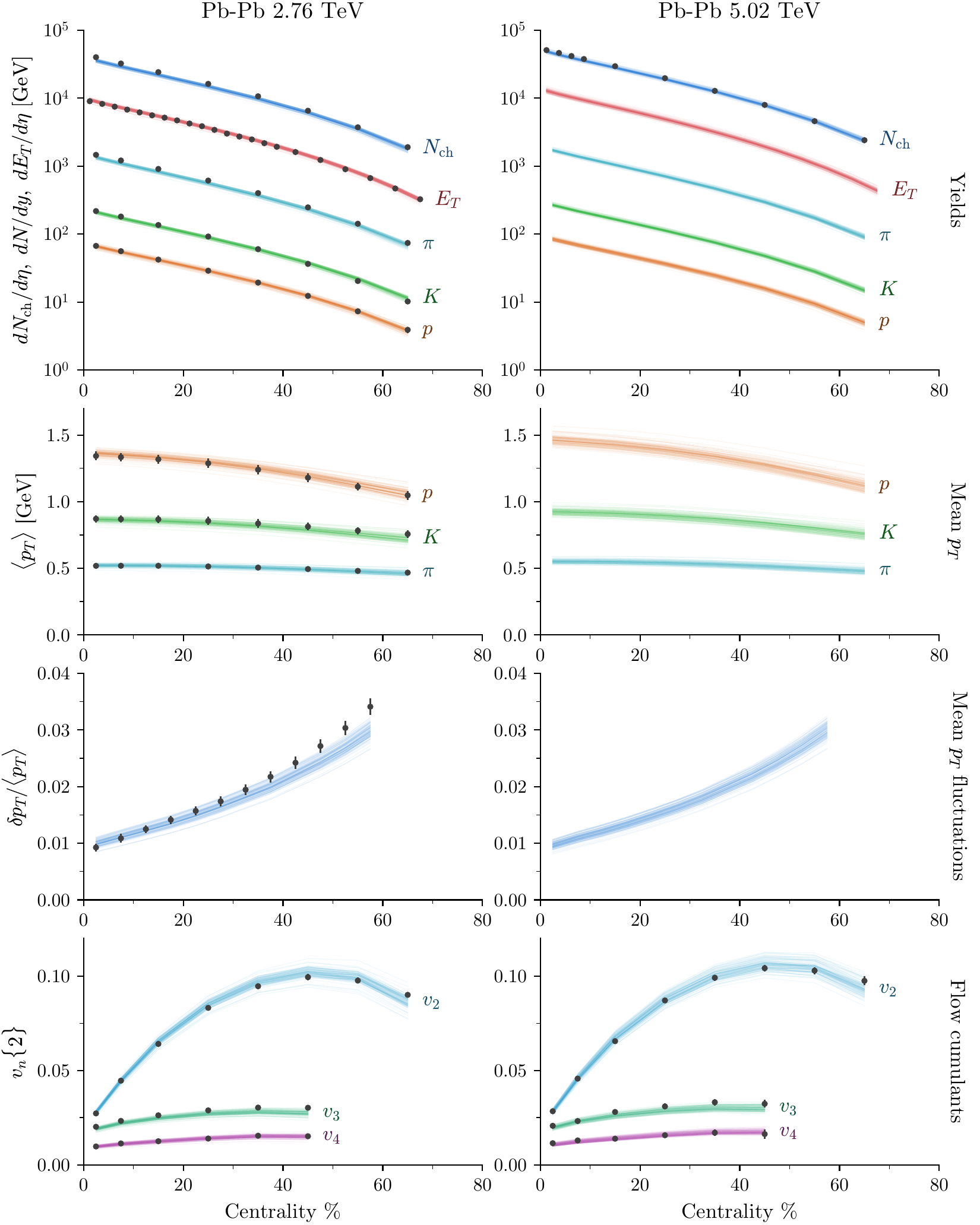}
  }
  \caption{
    Emulator predictions at parameter points randomly drawn from the posterior distribution.
    Experimental data are the same as figure \ref{fig:v3-observables-design}.
  }
  \label{fig:v3-observables-posterior}
\end{figure}

Before examining the posterior distribution in more detail, observe also figure \ref{fig:v3-observables-design}, showing the model calculations at each design point compared to experimental data, and \ref{fig:v3-observables-posterior}, which is analogous but with emulator predictions at parameter points randomly drawn from the posterior distribution.
As depicted in the second figure, the calibrated model accurately describes almost all experimental data points---the notable exception being peripheral mean $p_T$ fluctuations, which I will address later.
Overall, the fit is superior to the previous version (figure \ref{fig:v2-observables-samples}), with less tension among the identified particle yields and improved centrality dependence of the mean $p_T$ and elliptic flow cumulant $\vnk 2 2$.

Let us now consider each model parameter, proceeding in order along the diagonal.
It will be useful to refer back to the posterior distribution from \emph{\nameref{sec:v2}}, figure \ref{fig:v2-posterior}.

\subsubsection{Initial condition}

The normalization factors are well-constrained by particle and energy production data.
Interestingly, the factor at 5.02 TeV is about 30\% larger than at 2.76 TeV, even though experimental particle production only increases by about 20\% between the two energies \cite{Adam:2015ptt}.
This occurred because only $d\Nch/d\eta$ data are available at 5.02 TeV, while at 2.76 TeV there is also transverse energy and identified particle yields.
Since there is still some tension among these observables---notice that the model samples of $d\Nch/d\eta$ are slightly low at 2.76 TeV---the normalization decreases.

\begin{figure}
  \centering
  \includegraphics{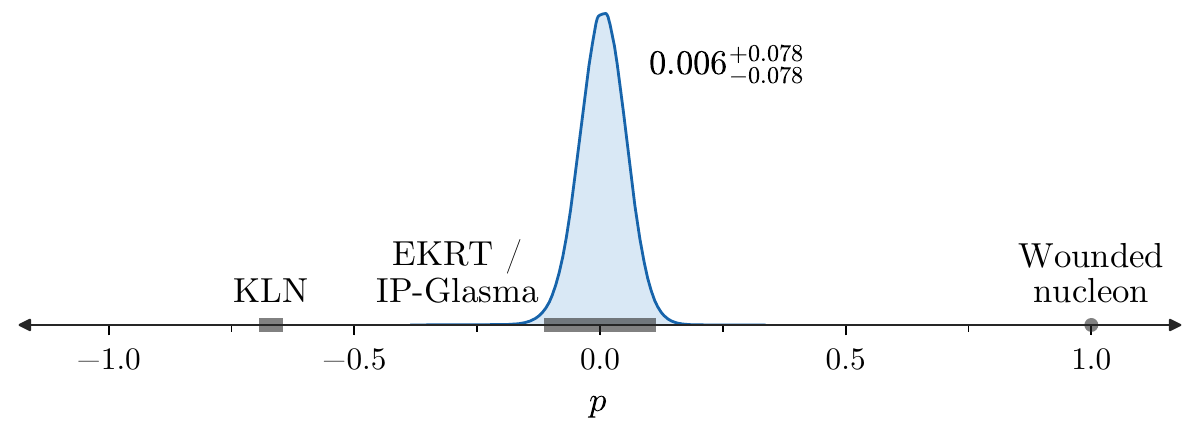}
  \caption{
    Marginal distribution of the \trento\ entropy deposition parameter $p$.
    Approximate $p$-values, established in subsection \ref{subsec:reproducing-ic}, are annotated for the KLN ($p \approx -0.67 \pm 0.01$), EKRT and IP-Glasma ($p \approx 0.0 \pm 0.1$), and wounded nucleon ($p = 1$) models.
  }
  \label{fig:v3-posterior-p}
\end{figure}

The \trento\ entropy deposition parameter $p$ has a narrow, approximately normal distribution centered at essentially zero, with half the uncertainty of the previous study, about $\pm0.08$ compared to $\pm0.17$.
This strongly corroborates that initial entropy deposition (or particle production) goes as the geometric mean of participant nuclear thickness,
\begin{equation}
  s \sim n \sim \sqrt{\T_A \T_B},
\end{equation}
see equations \eqref{eq:tr} and \eqref{eq:means}.
Although this does not tell us the physical mechanism driving entropy deposition---and many possibly theories could predict such general behavior---it does rule out models that do \emph{not} have this approximate scaling.
For example, consider figure \ref{fig:v3-posterior-p}, which shows an expanded view of the marginal distribution with the $p$-values of several existing models marked (compare to figure \ref{fig:v2-posterior-p-arrows}).
Clearly, the KLN and wounded nucleon models are excluded by this analysis, while EKRT and IP-Glasma are substantiated.

The standard deviation of nucleon multiplicity fluctuations has a strong peak around $\sigma_\text{fluct} \sim 1$, which corresponds to exponentially distributed fluctuations.
But the most compelling feature is simply that there are lower and upper bounds, meaning that some fluctuations, but not too much, are necessary to describe the data.
Of course, we already knew this from experimental proton-proton collision multiplicity distributions (see figure \ref{fig:nch}), but it is remarkable that the parameter estimation framework can extract this information from Pb-Pb data alone.
Note that the transformation from the gamma distribution shape parameter $k$ to the standard deviation $\sigma_\text{fluct} = 1/\sqrt k$ facilitated this inference, since zero fluctuations ($\sigma_\text{fluct} = 0$) corresponds to $k \rightarrow \infty$.

It appears that the design range for the Gaussian nucleon width $w$ was truncated on the upper end, although we do resolve a peak at $w \approx 0.98$ fm (a bit higher than the median).
This is reasonably close to the proton root-mean-square charge radius 0.88 fm \cite{Mohr:2015ccw}, but much larger than the previously estimated $\sim$0.5 fm (table \ref{tab:v2-posterior}).
Initially, this disparity seems contradictory, but the shift occurred because the previous model lacked pre-equilibrium free streaming, which generally increases radial flow;
smaller nucleons compensated by creating steep initial pressure gradients, thereby driving similar radial flow.
As for the apparent truncation:
It reflects our prior that $w < 1.0$ fm, which is perhaps justified.
After all, $w$ is the width of a Gaussian, not a hard radius, so $w = 1$ fm implies very large nucleons of diameter $\sim$4--6 fm.
If, based on physical considerations, we believe such large nucleons are unlikely, we could choose a prior with continuously decreasing probability as $w \rightarrow 1.0$ fm (instead of a sudden cutoff), and the posterior distribution would smoothly drop to zero.

The minimum inter-nucleon distance $\dmin$ enters the analysis as the volume $\dmin^3$, i.e.\ there is a uniform prior on $\dmin^3$ from 0 to 1.7 fm$^3$, and the visualized distribution is over the volume (note the nonuniform axis tick marks).
The distribution is more or less flat, suggesting that $\dmin$ does not influence the overall fit of the model to the present observables.
However, there is no doubt that this parameter does affect the model:
It modifies the initial eccentricity distributions and the final flow coefficients, and the emulator captures this dependence.
We can see a hint of this in the joint distribution between $\dmin$ and $\sigma_\text{fluct}$, which shows that increased multiplicity fluctuations correlate with increased minimum distance.
The interpretation:
A minimum distance prevents nucleons from piling up, but since only the beam-integrated thickness matters, increasing fluctuations---which scale the thickness of each nucleon---easily negates this effect.
Evidently, $\dmin$ only weakly affects model calculations of the present observables, but it's possible that calibrating to other data could reveal a nontrivial distribution for $\dmin$.

\subsubsection{Pre-equilibrium}

The sole free parameter related to pre-equilibrium evolution is the free-streaming time $\tfs$, whose distribution has a peak at ${\sim} 1.2 \pm 0.3\ \fmc$, consistent with the long-standing belief that hydrodynamic evolution begins early, around $\order{1\ \fmc}$.
Although free streaming is not the most realistic model, the existence of a peak means that a brief weakly-coupled pre-equilibrium stage is necessary to describe to the data.

Note that, in the present analysis, $\tfs$ is required to be the same at both beam energies, which may not be the case.
Future studies could seek to estimate independent values at different energies.

\subsubsection{QGP medium}

The most salient QGP medium parameters are those controlling its transport coefficients, namely, the temperature dependence of the specific shear and bulk viscosity, $(\eta/s)(T)$ and $(\zeta/s)(T)$, the determination of which is a primary goal of this work.

The marginal distribution for the minimum value of $\eta/s$ is approximately normal with a peak at $(\eta/s)_\text{min} \approx 0.085$ and 90\% credible interval 0.05--0.11, strikingly close to the conjectured lower bound $1/4\pi \simeq 0.08$ \cite{Danielewicz:1984ww,Policastro:2001yc,Kovtun:2004de}.
This of course does not prove the conjecture, but it is sensible to conclude that the QGP created in heavy-ion collisions behaves as a nearly-ideal fluid near the transition temperature.

Regarding the other $\eta/s$ parameters, there is a mild preference for a small positive slope, although zero slope (i.e.\ constant $\eta/s$) is not excluded.
We observe an anti-correlation between the slope and minimum, similar to the previous study, and another anti-correlation between the slope and curvature parameter, which itself has a broad marginal distribution with somewhat more density at negative curvature.
All of this points to $\eta/s$ likely increasing slowly with temperature, perhaps with a negative second derivative.

\begin{figure}[t]
  \makebox[\textwidth]{
    \includegraphics{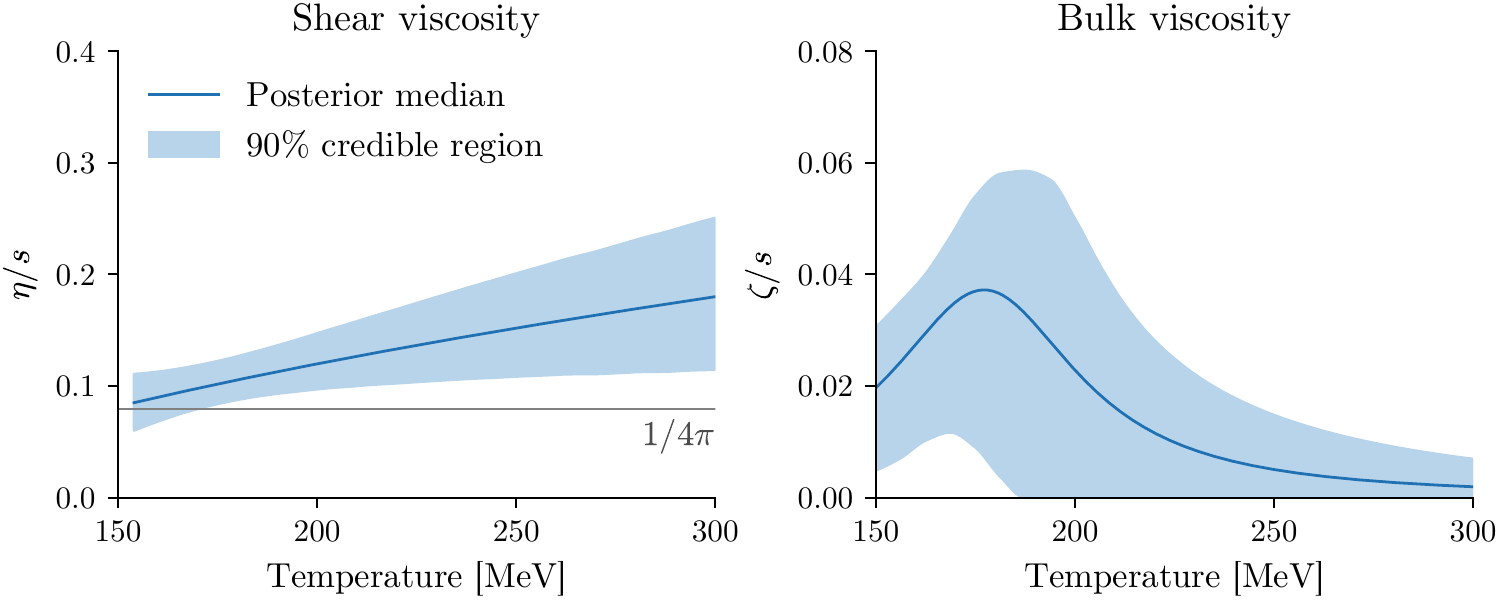}
  }
  \caption{
    Estimated temperature dependence of the specific shear and bulk viscosity, $(\eta/s)(T)$ and $(\zeta/s)(T)$.
    Lines are the parametrizations \eqref{v3:eta_s-T} and \eqref{v3:zeta_s-T} with the parameters set to their posterior median values;
    shaded regions are 90\% credible regions.
    The horizontal line in the shear viscosity plot indicates the conjectured lower bound $1/4\pi$ \cite{Danielewicz:1984ww,Policastro:2001yc,Kovtun:2004de}.
  }
  \label{fig:v3-region-shear-bulk}
\end{figure}

Figure \ref{fig:v3-region-shear-bulk}, left panel, visualizes the estimated temperature dependence of $\eta/s$ as the posterior median with a 90\% credible region (compare to figure \ref{fig:v2-etas-estimate}, note the $y$-axis range is different).
Similar to the previous study, there is a marked narrowing of the uncertainty at intermediate temperatures, although the narrowest region is now somewhat lower, $T \sim 175$ MeV compared to above 200 MeV before.
It is difficult to say why the range moved, but regardless, this characteristic suggests that the data have their greatest resolving power at intermediate temperatures, hence that is where $\eta/s$ is best constrained.
I emphasize that nothing about the $(\eta/s)(T)$ parametrization would impose such a narrowing---it arises naturally.

It is possible that $\eta/s$ does not reach its minimum value precisely at the transition temperature $T_c$, as the present parametrization requires.
Future work could add the location of the minimum as a degree of freedom and attempt to estimate it from the data.

Moving on to bulk viscosity:
The maximum value of $\zeta/s$ and the width of the peak both have skewed distributions, and their joint distribution shows that they trade off, i.e.\ the peak can be tall or wide, but not both.
This implies that it is the integral of $(\zeta/s)(T)$ that matters, not its specific form.
Meanwhile, the peak location ($T_0$) is not constrained, except for possibly ruling out a very narrow peak located at high temperature.
The right panel of figure \ref{fig:v3-region-shear-bulk} shows the estimated temperature dependence of $\zeta/s$, analogous to $\eta/s$.

Given the excellent performance of the model and the uncertainty quantification framework in the present analysis---which properly accounts for experimental statistical and systematic uncertainty and model uncertainty---we should take seriously the quantitative estimates shown in figure \ref{fig:v3-region-shear-bulk}, especially their credible regions.
Based on all the included information, and subject to the assumptions of the model, there is a 90\% chance that the true QGP $(\eta/s)(T)$ and $(\zeta/s)(T)$ curves lie within the pictured regions.

Finally, the particlization temperature $\Tsw$ has a narrow distribution located in the QCD crossover transition region.
As established in the previous study, $\Tsw$ is determined primarily by identified particle yield ratios, but where there was previously a discrepancy between the pion and kaon yields, there is now much less tension.
This is attributable to the inclusion of finite resonance mass width in the particlization model, which leads to increased production of resonances that feed down to pions.

\subsubsection{Systematic uncertainty}

The ``$\sigma$ model sys'' parameter is the model systematic uncertainty $\sigma_m\sys$ introduced in subsection \ref{subsec:likelihood-uq} to account for imperfections in the computational model.
As a reminder, it is defined relative to the overall variability of the model, e.g.\ $\sigma_m\sys = 0.1$ would mean that the model has systematic uncertainty equal to 10\% of its total variance.
The posterior distribution essentially equals the prior, so we cannot learn much about the model systematic uncertainty.
One interesting characteristic, present in most of the joint distributions with the physical parameters, is the widening of the posterior distribution with increasing $\sigma_m\sys$.
This reflects that, with large systematic error, the specific values of the model parameters don't matter as much.

\subsection{Maximum probability parameters}

\newcommand{\map}{{\small MAP}}

As a final verification of the calibrated model's performance, I calculated a large number\footnote{
  About $1 \times 10^6$ events, compared to $4 \times 10^4$ for the design points.
}
of events using the maximum a posteriori (\map) parameters, which are the mode of the posterior probability:
\begin{equation}
  \xv_\text{MAP} = \operatorname*{arg\,max}_{\xv} P(\xv|\D).
\end{equation}
The \map\ parameter values, determined by numerical optimization,\footnote{
  Starting the optimization algorithm from the posterior median.
}
are listed in table \ref{tab:v3-map}.

\begin{table}[ht]
  \centering
  \caption{Maximum a posteriori (\map) parameters.}
  \label{tab:v3-map}
  \small
  \begin{tabular}{llll}
    \toprule
    \multicolumn{2}{c}{Initial condition / Pre-eq} & \multicolumn{2}{c}{QGP medium}          \\
    \cmidrule(r){1-2} \cmidrule(l){3-4}
    \multirow{2}{*}{Norm} & 13.94 (2.76 TeV)       & $\eta/s$ min          & 0.081           \\
                          & 18.38 (5.02 TeV)       & $\eta/s$ slope        & 1.11 GeV$^{-1}$ \\
    $p$                   & 0.007                  & $\eta/s$ crv          & $-0.48$         \\
    $\sigma_\text{fluct}$ & 0.918                  & $\zeta/s$ max         & 0.052           \\
    $w$                   & 0.956 fm               & $\zeta/s$ width       & 0.022 GeV       \\
    $\dmin$               & 1.27 fm                & $\zeta/s$ $T_0$       & 183. MeV        \\
    $\tfs$                & 1.16 \fmc              & $\Tsw$                & 151. MeV        \\
    \bottomrule
  \end{tabular}
\end{table}

Using the \map\ events, I computed the usual calibration observables listed in subsection \ref{subsec:v3-observables}, which should approximate a ``best-fit'' of the model to data.
Further, if the calibrated model is a realistic representation of reality, it should be able to describe other observables that were \emph{not} used for calibration, and that potentially contain more information about the physical system.
To check this, I computed several higher-order flow observables that are too noisy for calibration, but are stable given the larger quantity of \map\ events.

I emphasize that the following is a secondary result of the analysis;
the primary result is the full posterior distribution, which a single model calculation cannot capture.
Ideally, one would perform model calculations at a number of parameter points sampled from the posterior distribution, but doing so would require a prohibitive amount of computation time.

\begin{figure}[p]
  \makebox[\textwidth]{
    \includegraphics[width=1.2\textwidth]{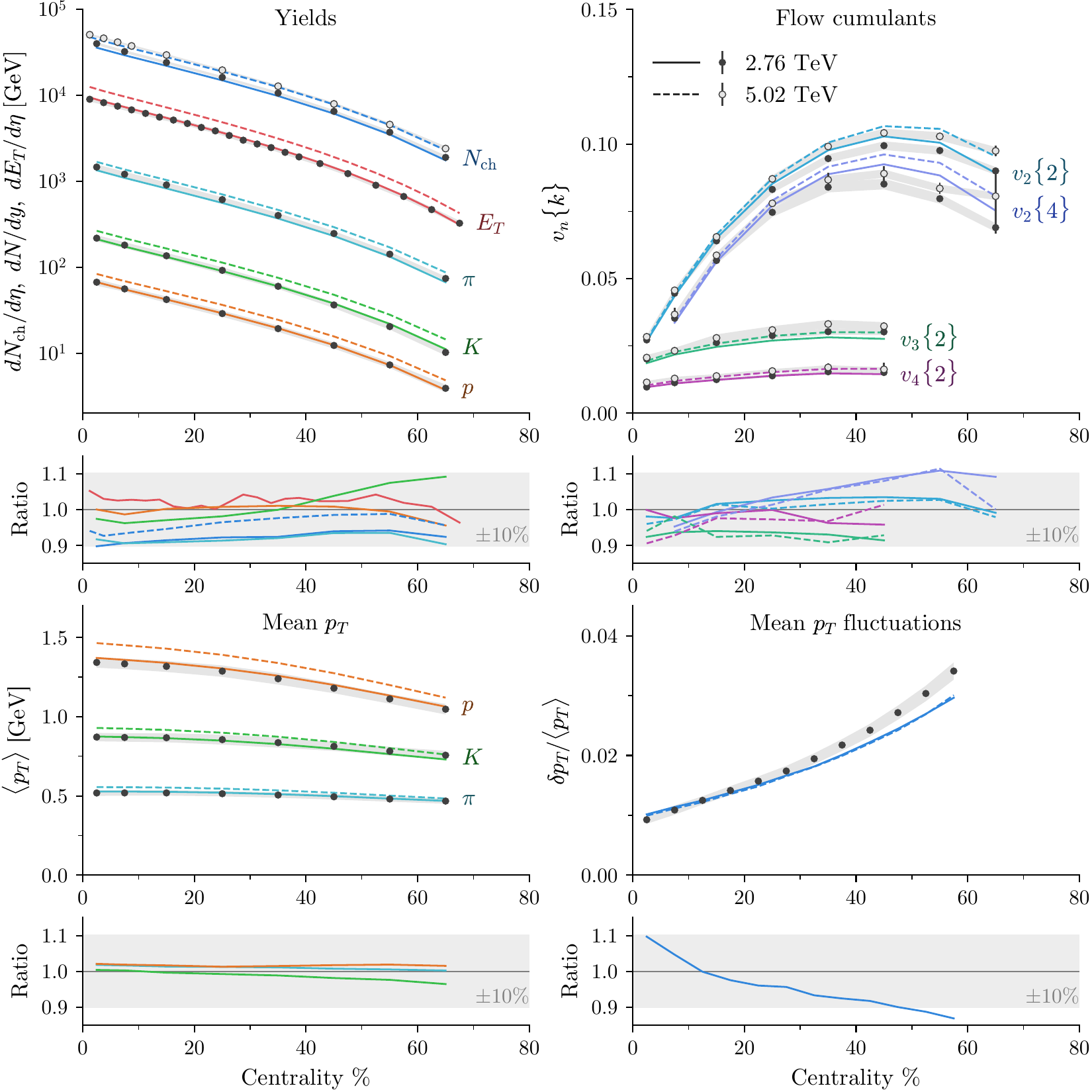}
  }
  \caption{
    Model calculations using the \map\ parameters listed in table \ref{tab:v3-map}.
    Solid lines are calculations at $\sqrt s = 2.76$ TeV; dashed 5.02 TeV.
    Filled symbols are ALICE data at 2.76 TeV \cite{Aamodt:2010cz,Adam:2016thv,Abelev:2013vea,Abelev:2014ckr,ALICE:2011ab}; empty 5.02 TeV (where available) \cite{Adam:2015ptt,Adam:2016izf}.
    The ratio axes show the ratio of the model calculations to data, where the gray band indicates $\pm 10$\%.
  }
  \label{fig:v3-observables-map}
\end{figure}

Figure \ref{fig:v3-observables-map} shows model calculations of the calibration observables at the \map\ point compared to experimental data.
The upper right ``flow cumulants'' panel also shows the four-particle elliptic flow $\vnk 2 4$, which was not a calibration observable.
The overall fit is superb, with almost every data point described within 10\%.
Arguably the worst fit is to the mean $p_T$ fluctuations, where the model calculations do not increase rapidly enough as a function of centrality.
We can likely attribute this to the lack of nucleon substructure in the initial condition model;
a model with quark and/or gluon constituents would have smaller hotspots, creating larger relative $p_T$ fluctuations in peripheral collisions \cite{Bozek:2017elk}.

As an additional cross-check observable, I computed symmetric cumulants $\SC(m, n)$, which quantify the correlations between event-by-event fluctuations of flow harmonics $v_m$ and $v_n$ \cite{Bilandzic:2013kga,ALICE:2016kpq}.
They are defined as the four-particle observable
\begin{equation}
  \begin{aligned}
    \SC(m, n)
      &= \davg{\cos[m(\phi_1 - \phi_3) + n(\phi_2 - \phi_4)]} \\
      &\quad\qquad {} - \davg{\cos[(m(\phi_1 - \phi_2)]} \davg{\cos[n(\phi_1 - \phi_2)]} \\
      &\approx \avg{v_m^2 v_n^2} - \avg{v_m^2} \avg{v_n^2},
  \end{aligned}
  \label{eq:sc}
\end{equation}
where the double average is over particles and events, as usual for flow cumulants (see discussion on page \pageref{loc:flow-corr-functions}), and the second equality is only approximate due to nonflow effects.
Since the two-particle correlations for $v_m$ and $v_n$ are subtracted, $\SC(m,n)$ is zero if $v_m$ and $v_n$ are uncorrelated.
Empirically, symmetric cumulants calculated from hydrodynamic models are highly sensitive to the temperature dependence of $\eta/s$ \cite{ALICE:2016kpq}.

Symmetric cumulants may be computed using $Q$-vectors;
the single-event two-particle correlation is
\begin{equation}
  \avg{\cos[n(\phi_1 - \phi_2)]} = \frac{1}{P_{M,2}} (|Q_n|^2 - M)
\end{equation}
and the four-particle mixed-harmonic correlation is \cite{ALICE:2016kpq}
\begin{equation}
  \begin{aligned}
    &\avg{\cos[m(\phi_1 - \phi_3) + n(\phi_2 - \phi_4)]} = \frac{1}{P_{M,4}} \Bigl\{
      |Q_m|^2|Q_n|^2 \\
      &\qquad {} - 2\Re[Q_{m+n}^{}Q_m^*Q_n^*] - 2\Re[Q_m^{}Q_{m-n}^*Q_n^*] + |Q_{m+n}|^2 + |Q_{m-n}|^2 \\
      &\qquad {} - (M-4)(|Q_m|^2 + |Q_n|^2) + M(M-6)
    \Bigr\},
  \end{aligned}
\end{equation}
where $P_{M,k}$ is the number of $k$-particle permutations, namely
\begin{equation}
  \begin{aligned}
    P_{M,2} &= M(M - 1), \\
    P_{M,4} &= M(M - 1)(M - 2)(M - 3).
  \end{aligned}
\end{equation}
The double averages in \eqref{eq:sc} are then obtained by averaging over events in a centrality class, weighting the single-event correlations by $P_{M,k}$.

\begin{figure}[t]
  \makebox[\textwidth]{
    \includegraphics{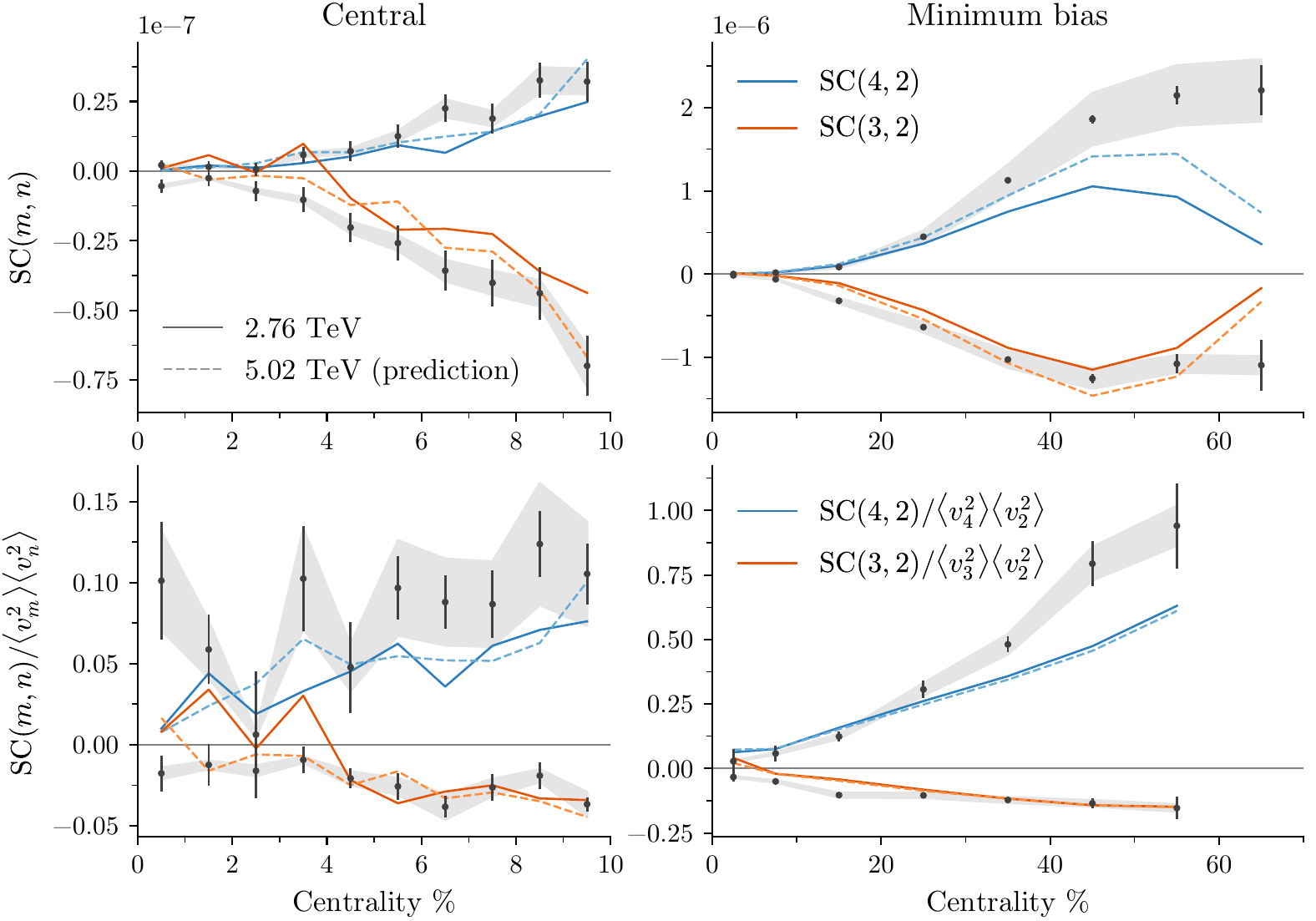}
  }
  \caption{
    Model calculations of symmetric cumulants using the MAP parameters.
    Solid lines are calculations at 2.76 TeV; dashed at 5.02 TeV (a prediction, as the data are not available).
    Data are from ALICE (2.76 TeV only) \cite{ALICE:2016kpq}.
  }
  \label{fig:v3-flow-corr}
\end{figure}

Figure \ref{fig:v3-flow-corr} shows model calculations of $\SC(4, 2)$ and $\SC(3, 2)$, as well as the normalized symmetric cumulants $\SC(m, n)/\avg{v_m^2} \avg{v_n^2}$, compared to experimental data \cite{ALICE:2016kpq}.
Considering that this is a sensitive observable and it did not enter into the calibration, the model provides a good description, with the correct signs and qualitative centrality trends.
But given this lone result, we can only speculate why the fit is imperfect or how it could be improved (unfortunately, a very large number of events, $\order{10^5}$, is required to compute symmetric cumulants with reasonable statistical noise).

\section{Future directions}
\label{sec:future-directions}

Although \emph{\nameref{sec:v3}} accomplished many of the salient goals of this work, there is always room for improvement.
The following is a non-exhaustive list of possible enhancements; a wish list.

\subsection{Computational model}

\paragraph{RHIC data}

In addition to the two LHC beam energies, we can calibrate on data from gold-gold collisions at $\sqrt s = 200$ GeV at the Relativistic Heavy-ion Collider, which should enhance constraining power.
In principle, there is no reason not to do this;
it is a matter of running the events and taking care to compute all observables correctly.

\paragraph{Nucleon substructure}

Implementing nucleon substructure in the initial condition model would permit calibration to data from small collision systems such as proton-lead, and could improve the performance of the model in peripheral nucleus-nucleus collisions.
In fact, this is already in progress, using an extension of \trento\ that models nucleons as superpositions of several smaller constituents \cite{Moreland:2017kdx}.

\paragraph{Full three-dimensional calculations}

Moving to full three-dimensional (not boost-invariant) initial conditions and hydrodynamics would enable calibration to new observables, such as particle rapidity distributions and rapidity-dependent flow.
A recent study \cite{Ke:2016jrd} applied Bayesian parameter estimation to constrain a 3D initial condition model, but did so without hydrodynamics, by mapping initial-state quantities directly to final-state observables.
This shortcut was necessary to avoid the great computational cost of 3+1D hydrodynamics, which indeed will be difficult to overcome.
One possible solution is to run hydrodynamic calculations on graphics processing units (GPUs);
such an implementation was recently developed \cite{Bazow:2016yra}.
GPUs can calculate single events much faster than CPUs, but the relative dearth of GPU computing resources inhibits running on a large scale.

\subsection{Parameter estimation method}

\paragraph{Beam energy dependence}

The only model parameter that I allowed to vary as a function of beam energy was the initial condition normalization factor.
But in principle, all parameters related to the initial condition or pre-equilibrium stages could be functions of energy, such as the nucleon size and free-streaming time.

\paragraph{Systematic uncertainty correlations}

I assumed a particular correlation structure for experimental systematic uncertainties, as described in subsection \ref{subsec:likelihood-uq}, specifically equation \eqref{eq:rho-ij-sys}.
This could certainly be improved, ideally with input from experimentalists.
It would also be interesting to test how much of an impact this has on parameter estimates.

\paragraph{Discrepancy model}

The complete formulation of Bayesian model calibration includes a discrepancy term which accounts for deviations between the model and reality \cite{Kennedy:2001bc,Higdon:2008cmc,Brynjarsdottir:2014imd}.
Adding this in, the relation between the model calculations and experimental data \eqref{eq:y-expt-model} becomes
\begin{equation}
  \yv_e = \yv_m(\xv) + \boldsymbol\delta + \epsv,
\end{equation}
where $\boldsymbol\delta$ is the discrepancy term, usually decomposed into some kind of basis functions.
The physical model parameters are then calibrated simultaneously with the discrepancy.

In practice, the simplified model systematic error parameter $\sigma_m\sys$, introduced in equation \eqref{eq:sigma-m-sys}, certainly subsumes some model discrepancy, but an explicit treatment of discrepancy would be preferable.

\paragraph{Sensitivity analysis}

A category of techniques for quantifying the relationships between model inputs and outputs, sensitivity analysis provides pertinent information such as which input parameters have the strongest effect on the outputs, which observables constrain each parameter, and which observables would benefit most from reduced uncertainty.
See, for example, sensitivity analysis applied to heavy-ion collisions \cite{Sangaline:2015isa} and galaxy formation \cite{Gomez:2013ida}.

\subsection{Other models and data}

The Bayesian parameter estimation method is not specific to the model and data used in this work;
it can be applied to other types of physical models and experimental data which describe different aspects of heavy-ion collisions, enabling inferences on new physical properties.
In particular, while this work focused on bulk properties and observables, there is already progress on quantifying properties related to hard processes, for example, a recent Bayesian analysis estimated the heavy-quark diffusion coefficient \cite{Xu:2017obm}, and the recently created \textsc{Jetscape} Collaboration \cite{Jetscape} is applying similar techniques to jets in heavy-ion collisions.

\chapter{Conclusion}
\label{ch:conclusion}

\lettrine{Q}{uark-gluon plasma} is one of the most exotic substances ever created, and one of the most extraordinarily difficult to characterize.
Produced as tiny fluid-like droplets in ultra-relativistic heavy-ion collisions, it almost instantly disintegrates into particles---the only observable evidence of the QGP's existence; the remnants, in essence, of a long-past explosion.

But all is not lost.
Although we can only observe the final state of heavy-ion collisions, we can infer the time evolution by matching the output of dynamical model calculations to corresponding experimental observations.
By encoding physical properties as model input parameters and tuning the parameters so that the model optimally describes the data, we can estimate the fundamental properties of the QGP and related characteristics of the collision.

This idea is not new, but prior to this work, its execution in heavy-ion physics had been limited.
Most previous studies considered only a single model parameter and observable, and reported rough estimates lacking meaningful uncertainties.
This is not to disparage earlier model-to-data analysis---it was integral to the progression of the field and informed many aspects of the present work---but the reality is that heavy-ion collision models have multiple interrelated parameters and there are a wide variety of experimental observables.
If we are to claim rigorous, quantitative estimates of QGP properties, we must account for all relevant sources of uncertainty and demand that the model describe as much data as possible.

In this dissertation, I have overcome previous limitations and produced the first estimates of QGP properties with well-defined quantitative uncertainties.
I developed a complete framework for applying Bayesian parameter estimation methods to heavy-ion collisions, calibrated a dynamical collision model to diverse experimental data, and derived a posterior probability distribution for numerous model parameters.

Several of these parameters directly connect to fundamental QGP transport coefficients, namely its temperature-dependent specific shear and bulk viscosity, $(\eta/s)(T)$ and $(\zeta/s)(T)$.
The final result is shown in figure \ref{fig:v3-region-shear-bulk}, reproduced here:
\begin{center}
  \vspace{.5em}
  \makebox[\textwidth]{
    \includegraphics{quant-qcd-props/v3/region_shear_bulk}
  }
\end{center}
The illustrated credible regions indicate quantitative 90\% uncertainties, accounting for experimental and model errors and subject to the assumptions of the model.
The estimated minimum value of the specific shear viscosity, $\eta/s = 0.085_{-0.025}^{+0.026}$, is conspicuously close to the conjectured lower bound $1/4\pi \sim 0.08$.

More important than the particular numerical values is that the estimates include quantitative measures of uncertainty;
not only do we now know the approximate $(\eta/s)(T)$ and $(\zeta/s)(T)$, we understand the precision of our knowledge.
Before this work, it was not possible to construct meaningful probability regions for temperature-dependent QGP transport coefficients.

These coefficients are fundamental physical properties;
their measurement a long-standing primary goal of heavy-ion physics.
Countless publications and presentations have studied and constrained $\eta/s$ and $\zeta/s$.
White papers have explicitly stated determination of transport coefficients as a principal objective.

In addition to transport properties, I simultaneously estimated characteristics of the initial state of heavy-ion collisions, including the scaling of initial state entropy deposition, the effective size of nucleons, and the duration of the pre-equilibrium stage that precedes QGP formation.
This was enabled in part by a flexible initial state model, \trento, developed by myself and fellow graduate student J.\ Scott Moreland specifically for this purpose.
Since the exact physical mechanisms governing the initial state are not known, the model parametrizes the relevant degrees of freedom so that they can be estimated and their remaining uncertainties propagated into the uncertainties on the transport coefficients.

This broad philosophy is key to achieving unbiased parameter estimates with faithful assessments of uncertainty.
We should, whenever possible, avoid imposing particular assumptions on the model, and instead use the parameter estimation method and the data to enrich our knowledge.
In other words, our Bayesian prior must embody what we actually know.
The prior comprises not only the prior probability distribution on the parameters, but also the basic design of the model;
if we impose an unfounded assumption, we effectively assert a prior including ``knowledge'' that we \emph{do not have}, and the posterior will be artificially narrow as a result.
Bayesian inference makes explicit how posterior results depend on prior knowledge.

\begin{center}
  \raisebox{.18\baselineskip}{\rule{.5\textwidth}{.4pt}}
\end{center}

\noindent
It is quite astonishing that we are capable of creating quark-gluon plasma and characterizing it with any precision.
In all likelihood, QGP does not presently exist anywhere else in the natural universe.
Only by colliding nuclei at ultra-relativistic speeds can we compress and heat matter enough to overcome the strong force and liberate quarks and gluons.
In the first moments after the Big Bang, similar temperature and density may have created a single large QGP from which everything originated.
We are, quite possibly, studying the source material of the universe itself.

There is, of course, more to be done.
The estimate of the minimum value of $\eta/s$, while more precise than previous results, still has 30\% uncertainty, which would not be considered particularly precise for many other measurements.
In section \ref{sec:future-directions}, I outlined some possible improvements to both the computational model and parameter estimation method which could reduce uncertainty and provide insights on new physical properties.
Other extensions of the analysis may inform pivotal decisions such as which experiments to run and which observables to measure.

Finally, the Bayesian parameter estimation method is not specific to the model used in this work.
There are entire other classes of models and data related to different physical phenomena in heavy-ion collisions.
Work is already underway applying the developed methodology in these areas.

Hopefully, this is only the beginning.

\backmatter

\chapter{Developed software}

\begin{center}
  \em Physics models and analysis tools that I have developed in my research
\end{center}

\section*{Original code}

\begin{description}[leftmargin=0pt,labelsep=1em]
  \item[trento] Initial condition model \\
    Relevant section: \ref{sec:ic} \\
    Source code: \url{https://github.com/Duke-QCD/trento} \\
    Documentation: \url{http://qcd.phy.duke.edu/trento}

  \item[freestream] Pre-equilibrium free streaming \\
    Relevant section: \ref{sec:pre-eq} \\
    Source code: \url{https://github.com/Duke-QCD/freestream} \\
    Documentation: \url{https://github.com/Duke-QCD/freestream#readme}

  \item[frzout] Particlization model (Cooper-Frye sampler) \\
    Relevant section: \ref{sec:particlization} \\
    Source code: \url{https://github.com/Duke-QCD/frzout} \\
    Documentation: \url{http://qcd.phy.duke.edu/frzout}

  \item[hic-eventgen] Heavy-ion collision event generator \\
    Relevant section: \ref{sec:comparing-expt} \\
    Source code: \url{https://github.com/Duke-QCD/hic-eventgen} \\
    Documentation: \url{https://github.com/Duke-QCD/hic-eventgen#readme}

  \item[hic-param-est] Implementation of Bayesian parameter estimation \\
    Relevant chapter: \ref{ch:param-est} \\
    Source code: \url{https://github.com/jbernhard/hic-param-est} \\
    Documentation: \url{http://qcd.phy.duke.edu/hic-param-est}
\end{description}

\section*{Adapted and modified code}

\begin{description}[leftmargin=0pt,labelsep=1em]
  \item[osu-hydro] The Ohio State University viscous hydrodynamics code \\
    Original source: \url{https://u.osu.edu/vishnu} \\
    Relevant section: \ref{sec:hydro} \\
    Source code: \url{https://github.com/jbernhard/osu-hydro} \\
    Documentation: \url{https://github.com/jbernhard/osu-hydro#readme}

  \item[urqmd-afterburner] UrQMD tailored for use as a hadronic afterburner \\
    Original source: \url{https://urqmd.org} \\
    Relevant section: \ref{sec:boltzmann} \\
    Source code: \url{https://github.com/jbernhard/urqmd-afterburner} \\
    Documentation: \\ \hspace*{1em} \url{https://github.com/jbernhard/urqmd-afterburner#readme}
\end{description}

\chapter{Acknowledgments}

\noindent
This work was supported by:
\begin{itemize}[leftmargin=1.7em]
  \item U.S.\ Department of Energy (DOE) grant number DE-FG02-05ER41367.
  \item National Science Foundation (NSF) grant number ACI-1550225.
\end{itemize}

\vspace{1.25em}

\noindent
Computation time was provided by:
\begin{itemize}[leftmargin=1.7em]
  \item The National Energy Research Scientific Computing Center (NERSC), the primary scientific computing facility for the Office of Science in the DOE.
  \item The Open Science Grid (OSG), funded by DOE and NSF.
\end{itemize}

\vspace{1.25em}

\noindent
Open-source programming libraries used in this work:
\begin{itemize}[leftmargin=1.7em]
  \item \textsc{NumPy} \cite{numpy}
  \item \textsc{SciPy} \cite{scipy}
  \item \textsc{scikit-learn} \cite{scikit-learn}
  \item \textsc{h5py} \cite{h5py}
  \item \textsc{matplotlib} \cite{matplotlib}
  \item \textsc{emcee} \cite{ForemanMackey:2012ig}
\end{itemize}

{\raggedright\printbibliography[heading=bibintoc, title={References}]}

\end{document}